%% file: paper.tex
\documentclass[a4paper,11pt]{article}
\usepackage[utf8]{inputenc}
\usepackage{authblk} 
\usepackage{amsmath} 
\usepackage{amssymb} 
\usepackage{graphicx}
\usepackage{hyperref} 
\usepackage{cite} 
\usepackage{color}
\usepackage{adjustbox}
\usepackage{subfig}
\usepackage{floatrow}
\usepackage{array}
\usepackage{multirow}
\usepackage{arydshln}
\usepackage{cancel}
\usepackage{soul}
\usepackage{wrapfig}
\usepackage{feynmf}
\usepackage[table]{xcolor}
\usepackage{color}
\usepackage{colortbl}
\usepackage{multirow}
\usepackage{graphicx}
\usepackage{caption}
\usepackage{adjustbox}
\usepackage{dashrule}
\usepackage{arydshln}
\usepackage{url}
\usepackage[super]{nth}
\usepackage{enumitem}
\setlist[itemize]{noitemsep, topsep=0pt}

\usepackage[top=2cm, bottom=2cm, left=1.5cm, right=1.5cm]{geometry} 
\usepackage{abstract} 
\title{An extension of the SM with vector-like quarks and hypothetical heavy bosons: model independent parametrisation at NLO order}
\author[1]{Chahra Rekaik \thanks{Email: \href{mailto:rekaik.chahra@univ-jijel.dz}{rekaik.chahra@univ-jijel.dz}}}
\author[1]{Mohamed Sadek Zidi \thanks{Email: \href{mailto:mohamed.sadek.zidi@univ-jijel.dz}{mohamed.sadek.zidi@univ-jijel.dz}}}
\affil[1]{LPTh, Department of Physics, Faculty of Exact and Computer Sciences\newline
University of Jijel, B.P. 98 Ouled Aissa, 18000 Jijel, Algeria}

\newcommand{\hp}{\mbox{${H^{\prime}}$}}
\newcommand{\zp}{\mbox{${Z^{\prime}}$}}
\newcommand{\wps}{\mbox{${W^{\prime}}$}}
\newcommand{\bp}{\mbox{${B^{\prime}}$}}
\newcommand{\bpp}{\mbox{${B^{\prime\prime}}$}}
\newcommand{\vp}{\mbox{${V^{\prime}}$}}

\newcommand{\call}{\mbox{${\cal L}$}}
\newcommand{\kpt}{\mbox{$\kappa_{\scriptscriptstyle T}$}}
\newcommand{\kpq}{\mbox{$\kappa_{\scriptscriptstyle Q}$}}
\newcommand{\kpwp}{\mbox{$\kappa_{\scriptscriptstyle W^{\prime}}$}}
\newcommand{\kpzp}{\mbox{$\kappa_{\scriptscriptstyle Z^{\prime}}$}}

\newcommand{\kphp}{\mbox{$\kappa_{\scriptscriptstyle H^{\prime}}$}}
\newcommand{\vkvp}{\mbox{$\varkappa_{\scriptscriptstyle{V^{\prime}}}$}}

\newcommand{\vkvnsq}{\mbox{$\varkappa_{\scriptscriptstyle{V^{\scriptscriptstyle 0}}}^2$}}
\newcommand{\vkvpmsq}{\mbox{$\varkappa_{\scriptscriptstyle{V^{\scriptscriptstyle \pm}}}^2$}}

\newcommand{\vkssq}{\mbox{$\varkappa_{\scriptscriptstyle{S}}^2$}}

\newcommand{\khl}{\mbox{$\kappa^{\scriptscriptstyle{(H)}}_{\scriptscriptstyle{L}}$}}
\newcommand{\khr}{\mbox{$\kappa^{\scriptscriptstyle{(H)}}_{\scriptscriptstyle{R}}$}}
\newcommand{\khpl}{\mbox{$\kappa^{\scriptscriptstyle{(H^{\prime})}}_{\scriptscriptstyle{L}}$}}
\newcommand{\khpr}{\mbox{$\kappa^{\scriptscriptstyle{(H^{\prime})}}_{\scriptscriptstyle{R}}$}}
\newcommand{\kzl}{\mbox{$\kappa^{\scriptscriptstyle{(Z)}}_{\scriptscriptstyle{L}}$}}
\newcommand{\kzr}{\mbox{$\kappa^{\scriptscriptstyle{(Z)}}_{\scriptscriptstyle{R}}$}}
\newcommand{\kzpl}{\mbox{$\kappa^{\scriptscriptstyle{(Z^{\prime})}}_{\scriptscriptstyle{L}}$}}
\newcommand{\kzpr}{\mbox{$\kappa^{\scriptscriptstyle{(Z^{\prime})}}_{\scriptscriptstyle{R}}$}}
\newcommand{\kwl}{\mbox{$\kappa^{\scriptscriptstyle{(W)}}_{\scriptscriptstyle{L}}$}}
\newcommand{\kwr}{\mbox{$\kappa^{\scriptscriptstyle{(W)}}_{\scriptscriptstyle{R}}$}}
\newcommand{\kwpl}{\mbox{$\kappa^{\scriptscriptstyle{(W^{\prime})}}_{\scriptscriptstyle{L}}$}}
\newcommand{\kwpr}{\mbox{$\kappa^{\scriptscriptstyle{(W^{\prime})}}_{\scriptscriptstyle{R}}$}}
\newcommand{\kbpplr}{\mbox{$\kappa^{\scriptscriptstyle{(B^{\prime\prime})}}_{\scriptscriptstyle{R/L}}$}}
\newcommand{\kbtildlr}{\mbox{$\kappa^{\scriptscriptstyle{(\tilde{B})}}_{\scriptscriptstyle{R/L}}$}}
\newcommand{\kbppl}{\mbox{$\kappa^{\scriptscriptstyle{(B^{\prime\prime})}}_{\scriptscriptstyle{L}}$}}
\newcommand{\kbppr}{\mbox{$\kappa^{\scriptscriptstyle{(B^{\prime\prime})}}_{\scriptscriptstyle{R}}$}}

\newcommand{\vkzp}{\mbox{$\varkappa_{\scriptscriptstyle{Z^{\prime}}}$}}
\newcommand{\vtzp}{\mbox{$\vartheta_{\scriptscriptstyle{Z^{\prime}}}$}}
\newcommand{\vkwp}{\mbox{$\varkappa_{\scriptscriptstyle{W^{\prime}}}$}}
\newcommand{\vtwp}{\mbox{$\vartheta_{\scriptscriptstyle{W^{\prime}}}$}}
\newcommand{\vkhp}{\mbox{$\varkappa_{\scriptscriptstyle{H^{\prime}}}$}}
\newcommand{\vtkhp}{\mbox{$\tilde{\varkappa}_{\scriptscriptstyle{H^{\prime}}}$}}
\newcommand{\kbpl}{\mbox{$\kappa^{\scriptscriptstyle{(B^{\prime})}}_{\scriptscriptstyle{L}}$}}

\newcommand{\ksl}{\mbox{$\kappa^{\scriptscriptstyle{(S)}}_{\scriptscriptstyle{L}}$}}
\newcommand{\ksr}{\mbox{$\kappa^{\scriptscriptstyle{(S)}}_{\scriptscriptstyle{R}}$}}

\newcommand{\kvpml}{\mbox{$\kappa^{\scriptscriptstyle{(V^{\scriptscriptstyle \pm})}}_{\scriptscriptstyle{L}}$}}
\newcommand{\kvnl}{\mbox{$\kappa^{\scriptscriptstyle{(V^{\scriptscriptstyle 0})}}_{\scriptscriptstyle{L}}$}}

\newcommand{\kvpmlsq}{\mbox{$\kappa^{\scriptscriptstyle{(V^{\scriptscriptstyle \pm})\, 2}}_{\scriptscriptstyle{L}}$}}
\newcommand{\kwplsq}{\mbox{$\kappa^{\scriptscriptstyle{(W^{\prime})\, 2}}_{\scriptscriptstyle{L}}$}}
\newcommand{\kwlsq}{\mbox{$\kappa^{\scriptscriptstyle{(W)\, 2}}_{\scriptscriptstyle{L}}$}}
\newcommand{\kzplsq}{\mbox{$\kappa^{\scriptscriptstyle{(Z^{\prime})\, 2}}_{\scriptscriptstyle{L}}$}}
\newcommand{\kzlsq}{\mbox{$\kappa^{\scriptscriptstyle{(Z)\, 2}}_{\scriptscriptstyle{L}}$}}
\newcommand{\khplsq}{\mbox{$\kappa^{\scriptscriptstyle{(H^{\prime})\, 2}}_{\scriptscriptstyle{L}}$}}
\newcommand{\khlsq}{\mbox{$\kappa^{\scriptscriptstyle{(H)\, 2}}_{\scriptscriptstyle{L}}$}}

\newcommand{\kvlsq}{\mbox{$\kappa^{\scriptscriptstyle{(V^{_0})\, 2}}_{\scriptscriptstyle{L}}$}}
\newcommand{\kslsq}{\mbox{$\kappa^{\scriptscriptstyle{(S)\, 2}}_{\scriptscriptstyle{L}}$}}
\newcommand{\ksrsq}{\mbox{$\kappa^{\scriptscriptstyle{(S)\, 2}}_{\scriptscriptstyle{R}}$}}

\newcommand{\ktbpplr}{\mbox{$\tilde{\kappa}^{\scriptscriptstyle{(B^{\prime\prime})}}_{\scriptscriptstyle{R/L}}$}}
\newcommand{\ktwlr}{\mbox{$\tilde{\kappa}^{\scriptscriptstyle{(W)}}_{\scriptscriptstyle{L/R}}$}}
\newcommand{\ktzlr}{\mbox{$\tilde{\kappa}^{\scriptscriptstyle{(Z)}}_{\scriptscriptstyle{L/R}}$}}
\newcommand{\kthlr}{\mbox{$\tilde{\kappa}^{\scriptscriptstyle{(H)}}_{\scriptscriptstyle{L/R}}$}}
\newcommand{\ktwplr}{\mbox{$\tilde{\kappa}^{\scriptscriptstyle{(W^{\prime})}}_{\scriptscriptstyle{L/R}}$}}
\newcommand{\ktzplr}{\mbox{$\tilde{\kappa}^{\scriptscriptstyle{(Z^{\prime})}}_{\scriptscriptstyle{L/R}}$}}
\newcommand{\kthplr}{\mbox{$\tilde{\kappa}^{\scriptscriptstyle{(H^{\prime})}}_{\scriptscriptstyle{L/R}}$}}
\newcommand{\kwlr}{\mbox{$\kappa^{\scriptscriptstyle{(W)}}_{\scriptscriptstyle{L/R}}$}}
\newcommand{\kzlr}{\mbox{$\kappa^{\scriptscriptstyle{(Z)}}_{\scriptscriptstyle{L/R}}$}}
\newcommand{\khlr}{\mbox{$\kappa^{\scriptscriptstyle{(H)}}_{\scriptscriptstyle{L/R}}$}}
\newcommand{\khplr}{\mbox{$\kappa^{\scriptscriptstyle{(H^{\prime})}}_{\scriptscriptstyle{L/R}}$}}
\newcommand{\vkwplr}{\mbox{$\kappa^{\scriptscriptstyle{(W^{\prime})}}_{\scriptscriptstyle{L/R}}$}}
\newcommand{\vkzplr}{\mbox{$\kappa^{\scriptscriptstyle{(Z^{\prime})}}_{\scriptscriptstyle{L/R}}$}}
\newcommand{\ktthplr}{\mbox{$\tilde{\tilde{\kappa}}^{\scriptscriptstyle{(H^{\prime})}}_{\scriptscriptstyle{L/R}}$}}

\newcommand{\ktbplr}{\mbox{$\tilde{\kappa}^{\scriptscriptstyle{(B^{\prime})}}_{\scriptscriptstyle{R/L}}$}}
\newcommand{\kzplr}{\mbox{$\kappa^{\scriptscriptstyle{(Z^{\prime})}}_{\scriptscriptstyle{L/R}}$}}
\newcommand{\kwplr}{\mbox{$\kappa^{\scriptscriptstyle{(W^{\prime})}}_{\scriptscriptstyle{L/R}}$}}
\newcommand{\khp}{\mbox{$\kappa^{\scriptscriptstyle{(H^{\prime})}}$}}
\begin{document}

\maketitle

\begin{abstract}
A model independent parametrization of an extension of the Standard Model including vector-like quarks, new heavy gauge bosons and an extra scalar, is introduced. Theoretical constraints on the model couplings and hypothetical particle masses are established by making use of perturbative unitarity. The consistency of the model renormalization and implementation, within the complex mass scheme for narrow widths, is verified. The production of heavy charged and neutral gauge bosons which subsequently decay to vector-like quarks at the LHC are investigated at both leading-order and next-to-leading order. The predictions of the model are obtained by using three approaches: the narrow-width-approximation, the complex mass scheme and a mix of both, where the advantages and weaknesses of each method are emphasized. This study demonstrates that, for certain configurations, off-shell effects remain significant even when the width-to-mass ratios of the unstable particles are sufficiently small, rendering the narrow-width approximation unreliable and necessitating the use of the complex mass scheme. The predictions are compared with CMS {\tt run II} data, where lower bounds on the charged heavy gauge boson mass are set as function of the free parameter normalizing their coupling to quarks.

\vspace{0.25cm}
\end{abstract}

\tableofcontents

\input{introduction}

\input{model}

\input{prod_wp_zp}

\input{summary}

\appendix

\input{appendix_A}

\input{appendix_B}

\input{appendix_C}

\input{appendix_D}

\section*{Acknowledgments}
We are very thankful to J. Ph. Guillet, P. Aurenche and N. Ferkous for useful discussions and precious remarks.

\bibliographystyle{unsrt}
\bibliography{biblio}
\end{document}

%% file: introduction.tex
\section{Introduction}
\label{int}

\noindent
Over the past decades, the Standard Model (SM) is shown to be a successful framework of particle physics. The discovery of the famous Higgs boson in 2012, long predicted by the SM, was one of the greatest crowning achievement of this theory \cite{A, B}.
Despite its successes, the SM fails to explain several phenomena. Experimentally, it cannot explain dark matter, dark energy, neutrino mass or matter–antimatter asymmetry,~\ldots etc. Theoretically, it offers no explanation for the hierarchy problem, gauge couplings unification and strong CP problem \cite{Gildener:1976ai,Weinberg:1975gm,Georgi:1974yf,Georgi:1974sy,Peccei:1977hh,1,2}.
In order to solve these problems, several theories beyond the SM (BSM) have been proposed  \cite{Pati:1974yy,Fritzsch:1974nn,03,04,05,06,Han:2003wu,07,08,09,010,011,012,013,014}, many of which predict the existence of new hypothetical particles. These include both fermions, such as vector-like quarks (VLQs), and heavy bosons, such as the gauge bosons $\zp$ and $\wps$ and extra-scalars. The main objective of this paper is to construct a phenomenological framework, independent of the specific BSM model, to aid in the search for VLQs and new heavy bosons at the LHC.\\

\noindent
VLQs are hypothetical colored spin-half fermions. The fact that their left and right chiralities belong to the same gauge group representation, allows to add their mass term without need to the Higgs mechanism~\cite{Aguilar-Saavedra:2013qpa,aguil1}. They are predicted by many BSM theories, see for example~\cite{extra-dim-vlqs,014,03,composite-higgs-vlqs,Ferretti:2013kya,Banerjee:2022izw}. The primary motivation of introducing such particles is to address the naturalness problem and to stabilize the Higgs mass~\cite{016,017}. Many model independent approaches are introduced to study VLQs at leading-order (LO) and next-to-leading order (NLO), where they are assumed to mix with SM particles via Yukawa interaction~\cite{model-indep-vlqs-1,model-indep-vlqs-2,model-indep-vlqs-3,model-indep-vlqs-4,Cacciapaglia:2011fx,model-indep-vlqs-5,model-indep-vlqs-6,model-indep-vlqs-7}. VLQs can be produced in pairs via the strong interaction or singly via the weak interaction. Pair production generally dominates at lower masses, while single production becomes more significant at higher masses. ATLAS and CMS collaboration have performed many publications addressing the search of VLQs for LHC {\tt run~I} and {\tt run~II}, where the lower limits on the masses are set between $\sim1$~TeV and $\sim1.5$~TeV, cf. ~\cite{cms1, Atlas1, cms2, Atlas2, Atlas3,cms4,Atlas4,Atlas5,cms5,cms6,cms6pp,cms6p}. \\

\noindent
The $\zp$ and $\wps$ are hypothetical spin-1 gauge bosons predicted by many BSM theories like $E_6$ inspired grand unification models~\cite{Hewett:1988xc, Gursey:1975ki, e6}, left-right symmetric models~\cite{Senjanovic:1975rk} and models with extra $U(1)$ symmetry~\cite{F,u1}. They share the same properties of the SM gauge bosons $Z$ and $W$ except that they are match more heavier, since they arise in the initial stages of spontaneous symmetry breaking (SSB) of larger gauge groups down to the SM one.  The Sequential Standard Model (SSM)~\cite{altarelli,fuks1} provide a minimal framework including $\zp$ and $\wps$ as a heavy copy of the ordinary $Z$ and $W$ bosons with similar mixing to SM fermions up to normalization factors. Such model is dedicated to guide experimental
searches for heavy resonances independently of the specific BSM theory~\cite{C,D,E}. The search of these new heavy resonances has been a focal point of interest by ATLAS and CMS collaborations in the last few years. Depending on the specific theoretical model,  the lower bounds on these boson masses can extend up to several TeV, see for example~\cite{cms6,CMS2021,ATLAS2018,CMS2017avb,cms7,ATLAS2023vxg}.\\

\noindent
The existence of VLQs is not optional but necessary for several BSM models that predict hypothetical gauge bosons and scalars. They emerge naturally in $E_6$-inspired models~\cite{Hewett:1988xc,Gursey:1975ki} and are required for the cancellation of one-loop quadratic divergences in the Higgs mass, as in little Higgs models~\cite{05,06,Han:2003wu}. Many recent phenomenological models incorporate these particle are introduced, see for example~\cite{Kim2018mks,Benbrik2019zdp,Fox2015kla,36}. The objective of this work is to extend the SM to include a singlet VLQ, $\zp$ and $\wps$ gauge bosons and an additional neutral scalar $\hp$, so that their interactions with SM fermions are independent of the given BSM model. The construction of our model is inspired by two key frameworks: the model independent parametrization of VLQs outlined in ref.~\cite{model-indep-vlqs-4} and the SSM model introduced in ref. \cite{fuks1}. The key point of our parametrization is that the couplings of the new hypothetical particles to the SM particles are controlled by their branching ratios, their masses and some kinematic functions encoding the kinematics of their partial decays. More generally, we provide two versions of the parametrization: one that applies when VLQs are heavier than the hypothetical bosons, and another for the opposite case. This makes this model flexible and allows for the simultaneous study of the physics of these hypothetical fermions and bosons, across various mass hierarchies, while remaining independent of any specific BSM model.  A first application of this model is the study of the production of the hypothetical gauge bosons and their decay to VLQs and SM quarks, a phenomena which has attracted significant attention in some publications from CMS collaboration  \cite{cms6, cms6p, cms7}. This will be the pillar of the phenomenological part of the current article.\\

\noindent
 This paper is organized in the following manner. In section~\ref{sec2}, we present a general overview of the model, focusing on the parametrization of the mixing of the new heavy particles with SM fermions, discussing the theoretical constraints imposed by perturbative unitarity on the couplings and masses of the new heavy states and outlining the implementation and the validation of the model at both LO and NLO. In section~\ref{sec3}, we examine the production of $\wps$ and $\zp$ that then decay into VLQ and SM particles, where we provide the total cross sections and differential distributions at both LO and NLO merged with parton shower. The calculation, in this section, is based on three approaches: {\it(i)} The narrow-width approximation, where all virtual unstable particles (the extra-gauge bosons and the VLQ) are treated as on-shell states, this approach is called NWA$_2$. {\it(ii)} The famous complex mass (CM) scheme, which represents the full calculation approach. {\it(iii)} A mixture of these two approaches, where only the VLQ is treated as on-shell state. This approach is denoted NWA$_1$. Each of these methods has its own advantages and weaknesses, as we will see in detail in this article. We also derive lower limits on the mass of the heavy charged gauge boson. We conclude this article with a summary and prospects of future research directions.\\

 \noindent
 To simplify the readability of the paper, we moved the calculation details about the model construction, the perturbative unitarity constraints, the comparison of NWA and CM scheme predictions and the validation of the model to 4 appendices. These appendices, which we believe they will be useful for the reader, are designed to give a comprehensive and transparent understanding of the model and its underlying framework. In appendix~\ref{appA}, we derive the model independent parametrization of the mixing of a VLQ with SM fermions via the bosons $W, Z, H, W^{\prime}$, $Z^{\prime}$, and $H^{\prime}$. In appendix~\ref{appB}, we provide the polarized matrix elements for multiple scattering processes involving these extra-particles both as real or virtual and calculate the corresponding partial-wave amplitudes at high energies. In appendix~\ref{appC}, we discuss the validity (and invalidity) of the narrow-width-approximation at LO and NLO. In appendix~\ref{appD}, we give more information on the model validation at NLO.

%% file: model.tex
\section{Model independent parametrization}
\label{sec2} 

\noindent
In this section, we provide a model independent parametrization of an extension of the SM with a singlet VLQ ($Q$), a neutral gauge boson ($\zp$), a charged gauge boson ($\wps$) and a neutral scalar ($\hp$). The quark $Q$ can be a vector-like top quark partner ($Q=T$) or a vector-like bottom quark partner ($Q=B$) which is assumed to mix only with one quark generation. This model is based on the fact that the VLQ interact with the ordinary quarks via the Yukawa mixing, where the parameters normalizing the couplings are expressed in terms of the branching ratios and the masses of the new particles in which the latter ones are treated as free parameters \cite{model-indep-vlqs-1,model-indep-vlqs-2,model-indep-vlqs-3,Cacciapaglia:2011fx,model-indep-vlqs-4,model-indep-vlqs-5}. The new heavy bosons are assumed to have the same couplings to ordinary fermions as the SM bosons up to some coefficients normalizing the interaction according to the SSM, see refs.\cite{altarelli,Langacker_2009,fuks1}. All model dependent interactions are ignored like the mixing between ordinary bosons and the new heavy bosons for example. In top of that the CKM matrix is taken to be diagonal, so no mixing between different ordinary quark generations is considered. We push the detailed calculation of the model construction from the main text to appendix~\ref{appA} to ease its reading but we consider
this appendix very useful to the reader.

\subsection{Lagrangian of the model}

\noindent
The general form of the Lagrangian that must be added to the SM to incorporate the interaction of the new heavy particles with the SM fermions is written as:
\begin{align}
\call&=
\call_{\scriptscriptstyle Q}+ 
\call_{\scriptscriptstyle B^{\prime}}
\label{lag0}
\end{align}
where $\call_{\scriptscriptstyle Q}$ describes the mixing between the vector-like quark $Q$ and the ordinary quarks via SM bosons ($\tilde{B}=H, Z, W$) and the new heavy bosons ($\bp=\zp, \wps, \hp$), while $\call_{\scriptscriptstyle B^{\prime}}$ encodes the interaction between the SM fermions and the new heavy bosons. The explicit formula of $\call_{\scriptscriptstyle Q}$ is given by:
\begin{align}
\call_{\scriptscriptstyle Q}&=i\, \bar{Q}\not{\!\!}D Q-m_{\scriptscriptstyle{Q}}\bar{Q}Q\nonumber\\
&-\biggl[H\bar Q(\khl P_L+\khr P_R)q_{\scriptscriptstyle i}+
\hp\bar Q(\khpl P_L+\khpr P_R)q_{\scriptscriptstyle i}+\text{h.c.}\biggr]\nonumber\\
&+\frac{g}{2c_{\scriptscriptstyle w}}\biggl[\bar Q\not{\!\! Z}(\kzl P_L+\kzr P_R)q_{\scriptscriptstyle i}+
\bar Q\not{\!\! Z^{\prime}}(\kzpl P_L+\kzpr P_R)q_{\scriptscriptstyle i}+\text{h.c.}\biggr]\nonumber\\
&+\frac{g}{\sqrt{2}}\biggl[\bar Q\not{\!\! W}(\kwl P_L+\kwr P_R)q_{\scriptscriptstyle i}^{\prime}+
\bar Q\not{\!\! \wps}(\kwpl P_L+\kwpr P_R)q_{\scriptscriptstyle i}^{\prime}+\text{h.c.}\biggr]
\label{lQ}
\end{align}
where $q_{\scriptscriptstyle i}$ and $q_{\scriptscriptstyle i}^{\prime}$ represent up- or down-type quark fields of a given generation
$i$, depending on the nature of $Q$ (for instance, if $Q\equiv T$, then $q_{\scriptscriptstyle i}\equiv u_{\scriptscriptstyle i}$ and $q_{\scriptscriptstyle i}^{\prime}\equiv d_{\scriptscriptstyle i}$), $D_{\mu}$ is the
covariant derivative which encodes only the QCD interaction of the VLQ \footnote{In our approach, we exclude the EW terms in the covariant derivative, as they are model-dependent and primarily relevant only for $Q$ pair production, which is dominated by QCD effects.}, $m_{\scriptscriptstyle{Q}}$ is the mass of the VLQ, $g$ is the weak coupling constant, $c_{\scriptscriptstyle w}$ is the cosine of the Weinberg mixing angle and $P_{R/L}$ is the right and left-handed chirality projectors. The free parameters $\kbpplr$ characterize the strength of the mixing
between $Q$ and the quarks via the bosons $B^{\prime\prime}=W,Z,H,\wps,\zp,\hp$, respectively. 

\noindent
The Lagrangian $\call_{\scriptscriptstyle B^{\prime}}$ can be expressed as:
\begin{align}
\call_{\scriptscriptstyle B^{\prime}}&=\call^{\scriptscriptstyle\text{kin}}_{\scriptscriptstyle B^{\prime}}+\call^{\scriptscriptstyle\text{mass}}_{\scriptscriptstyle B^{\prime}}\nonumber\\
&+  \frac{g}{2c_{\scriptscriptstyle w}}\underset{i}{\sum}\{\vkzp\underset{q=u,d}{\sum}[\bar q_{\scriptscriptstyle i}\not{\!\! \zp}(g_{\scriptscriptstyle{V}}^{\scriptscriptstyle{q_i}}+g_{\scriptscriptstyle{A}}^{\scriptscriptstyle{q_i}}\gamma_5)q_{\scriptscriptstyle i}+\text{h.c.}]+\vtzp\underset{f=l, \nu}{\sum}[\bar f_{\scriptscriptstyle i}\not{\!\! \zp}(g_{\scriptscriptstyle{V}}^{\scriptscriptstyle{f_i}}+g_{\scriptscriptstyle{A}}^{\scriptscriptstyle{f_i}}\gamma_5)f_{\scriptscriptstyle i}+\text{h.c.}]\}
\nonumber\\
&+\frac{g}{\sqrt{2}}\underset{i}{\sum}\left\{\vkwp[\bar u_{\scriptscriptstyle i}\not{\!\! \wps}P_L d_{\scriptscriptstyle i}+\text{h.c.}]+\vtwp[\bar \nu_{l_{\scriptscriptstyle i}}\not{\!\! \wps}P_L\, l_{\scriptscriptstyle i}+\text{h.c.}]\right\}
-\vkhp\frac{m_{\scriptscriptstyle{t}}}{v}\hp\bar tt
\label{lBp}
\end{align}
where $i$ denotes the generation index. The quantities $g_{\scriptscriptstyle{V}}^{\scriptscriptstyle{f}}$ and $g_{\scriptscriptstyle{A}}^{\scriptscriptstyle{f}}$ are given by: $g_{\scriptscriptstyle{V}}^{\scriptscriptstyle{f}}=\frac{1}{2}I_3^{\scriptscriptstyle{f}}-Q^{\scriptscriptstyle{f}}s_{\scriptscriptstyle w}^2$ and $g_{\scriptscriptstyle{A}}^{\scriptscriptstyle{f}}=-\frac{1}{2}I_3^{\scriptscriptstyle{f}}$ for $f=q, l, \nu$, where $I_3^{\scriptscriptstyle{f}}$ stands for the 3\textsuperscript{rd} component of the isospin and $Q^{\scriptscriptstyle{f}}$ is the electric charge of the fermion~$f$. As mentioned above, the parameters $\varkappa_{{\scriptscriptstyle B^{\prime}}}$ and $\vartheta_{{\scriptscriptstyle B^{\prime}}}$ (for $B^{\prime}=\hp, \zp, W^{\prime}$) are introduced to normalize the interactions of the new bosons. In eq. (\ref{lBp}), we assume that when $\vkzp=\vkwp=\vtzp=\vtwp=\vkhp=1$, the SM fermions interact with the new heavy bosons $\bp$ exactly the same way as they do with SM bosons $W$, $Z$ and $H$, respectively. Unlike the SSM, we add an extra neutral scalar in which its {\it vev} ($v^{\prime}$) is hidden in the coefficient $\vkhp$ (i.e. $\vkhp=v/v^{\prime}\vtkhp$ cf. appendix \ref{appA}) to remain as model-independent as possible. We notice that the kinematic and the mass terms of the new heavy bosons ($\call^{\scriptscriptstyle\text{kin}}_{\scriptscriptstyle B^{\prime}}, \call^{\scriptscriptstyle\text{mass}}_{\scriptscriptstyle B^{\prime}}$) are not included for simplicity.

\subsection{Couplings parametrization}
\noindent
In this section, we show how to parameterize, in model-independent manner, the coefficients characterizing the strength of the electroweak interaction of $Q$ (i.e. $\kbpplr$), see appendix~\ref{appA} for further details. We consider benchmark models where $Q$ mixes only with one quark generation (1\textsuperscript{st}, 2\textsuperscript{nd} or 3\textsuperscript{rd}) via the SM bosons ($\tilde{B}=H, Z, W$) and the new heavy bosons ($\bp=\hp, \zp, \wps$). We assume that the right-handed couplings are negligible compared to the left-handed ones (i.e. $\kbppr=0$ and $\kbppl\neq0$) except for the Higgs where both are taken into account for relatively low $Q$ masses~\footnote{It is shown in refs. \cite{model-indep-vlqs-3,Cacciapaglia:2011fx,model-indep-vlqs-4} that only one of the mixing angle is significant, the other one being suppressed by factor of $m_{\scriptscriptstyle{q}}/m_{\scriptscriptstyle{Q}}$. Therefore, we take into account only the left-handed couplings of the gauge bosons and neglect the right-handed ones, as it is the case for BSM models with VLQ singlets, for all the quarks except the top quark mixing with the Higgs.}~\cite{model-indep-vlqs-4}.\\

\noindent
In the current work, the new bosons are supposed to be heavier than all the particles of the model, so they can decay to VLQ and ordinary fermions(other possibilities are discussed in appendix~\ref{appA}). The main idea of our parametrization is to express the free parameters $\kwpl$, $\kzpl$, $\khpl$ and $\khpr$ in terms of $m_{\scriptscriptstyle Q}$ and the branching ratios $\xi_{\scriptscriptstyle{Qq^{\prime}}}^{\scriptscriptstyle{W^{\prime}}}$, $\xi_{\scriptscriptstyle{Qq}}^{\scriptscriptstyle{Z^{\prime}}}$ and $\xi_{\scriptscriptstyle{Qq}}^{\scriptscriptstyle{H^{\prime}}}$ of the decays: $\wps\rightarrow Qq^{\prime}$, $\zp\rightarrow Qq$ and $\hp\rightarrow Qq$, which are treated as free parameters. The detailed calculation of these parameters is provided in subsection~\ref{appA2}. We have:
\begin{align}
\kwpl&=\kpwp \sqrt{\frac{\xi^{\scriptscriptstyle{W^{\prime}}}_{\scriptscriptstyle{Qq'}}}{\Gamma_{\scriptscriptstyle{Qq^{\prime}}}^{\scriptscriptstyle{W^{\prime}}}}}&
\kzpl&=\kpzp \sqrt{\frac{\xi^{\scriptscriptstyle{Z^{\prime}}}_{\scriptscriptstyle{Qq}}}{\Gamma_{\scriptscriptstyle{Qq}}^{\scriptscriptstyle{Z^{\prime}}}}}&
\khpl&=\kphp \frac{m_{\scriptscriptstyle Q}}{v}\sqrt{\frac{\xi^{\scriptscriptstyle{H^{\prime}}}_{\scriptscriptstyle{Qq}}}{\Gamma_{\scriptscriptstyle{Qq}}^{\scriptscriptstyle{H^{\prime}}}}}.
\label{kzpwpcase1} 
\end{align}
 where $\kpwp$, $\kpzp$ and $\kphp$ are new parameters which have been introduced to hide the couplings of the new heavy bosons to ordinary fermions, they are related to the SSM parameters $\varkappa_{{\scriptscriptstyle B^{\prime}}}$ and $\vartheta_{{\scriptscriptstyle B^{\prime}}}$. The quantities $\Gamma_{\scriptscriptstyle{Qq/q^{\prime}}}^{\scriptscriptstyle{B^{\prime}}}$ are the kinematic functions, associated to the partial decay rate of $B^{\prime}\rightarrow Qq/q^{\prime}$, cf. eqs.~(\ref{xiwpQq},~\ref{xizpQq},~\ref{xihpQq}).\\
 
 \noindent
 Since the VLQ is allowed to decay to SM bosons, the parameters $\kbtildlr$ can be expressed as (cf. sub.sec~\ref{appA1}):  
\begin{align}
\kwl&=\kpq\sqrt{\frac{\xi_{\scriptscriptstyle{W}}}{\Gamma_{\scriptscriptstyle{W}}}}&
\kzl&=\kpq\sqrt{\frac{\xi_{\scriptscriptstyle{Z}}}{\Gamma_{\scriptscriptstyle{Z}}}}&
\khl&=\kpq\frac{m_{\scriptscriptstyle{Q}}}{v}\sqrt{\frac{\xi_{\scriptscriptstyle{H}}}{\Gamma_{\scriptscriptstyle{H}}}}&
\khr&=\kpq\frac{m_{\scriptscriptstyle{q}}}{v}\sqrt{\frac{\xi_{\scriptscriptstyle{H}}}{\Gamma_{\scriptscriptstyle{H}}}}
\label{kzpwpcase2}
\end{align}
$\xi_{\scriptscriptstyle \tilde{B}}$ represents the branching ratio of $Q$ into one of the bosons $\tilde{B}\equiv W,Z,H$ and a quark $q$, and must satisfies $\sum_{\scriptscriptstyle{\tilde{B}}}\xi_{\scriptscriptstyle{\tilde{B}}}=1$. Additionally, $v$ denotes the SM {\it vev} and $\Gamma_{\scriptscriptstyle{\tilde{B}}}$ are the kinematic functions provided in eq. (\ref{kinematic1}).\\

\noindent
The number of independent parameters is related to the studied benchmark model. In the CSSM for example and in the benchmark model where only $\wps$, $Z$ and $H$ bosons are involved and $Q$ mixes with 3\textsuperscript{rd} quark generation, designated by $\textbf{Q}^{\scriptscriptstyle\{3\}}_{\scriptscriptstyle\{W^{\prime}, Z, H\}}$, the number of free parameters is 5 ($\xi_{\scriptscriptstyle Z}$, $\xi^{\scriptscriptstyle{W^{\prime}}}_{\scriptscriptstyle{Qq}}$, $m_{\scriptscriptstyle Q}$, $m_{\scriptscriptstyle W^{\prime}}$  and $\kpq$). We note that in SSM, $\kpwp$, $\kpzp$ and $\kphp$ are not free parameters, see appendix~\ref{appA} for more discussion.\\

\noindent
We recall that the couplings of the VLQ can be large in this model, especially those associated with the new gauge bosons $Z^{\prime}$, $W^{\prime}$, $H^{\prime}$ and the ordinary Higgs. This is shown in figures.~\ref{coulingscase1} and \ref{coulingscase2}. Therefore, one has to impose some constraints on the free parameters and the masses of the model to ensures that perturbative limit is satisfied, at least at the LHC energy scale, which we address in the next subsection.
\subsection{Validity of the perturbative treatment}
\label{pertun}

\noindent
The aim of this paragraph is to find upper bounds on the couplings of the VLQs, given in eqs.~(\ref{kzpwpcase1}) and (\ref{kzpwpcase2}), such that we remain within perturbative limit in the mass and energy range considered in this work.
The validity of the perturbation expansion requires that the partial-wave amplitude (PWA) must be smaller than unity~\cite{prtunit1,prtunit2,prtunit3}. We have shown, in appendix~\ref{appB}, that the $J=0$ PWA ($a_{_0}$) has the dominant contribution for reactions of interest in our study (i.e. higher PWA can be ignored).\\

\noindent
According to eqs.~(\ref{kzpwpcase1}) and (\ref{kzpwpcase2}), the coupling constants are the largest when $\xi^{\scriptscriptstyle{B^{\prime}}}_{\scriptscriptstyle{Qq}}, \xi_{\scriptscriptstyle{\tilde{B}}}\approx1$. This condition holds for benchmark models involving only one SM boson or $B^{\prime}$ decays to $Qq$ \footnote{For benchmark models involving more than one boson, $\xi_{\scriptscriptstyle{B^{\prime\prime}}}<1$, as other decay modes of $Q$ become possible.}. Hence, it is sufficient to consider only the following six benchmark scenarios: ${\bf Q}^{\scriptscriptstyle\{3\}}_{\scriptscriptstyle\{W\}}$, ${\bf Q}^{\scriptscriptstyle\{3\}}_{\scriptscriptstyle\{Z\}}$, ${\bf Q}^{\scriptscriptstyle\{3\}}_{\scriptscriptstyle\{H\}}$, ${\bf Q}^{\scriptscriptstyle\{3\}}_{\scriptscriptstyle\{W^{\prime}\}}$, ${\bf Q}^{\scriptscriptstyle\{3\}}_{\scriptscriptstyle\{Z^{\prime}\}}$ and ${\bf Q}^{\scriptscriptstyle\{3\}}_{\scriptscriptstyle\{H^{\prime}\}}$ to set upper bounds on the couplings.
For the polarized elastic process $q\bar Q\rightarrow q\bar Q$\footnote{Other $2\rightarrow 2$ processes have been examined, where some of them lead to the same restrictions while others cannot be used to constrain the model. See appendix~\ref{appB} for more details.}, we find that the absolute values of the highest eigenvalues of the non-vanishing PWAs in the high-energy limit (cf. eq.~(\ref{lam0})) are given by:
\begin{align}
\left|\lambda_{0}^{\scriptscriptstyle\{V^{_\pm}\}}\right|&=\frac{g^2\kvpmlsq}{\pi}\frac{3}{32}\frac{m_{\scriptscriptstyle Q}^2}{m_{_{V^{_\pm}}}^2},&
\left|\lambda_{0}^{\scriptscriptstyle\{V^{_0}\}}\right|&=\frac{g^2 \kvlsq}{c_{\scriptscriptstyle w}^2\, \pi}\frac{3}{64}\frac{m_{\scriptscriptstyle Q}^2}{m_{\scriptscriptstyle V^{_0}}^2},&
\left|\lambda_{0}^{\scriptscriptstyle\{S\}}\right|&=\frac{\kslsq}{\pi}\frac{3}{16}
\label{lam1}
\end{align}
\noindent
with ($V^{\pm}=W^{\prime},W$), ($V^{0}=Z^{\prime},Z$) and ($S=H^{\prime},H$). 

\vspace{0.25cm}
\noindent
Let us begin by constraining the parameters controlling the mixing between the VLQ and SM bosons ($\kwl$, $\kzl$ and $\khl$). By imposing the conditions $|\lambda_0^{\scriptscriptstyle\{W\}}|<1$, $|\lambda_0^{\scriptscriptstyle\{Z\}}|<1$ and $|\lambda_0^{\scriptscriptstyle\{H\}}|<1$, we find that these couplings are upper bounded by:
\begin{align}
\kwl&\leq \left(\frac{8}{3}\, \frac{s_w^2}{\alpha}\frac{m_{\scriptscriptstyle{W}}^2}{m_{\scriptscriptstyle{Q}}^2}\right)^{1/2},&
\kzl&\leq \left(\frac{16}{3}\, \frac{s_w^2\, c_w^2}{\alpha}\frac{m_{\scriptscriptstyle{Z}}^2}{m_{\scriptscriptstyle{Q}}^2}\right)^{1/2},&
\khl&\leq \left(\frac{16}{3}\, \pi\right)^{1/2}. 
\label{conwzh1} 
\end{align}
where $\alpha$ is the electromagnetic coupling (related to the weak coupling by: $\alpha_w=g^2/(4\pi)\equiv \alpha/s_w^2$).
From  eq. (\ref{kzpwpcase1}) and the constraints given in eq. (\ref{conwzh1}), we find that, to ensure perturbative unitarity, the free parameter $\kpq$ and $m_{\scriptscriptstyle Q}$ must be bounded from above by: 
\begin{align}
&{\bf Q}^{\scriptscriptstyle\{3\}}_{\scriptscriptstyle\{\tilde{B}\}}:\quad \kpq\, m_{\scriptscriptstyle Q} \lesssim \left(\frac{8}{3}\, \gamma_{\scriptscriptstyle{\tilde{B}}}\, m_{\scriptscriptstyle{W}}^2\, \frac{s_w^2}{\alpha}\, \frac{\Gamma_{\scriptscriptstyle \tilde{B}}^{\scriptscriptstyle 0}}{\xi_{\scriptscriptstyle \tilde{B}}}\right)^{1/2} \approx \left(\frac{8}{3}\, m_{\scriptscriptstyle{W}}^2\, \frac{s_w^2}{\alpha}\right)^{1/2}
&\text{with}&& 
\gamma_{\scriptscriptstyle{W}}&=1,& 
\gamma_{\scriptscriptstyle{Z}}&=\gamma_{\scriptscriptstyle{H}}=2.
\label{conwzh2} 
\end{align}
where $\Gamma_{\scriptscriptstyle \tilde{B}}^{\scriptscriptstyle 0}$ are the kinematic functions taken in the asymptotic limit $m_{\scriptscriptstyle Q}\rightarrow\infty$, which is justified since $m_{\scriptscriptstyle Q}\gg m_{\scriptscriptstyle \tilde{B}}$.
To obtain these formulae, we have made the substitutions: $m_{\scriptscriptstyle Z}=m_{\scriptscriptstyle W}/c_w$ and $v=2\, m_{\scriptscriptstyle W}/g$.

\noindent
At the $Z$-pole, i.e. $1/\alpha=128.98$ and $s_w^2=0.2312$, and for $m_{\scriptscriptstyle Q}$ ranging between $1200\, \text{GeV}$ and $7000\, \text{GeV}$, the upper bounds on the $\kpq$ range approximately between $0.6$ and $0.1$. Thus, for $m_{\scriptscriptstyle Q}$ accessible within the LHC's energy range, the value of the free parameter $\kpq$ can be set around 0.1, which is our first conclusion. This is, in fact, in agreement with the experimental limits discussed in refs.~\cite{model-indep-vlqs-4,model-indep-vlqs-6}. \\

\noindent
Now, let's turn to the couplings $\kwpl$, $\kzpl$ and $\khpl$. The perturbative unitarity condition applied on the highest eigenvalues of the PWAs, given on the right-hand side of eq.~(\ref{lam1}), shows that the couplings of the new heavy bosons are upper bounded by:
\begin{align}
\kwpl&\leq \left(\frac{8}{3}\, \frac{s_w^2}{\alpha}\frac{m_{\scriptscriptstyle{W^{\prime}}}^2}{m_{\scriptscriptstyle{Q}}^2}\right)^{1/2},&
\kzpl&\leq \left(\frac{16}{3}\, \frac{s_w^2\, c_w^2}{\alpha}\frac{m_{\scriptscriptstyle{Z^{\prime}}}^2}{m_{\scriptscriptstyle{Q}}^2}\right)^{1/2},&
\khpl&\leq \left(\frac{16}{3}\, \pi\right)^{1/2}. 
\label{conwpzphp1} 
\end{align}

\noindent
Substituting the couplings from eq.~(\ref{kzpwpcase2}) in the constraints (\ref{conwpzphp1}), we show that the parameters $\kpwp$, $\kpzp$, $\kphp$ must be bounded from above by:
\begin{align}
&{\bf Q}^{\scriptscriptstyle\{3\}}_{\scriptscriptstyle\{W^{\prime}\}}:\quad&& \kpwp \leq \left[\frac{8}{3}\, \frac{s_w^2}{\alpha}\, \frac{1}{x_{\scriptscriptstyle W^{\prime}}^2}\, \frac{\Gamma^{\scriptscriptstyle W^{\prime}}_{\scriptscriptstyle Qq^{\prime}}}{\xi_{\scriptscriptstyle Qq^{\prime}}^{\scriptscriptstyle W^{\prime}}}\right]^{1/2}\equiv\left[\frac{8}{3}\, \frac{s_w^2}{\alpha}\, \left(1/x_{\scriptscriptstyle W^{\prime}}^2-3/2+x_{\scriptscriptstyle W^{\prime}}^4/2\right)\right]^{1/2}, \nonumber\\
&{\bf Q}^{\scriptscriptstyle\{3\}}_{\scriptscriptstyle\{Z^{\prime}\}}:\quad &&\kpzp  \leq \left[\frac{16}{3}\, \frac{s_w^2\, c_w^2}{\alpha}\, \frac{1}{x_{\scriptscriptstyle Z^{\prime}}^2}\, \frac{\Gamma^{\scriptscriptstyle Z^{\prime}}_{\scriptscriptstyle Qq}}{\xi^{\scriptscriptstyle Z^{\prime}}_{\scriptscriptstyle Qq}}\right]^{1/2}\equiv\left[\frac{4}{3}\, \frac{s_w^2\, c_w^2}{\alpha}\, \left(1/x_{\scriptscriptstyle Z^{\prime}}^2-3/2+x_{\scriptscriptstyle Z^{\prime}}^4/2\right)\right]^{1/2}, \nonumber\\
%
&{\bf Q}^{\scriptscriptstyle\{3\}}_{\scriptscriptstyle\{H^{\prime}\}}:\quad &&\kphp \leq \left[\frac{16}{3}\,\frac{m_{\scriptscriptstyle W}^2\, s_w^2}{\alpha}\, \frac{1}{x_{\scriptscriptstyle H^{\prime}}^2}\, \frac{\Gamma^{\scriptscriptstyle H^{\prime}}_{\scriptscriptstyle Qq}}{\xi^{\scriptscriptstyle H^{\prime}}_{\scriptscriptstyle Qq}}\right]^{1/2}\!\!\!\!\!\!\!\!/m_{\scriptscriptstyle H^{\prime}}\equiv \left[\frac{16}{3}\,\frac{m_{\scriptscriptstyle W}^2\, s_w^2}{\alpha}\, \left(1/x_{\scriptscriptstyle H^{\prime}}^2-2+x_{\scriptscriptstyle H^{\prime}}^2\right)\right]^{1/2}\!\!\!\!\!\!\!\!/m_{\scriptscriptstyle H^{\prime}} .  
\label{conwpzphp2} 
\end{align}
where we introduced $x_{\scriptscriptstyle B^{\prime}}=m_{\scriptscriptstyle Q}/m_{\scriptscriptstyle B^{\prime}}$ and set $\xi^{\scriptscriptstyle B^{\prime}}_{\scriptscriptstyle Qq/q^{\prime}}\equiv 1$ (for $B^{\prime}\equiv W^{\prime}, Z^{\prime}, H^{\prime}$)\footnote{
Unlike the previous case, the branching ratios $\xi^{\scriptscriptstyle B^{\prime}}_{\scriptscriptstyle Qq/q^{\prime}}$ are not known in the asymptotic limit $x_{\scriptscriptstyle B^{\prime}}\rightarrow0$, since the $B^{\prime}$ can also decay to SM fermions according to the SSM. Thus, setting $\xi^{\scriptscriptstyle B^{\prime}}_{\scriptscriptstyle Qq/q^{\prime}}\equiv 1$ is merely a choice to derive upper limits on the parameters $\kappa_{\scriptscriptstyle B^{\prime}}$ such that perturbative unitarity is satisfied. The obtained limits are necessary valid for smaller values of $\xi^{\scriptscriptstyle B^{\prime}}_{\scriptscriptstyle Qq/q^{\prime}}$.}.\\

\noindent
One has to distinguish between 3 regimes. In the near-threshold regime (i.e. $x_{\scriptscriptstyle B^{\prime}}\approx 1$), the parameters $\kappa_{\scriptscriptstyle B^{\prime}}$ must be extremely small ($\kappa_{\scriptscriptstyle B^{\prime}}\equiv\delta\rightarrow 0$) to ensure perturbative unitarity. Therefore, we will not consider this regime because the interaction of the new heavy bosons must be negligible to remain within the perturbative framework. In the asymptotic region (i.e. $x_{\scriptscriptstyle B^{\prime}}\approx 0$), the upper limits on the couplings are given by:
\begin{align}
\kpwp\, &\leq \frac{2\sqrt{2}}{\sqrt{3\, \alpha}}\, \frac{s_w}{x_{\scriptscriptstyle W^{\prime}}}\underset{{x_{\scriptscriptstyle W^{\prime}}\rightarrow 0}}{\longrightarrow}\infty ,&
\kpzp\, &\leq \frac{2}{\sqrt{3\, \alpha}}\, \frac{s_w\, c_w}{x_{\scriptscriptstyle Z^{\prime}}}\underset{{x_{\scriptscriptstyle Z^{\prime}}\rightarrow 0}}{\longrightarrow}\infty ,&
\kphp\,  &\leq \frac{4\, s_w}{\sqrt{3\, \alpha}} \frac{m_{\scriptscriptstyle W}}{m_{\scriptscriptstyle Q}}.
\label{uppercas2}
\end{align}
From the first two inequalities in eq.~(\ref{uppercas2}), the parameters $\kpwp$  and $\kpzp$ can be very large in the asymptotic limit without disrupting perturbative treatment, regardless of mass of the VLQ. However $\kphp$ depend on the mass $m_{\scriptscriptstyle Q}$ (cf. last inequality in eq.~(\ref{uppercas2})). For instance, when $m_{\scriptscriptstyle Q}= 7$ TeV, we find that $\kphp\leq 0.14$.
In the intermediate regime (i.e. $x_{\scriptscriptstyle B^{\prime}}<1$), which is more realistic and the focus of this work, we observe that the parameters $\kpwp$ and $\kpzp$ depend on the fraction $x_{\scriptscriptstyle B^{\prime}}$. The values of the couplings $\kpwp$, $\kpzp$ and $\kphp$ for $x_{\scriptscriptstyle B^{\prime}}=1/2,\, 2/3,\, 3/4$ and $5/6$ are provided in table~\ref{Tab2}.

\begin{table*}[h!]
\centering
 \renewcommand{\arraystretch}{1.40}
 \setlength{\tabcolsep}{10pt}
  \boldmath
 \begin{adjustbox}{width=8cm,height=1.0cm}
 \begin{tabular}{!{\vrule width 1pt}l!{\vrule width 2pt}c!{\vrule width 1pt}c!{\vrule width 2pt}c!{\vrule width 1pt}c!{\vrule width 1pt}}
  \noalign{\hrule height 1pt}
{\Large$\bf m_{\scriptscriptstyle Q}/m_{\scriptscriptstyle B^{\prime}}$} & {\Large$\bf \kpwp\left[\bf Q^{\scriptscriptstyle\{3\}}_{\scriptscriptstyle\{W'\}}\right]$} & {\Large$\bf \kpzp\left[\bf Q^{\scriptscriptstyle\{3\}}_{\scriptscriptstyle\{Z'\}}\right]$} &\multicolumn{1}{c!{\vrule width 1pt}}{{\Large$\bf \kphp\left[\bf Q^{\scriptscriptstyle\{3\}}_{\scriptscriptstyle\{H'\}}\right]$}}\\
\noalign{\hrule height 1pt}
 \cline{1-4}
{\bf {\large $1/2$}}&$14.2$ &$8.7$ &$0.25$ \\
{\bf {\large $2/3$}}& $8.2$& $5.0$&$0.14$ \\
{\bf {\large $3/4$}}&$5.8$ &$3.6$ &$0.10$ \\
{\bf {\large $5/6$}}&$3.7$ &$2.3$ &$0.06$ \\
\noalign{\hrule height 1pt}
\end{tabular}
\end{adjustbox}
  \caption{\small Upper bounds on $\kpwp$, $\kpzp$ and $\kphp$ (for $\kphp$, we set $m_{\scriptscriptstyle H^{\prime}}=6$ TeV).}
   \label{Tab2}
  \end{table*}

\noindent
It is very important to recall that the bound on the PWA, in order to remain within the perturbative regime, is usually $|a_{J}|\leq 1$ or $|\text{Re}\left(a_{J}\right)|\leq 1/2$ for elastic scatterings. Since the tree-level amplitude is real, one can adopt the latter condition instead of the former one, as it is more restrictive and ensures that we stay away from the non-perturbative region. This leads to tighter upper bounds on the couplings, where one has to divide all the bounds obtained previously by $\sqrt{2}$. For example, the upper limit on $\kpq$, for $m_{\scriptscriptstyle Q}=7$ TeV, becomes 0.07 instead of 0.1 and so on.\\

\noindent
In appendix~\ref{appA}, we considered a more general analysis by treating two cases: {\it case 1}, where $m_{\scriptscriptstyle Q}$ is considered as the heaviest particle in the model, and {\it case 2}, where it is assumed to be lighter than the $\bp$'s.
In our phenomenological study, we assume that $m_{\scriptscriptstyle W^{\prime}, Z^{\prime}}>m_{\scriptscriptstyle Q}$, which is more realistic and corresponds to the regime explored in many CMS publications (cf. refs.~\cite{cms6,cms7}). Therefore, the couplings are parameterized according to \textit{case 2}. We recall that $\kpq$ is constrained only from the mixing to ordinary gauge bosons, we will fix it to $0.1$. This choice ensures perturbative unitarity up to 7 TeV for the VLQ masses and satisfies the experimental constraints from atomic parity violation \cite{model-indep-vlqs-4}, assuming mixing with quarks of the 3\textsuperscript{rd} generation.
\subsection{Renormalization and validation of the model}
\label{se24}
\noindent
This model support computations at both LO and QCD NLO in the complex mass scheme, provided that the unstable virtual particles have narrow widths. In this subsection, we present numerical validation of the model under these conditions, see appendix~\ref{appD} for further details. The Feynman rules and all necessary components for calculating one-loop QCD radiative corrections, such as ultraviolet counterterms and one-loop reduction rational terms, are generated automatically using {\tt FeynRules}, with the support of the {\tt NLOCT} and {\tt FeynArts} packages \cite{ref231,ref232,ref4,ref5}. We have generated 2 versions of the model in the {\tt UFO} format (Universal FeynRules Output), corresponding to the 4- and 5-flavor schemes (respectively 4FS and 5FS), which are publicly available in \url{https://github.com/mszidi/VLQ_Wp_Zp_Hp_NLO}. \\

\noindent
It is important to note that to renormalize the model in QCD, we closely follow the strategy adopted in ref.~\cite{model-indep-vlqs-6}. Specifically, we treat the parameters controlling the coupling strengths ($\kbpplr$) as free parameters to avoid their dependence on the particle masses. In other words, we do not implement their explicit expressions (cf.~eqs~(\ref{kzpwpcase2}) and (\ref{kzpwpcase1})) in {\tt FeynRules} input files, but we retain the compact form of the Lagrangian $\call_{\scriptscriptstyle Q}$ as given in eq.~(\ref{lQ}). This approach prevents conflicts arising from the relation between the mass and coupling counterterms, thereby ensuring the cancellation of ultraviolet divergences at the one-loop level.\\

\noindent
The two key ingredients required to construct the QCD one-loop renormalized model are the {\it renormalization counterterms}, needed to cancel the UV divergences, and the {\it $R_2$ rational terms}. The latter ones are finite quantities arising from the space-time $d$-$4$ component of the one-loop amplitude numerator after reducing it to Passarino-Veltman scalar functions~\cite{Passarino:1978jh}. These ingredients are automatically calculated by {\tt FeynRules/NLOCT} with the help of {\tt FeynArts}. The presence of the VLQ modifies the gluon field wave-function renormalization countertem, while those associated to the SM quarks remain the same as in the SM. In the on-shell renormalization scheme, the renormalization constants at first order in $\alpha_s$ for the gluon wave-function, the left- and right-handed $T$-quark field wave-functions and mass are given by:     
\begin{align}
\delta Z_{\scriptscriptstyle G}&=\delta Z_{\scriptscriptstyle G}^{\scriptscriptstyle \text{SM}}-\frac{g_s^2}{24\, \pi^2}\sum_{q=t, Q}\left[-1/3+B_{0}(0, m_{\scriptscriptstyle q}^2,m_{\scriptscriptstyle q}^2)+2\, m_{\scriptscriptstyle q}^2\, B_{0}^{\prime}(0, m_{\scriptscriptstyle q}^2,m_{\scriptscriptstyle q}^2)\right]\label{delG}\\
\delta Z_{\scriptscriptstyle Q}^{\scriptscriptstyle L/R}&= \frac{\alpha_s}{3\, \pi}\, \left[1+2\, B_{1}\left(m_{\scriptscriptstyle Q}^2,m_{\scriptscriptstyle Q}^2,0\right)+4\, m_{\scriptscriptstyle Q}^2\, \left(2\, B_{0}^{\prime}\left(m_{\scriptscriptstyle Q}^2,0,m_{\scriptscriptstyle Q}^2\right)+B_{1}^{\prime}\left(m_{\scriptscriptstyle Q}^2,m_{\scriptscriptstyle Q}^2,0\right)\right)\right] \\
\delta m_{\scriptscriptstyle Q}& = \frac{\alpha_s}{3\, \pi}\, m_{\scriptscriptstyle Q}\, \left[1-4\, B_{0}\left(m_{\scriptscriptstyle Q}^2,0,m_{\scriptscriptstyle Q}^2\right)-2\, B_{1}\left(m_{\scriptscriptstyle Q}^2,m_{\scriptscriptstyle Q}^2,0\right)\right]
\end{align}
Here, $B_{0,1}$ and $B_{0,1}^{\prime}$ denote the Passarino-Veltman 2-point functions and their derivatives, while $\delta Z_{\scriptscriptstyle G}^{\scriptscriptstyle \text{SM}}$ (with the UV divergence $\frac{5\, g_s^2}{24 \pi^2\varepsilon}$) is the SM gluon field wave-function counterterm for five active light quark flavors. The strong coupling constant is renomalized in the zero-momentum scheme by subtracting the contributions of the massive quark ($t$ and $Q$ quarks in 5FS) from the gluon self-energy. For further details, see refs.~\cite{ref4, Beenakker:2002nc}.\\ 

\noindent
The $R_2$ rational terms, involving the extra-gauge bosons and the $T$ quark in the case of mixing with 3-generation quarks, are provided in table~\ref{R2}. In this table, we use the shorthand notations: $V^{\scriptscriptstyle 0}\equiv Z, \zp$, $V^{\scriptscriptstyle \pm}\equiv W, \wps$, $S\equiv H, \hp$, $\varkappa_{\scriptscriptstyle Z}=\varkappa_{\scriptscriptstyle W}=1$, $\varkappa_{\scriptscriptstyle H}=y_t/\sqrt{2}$, $\beta_{q_u}=4$ and $\beta_{q_d}=2$. The Kronecker delta-functions $\delta_{ij}$ and $\delta_{ab}$ enforce the quark and gluon color conservation, respectively. The tensor $V_{\mu_1\mu_2\mu_3\mu_4}$ is defined as: $V_{\mu_1\mu_2\mu_3\mu_4}=g_{\mu_1\mu_2}g_{\mu_3\mu_4}+g_{\mu_1\mu_3}g_{\mu_2\mu_4}+g_{\mu_1\mu_4}g_{\mu_2\mu_3}$, where $g_{\mu\nu}$ represents the metric tensor.   
\begin{table*}[h!]

\centering
 \renewcommand{\arraystretch}{1.40}
 \setlength{\tabcolsep}{10pt}
 \begin{adjustbox}{width=18cm,height=2.25cm}
 \boldmath
 \begin{tabular}{!{\vrule width 1pt}l!{\vrule width 1pt}l!{\vrule width 2pt}l!{\vrule width 1pt}l!{\vrule width 2pt}l!{\vrule width 1pt}l!{\vrule width 1pt}}
  \noalign{\hrule height 1pt}
$R_2$ & \textbf{Expression} & $R_2$ & \textbf{Expression} & $R_2$ & \textbf{Expression}  \\
\noalign{\hrule height 1pt}
$\bar{T}-T$ &
$ \frac{-i\, g_s^2}{12\pi^2}\left(\not{\!p}-2\, m_{\scriptscriptstyle T}\right)\, \delta_{ij}$ &
$G-G-V^{\scriptscriptstyle 0}-V^{\scriptscriptstyle 0}$ & $\frac{i\, g^2\, g_s^2}{288\, \pi^2\, c_{\scriptscriptstyle{w}}^2}\, \left[3\, \kvnl^2+\vkvnsq\left(9-18 s_{\scriptscriptstyle w}^2+20 s_{\scriptscriptstyle w}^4\right)\right]\, V_{\mu_1\mu_2\mu_3\mu_4}\, \delta_{ab}$&
$G-G-H$ & $\frac{-i\, g_s^2\, m_{\scriptscriptstyle T} y_t}{8\sqrt{2}\pi^2}\, g_{\mu_1\mu_2}\, \delta_{ab}$\\
\cline{1-6}
$\bar{T}-t-S$ & 
$ \frac{i\, g_s^2}{3\pi^2}\left(\ksl P_{\scriptscriptstyle L}+\ksr P_{\scriptscriptstyle R} \right)\, \delta_{ij}$ &
$G-G-Z-\zp$ &
$\frac{i\, g^2\, g_s^2}{288\, \pi^2\, c_{\scriptscriptstyle{w}}^2}\, \left[3\, \kzl\, \kzpl+\vkzp\left(9-18 s_{\scriptscriptstyle w}^2+20 s_{\scriptscriptstyle w}^4\right)\right]\, V_{\mu_1\mu_2\mu_3\mu_4}\, \delta_{ab}$&
$G-G-\hp$& $\frac{-i\, g_s^2\, m_{\scriptscriptstyle T}\, \vkhp}{8\, \pi^2}\, g_{\mu_1\mu_2}\, \delta_{ab}$\\
\cline{1-6}
$\bar{t}-T-S$ & 
$ \frac{i\, g_s^2}{3\pi^2}\left(\ksl P_{\scriptscriptstyle R}+\ksr P_{\scriptscriptstyle L} \right)\, \delta_{ij}$ &
$G-G-V^{\scriptscriptstyle\pm}-V^{\scriptscriptstyle\pm}$& 
$\frac{i\, g^2\, g_s^2}{96\, \pi^2}\, \left[\kvpmlsq+3\, \vkvpmsq\right]\, V_{\mu_1\mu_2\mu_3\mu_4}\, \delta_{ab}$ &
$q_{\scriptscriptstyle u}-q_{\scriptscriptstyle d}-W$&
$\frac{-i\, g\, g_s^2}{6\, \sqrt{2}\, \pi^2}\, \gamma_{\mu}\, P_L\, \delta_{ij}$\\
\cline{1-6}
$\bar{T}-t-V^{\scriptscriptstyle 0}$ & 
$ \frac{-i\, g\, g_s^2}{12\pi^2\, c_{\scriptscriptstyle w}}\, \kvnl \gamma_{\mu}\, P_{\scriptscriptstyle L}\, \delta_{ij}$ &
$G-G-W-\wps$ &  
$\frac{i\, g^2\, g_s^2}{96\, \pi^2}\, \left[\kwl\,\kwpl +3\, \vkwp\right]\, V_{\mu_1\mu_2\mu_3\mu_4}\, \delta_{ab}$&
$q_{\scriptscriptstyle u}-q_{\scriptscriptstyle d}-\wps$&
$\frac{-i\, g\, g_s^2}{6\, \sqrt{2}\, \pi^2}\, \vkwp \gamma_{\mu}\, P_L\, \delta_{ij}$\\
\cline{1-6}
$\bar{T}-b-V^{\scriptscriptstyle \pm}$ &
$ \frac{-i\, g\, g_s^2}{6\sqrt{2}\pi^2}\, \kvpml \gamma_{\mu}\, P_{\scriptscriptstyle L}\, \delta_{ij}$ & 
$G-G-S-S$&
$\frac{-i\, g_s^2}{16\, \pi^2}\, \left[2 \vkssq+2\left(\kslsq+\ksrsq\right)\right]\, g_{\mu_1\mu_2}\, \delta_{ab}$&
$q-q-Z$&
$\frac{-i\, g\, g_s^2}{36\, \pi^2\, c_{\scriptscriptstyle w}}\, \gamma_{\mu} \left[(3-\alpha_{q}s_{\scriptscriptstyle w}^2)\, P_L-\alpha_{q}s_{\scriptscriptstyle w}^2 P_{R}\right]\delta_{ij}$\\
\cline{1-6}
$\bar{T}-T-G$ & 
$ \frac{-i\, g_s^3}{6\pi^2}\, \gamma_{\mu}\, (T^{a})_{ij}$ & 
$G-G-H-\hp$& 
$\frac{-i\, g_s^2}{16\, \pi^2}\, \left[\sqrt{2}\, \vkhp y_t+2\left(\khl\, \khpl+\khr\, \khpr\right)\right]\, g_{\mu_1\mu_2}\, \delta_{ab}$&
$q-q-\zp$&
$\frac{-i\, g\, g_s^2}{36\, \pi^2\, c_{\scriptscriptstyle w}}\, \vkzp\,\gamma_{\mu} \left[(3-\beta_{q}s_{\scriptscriptstyle w}^2)\, P_L-\beta_{q}s_{\scriptscriptstyle w}^2 P_{R}\right]\delta_{ij}$\\
\noalign{\hrule height 1pt}
\end{tabular}
\end{adjustbox}
  \caption{\small   Some $R_2$ rational terms associated with the $T$ quark and the heavy bosons of the model.}
   \label{R2}
  \end{table*}
  
\noindent
The model is validated at both LO and NLO in the CM scheme for relatively small widths in multiple ways\footnote{The model should be improved for broad widths in NLO calculations, but we leave this extension for future work.}: 

\vspace{0.25cm}
\noindent
{\bf 1- Comparison with SM counterparts:} We compared its predictions for several reactions with their  corresponding counterparts in the SM, which are expected to be the same if the hypothetical particles $T$, $\wps$ and $\zp$ are constrained to have the same masses, couplings and total width as their SM partners $t$, $W$ and $Z$, provided that these processes are described by the same Feynman diagrams at the Born, loop and real-emission levels. For example, the single production of a $T$ quark in association with a $\bar{b}$ quark via virtual $\wps$ exchange in the $s$-channel ($pp\rightarrow \{\wps\} \rightarrow T \bar{b}$) is expected to have exactly the same LO and NLO cross sections as its SM counterpart process, $pp\rightarrow \{W\} \rightarrow t \bar{b}$, provided that: $m_{\scriptscriptstyle T}=m_{\scriptscriptstyle t}$, $m_{\scriptscriptstyle W^{\prime}}=m_{\scriptscriptstyle W}$, $\vkwp=\kwpl=1$ and $\Gamma^{\text{ToT}}_{\scriptscriptstyle W^{\prime}}=\Gamma^{\text{ToT}}_{\scriptscriptstyle W}$ (where $\Gamma^{\text{ToT}}_{\scriptscriptstyle W^{\prime},W}$ denotes the total width). This agreement arise because the 2 processes share identical Feynman diagrams at both LO and NLO (including Born, virtual and real-emission), and the parameters ($\kbpplr$) are treated as free ones when generating the counterterms. \\

\noindent
To cover a wide range of one-loop typologies, we consider both $2\rightarrow2$ and $2\rightarrow3$ processes listed in table~\ref{TestProcess}, where some involve the VLQ as virtual and/or real, while others include 3- and 4-gluon vertices. We note that the processes 1 are of pure QCD origin while the remaining are mixed weak-QCD processes. For certain reactions, we need to carefully select their associated Feynman diagrams (applying appropriate filters when necessary) to ensure that each process in our model (labeled with {\tt vlQBp} in table~\ref{TestProcess}) has identical Feynman diagrams to its SM counterpart. For instance, in the SM process $pp\rightarrow \{W, t\} \rightarrow \bar{b}tH$, we must exclude all Feynman diagrams containing $W-W-H$ and $b-b-H$ vertices, as their counterparts do not appear in the corresponding {\tt vlQBp} reaction $pp\rightarrow \{\wps, T\}\rightarrow \bar{b} T \rightarrow \bar{b}tH$ (cf. processes 4). It is important to note that discarding some diagrams renders the SM amplitude incomplete. However, this does not affect the validation technique in any way, as it is done purely for comparative analysis.  \\

\noindent
The results, presented in table~\ref{TestProcess}, are obtained in the 4FS (i.e. $t$ and $b$ are treated as massive). We numerically compute the hadronic cross sections using {\tt MadGraph5\_v2\_9\_6}. We employed the parton distribution functions {\tt NNPDF\_3.0}~\cite{NNPDF:2014otw} and set the renormalization and factorization scales equal ($\mu_R=\mu_F=\mu$), where $\mu$ is a dynamical scale defined as the sum of the transverse momentum of the final-state particles divided by~2. The SM predictions are generated using the {\tt UFO} model {\tt loop\_qcd\_qed\_sm\_Gmu} (supporting CM scheme~\cite{cmssch1}) provided by {\tt MadGraph5} collaboration. For the {\tt vlQBp} model processes, we imposed the following assumptions:
\begin{align}
\kzpl&=\kzl=2\, (I_3^{t}-Q^{ t}\, s_{\scriptscriptstyle w}^2)
&
\vkwp&=\vkzp=1
&
m_{\scriptscriptstyle Z^{\prime}}&=m_{\scriptscriptstyle Z}
&
m_{\scriptscriptstyle T}&=m_{\scriptscriptstyle t}
&
\Gamma^{\text{ToT}}_{\scriptscriptstyle Z^{\prime}}&=\Gamma^{\text{ToT}}_{\scriptscriptstyle Z}
\notag\\
\kzpr&=\kzr=- 2\, Q^{ t}\, s_{\scriptscriptstyle w}^2
&
\khl&=\khr=m_{\scriptscriptstyle T}/v
&
m_{\scriptscriptstyle W^{\prime}}&=m_{\scriptscriptstyle W}
&
\kwpl&=1
&
\Gamma^{\text{ToT}}_{\scriptscriptstyle W^{\prime}}&=\Gamma^{\text{ToT}}_{\scriptscriptstyle W}
 \label{KinematicAssumption}
\end{align}
where the masses and widths are set to their SM values (we also enforce $\Gamma^{\text{ToT}}_{\scriptscriptstyle T}=\Gamma^{\text{ToT}}_{t}$).
\begin{table*}[h!]

\centering
 \renewcommand{\arraystretch}{1.40}
 \setlength{\tabcolsep}{10pt}
 \begin{adjustbox}{width=17cm,height=3.5cm}
 \boldmath
 \begin{tabular}{!{\vrule width 2pt}l!{\vrule width 2pt}l!{\vrule width 2pt}l!{\vrule width 2pt}l!{\vrule width 1pt}l!{\vrule width 2pt}l!{\vrule width 1pt}l!{\vrule width 2pt}}
\cline{4-7}
 \multicolumn{1}{c!}{{}}& \multicolumn{1}{c!}{{}}& 
 &\multicolumn{2}{c!{\vrule width 2.0pt}}{{{\Large$\mathbf{\sigma_{\scriptscriptstyle LO}[pb]}$}}} 
 &\multicolumn{2}{c!{\vrule width 2.0pt}}{{{\Large$\mathbf{\sigma_{\scriptscriptstyle NLO}[pb]}$}}} \\
 \cline{2-7}
 \multicolumn{1}{c!{\vrule width 2.0pt}}{{}}&\multicolumn{1}{c!{\vrule width 2.0pt}}{\Large\textbf{Model}} &\multicolumn{1}{c!{\vrule width 2.0pt}}{\Large\textbf{Process}}  & {\Large\textbf{zero-width}} &{\Large\textbf{finite-width}} & {\Large\textbf{zero-width}}&{\Large\textbf{finite-width}}\\
\noalign{\hrule height 1pt}
{\Large$1$} 
&\textbf{vlQBp}&  $p\, p \overset{\textbf{pure QCD}}{\xrightarrow{\hspace{2.0cm}}} T\, \bar{T}$ 
&$459.3 \pm 0.51$& $459.5 \pm 0.50$ 
&$678.6 \pm 0.50$& $678.2 \pm 0.54$\\
&\textbf{SM}&  $p\, p \overset{\textbf{pure QCD}}{\xrightarrow{\hspace{2.0cm}}} t\, \bar{t}$ 
& $459.5 \pm 0.50$& $459.3 \pm 0.51$
&$678.9 \pm 0.49$& $679.3\pm0.50$\\
\cline{1-7}
{\Large$2$} 
&\textbf{vlQBp}&  $p\, p \overset{\textbf{virtual}\,\, W^{\prime}}{\xrightarrow{\hspace{2.0cm}}} T\, \bar{b}$ 
&$4.853 \pm 5.2\times 10^{-3}$& $4.837 \pm 5.2\times 10^{-3}$ 
&$106.7 \pm 7.1\times 10^{-2}$& $106.7 \pm 7.6\times 10^{-2}$  \\
&\textbf{SM}&  $p\, p \overset{\textbf{virtual}\,\, W}{\xrightarrow{\hspace{2.0cm}}} t\, \bar{b}$ 
&$4.850 \pm 5.2\times 10^{-3}$&$4.848 \pm 5.3\times 10^{-3}$  
&$106.7 \pm 7.2\times 10^{-2}$&$106.9 \pm 6.4\times 10^{-2}$ \\
\cline{1-7}
{\Large$3$} 
&\textbf{vlQBp}&  $p\, p \overset{\textbf{virtual}\,\, Z^{\prime}}{\xrightarrow{\hspace{2.0cm}}} T\, \bar{t}$ 
&$0.1452 \pm 8.9\times 10^{-5}$& $0.1454 \pm 9.2\times 10^{-5}$ 
&$2.566 \pm 1.9\times 10^{-3}$ & $2.566 \pm 1.9\times 10^{-3}$ \\
&\textbf{SM}&  $p\, p \overset{\textbf{virtual}\,\, Z}{\xrightarrow{\hspace{2.0cm}}} t\, \bar{t}$ 
& $0.1454 \pm 9.4\times 10^{-5}$&$0.1457 \pm 9.2\times 10^{-5}$ 
& $2.569 \pm 1.6\times 10^{-3}$& $2.569 \pm 1.7\times 10^{-3}$\\
\cline{1-7}
{\Large$4$} 
&\textbf{vlQBp}&  $p\, p \overset{\textbf{virtual}\,\, W^{\prime}, T}{\xrightarrow{\hspace{2.0cm}}} \bar{b}\, t\, H$ 
&$1.131\times 10^{-3} \pm 1.2\times 10^{-6}$& $1.128\times 10^{-3} \pm 1.2\times 10^{-6}$ 
&$1.012\times 10^{-1} \pm 6.9\times 10^{-5}$&$1.010\times 10^{-1} \pm 7.8\times 10^{-5}$\\
&\textbf{SM}&  $p\, p \overset{\textbf{virtual}\,\, W, t}{\xrightarrow{\hspace{2.0cm}}} \bar{b}\, t\, H$ 
& $1.130\times 10^{-3} \pm 1.1\times 10^{-6}$& $1.131\times 10^{-3} \pm 1.1\times 10^{-6}$ 
& $1.009\times 10^{-1} \pm 6.6\times 10^{-5}$& $1.010\times 10^{-1} \pm 7.5\times 10^{-5}$\\
\cline{1-7}
{\Large$5$} 
&\textbf{vlQBp}&  $p\, p \overset{\textbf{virtual}\,\, W^{\prime}, T}{\xrightarrow{\hspace{2.0cm}}} \bar{b}\, t\, Z$ 
&$1.874\times 10^{-3} \pm 1.3\times 10^{-6}$& $1.873\times 10^{-3} \pm 1.3\times 10^{-6}$  
& $2.370\times 10^{-1} \pm 1.9\times 10^{-4}$ & $2.372\times 10^{-1} \pm 2.4\times 10^{-4}$\\
&\textbf{SM}&  $p\, p \overset{\textbf{virtual}\,\, W, t}{\xrightarrow{\hspace{2.0cm}}} \bar{b}\, t\, Z$ 
& $1.875\times 10^{-3} \pm 1.3\times 10^{-6}$& $1.877\times 10^{-3} \pm 1.3\times 10^{-6}$ 
& $2.375\times 10^{-1} \pm 2.4\times 10^{-4}$& $2.376\times 10^{-1} \pm 1.9\times 10^{-4}$\\
\cline{1-7}
{\Large$6$} 
&\textbf{vlQBp}&  $p\, p \overset{\textbf{virtual}\,\, Z^{\prime}, T}{\xrightarrow{\hspace{2.0cm}}} \bar{t}\, t\, H$ 
&$2.245\times 10^{-4} \pm 1.3\times 10^{-7}$& $2.246\times 10^{-4} \pm 1.3\times 10^{-7}$
&$9.755\times 10^{-3} \pm 5.4\times 10^{-6}$& $9.742\times 10^{-3} \pm 5.0\times 10^{-6}$\\
&\textbf{SM}&  $p\, p \overset{\textbf{virtual}\,\, Z, t}{\xrightarrow{\hspace{2.0cm}}} \bar{t}\, t\, H$ 
&$2.246\times 10^{-4} \pm 1.3\times 10^{-7}$& $2.249\times 10^{-4} \pm 1.3\times 10^{-7}$
&$9.713\times 10^{-3} \pm 6.3\times 10^{-6}$& $9.714\times 10^{-3} \pm 4.3\times 10^{-6}$ \\
\cline{1-7}
{\Large$7$} 
&\textbf{vlQBp}&  $p\, p \overset{\textbf{virtual}\,\, Z^{\prime}, T}{\xrightarrow{\hspace{2.0cm}}} T\, \bar{t}\, Z$ 
&$9.071\times 10^{-4} \pm 6.5\times 10^{-7}$& $9.069\times 10^{-4} \pm 5.1\times 10^{-7}$ 
&$7.786\times 10^{-3} \pm 3.6\times 10^{-6}$& $7.786\times 10^{-3} \pm 4.8\times 10^{-6}$ \\
&\textbf{SM}&  $p\, p \overset{\textbf{virtual}\,\, Z, t}{\xrightarrow{\hspace{2.0cm}}} t\, \bar{t}\, Z$ 
&$9.100\times 10^{-4} \pm 7.6\times 10^{-7}$& $9.101\times 10^{-4} \pm 7.8\times 10^{-7}$
&$7.783\times 10^{-3} \pm 4.1\times 10^{-6}$& $7.793\times 10^{-3} \pm 3.8\times 10^{-6}$\\
\noalign{\hrule height 1pt}
\end{tabular}
\end{adjustbox}
  \caption{\small Comparison between $\sigma_{\scriptscriptstyle \text{LO}/\text{NLO}}$ of some processes in the model vlQBp with their SM counterparts.}
   \label{TestProcess}
\end{table*}

\noindent

\noindent
We recall that these results are computed in the CM scheme, an option available in {\tt MadGraph5} (enabled via {\tt set complex\_mass\_scheme}). This table shows an excellent agreement between our model predictions and the SM counterparts, both at LO and NLO in QCD. We note that we considered also configurations, where the unstable particles have masses of TeV order and the width-to-mass ratios: $1\%$, $3\%$, $6\%$, $10\%$ and $30\%$. We also, found a good agreement between vlQBp and SM counterparts cross sections for width-to-mass ratio less than $10\%$, some results are provided in appendix~\ref{appD}. This strongly confirms the validity and consistency of our model within the CM scheme, at least for the processes considered in this article and for width-to-mass ratios less than $10\%$. In other words, the counterterms and the rational terms generated in this scheme by {\tt FeynRules/FeynArts/NLOCT} (enabled by setting {\tt ComplexMass -> True} when generating the counterterms at 1st order in $\alpha_s$ by {\tt NLOCT}) are correct, at least for the processes and configuration considered here.
Further discussion on the CM scheme follows in the next section.
For full transparency, all analysis scripts, configuration cards and additional materials required to reproduce these results are publicly available at \url{https://github.com/mszidi/VLQ_Wp_Zp_Hp_NLO}.\\

\noindent
{\bf 2- {\tt MadGraph5} CM scheme automatic check:} We checked the consistency of the CM scheme implementation by the {\tt MadGraph5} option {\tt check cms} described in \url{https://cp3.irmp.ucl.ac.be/projects/madgraph/wiki/ComplexMassScheme}. We found that all processes (Born and virtual) in the configurations considered in this paper passed the test, some examples are given in appendix~\ref{appD}. This again shows that this model can be used for computation at both LO and NLO in the presence of unstable particles with narrow widths. \\ 

\noindent
{\bf 3- Reproducing the results with an extended version {\tt loop\_qcd\_qed\_sm\_Gmu}:} We extend the SM {\tt UFO} model {\tt loop\_qcd\_qed\_sm\_Gmu}, which supports the CM scheme~\cite{cmssch1}, by manually implementing all necessary UV counterterms and $R_2$ rational terms associated to the VLQ $T$. The structure of these terms is analogous to those of the top quark but with specific modifications for the $T$ partner. We have calculated the LO and NLO cross sections of the processes $pp\rightarrow\{W^{\prime}\}\rightarrow T\bar{b}+\bar{T}b$ and $pp\rightarrow\{W^{\prime}, T\}\rightarrow t\bar{b}H+\bar{t}bH$ (involving virtual $W^{\prime}$ and/or $T$ propagators) using both our dedicated model and the modified {\tt loop\_qcd\_qed\_sm\_Gmu} model. The results show excellent agreement. We note that the $W^{\prime}$ is not implemented in the modified SM version, instead, the SM $W$ boson was used as its proxy. The results and further details on this approximation are provided in appendix~\ref{appD}.\\  

\noindent
{\bf 4- Comparison with existing models:} We note that we have also compared both single production and pair production of VLQs (in the absence of $\wps$ and $\zp$) with some results reported in refs.~\cite{model-indep-vlqs-6,model-indep-vlqs-7}. Additionally, in scenarios without VLQs, we compared the cross sections for leptons and neutrinos production via $\zp$ and $\wps$ exchange with those obtained using the model introduced in ref.~\cite{fuks1}. \\ 

\noindent
In the next section, we use the model {\tt vlQBp} to study the production of $\wps$ and $\zp$ and their decay to VLQ at the LHC. The width-to-mass ratio of the unstable particles for the chosen configurations in this study range between $0.19\%$ and $5.88\%$, which means that this model can be used safely for NLO calculations even for Feynman diagrams involving virtual unstable particles. Despite this advantage, we do not rely solely on the full calculation within the CM scheme but adopt the NWA as a precautionary measure.

%% file: prod_wp_zp.tex
\section{Production of new heavy gauge bosons decaying to vector-like quarks}
\label{sec3}

In this section, we examine the production and decay of the hypothetical heavy gauge bosons $\zp$ and $\wps$ at the LHC. We focus on Drell-Yan-like reactions, which dominate the production of these resonances in comparison to vector-boson fusion processes\footnote{This dominance is shown in various studies, see for example \cite{wpth1}. {We have also confirmed this numerically within our model}.}. We assume that they decay into a vector-like top quark partner~$T$, which subsequently decays into $t H$ and/or $t Z$, considering fully hadronic decays of each resulting particle. Those new heavy bosons would manifest as a narrow peak in the invariant mass distribution of the produced jets, where one distinguishing feature of these jets is their significant Lorentz boost resulting from the large masses of the hypothetical particles, making them notably more energetic than those typically observed in SM events~\cite{cms6,cms7}.

\begin{figure}[h!]
\includegraphics[width=12.0cm,height=3.15cm]{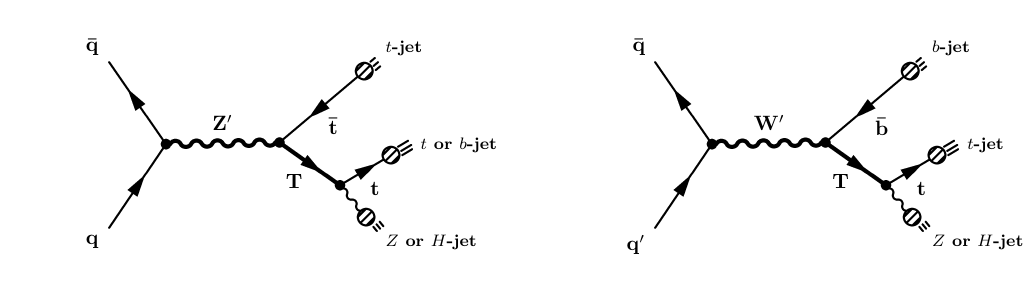}
 \caption{\small Production of $\zp$ and $\wps$ and their subsequent decays into $T$ and jets at the LHC.}
 \label{pp-zp-wp}
\end{figure}

\noindent
The production and decay mechanism of these particles is illustrated in figure~\ref{pp-zp-wp}. We depict the  Feynman diagrams showing the production of the $\zp$ and $\wps$ gauge bosons and their subsequent decays into VLQ and jets. The purpose of this section is to calculate the production cross section for heavy gauge bosons through Drell-Yan processes, employing both the narrow width approximation (NWA) and the complex mass scheme (CM). These computations will be performed at LO and NLO in QCD within the 4-flavor scheme 4FS (all SM quarks are assumed to be massless, except the top and bottom quarks). Additionally, we will provide differential distributions in the invariant mass for the final-state jets at LO+PS and NLO+PS levels, where PS indicates the parton shower {\tt pythia8}~\cite{Sjostrand:2014zea}.

\noindent
In the whole paper, we use {\tt MadGraph5\_v2\_9\_6} to numerically compute the hadronic cross sections. We employ the parton distribution functions {\tt NNPDF\_3.0}~\cite{NNPDF:2014otw} and set the renormalization and factorization scales equal ($\mu_R~=~\mu_F~=~\mu$), where $\mu$ is a dynamical scale defined as the sum of the transverse momentum of the final-state particles divided by 2.


\subsection{Brief overview of NWA and CM schemes}
One of the main objectives on this work is to compare the CM and NWA predictions, for relatively small widths, at both LO and NLO for the reactions under study in this work. To that end, we would like to begin by giving a brief overview of each method.

\noindent
$\bullet$ In the NWA, the cross section associated to the processes described by the Feynman diagrams, depicted in figure~\ref{pp-zp-wp}, can be expressed in the following simple form:
\begin{align} 
\sigma_{\scriptscriptstyle{V^{\prime}}} &= 
\sigma^{\scriptscriptstyle{pp \rightarrow V^{\prime}}}\, \times\, \text{Br}[V^{\prime} \rightarrow qT] \, \times\! \sum_{\scriptscriptstyle \tilde{B}=W,Z,H}\!\!\!\!\!\!\text{Br}[T \rightarrow q^{\prime} \tilde{B}] & \text{with}&&
\sigma^{\scriptscriptstyle{pp \rightarrow V^{\prime}}}&= \left[\vkvp\right]^2 \sigma^{\scriptscriptstyle{pp \rightarrow V^{\prime}}}_{\scriptscriptstyle{0}}
\label{signwa}
\end{align} 
where $\sigma^{\scriptscriptstyle{pp \rightarrow V^{\prime}}}_{\scriptscriptstyle{0}}$ is the Drell-Yan cross section for $\vkvp = 1$. The sum over $\tilde{B}$ in eq.~(\ref{signwa}) covers all possible decay channels of $T$ in a specific benchmark scenario. Since the branching ratios $\text{Br}[T \rightarrow q^{\prime} \tilde{B}] \equiv \xi_{\scriptscriptstyle{\tilde{B}}}$ are fixed in the asymptotic limit and always sum to one\footnote{Since we are looking for all final state jets, one has to consider all possible decay modes of $T$ in the given benchmark scenario.}, cf. section \ref{sec2}, the behavior of the cross section $\sigma_{\scriptscriptstyle{V^{\prime}}}$ in eq.~(\ref{signwa}) is controlled by only the following three free parameters: $\vkvp$, the branching ratio $\text{Br}[V^{\prime} \rightarrow qT] \equiv \xi_{\scriptscriptstyle{Tq}}^{\scriptscriptstyle{V^{\prime}}}$ and the mass of the hypothetical gauge boson $m_{\scriptscriptstyle{V^{\prime}}}$ (with $\vp=\wps, \zp$).  \\

 \noindent
 The approximate equalities in eq.~(\ref{signwa}) can be proven by replacing the denominators of the squared propagators of the unstable particles ($\vp$ and $T$) in the squared amplitude with their Breit-Wigner form:
 \begin{align}
 \frac{1}{|q^2-m_{\scriptscriptstyle A}^2+i\, \Gamma^{\scriptscriptstyle\text{ToT}}_{\scriptscriptstyle A}\, m_{\scriptscriptstyle A}|^2}&
 \approx \frac{\pi}{\Gamma^{\scriptscriptstyle\text{ToT}}_{\scriptscriptstyle A}\, m_{\scriptscriptstyle A}}\, \delta(q^2-m_{\scriptscriptstyle A}^2)+\mathcal{P}\left[\frac{1}{(q^2-m_{\scriptscriptstyle A}^2)^2}\right]+\mathcal{O}\left(\frac{\Gamma^{\scriptscriptstyle\text{ToT}}_{\scriptscriptstyle A}}{m_{\scriptscriptstyle A}}\right)
\end{align}
where $\Gamma^{\scriptscriptstyle\text{ToT}}_{\scriptscriptstyle A}$ and $m_{\scriptscriptstyle A}$ are, respectively, the total width and the mass of the particle $A$ (with $A\equiv \wps, \zp, T$ in this case).
 This involves approximating the denominator of the squared propagator as a delta function (first term) plus a principal value term ($\mathcal{P}$ stands for the principle value operator), neglecting higher-order terms in the width-to-mass ratio. A detailed proof of this approach, at LO, is shown in appendix \ref{appC}.\\

\noindent
It is very important to notice that one of the crucial requirements for the validity of this approach is that the width-to-mass ratio of the unstable particles, $\wps$, $\zp$ and $T$, must remain sufficiently small. In fact, this condition is necessary but not sufficient for the validity of the NWA. Additional conditions must also be satisfied, such as: the propagator can be factorized from the matrix element, interference with other contributions must vanish or be negligible and the production and subsequent decays must occur far from thresholds. For more detail on this approach, see appendix~\ref{appC} and refs. \cite{width2,width1,Schwartz2014sze,Fuchs:2014ola}.  \\

\noindent 
$\bullet$ In the CM approach \cite{cmssch1,cmssch2}, which provides greater accuracy, the mass of any unstable particle in the model is replaced by $(m_{_A}^2-i\, m_{_A}\, \Gamma^{\scriptscriptstyle\text{ToT}}_{_A})^{1/2}$ for each $A=\wps, T, W, Z, H, t$. This substitution redefines all quantities that depend on particle masses, including, for example, the propagator of unstable fermions ($\mathcal{P}_{_f}$) and the renormalized Weinberg angle cosine ($c_{_w}$), which are now expressed as follows:
\begin{align}
  \mathcal{P}_{_f}&=\frac{\not{\!p}+\mu_{_f}}{p^2-\mu_{_f}^2}\quad \text{with}\quad  \mu_{_f}^2=m_{_f}^2-i\, m_{_f}\,\Gamma^{\scriptscriptstyle\text{ToT}}_{_f}&
 c_{_w}^2&=\frac{m_{_W}^2-i\, m_{_W}\, \Gamma^{\scriptscriptstyle\text{ToT}}_{_Z}}{m_{_Z}^2-i\, m_{_Z}\, \Gamma^{\scriptscriptstyle\text{ToT}}_{_Z}}
\end{align}
The ability to perform calculations (for narrow widths) in the CM framework is achieved in our model, thanks to {\tt FeynRules}, {\tt NLOCT} ( enabled via {\tt ComplexMass -> True}) and {\tt MadGraph5}.
In this case, the cross section depends generally on five free parameters which are: $\vkvp$, $\xi_{\scriptscriptstyle{Tq}}^{\scriptscriptstyle{V^{\prime}}}$, $m_{_{V^{\prime}}}$, $\kpt$ and $m_{_T}$. However, for configurations where NWA and CM give approximately the same results (at LO and NLO), the dependence on the last two parameters ($\kpt$ and $m_{_T}$) is significantly reduced making the cross section effectively depends only on the 3 remaining free parameters even in this approach. We recall that the LO and NLO predictions in the CM scheme for narrow widths are validated in section~\ref{sec2} and appendix~\ref{appD}, cf. tables~\ref{TestProcess}, \ref{tabl1AppD}, \ref{tabl2AppD} and \ref{tabloopqcd}.\\

\noindent
In what follows, we employ 3 approaches to compute the LO and NLO cross sections: the CM scheme (full calculation method), the NWA$_2$ method (both unstable particles are treated as on-shell states, cf.~ eq.~(\ref{signwa}), i.e. NWA is applied twice), and the the NWA$_1$ method, which is a hybrid between NWA and CM ($T$ is considered as on-shell state, i.e. NWA is applied only once). 
\subsection{Search for $\wps$ boson decaying into vector-like quark}
\label{prod_wp}
\noindent
The first search for heavy vector charged boson decaying to VLQ at the LHC is performed by CMS collaboration in refs. \cite{cms6, cms6p}, where exclusion limits ranging between $0.01-0.43$ pb are set on the production cross section.
In this subsection, we focus on studying the production and decay of $\wps$ into a VLQ $T$ and a bottom quark, the $T$ subsequently decays into $tH$ and/or $tZ$, according to the model presented in section \ref{sec2}, where we emphasis the all-jets final state. 
We compute the inclusive cross-section and differential distributions at LO and NLO in QCD. The full NLO QCD cross-section is obtained by combining one-loop QCD corrections and Born amplitudes with UV counterterms and real emission contributions, employing the FKS subtraction framework \cite{fks1}. The calculations are fully automated through {\tt MadGraph5}, which utilizes {\tt MadLoop} and {\tt MadFKS} to compute the one-loop and real emission amplitudes, as well as the IR subtraction counterterms \cite{madloop, fks2, madgraph}.\\

\noindent
Let's investigate the 3 following benchmark models\footnote{We can take the branching ratios of $T$ in their asymptotic limit, since $m_{\scriptscriptstyle T}$ is sufficiently larger than SM particle masses. }:
\begin{align}
&{\bf T^{\scriptscriptstyle\{3\}}_{\scriptscriptstyle\{H,W'\}}:}  & \text{with}&& &\text{Br}(T\rightarrow tH)=100\% \notag\\
&{\bf T^{\scriptscriptstyle\{3\}}_{\scriptscriptstyle\{Z,W'\}}:}  & \text{with}&& &\text{Br}(T\rightarrow tZ)=100\%\notag\\
&{\bf T^{\scriptscriptstyle\{3\}}_{\scriptscriptstyle\{Z,H,W'\}}:}& \text{with}&& &\text{Br}(T\rightarrow tH)=\text{Br}(T\rightarrow tZ)=50\%
\label{scenarios}
\end{align}

\noindent
The lowest-order $\mathcal{O}(\alpha^3)$ subprocesses yielding the final states $btH$ and/or $btZ$, via virtual $\wps$ and $T$, are:
\begin{align}
q\bar q^{\prime}&\rightarrow\{W^{\prime}, T\}\rightarrow b\bar t H+\bar bt H, & q\bar q^{\prime}&\rightarrow\{W^{\prime}, T\}\rightarrow b\bar t Z+\bar bt Z. \end{align}

\begin{figure}[h!]
\subfloat[\label{qqWp1}]{\includegraphics[width=4.5cm,height=3.25cm]{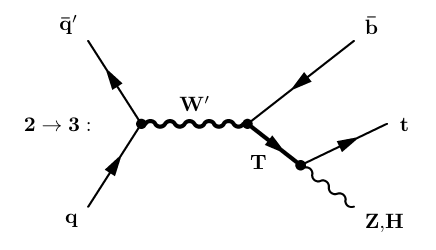}}\quad\quad
\subfloat[\label{qqWp2}]{\includegraphics[width=4.5cm,height=3.25cm]{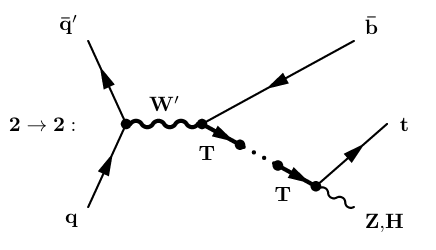}}\quad\quad
\subfloat[\label{qqWp3}]{\includegraphics[width=4.5cm,height=3.25cm]{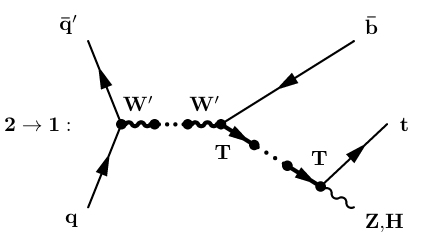}}
 \caption{\small Production and decay of $\wps$ at the LHC. In (a), the diagram illustrates the process $q\bar{q}^{\prime} \rightarrow \bar{b}tH / \bar{b}tZ$ (CM), while (b) shows the production of a single resonant $T$ decaying into $tH/tZ$ i.e. $q\bar{q}^{\prime} \rightarrow \bar{b}T$, followed by $T \rightarrow tH / tZ$ (NWA$_1$). Diagram (c) depicts the production of a resonant $W^{\prime}$, which decays into $\bar{b}$ and resonant $T$ decaying into $tH/tZ$ i.e. $q\bar{q}^{\prime} \rightarrow W^{\prime}$, then $W^{\prime} \rightarrow \bar{b}T$ with $T \rightarrow tH / tZ$ (NWA$_2$).}
 \label{pp-wp}
\end{figure}

\noindent
In these benchmark scenarios, the leading order weak production mechanism involves $W^{\prime}$ boson exchange in the $s$-channel, see the tree-level Feynman diagrams depicted in sub-figure \ref{qqWp1}. We select 3 different values for $m_{_T}$ for each fixed $m_{_{W^{\prime}}}$ which are: $m_{\scriptscriptstyle{T}}= 1/2\,  m_{\scriptscriptstyle{W^{\prime}}}, 2/3\, m_{\scriptscriptstyle{W^{\prime}}}, 3/4\, m_{\scriptscriptstyle{W^{\prime}}}$, as adopted in refs. \cite{cms6, cms6p}.
It's important to note that the production cross-section is primary influenced by the free parameters $\kpt$, $\vkwp$ (or equivalently by $\kpwp$), $\xi_{_{Tb}}^{_{W^{\prime}}}$, and the masses of $\wps$ and $T$ within the specified benchmark scenarios. The parameter $\kpt$ and $m_{\scriptscriptstyle T}$ play a marginal role when the NWA and the CM schemes give approximately the same predictions, as we will show later on. The graphs in sub-figures \ref{qqWp2} and \ref{qqWp3} depict the same reaction, with the difference that they show cases where the $T$ and/or $\wps$ particles are on-shell (according to NWA$_1$ and NWA$_2$). Specifically, this occurs when their total width-to-mass ratio is relatively small (below a few percent), allowing for the application of the NWA, if other conditions are satisfied, see appendix~\ref{appC}.

\noindent 
Before performing the LO and NLO calculation of the production cross sections and confronting our predictions to experimental data and since we want to compare the NWA and CM schemes, let's take a look on the total width over mass of the heavy particles $\wps$ and $T$ (denoted respectively $\Gamma_{\scriptscriptstyle{W^{\prime}}}^{\scriptscriptstyle{\text{ToT}}}/m_{\scriptscriptstyle W^{\prime}}$ and $\Gamma_{\scriptscriptstyle{T}}^{\scriptscriptstyle{\text{ToT}}}/m_{\scriptscriptstyle T}$). They can be expressed as the following:  
\begin{align}
 \frac{\Gamma_{\scriptscriptstyle{W^{\prime}}}^{\scriptscriptstyle{\text{ToT}}}}{m_{\scriptscriptstyle W^{\prime}}}&=\frac{\Gamma[W^{\prime}\rightarrow Tb]/\xi_{\scriptscriptstyle{Tb}}^{\scriptscriptstyle{W^{\prime}}}}{m_{\scriptscriptstyle W^{\prime}}}=\frac{g^2}{4\, \pi}\frac{\vkwp^2}{1-\xi_{\scriptscriptstyle{Tb}}^{\scriptscriptstyle{W^{\prime}}}}.
 &
 \frac{\Gamma_{\scriptscriptstyle{T}}^{\scriptscriptstyle{\text{ToT}}}}{m_{\scriptscriptstyle T}}&=\frac{\Gamma[T\rightarrow tH/tZ]/\xi_{\scriptscriptstyle{Z/H}}}{m_{\scriptscriptstyle T}}=\frac{\kpt^2\, g^2\, m_{\scriptscriptstyle T}^2}{64\, \pi\, m_{\scriptscriptstyle W}^2}
 \label{GamTpOvM}
\end{align}
We observe that $\Gamma_{\scriptscriptstyle{W^{\prime}}}^{\scriptscriptstyle{\text{ToT}}}/m_{\scriptscriptstyle W^{\prime}}$ is solely determined by $\vkwp$ and $\xi_{\scriptscriptstyle{Tb}}^{\scriptscriptstyle{W^{\prime}}}$, independent of the hypothetical particle masses, due to our parametrization. Regarding the ratio $\Gamma_{\scriptscriptstyle{T}}^{\scriptscriptstyle{\text{ToT}}}/m_{\scriptscriptstyle T}$, we find that it remains the same in the 3 benchmark scenarios considered above, as the branching ratios ($\xi_{\scriptstyle Z}$ and $\xi_{\scriptstyle H}$) sum to unity, cf. eq.~(\ref{scenarios}), but it does depend on the mass of $T$ and the parameter $\kpt$.
\begin{figure}[h!]
\begin{center}
\subfloat[\label{width1}]{\includegraphics[width=6.0cm,height=4.0cm]{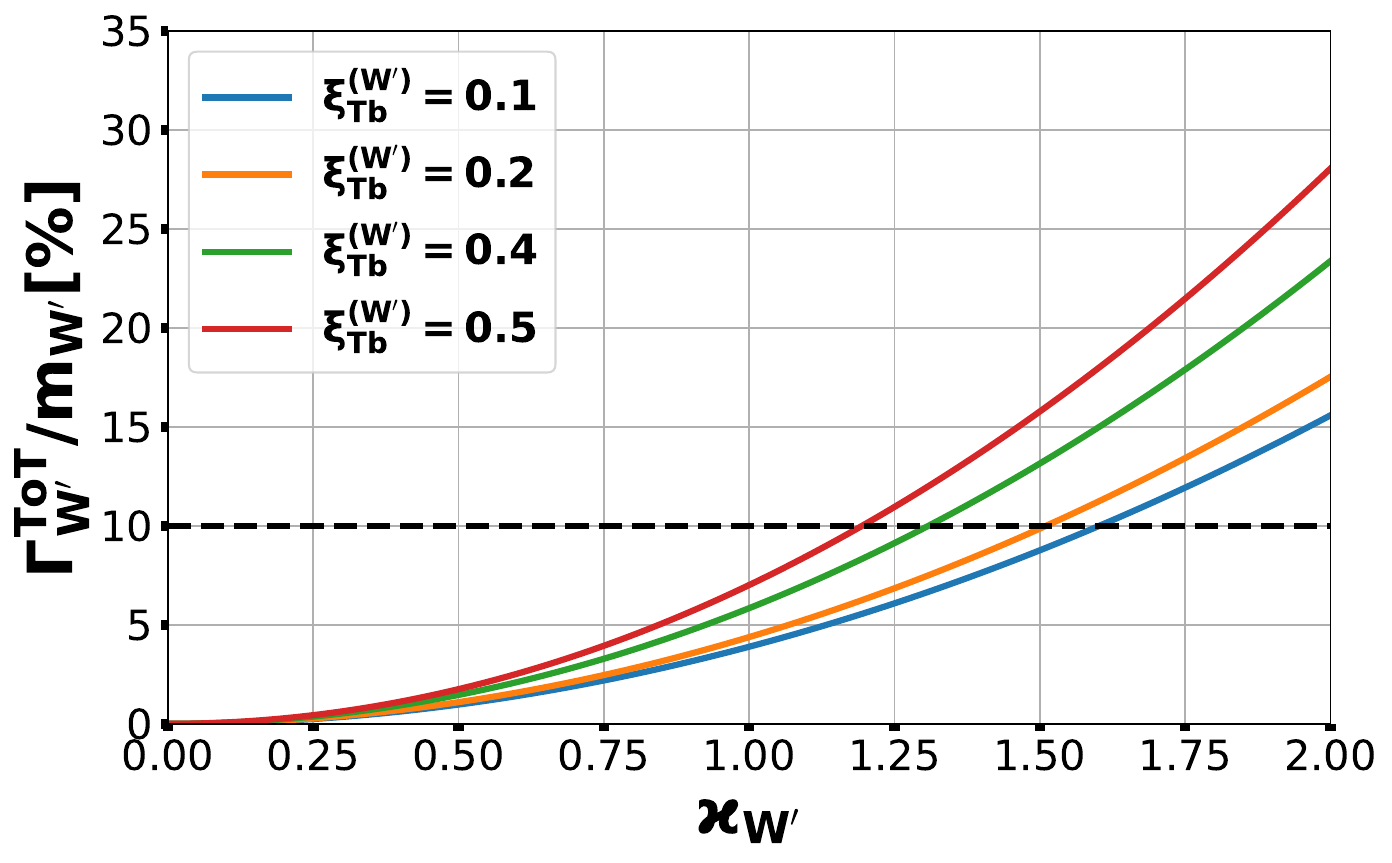}}
\subfloat[\label{width2}]{\includegraphics[width=6.0cm,height=4.0cm]{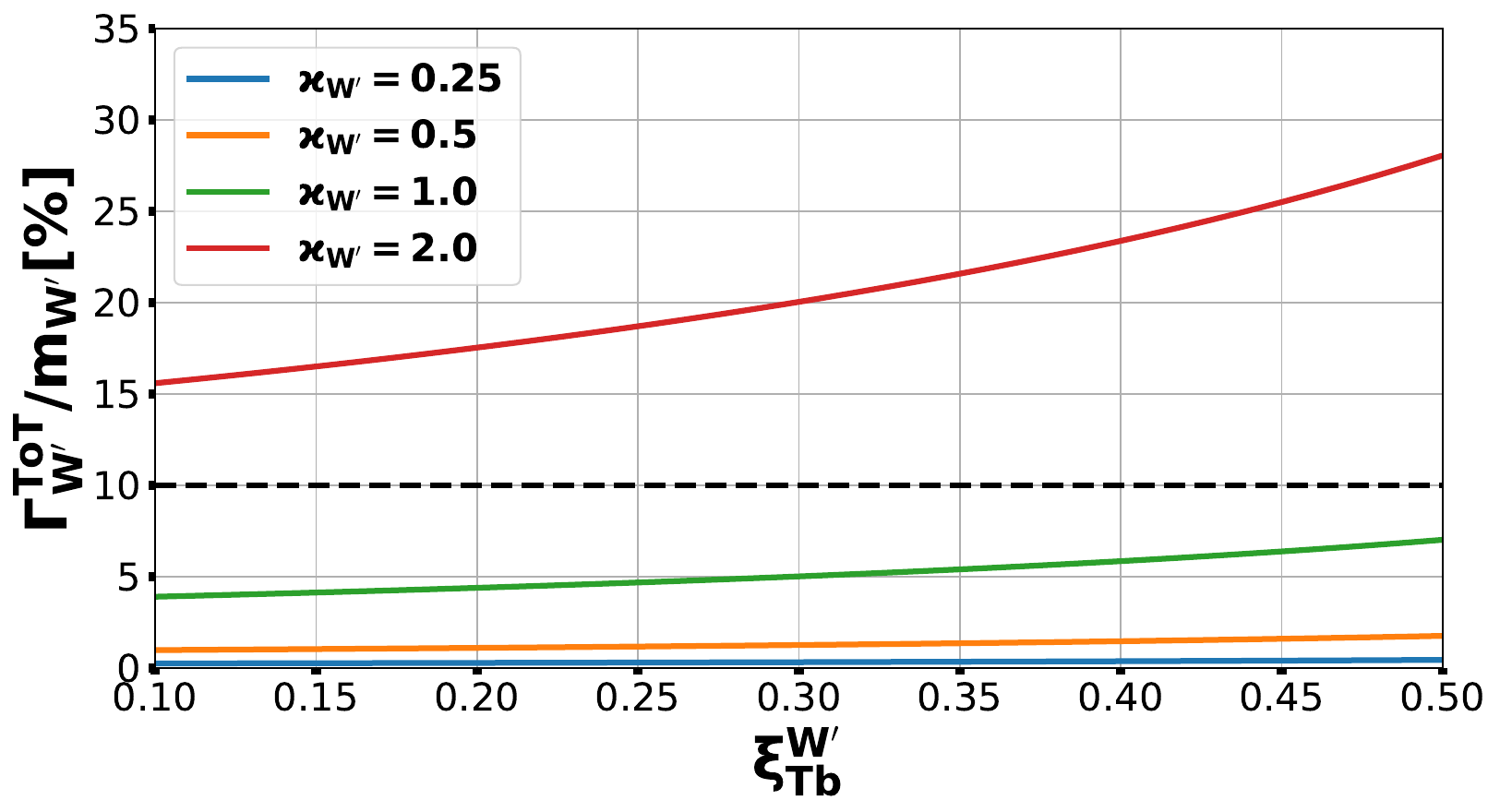}}
\subfloat[\label{width3}]{\includegraphics[width=6.0cm,height=4.0cm]{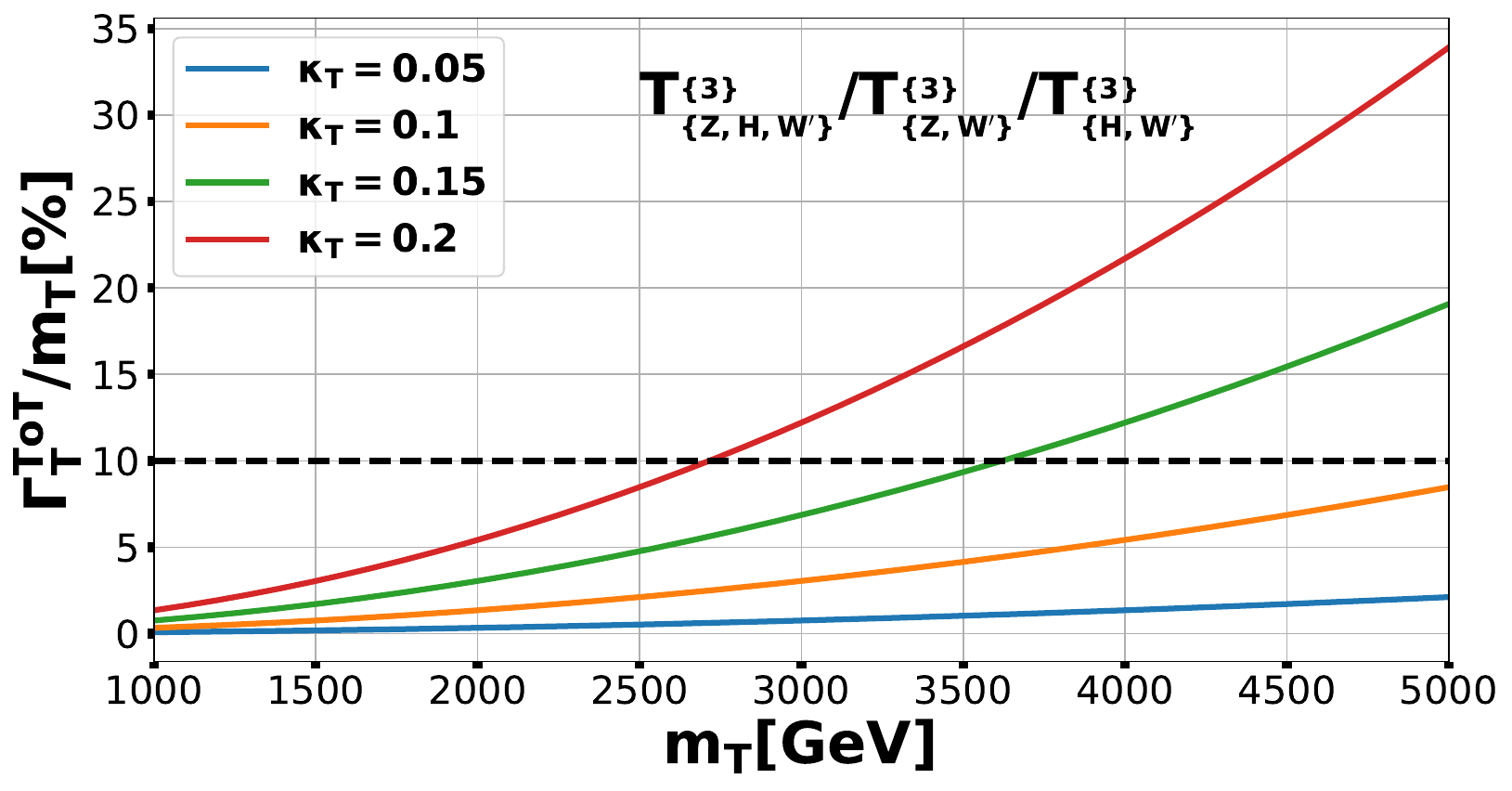}}
\end{center}
\vspace{-0.75cm}
\caption{\small Variation of the total width over mass ratio of $\wps$ and $T$ in terms of $\vkwp$, $\xi_{\scriptscriptstyle{Tb}}^{\scriptscriptstyle{W^{\prime}}}$ and $m_{_T}$ (and $\kpt$).}
\label{WidOverMass}
\end{figure}

\noindent
In figure~\ref{WidOverMass}, we show the variation of the total width over mass of both $\wps$ and $T$ (i.e. $\Gamma_{\scriptscriptstyle W^{\prime}}^{\scriptscriptstyle \text{ToT}}/m_{\scriptscriptstyle W^{\prime}}$ cf. sub-figures~\ref{width1},~\ref{width2} and $\Gamma_{\scriptscriptstyle T}^{\scriptscriptstyle \text{ToT}}/m_{\scriptscriptstyle T}$ cf. sub-figure~\ref{width3}), expressed as a percentage.
 We observe that for $\xi_{{Tb}}^{{W^{\prime}}} \leq 0.5$ and $\vkwp \leq 1$, the total width-to-mass ratio of $\wps$ stays below $10\%$ (below the black dashed line). For the particle $T$, this ratio remains always under $10\%$ for $\kpt \leq 0.1$ and for LHC-accessible masses, which can be significantly large. This indicates that the total widths over mass for both $\wps$ and $T$ are narrow in these configurations. This allows the use of both the NWA and CM scheme (which is validated for small widths). 
 
\begin{table*}[h!]
\centering
 \renewcommand{\arraystretch}{1.40}
 \setlength{\tabcolsep}{12.5pt}
 \begin{adjustbox}{width=15cm,height=1.00cm}
 \boldmath
 \begin{tabular}{!{\vrule width 1.5pt}l!{\vrule width 1.5pt}l!{\vrule width 1pt}l!{\vrule width 1pt}l!{\vrule width 1.5pt}l!{\vrule width 1pt}l!{\vrule width 1pt}l!{\vrule width 1.5pt}l!{\vrule width 1pt}l!{\vrule width 1pt}l!{\vrule width 1.5pt}l!{\vrule width 1pt}l!{\vrule width 1.5pt}}
 \cline{2-10}
\multicolumn{1}{c!{\vrule width 1.5pt}}{{}}
&\multicolumn{3}{c!{\vrule width 1.5pt}}{{$\bf \vkwp=0.5$}}
&\multicolumn{3}{c!{\vrule width 1.5pt}}{{$\bf \vkwp=1.0$}}
&\multicolumn{3}{c!{\vrule width 1.5pt}}{{$\bf \vkwp=2.0$}}\\
\noalign{\hrule height 1pt}
$\bf \xi^{\scriptscriptstyle{W^{\prime}}}_{\scriptscriptstyle{Tb}}$&\bf 0.10 &\bf  0.20 &\bf  0.40 &\bf 0.10 &\bf 0.20 &\bf 0.40 &\bf 0.10 &\bf 0.20 &\bf 0.40 \\
\cdashline{1-10}
$\bf \kpwp$&\bf 1.05&\bf 1.12& \bf 1.29& \bf 2.11 &\bf 2.24 &\bf 2.58&\bf 4.22 &\bf 4.47 & \bf 5.16\\
\cdashline{1-10}
$\bf \frac{\Gamma_{\scriptscriptstyle{W^{\prime}}}^{\scriptscriptstyle{\text{ToT}}}}{m_{\scriptscriptstyle W^{\prime}}}[\%]$&\bf 0.97 &\bf 1.09 &\bf 1.46 &\bf 3.89 &\bf 4.38 &\bf 5.84 &\bf 15.58 &\bf 17.52 &\bf  23.37 \\
\noalign{\hrule height 1pt}
\end{tabular}
\end{adjustbox}
  \caption{\footnotesize Numerical values of $\kpwp$ for $\vkwp=0.5, 1, 2$ and $\xi^{\scriptscriptstyle{W^{\prime}}}_{\scriptscriptstyle{Tb}}=0.1, 0.2, 0.4$.}
   \label{tabkwpPerturba}
  \end{table*}

\noindent 
It is very important to note that for these configurations ($\xi_{{Tb}}^{{W^{\prime}}} \leq 0.5$ and $\vkwp \leq 1$), perturbative unitarity is always preserved regardless $m_{\scriptscriptstyle T}$ and $m_{\scriptscriptstyle W^{\prime}}$. In table~\ref{tabkwpPerturba}, we show that the parameter $\kpwp$, given by $\kpwp=2\vkwp/(1-\xi_{{Tb}}^{{W^{\prime}}})^{1/2}$, is always below the upper bounds given in table~\ref{Tab2}, cf.~section~\ref{sec2}.\\

\noindent
Now, let’s focus on two scenarios: one we call the {\it Canonical Sequential SM} (abbreviated by CSSM) with $\vkwp=1$, and another we call {\it Democratic Decay} (denoted DD) with $\xi_{{Tb}}^{{W^{\prime}}}=0.2$. The latter is referred to as {\it democratic} because the $W^{\prime}$ decays equally to quarks of different types (both VLQs and ordinaries), i.e. $\xi_{{Tb}}^{{W^{\prime}}} = \xi_{{u_id_i}}^{{W^{\prime}}} = 20\%$. To improve accuracy and obtain more reliable predictions with reduced dependence on unphysical scales (factorization and renormalization scales), the calculation should be performed at NLO. To proceed, we select the following two configurations:
\begin{itemize}
\item {\bf DD scenario}: $\xi_{{Tb}}^{{W^{\prime}}}=20\%$ with $\vkwp = 0.5$ ($\Gamma_{\scriptscriptstyle{W^{\prime}}}^{\scriptscriptstyle{\text{ToT}}}/m_{\scriptscriptstyle W^{\prime}}=1.09\%$).
\item {\bf CSSM scenario} : $\vkwp=1$ with $\xi_{{Tb}}^{{W^{\prime}}} = 5\%$ ($\Gamma_{\scriptscriptstyle{W^{\prime}}}^{\scriptscriptstyle{\text{ToT}}}/m_{\scriptscriptstyle W^{\prime}}=3.69\%$). 
\end{itemize} 

\begin{figure}[h!]
\centering
\includegraphics[width=5.5cm,height=4.0cm]{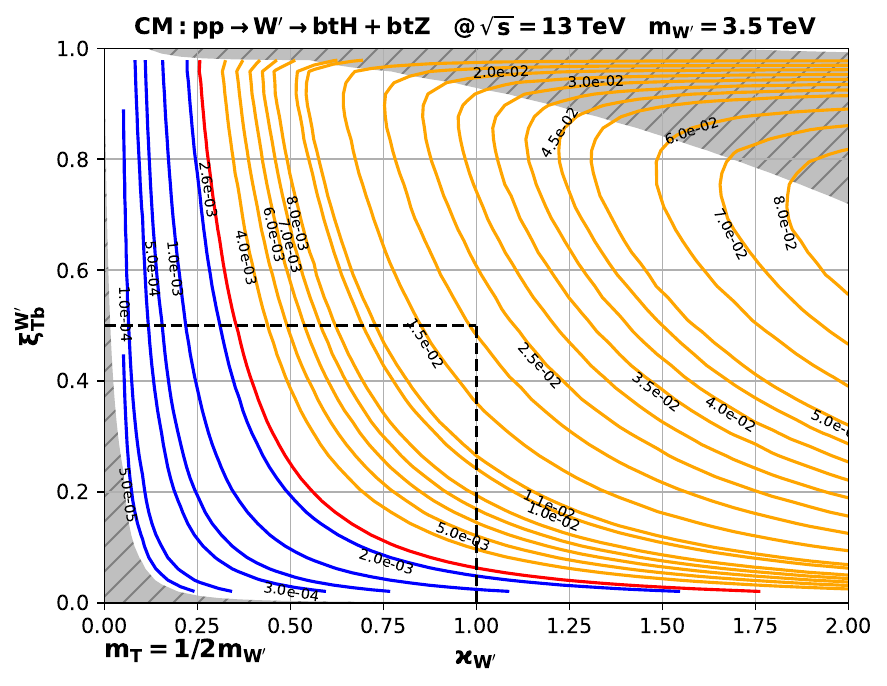}
\includegraphics[width=5.5cm,height=4.0cm]{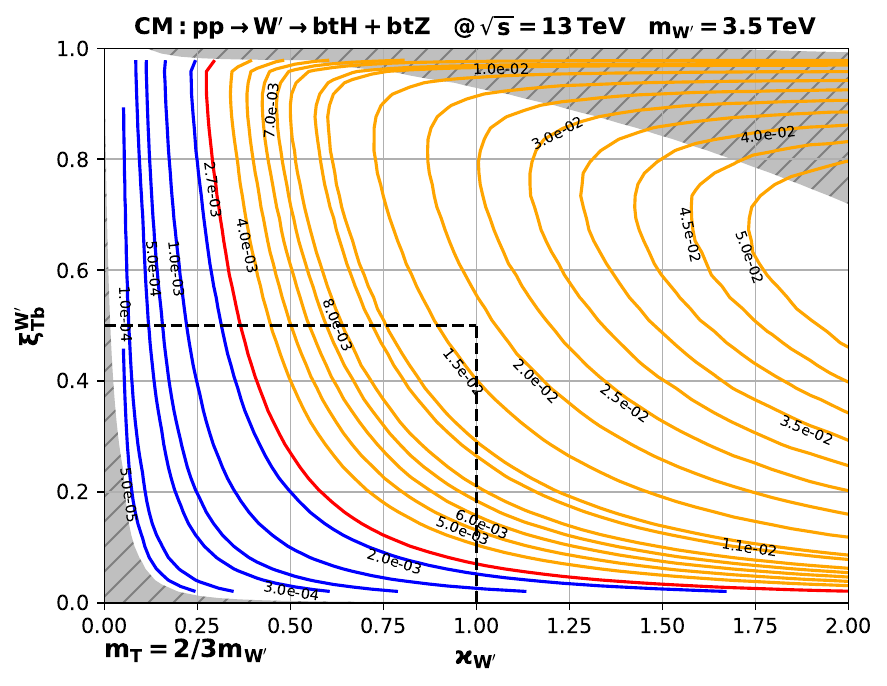}
\includegraphics[width=5.5cm,height=4.0cm]{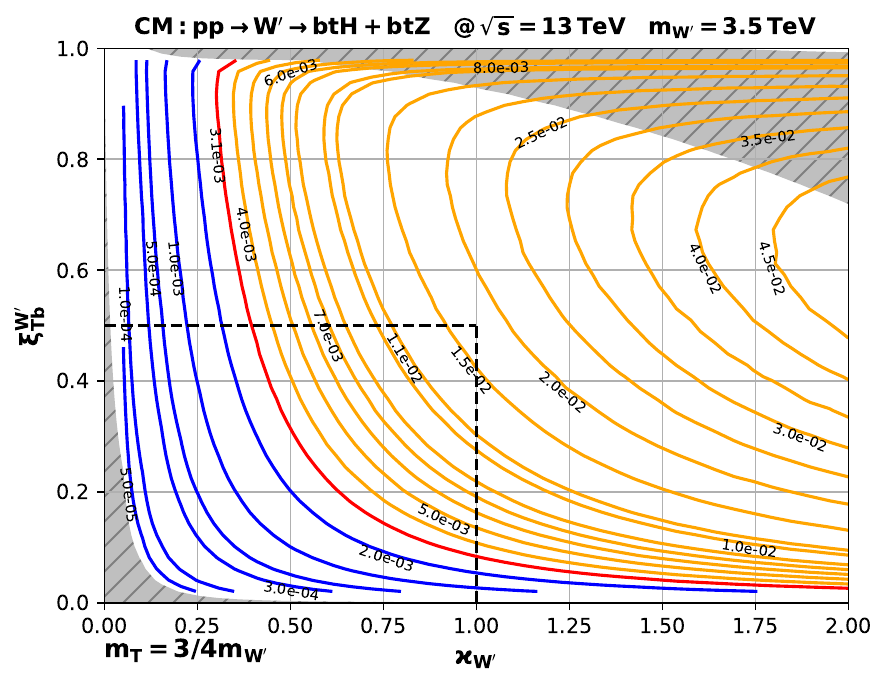}
 \caption{\footnotesize Parameter scan of the LO cross sections $\sigma^{\scriptscriptstyle pp\rightarrow tbH+tbZ}$ (CM), for $m_{\scriptscriptstyle W^{\prime}}=3.5$~TeV, $m_{\scriptscriptstyle T}=1/2, 2/3, 3/4 m_{\scriptscriptstyle W^{\prime}}$ at center-of-mass energy $\sqrt{s}=13\, \text{TeV}$. Experimental upper limits provided by the CMS collaboration (red curves), available on the HEPDATA website: \url{https://www.hepdata.net/record/ins2039384}; see also refs.~\cite{cms6,cms6pp}. The blue and orange curves represent the non-excluded and excluded predictions of the dedicated model {\tt vlQBp}, respectively.}
\label{scanXsec}
\end{figure}

\noindent
We note that in these two configurations, the cross sections are not excluded experimentally (see figure~\ref{scanXsec}). Moreover, the NLO calculations in the CM scheme are available in the model, since $\wps$ and $T$ widths are narrow. We recall that the $T$ width-to-mass ratio range between $0.19\%$ and $4.77\%$ over the considered masses.

\begin{figure}[h!]
    \raggedright  
    \hspace{0.5cm} 
    \boxed{\includegraphics[width=8.5cm,height=2.5cm]{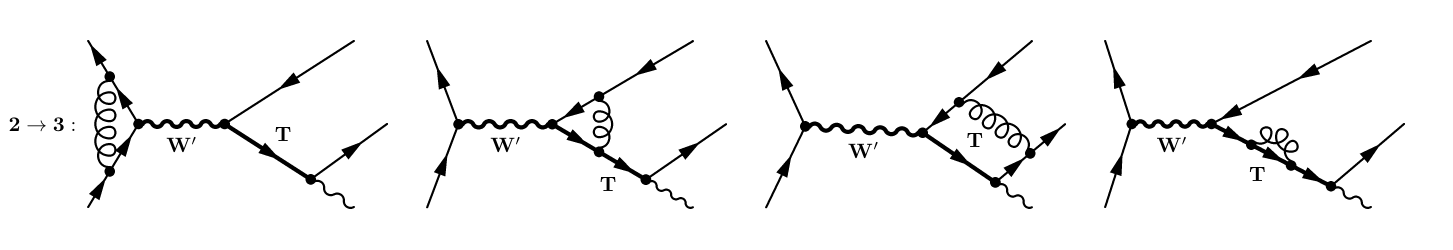}}
    \boxed{\includegraphics[width=4.75cm,height=2.5cm]{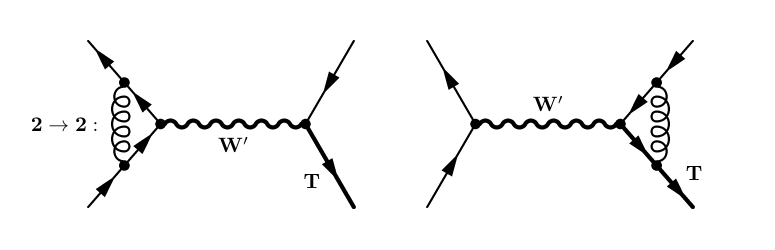}}
    \boxed{\includegraphics[width=2.25cm,height=2.5cm]{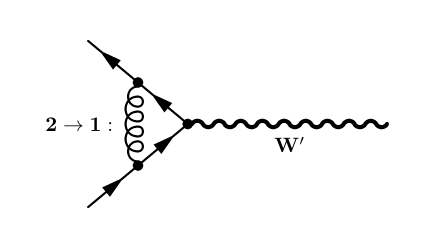}}
    \caption{\small Some virtual correction diagrams: CM (left), $\text{NWA}_{1}$ (middle) and $\text{NWA}_{2}$ (right).}
    \label{loops}
\end{figure}

\noindent
The one-loop virtual amplitude at order $\mathcal{O}(\alpha^3\alpha_s)$ is derived by combining the weak Born-level Feynman diagrams (sub-figure~\ref{qqWp1}) with their one-loop QCD corrections, which include diagrams containing one virtual gluon. These corrections consist of bubble, triangle and box diagrams with two strong vertices (the pentagons and other boxes are not included because they are color-suppressed). Some examples are shown on the left panel of figure~\ref{loops}. All necessary UV counterterms are included to suppress the ultraviolet divergences in this amplitude. {\tt MadGraph5} can manage this automatically once the renormalized model in {\tt UFO} format is implemented, which we have generated automatically using {\tt FeynRules}. \\

\noindent
The loop diagrams shown in the middle (2-to-2 process) and right (2-to-1 process) panels of figure \ref{loops} represent some virtual correction graphs to the Born diagrams in sub-figures \ref{qqWp2} and \ref{qqWp3}, respectively. These diagrams are needed to obtain the NLO predictions in NWA$_1$ and NWA$_2$ approaches. We emphasize that the branching ratios remain unchanged in this case, specifically $\xi_{\scriptscriptstyle{Z}}=\xi_{\scriptscriptstyle{H}}=0.5$ and $\xi_{{Tb}}^{{W^{\prime}}}=0.2$ and $0.05$ for the DD and CSSM scenarios, respectively. Although it might seem necessary to compute the widths of $T$ and $\wps$ at NLO, this is actually not required, as the branching ratios of these particles are treated as free parameters within our model, serving as inputs rather than quantities derived from the calculated widths, cf.~eq.(\ref{GamTpOvM}).

\begin{figure}[h!]
\raggedright
\includegraphics[width=17cm,height=2.0cm]{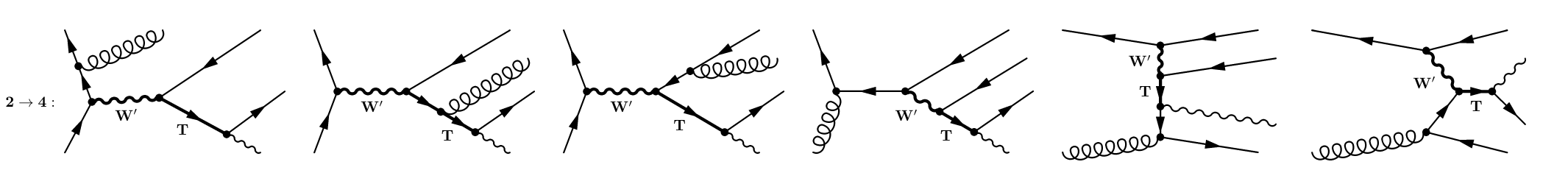}

\raggedright
\includegraphics[width=14.17cm,height=2.0cm]{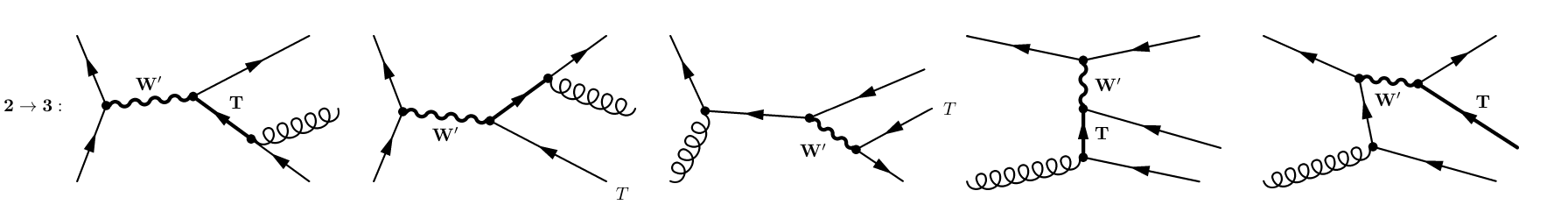}

\raggedright
\includegraphics[width=8.50cm,height=2.0cm]{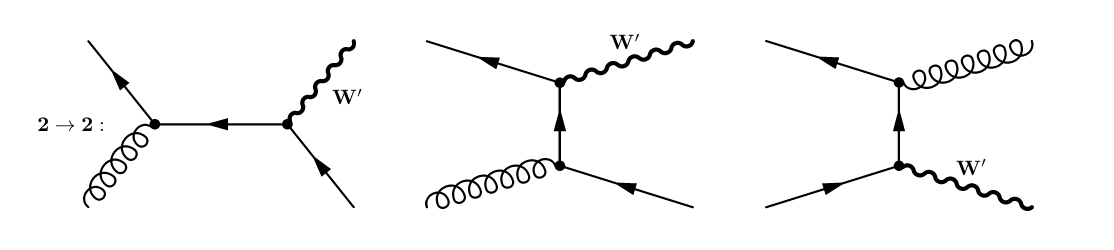}
\caption{\small Real emission diagrams: CM (up), $\text{NWA}_{1}$ (middle) and $\text{NWA}_{2}$ (down).}
\label{reals}
\end{figure}

\noindent
The interference of Born and virtual diagrams has infrared (soft/collinear) divergences. To preserve perturbative consistency, diagrams with the real emission of an additional parton (order $\mathcal{O}(\alpha^3\alpha_s)$) must be included. The real emission processes that we have to consider are: \begin{align} 
q\bar{q}^{\prime}&\rightarrow\left[b \bar t H/Z+\bar b t H/Z\right]\, g, &
qg&\rightarrow \left[b \bar t H/Z+\bar b t H/Z\right]\, q, &
\bar q^{\prime} g&\rightarrow \left[b \bar t H/Z+\bar b t H/Z\right]\, \bar q^{\prime}. 
\label{real} 
\end{align}

\noindent
 In figure~\ref{reals}, we depict some of the real emission Feynman diagrams associated with the Born process shown in figure~\ref{pp-wp}. The first process in eq.~(\ref{real}) arises from the Born diagrams with the emission of an extra gluon from the initial or final state quarks, cf. the first 3 diagrams of the first row of figure~\ref{reals}. Since both initial state partons are massless, then initial state emission can be soft and collinear divergent. The same thing is true for the emission from the final sate, except if the emitting parton is massive (i.e. $T$ or $t$), where only soft divergences are encountered. The soft divergences of the squared amplitude of these diagrams are canceled once we integrate over gluon momentum and combine it with virtual contribution, thanks to the Kinoshita-Lee-Nauenberg theorem (KLN) \cite{kln1, kln2}. As for real emission processes initiated by gluon, they must be added because they contribute to quark structure functions (PDFs) scaling violation, they might contain initial state collinear singularities. After combining virtual and real emission amplitudes, only initial states collinear singularities remain, which are factorized in the NLO parton distribution functions, thanks to the factorization theorem \cite{fact1, fact2}. Rows 2 and 3 of figure~\ref{reals} show real emission diagrams corresponding, respectively, to $2\rightarrow2$ and $2\rightarrow1$ Born diagrams in figure~\ref{pp-wp}, i.e. $\text{NWA}_{1}$ and $\text{NWA}_{2}$ approximations, respectively.\\

\begin{figure}[h!]
 \centering
\includegraphics[width=8.0cm,height=4.0cm]{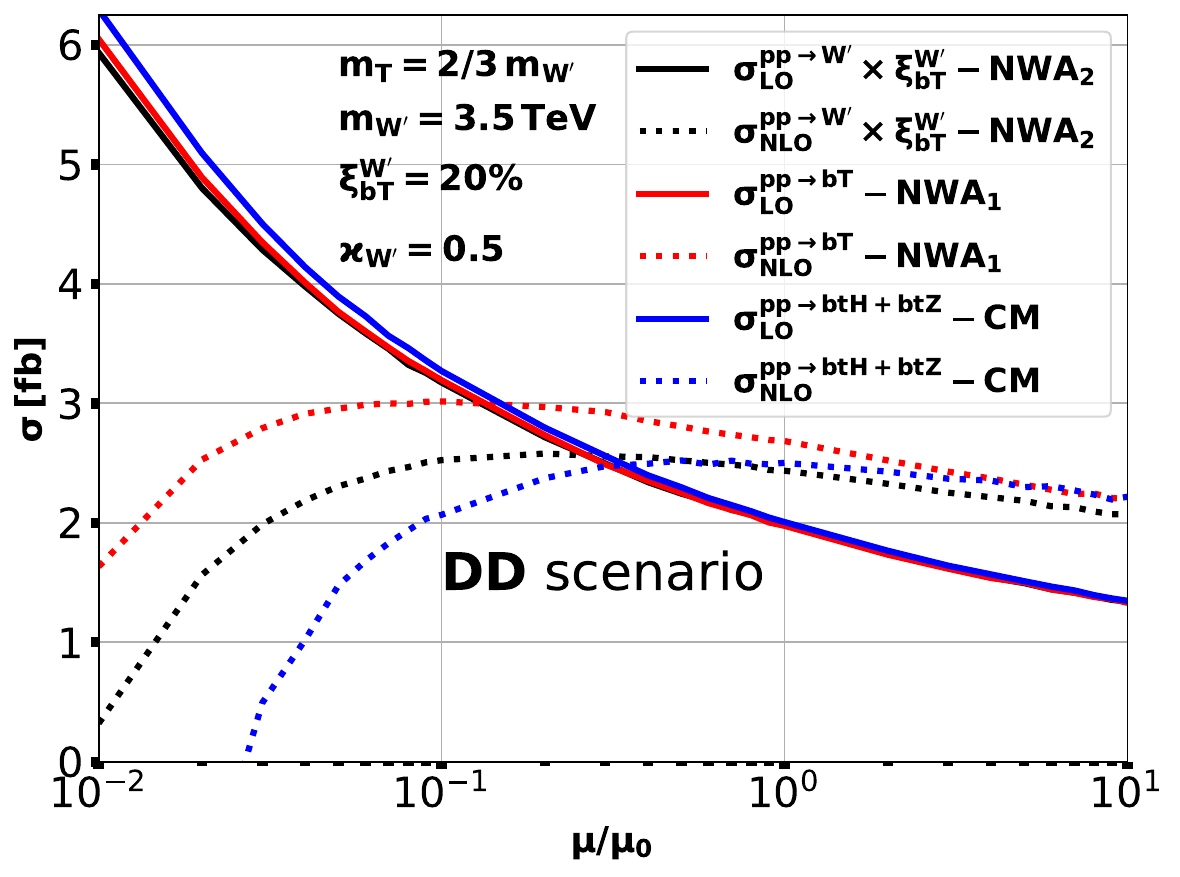}\quad\quad\quad
\includegraphics[width=8.0cm,height=4.0cm]{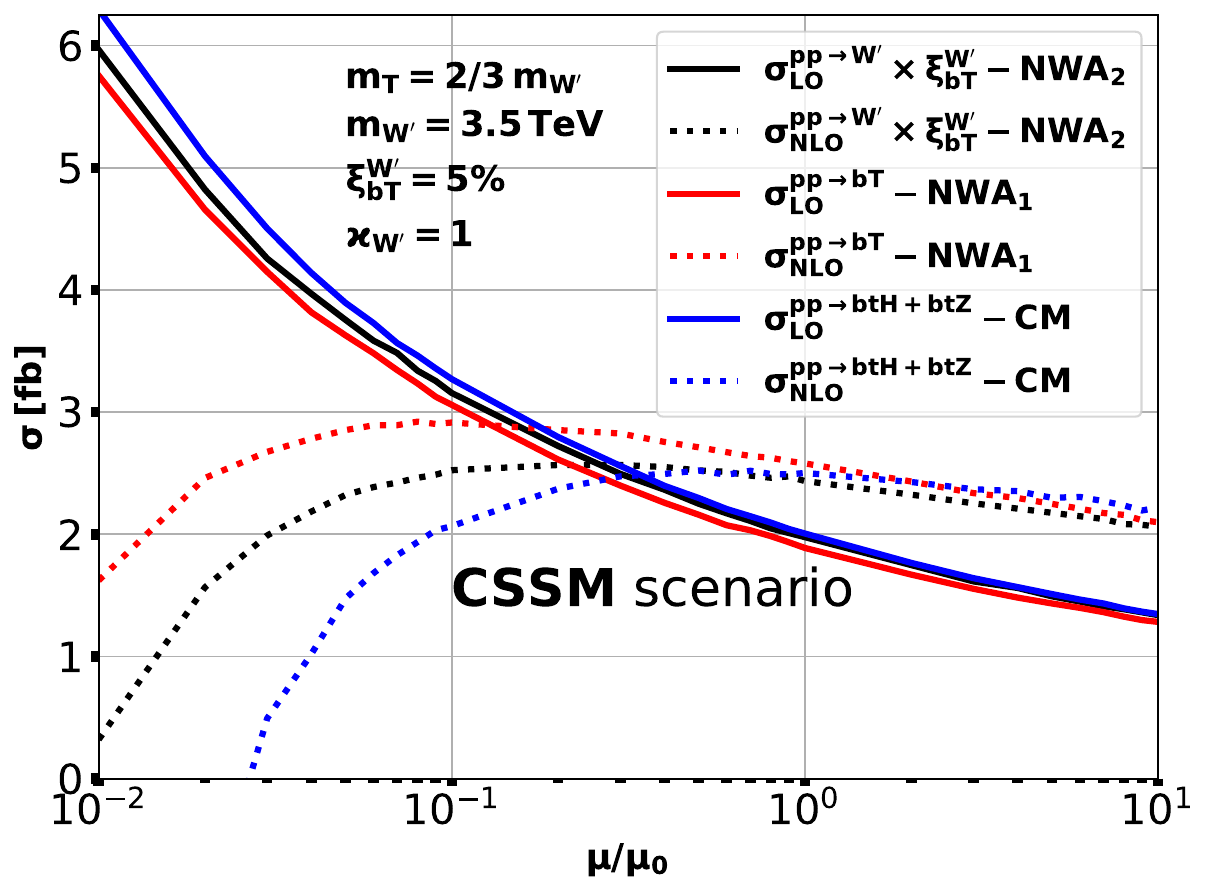}
\caption{\small Scale variation of LO and NLO cross sections in the 3 approximations: CM, $\text{NWA}_{1}$ and $\text{NWA}_{2}$.}
\label{scalVars}
\end{figure}

\noindent
As we have said above, one of the major motivations for considering NLO calculations is the reduction of the dependency on the non-physical scales of the cross section\footnote{Due the importance of the automation of the NLO and the next-to-next-leading order (NNLO) calculations \cite{Guillet:2021qmg,Guillet:2019hfo,Guillet:2018fsc,Guillet:2018skp,Guillet:2018cdm,Guillet:2013mta}}. To explicitly demonstrate this, we plot the variation of the inclusive cross sections, for $m_{\scriptscriptstyle W^{\prime}}=3/2\, m_{\scriptscriptstyle T} = 3.5\, \text{TeV}$, in the 3 approximations: CM ($\sigma_{\scriptscriptstyle\text{LO/NLO}}^{\scriptscriptstyle pp\rightarrow btH/Z}$), $\text{NWA}_{1}$ ($\sigma_{\scriptscriptstyle\text{LO/NLO}}^{\scriptscriptstyle pp\rightarrow bT}$) and $\text{NWA}_{2}$ ($\sigma_{\scriptscriptstyle\text{LO/NLO}}^{\scriptscriptstyle pp\rightarrow W^{\prime}}\times\xi_{\scriptscriptstyle Tb}^{\scriptscriptstyle W^{\prime}}$) in term of $\mu/\mu_0$, where $\mu_F=\mu_R=\mu$ and the central scale $\mu_0$ is the sum of transverse mass of final state particles divided by 2 ($\mu_0=H_T/2$), see figure.~\ref{scalVars}.

\noindent
We observe that, for the 3 approximations, the LO cross sections ($\sigma_{\scriptscriptstyle\text{LO}}^{\scriptscriptstyle pp\rightarrow btH/Z}$, $\sigma_{\scriptscriptstyle\text{LO}}^{\scriptscriptstyle pp\rightarrow bT}$ and $\sigma_{\scriptscriptstyle\text{LO}}^{\scriptscriptstyle pp\rightarrow W^{\prime}}\times\xi_{\scriptscriptstyle Tb}^{\scriptscriptstyle W^{\prime}}$) exhibit a stronger sensitivity to scale variation, behaving as monotonically decreasing functions on $\mu/\mu_0$. In contrast, the NLO cros sections ($\sigma_{\scriptscriptstyle\text{NLO}}^{\scriptscriptstyle pp\rightarrow btH/Z}$, $\sigma_{\scriptscriptstyle\text{NLO}}^{\scriptscriptstyle pp\rightarrow bT}$ and $\sigma_{\scriptscriptstyle\text{NLO}}^{\scriptscriptstyle pp\rightarrow W^{\prime}}\times\xi_{\scriptscriptstyle Tb}^{\scriptscriptstyle W^{\prime}}$) show significantly reduced dependence, particularly at higher scales. Thus, the scale dependence is extremely reduced for NWA$_1$, NWA$_2$ and CM scheme, at NLO. It should be noted, however, that for very small scales, the LO and NLO predictions become very sensitive to scale variations. In this regime, the reliability of the predictions is compromised, as the calculations enter the non-perturbative QCD domain where the perturbative framework is no longer valid.

\begin{figure}[h!]
\centering
\includegraphics[width=5.75cm,height=3.75cm]{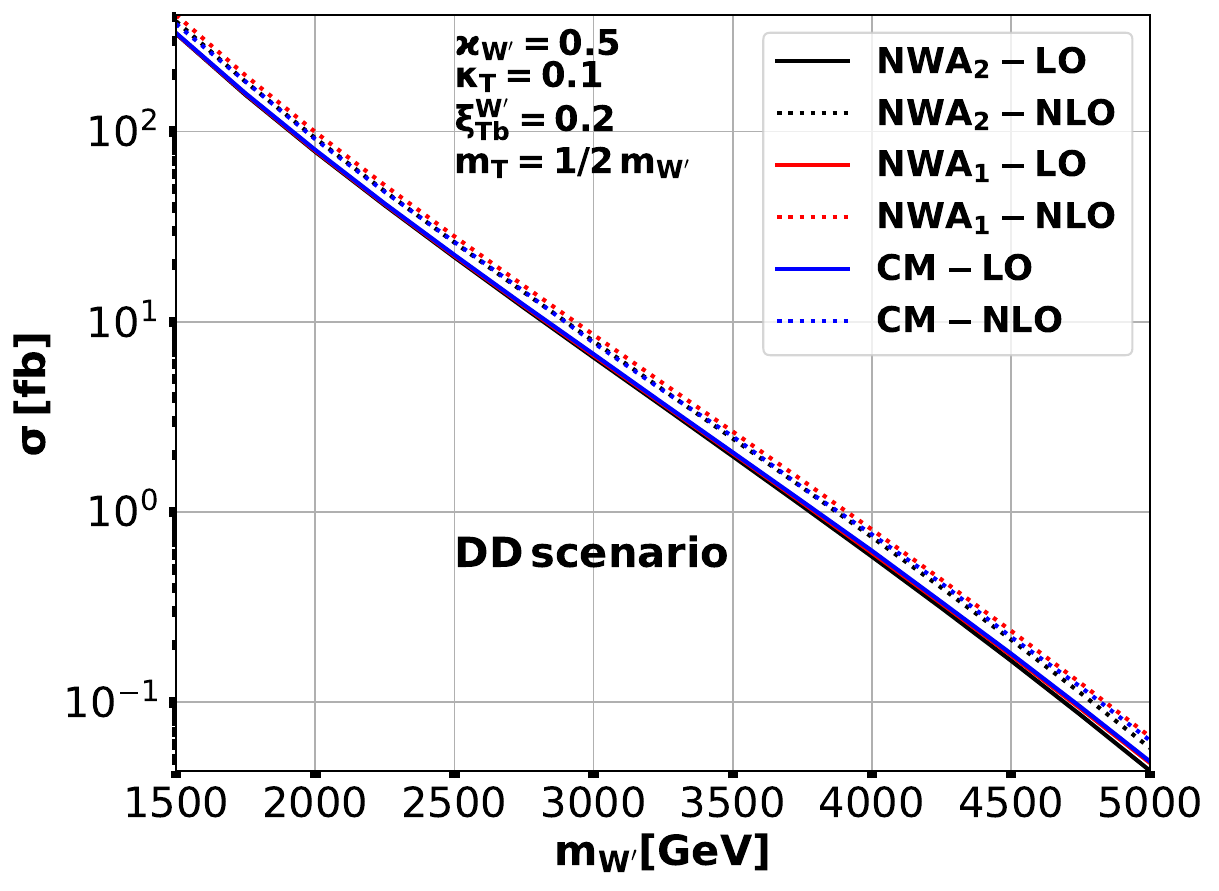}
\includegraphics[width=5.75cm,height=3.75cm]{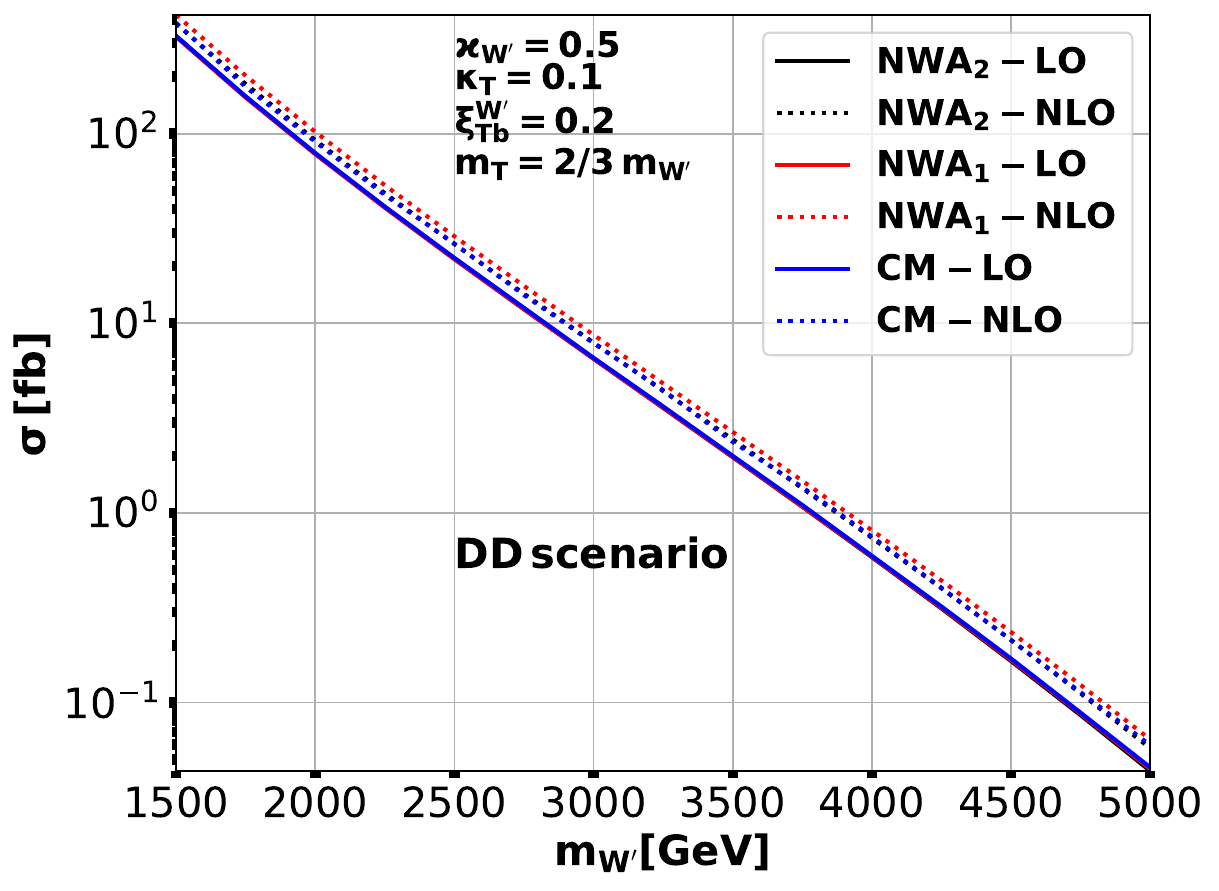}
\includegraphics[width=5.75cm,height=3.75cm]{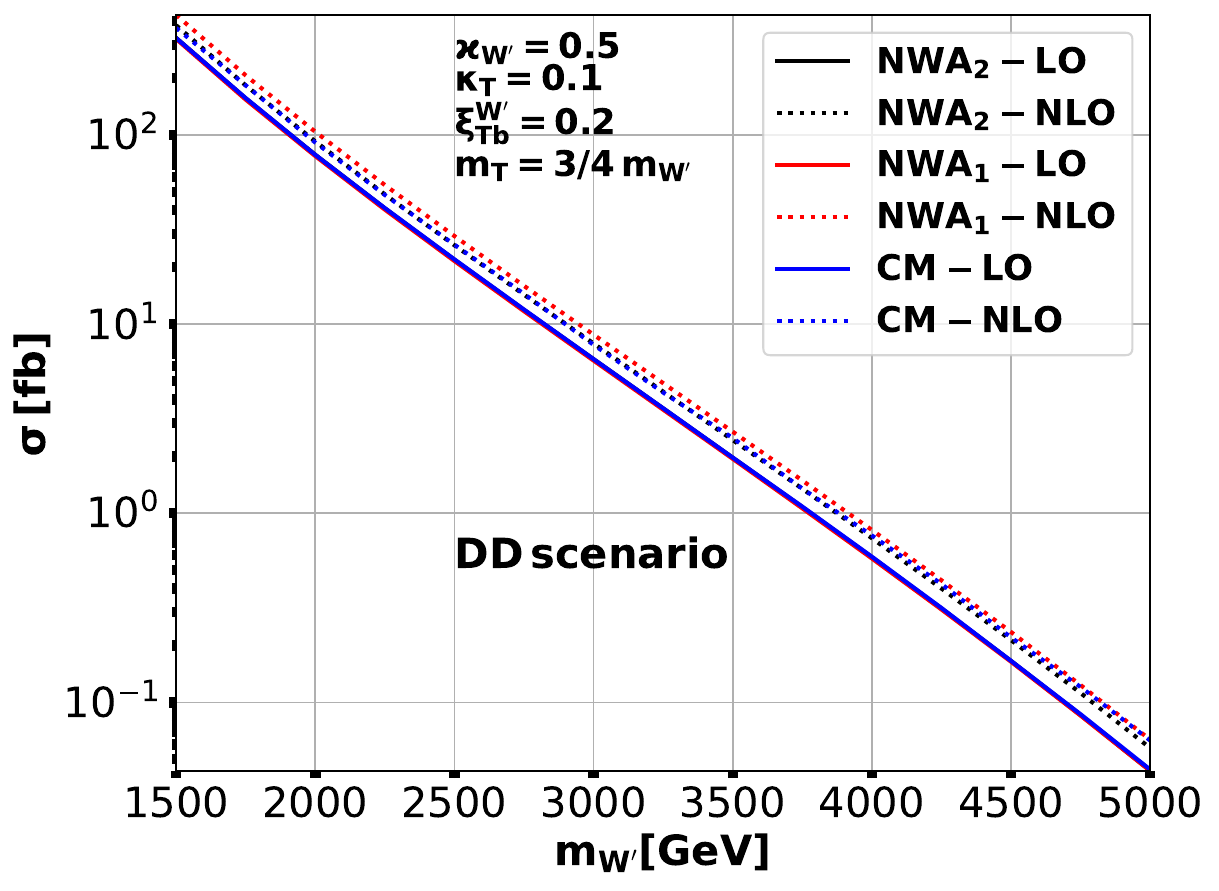}

\includegraphics[width=5.75cm,height=3.75cm]{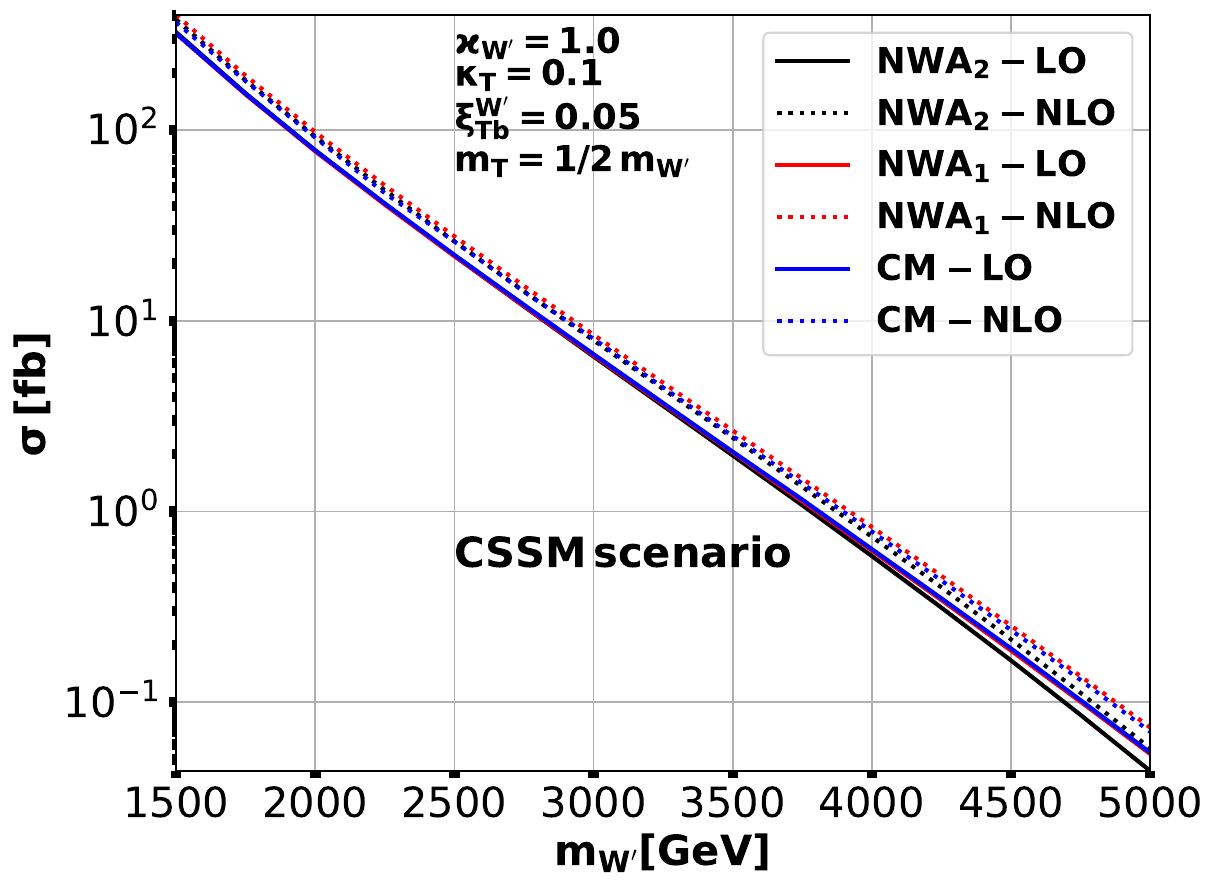}
\includegraphics[width=5.75cm,height=3.75cm]{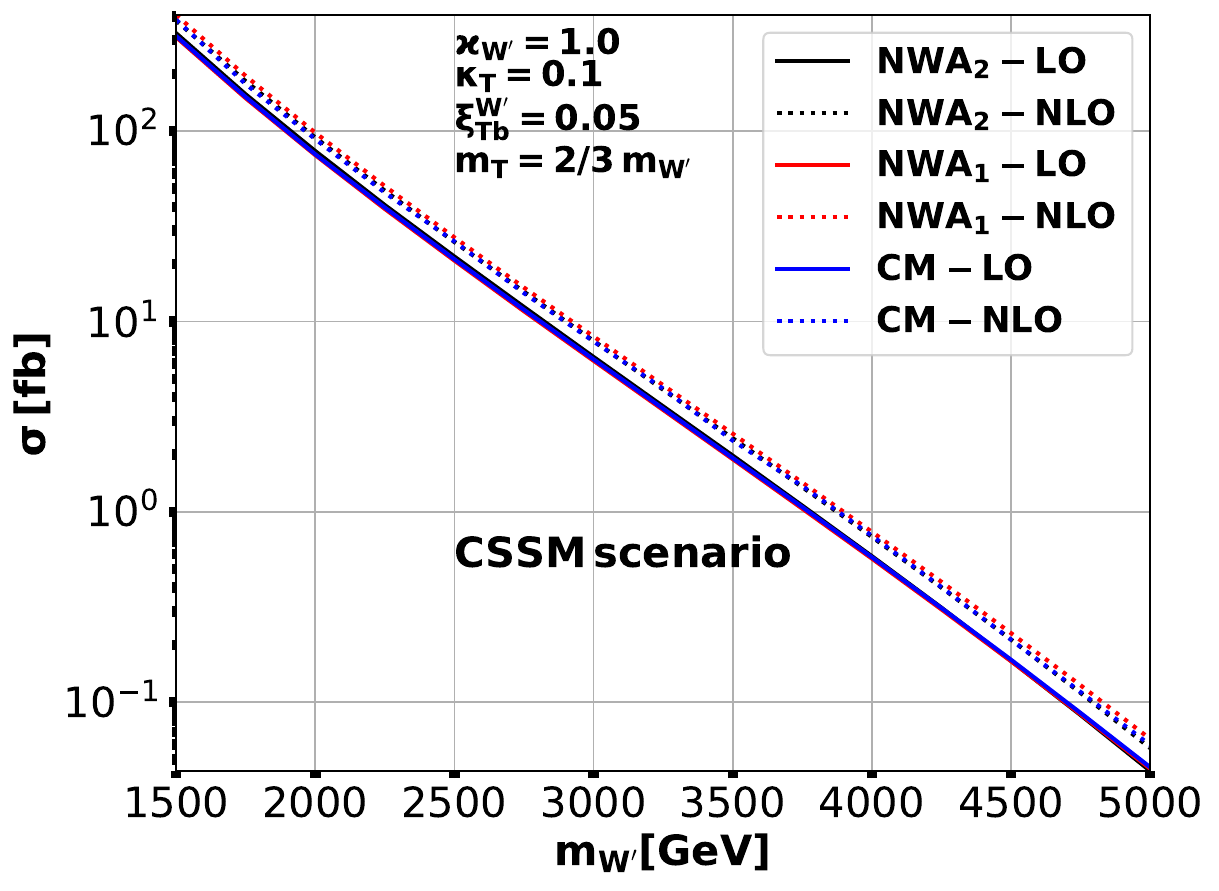}
\includegraphics[width=5.75cm,height=3.75cm]{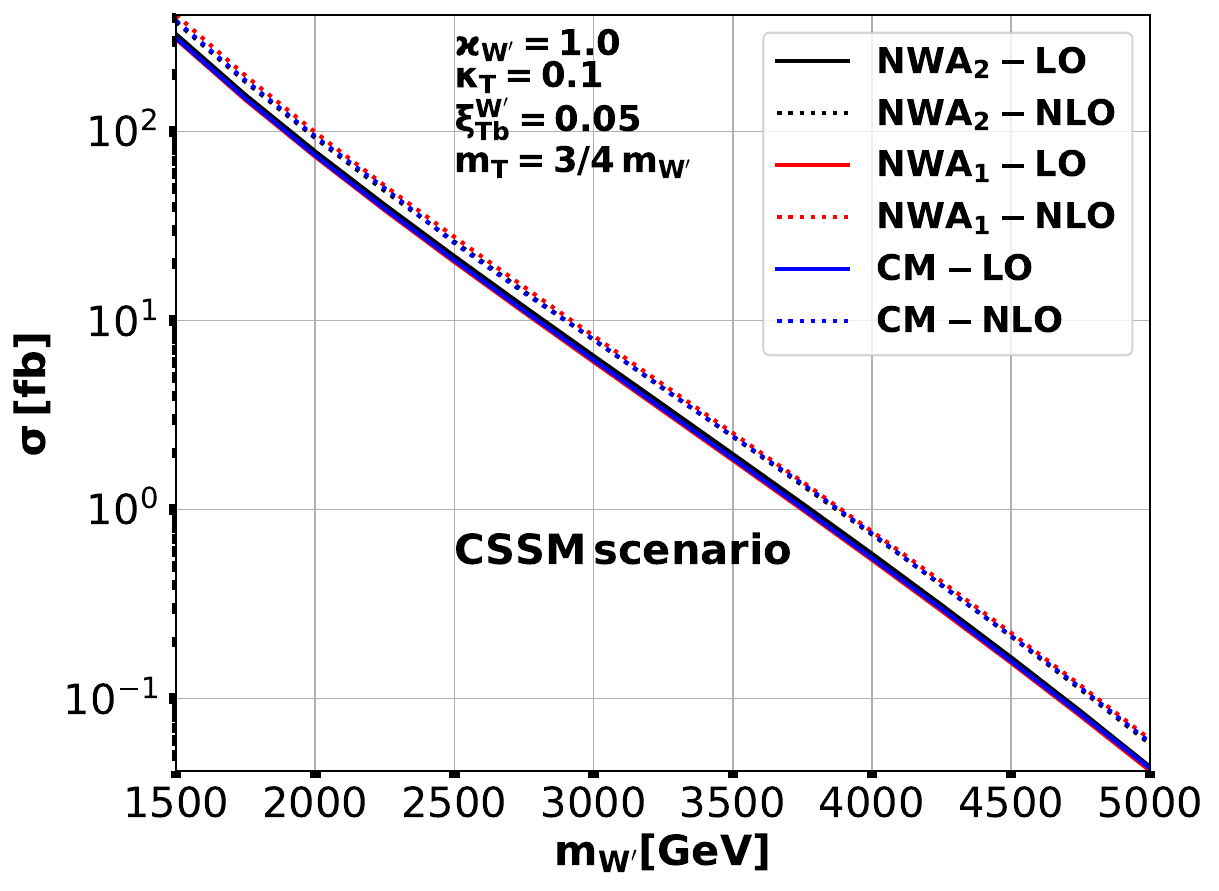}
 \caption{\small Variation of LO and NLO cross sections in term of $m_{\scriptscriptstyle W^{\prime}}$ for DD scenario (up) and CSSM scenario (down).}
\label{mWpVarsXsec}
\end{figure}

\vspace{0.25cm}
\noindent
In figure~\ref{mWpVarsXsec}, we illustrate the variation of the LO and NLO cross sections in CM, $\text{NWA}_{1}$ and $\text{NWA}_{2}$ approaches as functions of $m_{W^{\prime}}$ for $m_{_T}=\frac{1}{2}$, $\frac{2}{3}$ and $\frac{3}{4}\, m_{\scriptscriptstyle W^{\prime}}$ in the  DD and CSSM scenarios. The first observation that draws our attention is that all the LO and NLO cross sections behave similarly (all increase with $m_{\scriptscriptstyle W^{\prime}}$\footnote{They decrease when the masses increase because the phase space becomes more restricted.}) and are roughly comparable for the 3 employed approximations (CM, NWA$_1$ and NWA$_2$) and the 3 adopted $m_{\scriptscriptstyle T}$ hierarchies. To examine this more closely, we provide some numerical values for those cross sections, in DD scenario, associated with their scale and PDF uncertainties in table~\ref{tabVp2Tev}. We mention that the numerical values of the cross section are displayed in the format $\sigma\, _{-\text{scal}\%}^{+\text{scal}\%}\, _{-\text{pdf}\%}^{+\text{pdf}\%}$, where $\pm \text{scal}\%$ and $\pm \text{pdf}\%$ represent the scale and PDF uncertainties in percentage, respectively. 

\noindent
From this table, we find that the scale uncertainties are extremely reduced in NLO compared to LO, confirming the conclusion drawn from figure~\ref{scalVars}. Regarding the PDF uncertainties, they are almost the same for LO and NLO, however, they increase remarkably for higher $\wps$ masses (from $5\%$ for $m_{\scriptscriptstyle W^{\prime}}=2.5$ TeV to $31\%$  $m_{\scriptscriptstyle W^{\prime}}=4.0$ TeV). This is due to the lack of PDF constraints at high energy scales. The predictions of the NWA$_2$ are completely independent of $m_{\scriptscriptstyle T}$, which is expected, as the cross section in this case is given by $\sigma_{\scriptscriptstyle\text{LO/NLO}}^{\scriptscriptstyle pp\rightarrow W^{\prime}}\times\xi_{\scriptscriptstyle Tb}^{\scriptscriptstyle W^{\prime}}$ and $\xi_{\scriptscriptstyle Tb}^{\scriptscriptstyle W^{\prime}}$ is treated as a free parameter of the model. The LO predictions from CM scheme, NWA$_1$ and NWA$_2$ are nearly identical, with relative difference not exceeding $2\%$ in all the cases. However, the NLO predictions for NWA$_1$ and NWA$_2$ are not as close to CM scheme predictions in all configurations. The largest relative differences are rerecorded for the high-mass case ($m_{\scriptscriptstyle T}=3/4\, m_{\scriptscriptstyle W^{\prime}}$), about $8\%$ for NWA$_1$ and $3\%$ for NWA$_2$, while the smallest differences are recorded for the low-mass case ($m_{\scriptscriptstyle T}=1/2\, m_{\scriptscriptstyle W^{\prime}}$), about $4.75\%$ for NWA$_1$ and $2.75\%$ for NWA$_2$. This is due to the fact that in the former case, we are closer to the decay threshold than in the latter case, which makes the off-shell effect more important, see appendix~\ref{appC} for more details. 

\begin{table*}[h!]
\boldmath
\centering
 \renewcommand{\arraystretch}{1.40}
 \setlength{\tabcolsep}{10pt}
 \begin{adjustbox}{width=18cm,height=3.0cm}
 \begin{tabular}
 {!{\vrule width 2pt}l!{\vrule width 2pt}l!{\vrule width 2pt}c:c!{\vrule width 2pt}c:c!{\vrule width 2pt}c:c!{\vrule width 2pt}c!{\vrule width 2pt}}
  \cline{1-8}
$\bf m_{\scriptscriptstyle W^{\prime}}$& $\frac{\Gamma_{\scriptscriptstyle T}^{\scriptscriptstyle\text{ToT}}}{m_{\scriptscriptstyle T}}$ &\multicolumn{2}{c!{\vrule width 2pt}}{{\bf CM}}
&\multicolumn{2}{c!{\vrule width 2pt}}{{\bf NWA$_1$}}
&\multicolumn{2}{c!{\vrule width 2pt}}{{\bf NWA$_2$}}&\multicolumn{1}{c}{{}}\\
\cline{3-9}
 \textbf{[TeV]}& $[\%]$ &$\bf\sigma_{\scriptscriptstyle{\textbf{LO}}}[\textbf{fb}]$& $\bf\sigma_{\scriptscriptstyle{\textbf{NLO}}}[\textbf{fb}]$&$\bf\sigma_{\scriptscriptstyle{\textbf{LO}}}[\textbf{fb}]$& $\bf\sigma_{\scriptscriptstyle{\textbf{NLO}}}[\textbf{fb}]$&$\bf\sigma_{\scriptscriptstyle{\textbf{LO}}}[\textbf{fb}]$& $\bf\sigma_{\scriptscriptstyle{\textbf{NLO}}}[\textbf{fb}]$& $\frac{m_{\scriptscriptstyle T}}{m_{\scriptscriptstyle W^{\prime}}}$\\
  \noalign{\hrule height 1pt}
  $2.5$ & $0.53$& $22.52\, ^{+11.1\%}_{-9.4\%}\, ^{+5.2\%}_{-5.2\%}$ & $26.84\, ^{+2.7\%}_{-3.3\%}\, ^{+5.1\%}_{-5.1\%}$
  & $22.25\, ^{+11.0\%}_{-9.3\%}\, ^{+5.3\%}_{-5.3\%}$ & $28.15\, ^{+3.8\%}_{-4.0\%}\, ^{+5.2\%}_{-5.2\%}$
  & $21.88\, ^{+10.8\%}_{-9.2\%}\, ^{+5.3\%}_{-5.3\%}$ & $26.10\, ^{+2.7\%}_{-3.2\%}\, ^{+5.2\%}_{-5.2\%}$&\\
\cdashline{2-8}
   $3.5$& $1.04$ & $2.073\, ^{+14.5\%}_{-11.9\%}\, ^{+16.2\%}_{-16.2\%}$ & $2.520\, ^{+3.3\%}_{-4.3\%}\, ^{+15.5\%}_{-15.5\%}$
   & $2.031\, ^{+14.3\%}_{-11.8\%}\, ^{+16.3\%}_{-16.3\%}$ & $2.648\, ^{+4.6\%}_{-5.2\%}\, ^{+15.3\%}_{-15.3\%}$
   & $1.971\, ^{+14.1\%}_{-11.7\%}\, ^{+16.7\%}_{-16.7\%}$ & $2.434\, ^{+3.6\%}_{-4.4\%}\, ^{+15.7\%}_{-15.7\%}$&$\frac{1}{2}$\\
\cdashline{2-8}
  $4.0$& $1.36$ & $0.624\, ^{+16.0\%}_{-13.0\%}\, ^{+33.0\%}_{-33.0\%}$ & $0.775\, ^{+4.0\%}_{-5.0\%}\, ^{+30.5\%}_{-30.5\%}$
  & $0.611\, ^{+15.8\%}_{-12.8\%}\, ^{+33.1\%}_{-33.1\%}$ & $0.809\, ^{+5.2\%}_{-5.8\%}\, ^{+30.6\%}_{-30.6\%}$
  & $0.584\, ^{+15.7\%}_{-12.8\%}\, ^{+34.4\%}_{-34.4\%}$ & $0.738\, ^{+4.1\%}_{-5.0\%}\, ^{+31.5\%}_{-31.5\%}$&\\
 \noalign{\hrule height 1pt}
  $2.5$&$0.94$ & $22.07\, ^{+11.1\%}_{-9.4\%}\, ^{+5.2\%}_{-5.2\%}$ & $26.68\, ^{+3.0\%}_{-3.5\%}\, ^{+5.1\%}_{-5.1\%}$
  & $21.75\, ^{+10.9\%}_{-9.3\%}\, ^{+5.3\%}_{-5.3\%}$ & $28.77\, ^{+4.4\%}_{-4.5\%}\, ^{+5.2\%}_{-5.2\%}$
  & $21.84\, ^{+10.8\%}_{-9.2\%}\, ^{+5.3\%}_{-5.3\%}$ & $26.14\, ^{+2.7\%}_{-3.3\%}\, ^{+5.2\%}_{-5.2\%}$ &\\
\cdashline{2-8}
    $3.5$& $1.85$ & $1.999\, ^{+14.5\%}_{-11.9\%}\, ^{+16.6\%}_{-16.6\%}$ & $2.501\, ^{+3.9\%}_{-4.7\%}\, ^{+15.1\%}_{-15.1\%}$
    & $1.971\, ^{+14.2\%}_{-11.7\%}\, ^{+16.6\%}_{-16.6\%}$ & $2.672\, ^{+5.3\%}_{-5.6\%}\, ^{+15.8\%}_{-15.8\%}$
    & $1.971\, ^{+14.1\%}_{-11.7\%}\, ^{+16.8\%}_{-16.8\%}$ & $2.436\, ^{+3.6\%}_{-4.4\%}\, ^{+15.6\%}_{-15.6\%}$ &$\frac{2}{3}$\\
\cdashline{2-8}
  $4.0$& $2.41$ & $0.598\, ^{+16.1\%}_{-13.0\%}\, ^{+33.1\%}_{-33.1\%}$ & $0.7616\, ^{+4.5\%}_{-5.4\%}\, ^{+31.5\%}_{-31.5\%}$
  & $0.588\, ^{+15.8\%}_{-12.8\%}\, ^{+33.8\%}_{-33.8\%}$ & $0.811\, ^{+5.9\%}_{-6.2\%}\, ^{+31.4\%}_{-31.4\%}$
  & $0.585\, ^{+15.7\%}_{-12.8\%}\, ^{+34.3\%}_{-34.3\%}$ & $0.738\, ^{+4.2\%}_{-5.1\%}\, ^{+31.6\%}_{-31.6\%}$ &\\
 \noalign{\hrule height 1pt}
  $2.5$& $1.19$ & $22.00\, ^{+11.1\%}_{-9.5\%}\, ^{+5.2\%}_{-5.2\%}$ & $26.99\, ^{+3.5\%}_{-3.9\%}\, ^{+5.1\%}_{-5.1\%}$
  & $21.56\, ^{+10.9\%}_{-9.3\%}\, ^{+5.4\%}_{-5.4\%}$ & $29.19\, ^{+4.8\%}_{-4.8\%}\, ^{+5.2\%}_{-5.2\%}$
  & $21.90\, ^{+10.8\%}_{-9.2\%}\, ^{+5.3\%}_{-5.3\%}$ & $26.10\, ^{+2.7\%}_{-3.3\%}\, ^{+5.2\%}_{-5.2\%}$ &\\
\cdashline{2-8}
    $3.5$& $2.34$ & $1.968\, ^{+14.5\%}_{-11.9\%}\, ^{+16.8\%}_{-16.8\%}$ & $2.507\, ^{+4.2\%}_{-4.9\%}\, ^{+15.8\%}_{-15.8\%}$
    & $1.943\, ^{+14.2\%}_{-11.7\%}\, ^{+16.8\%}_{-16.8\%}$ & $2.704\, ^{+5.8\%}_{-5.9\%}\, ^{+15.8\%}_{-15.8\%}$
    & $1.968\, ^{+14.1\%}_{-11.7\%}\, ^{+16.8\%}_{-16.8\%}$ & $2.438\, ^{+3.6\%}_{-4.4\%}\, ^{+15.7\%}_{-15.7\%}$ & $\frac{3}{4}$\\
\cdashline{2-8}
  $4.0$& $3.05$ & $0.585\, ^{+16.1\%}_{-13.1\%}\, ^{+34.4\%}_{-34.4\%}$ & $0.750\, ^{+4.5\%}_{-5.4\%}\, ^{+31.6\%}_{-31.6\%}$
  & $0.578\, ^{+15.8\%}_{-12.8\%}\, ^{+34.4\%}_{-34.4\%}$ & $0.816\, ^{+6.4\%}_{-6.5\%}\, ^{+31.7\%}_{-31.7\%}$
  & $0.584\, ^{+15.7\%}_{-12.8\%}\, ^{+34.3\%}_{-34.3\%}$ & $0.739\, ^{+4.1\%}_{-5.0\%}\, ^{+31.4\%}_{-31.4\%}$ &\\
 \noalign{\hrule height 1pt}
\end{tabular}
\end{adjustbox}
  \caption{\footnotesize Hadronic cross section for the reaction $pp\rightarrow\{W^{\prime}, T\}\rightarrow t\bar{b}Z+t\bar{b}H+\text{h.c.}$ in the benchmark scenario ${\bf T^{\scriptscriptstyle\{3\}}_{\scriptscriptstyle\{Z,H,W'\}}}$ with $\Gamma_{\scriptscriptstyle W^{\prime}}^{\scriptscriptstyle\text{ToT}}/m_{\scriptscriptstyle W^{\prime}}=1.09\%$, $\xi_{{Tb}}^{{W^{\prime}}}=20\%$ and $\vkwp = 0.5$ (DD scenario).}
   \label{tabVp2Tev}
  \end{table*}

\noindent
 The discussion above suggests that the NWA$_2$ approximation remains valid at NLO order, as it does at LO, for the DD scenario with $\vkwp=0.5$, with about an accuracy of about $3\%$ compared to the full approach (i.e. the CM scheme) for all the considered VLQ mass hierarchies.  \\

\noindent
A natural question arises: which of the NLO approximations is the most suitable to adopt? The ideal choice would be the one that is least dependent on nonphysical scales, the most accurate and the lowest computational cost. Regarding scale dependence, the 3 NLO approximations (CM, NWA$_1$ and NWA$_2$) exhibit similar behavior, as their results converge at higher scales, as illustrated in figure~\ref{scalVars} and exposed in table~\ref{tabVp2Tev}. In terms of accuracy, the CM scheme is the most precise since it fully accounts the finite width and off-shell effects of all the virtual unstable particles.
With respect to computational efficiency, NWA$_2$ is the fastest and the least time-consuming option. It shows a significant reduction in computation cost, especially compared to CM option\footnote{At NLO, the $\text{NWA}_{2}$ completes in minutes, whereas the CM approach requires several hours for certain kinematic configurations when a precision of 0.001 is requested in {\tt MadGraph5}.}. This is because it involves fewer loop diagrams and simpler topologies compared to the CM approach.\\

\noindent
Therefore, for the specific case of DD scenario, where  the predictions of $\text{NWA}_{2}$ and CM scheme are in reasonable agreement (within uncertainties), $\text{NWA}_{2}$ is more advantageous for quick estimation of the NLO cross section, as it not only simplifies the calculations but also reduces the number of free parameters to just two: $\vkwp$ and the mass of $W^{\prime}$, which suffice to control the cross section. We strongly emphasize that this conclusion is only valid for the DD scenario. For other scenarios the discrepancy between NWA$_2$ and CM scheme may be important\footnote{We show, in appendix~\ref{appC} that NWA$_2$ and CM discrepancies may exceed $7\%$ for scenarios other than those considered here.}. In general, the CM scheme remains the most reliable and accurate method, despite its high computation cost, cf. appendix~\ref{appC}.

\begin{figure}[h!]
\centering
\includegraphics[width=5.75cm,height=4.5cm]{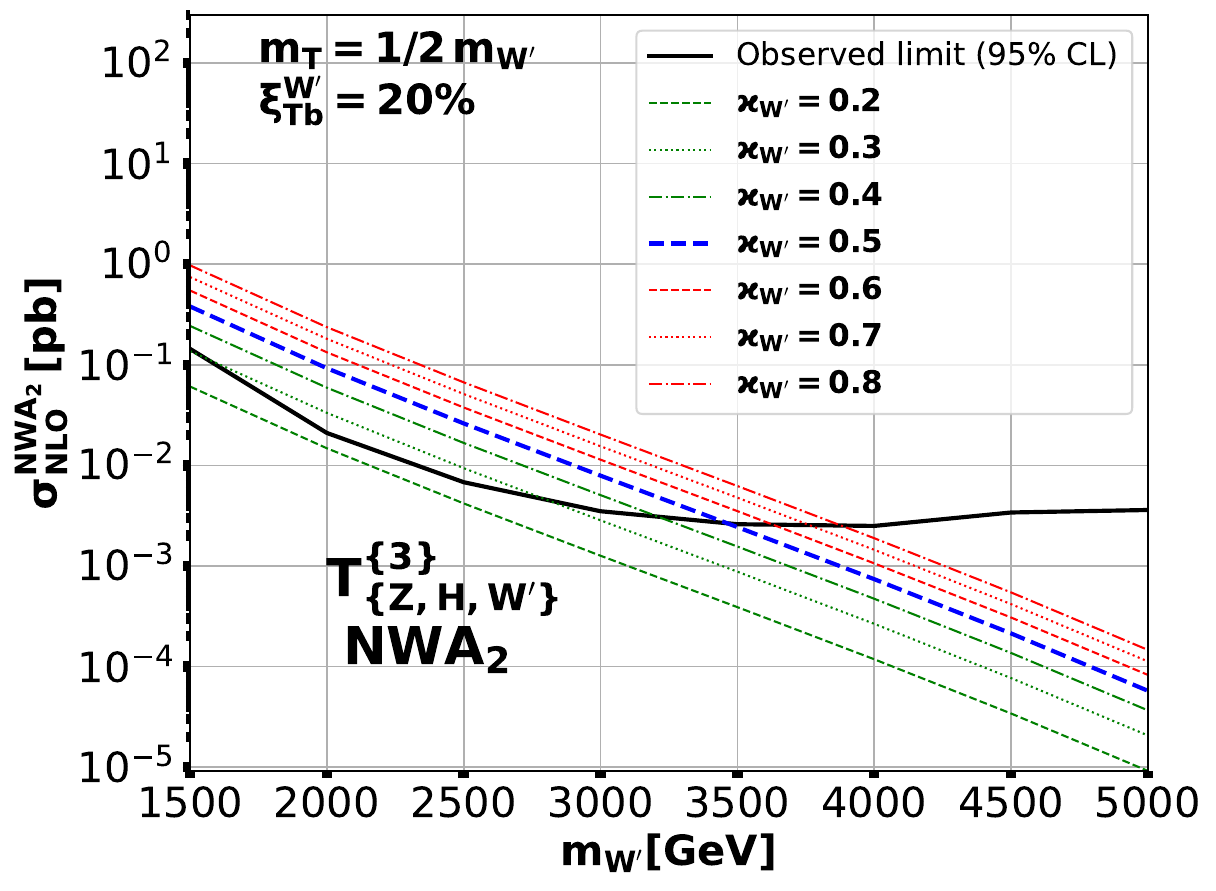}
\includegraphics[width=5.75cm,height=4.5cm]{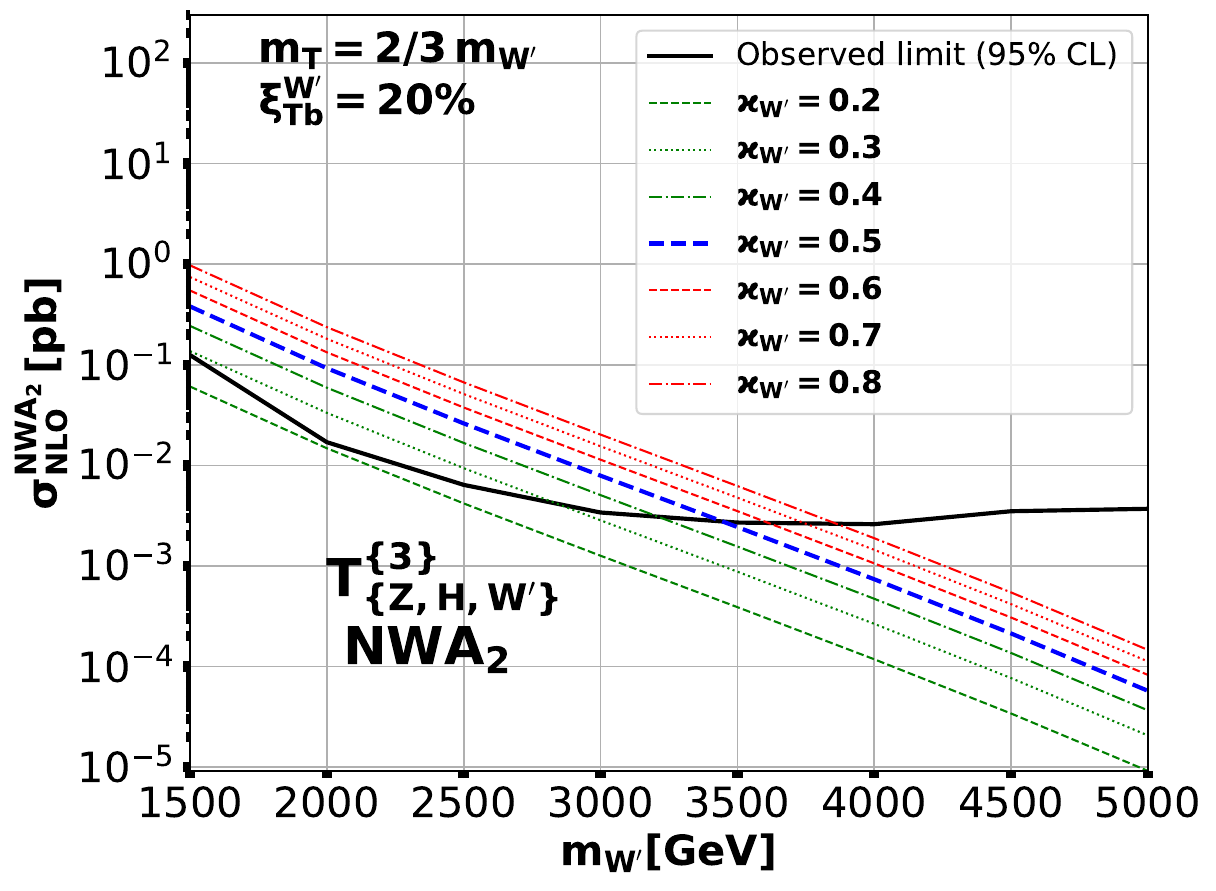}
\includegraphics[width=5.75cm,height=4.5cm]{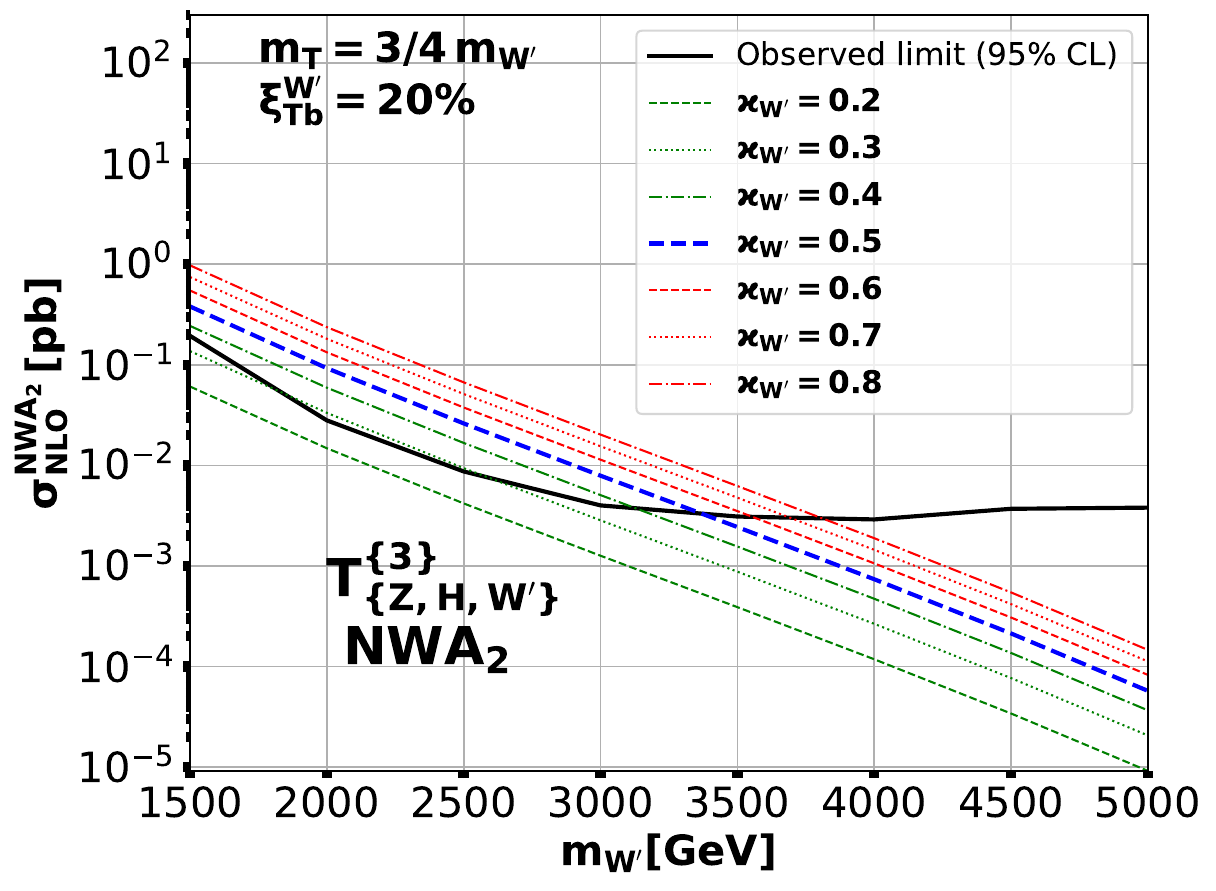}
\includegraphics[width=5.75cm,height=4.5cm]{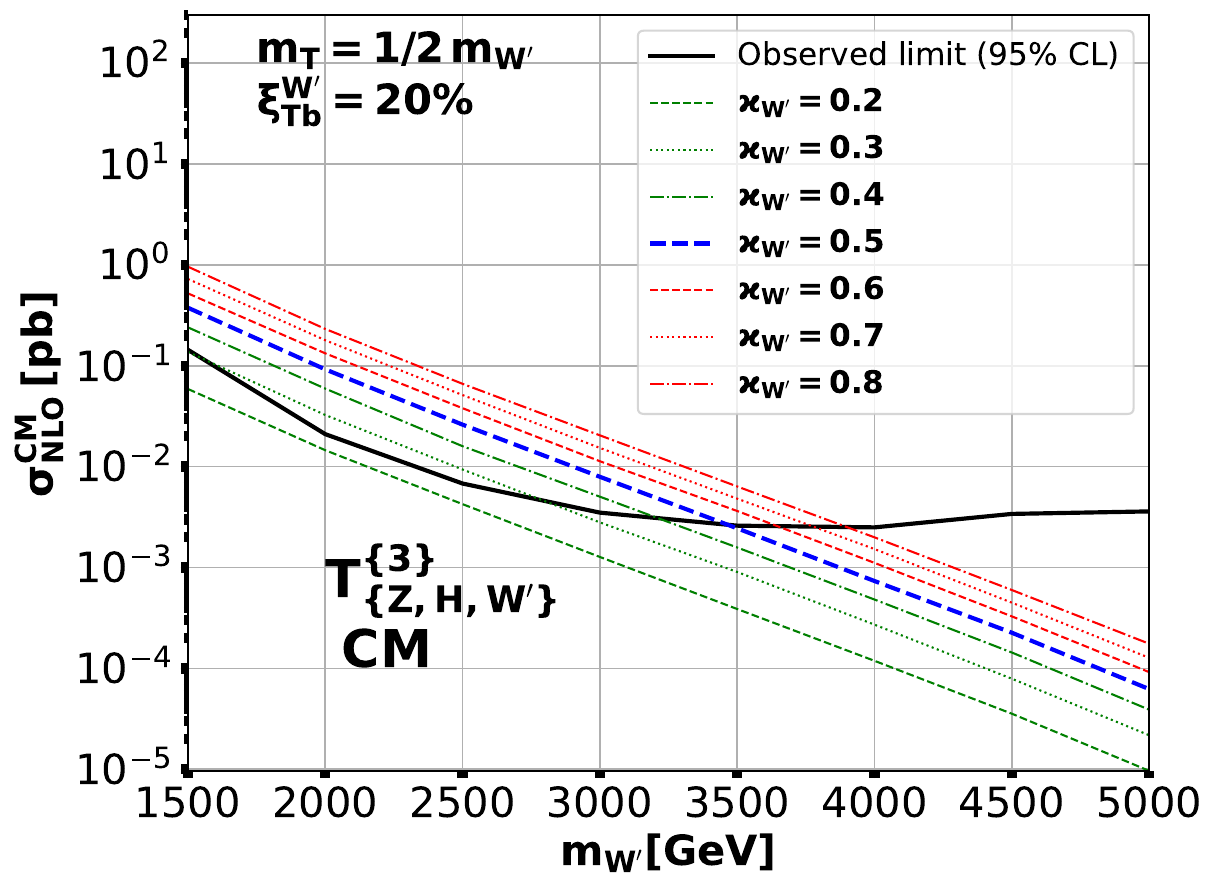}
\includegraphics[width=5.75cm,height=4.5cm]{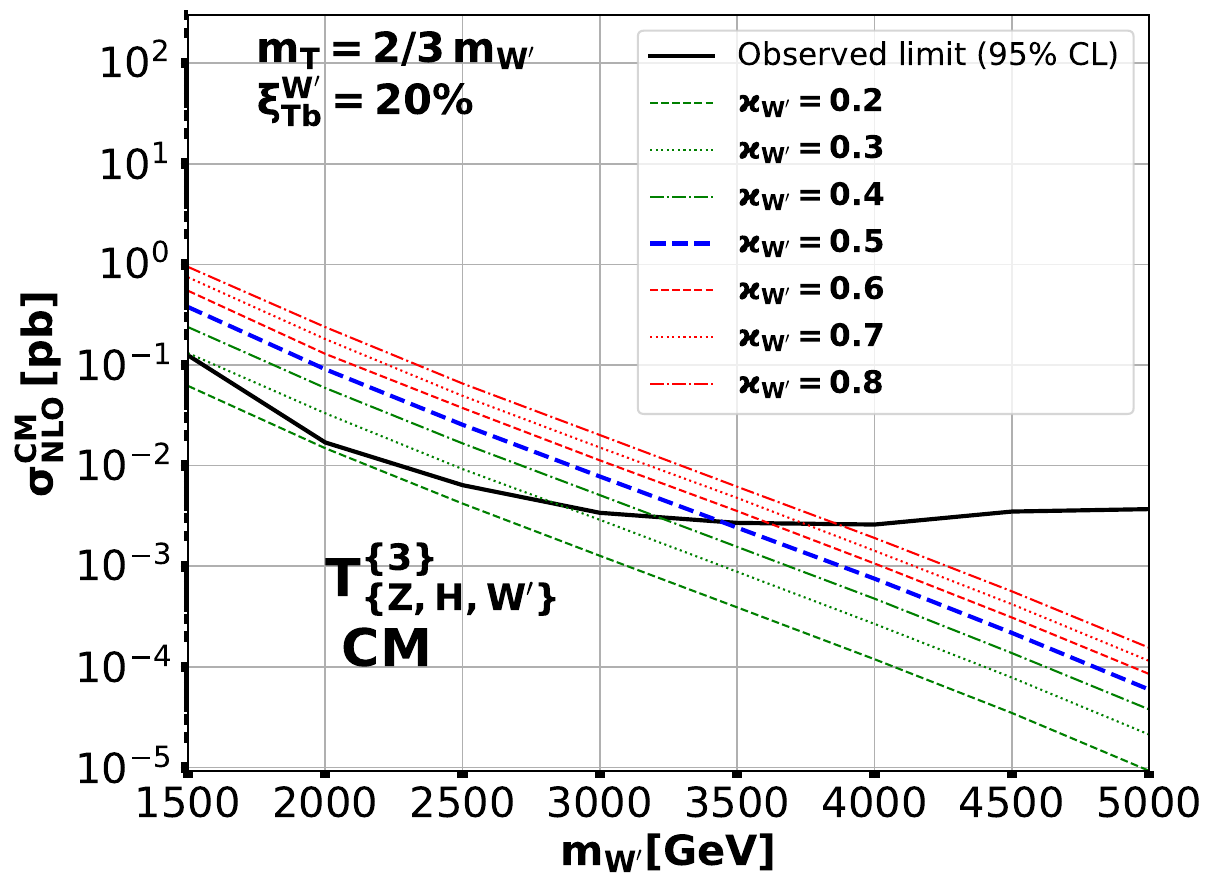}
\includegraphics[width=5.75cm,height=4.5cm]{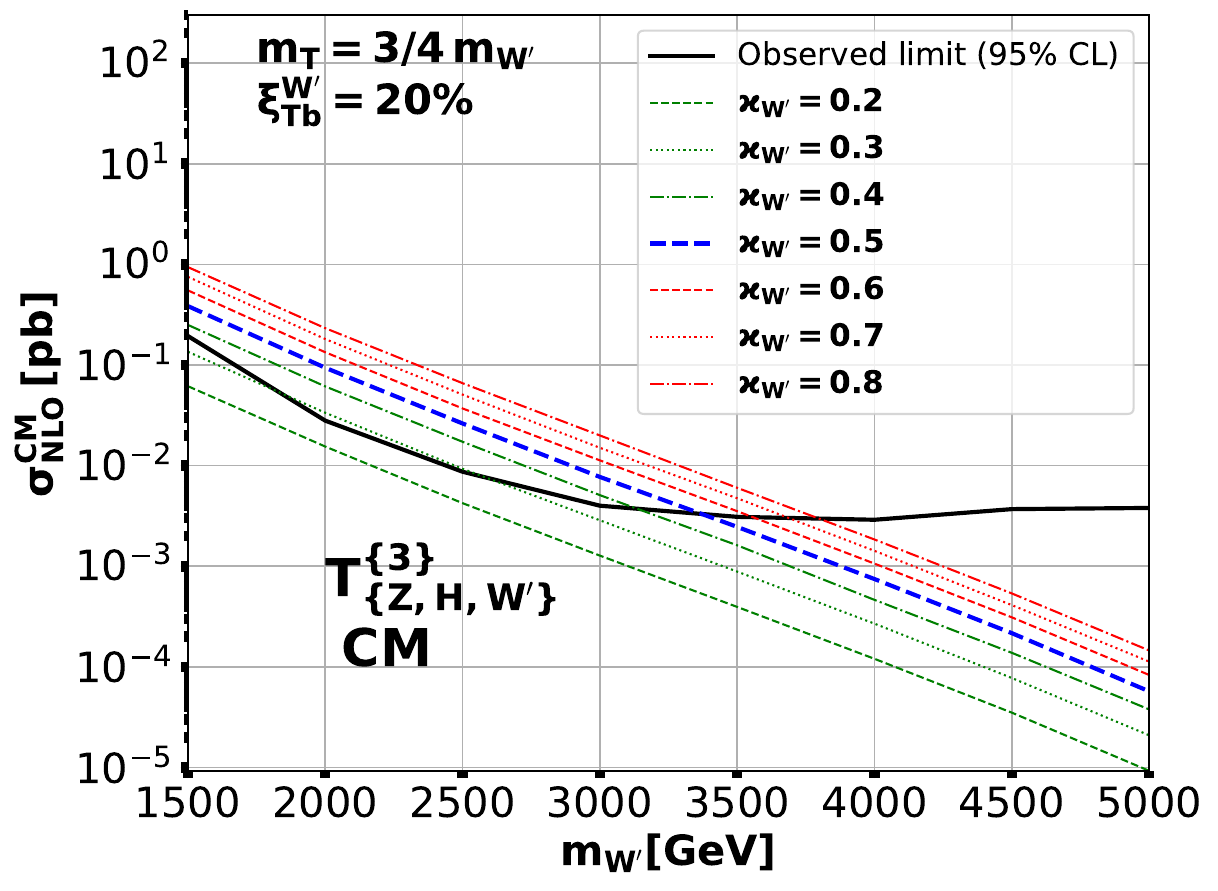}
 \caption{\small NLO predictions versus data: $m_{\scriptscriptstyle T}=1/2\, m_{\scriptscriptstyle W^{\prime}}$ (left),  $m_{\scriptscriptstyle T}=2/3\, m_{\scriptscriptstyle W^{\prime}}$ (middle) and $m_{\scriptscriptstyle T}=3/4\, m_{\scriptscriptstyle W^{\prime}}$ (right).}
\label{datavsth}
\end{figure}
\noindent
Now, we are ready to compare our NLO predictions to CMS experimental data available at \url{https://www.hepdata.net/record/ins2039384}. In figure~\ref{datavsth}, we present the upper observed limit at $95\%$ confidence level (black curve) and the predictions of our model at NLO for various values of $\vkwp$ (dotted, dashed, dotted-dashed green, blue and red curves), in the DD scenario (i.e. the $W^{\prime}$ boson decays equally to all type of quarks), in the NWA$_2$ (upper panels) and the CM scheme (lower panels). Here, the lower limit on the $W^{\prime}$ exhibits a strong dependence on only one free parameter which is $\vkwp$. Specifically, the lower mass limit is approximately $2.8$ TeV for $\vkwp=0.3$ and increases to $3.9$ TeV for $\vkwp=0.8$. Regarding the case where $\vkwp=0.2$, the $W^{\prime}$ mass is not constrained by this data. We note that the difference between the predictions of the NWA$_2$ and CM schemes is not clearly visible in these plots due to the logarithmic scale on the Y-axis. In practice, the 2 approaches yield approximately the same $\wps$ mass lower limits (differing by barely a few GeV), making NWA$_2$ and CM scheme effectively equivalent in this context, for the DD scenario.\\

\begin{figure}[h!]
\includegraphics[width=8cm,height=5cm]{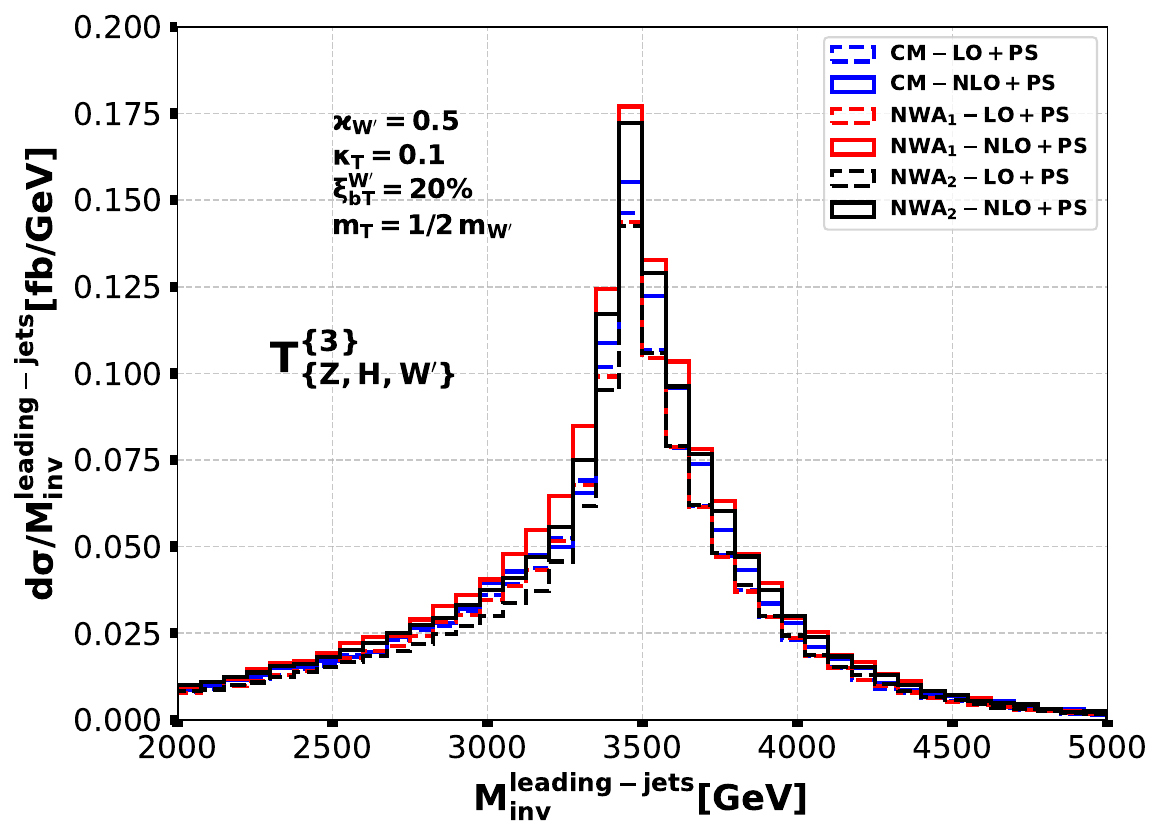}
\includegraphics[width=8cm,height=5cm]{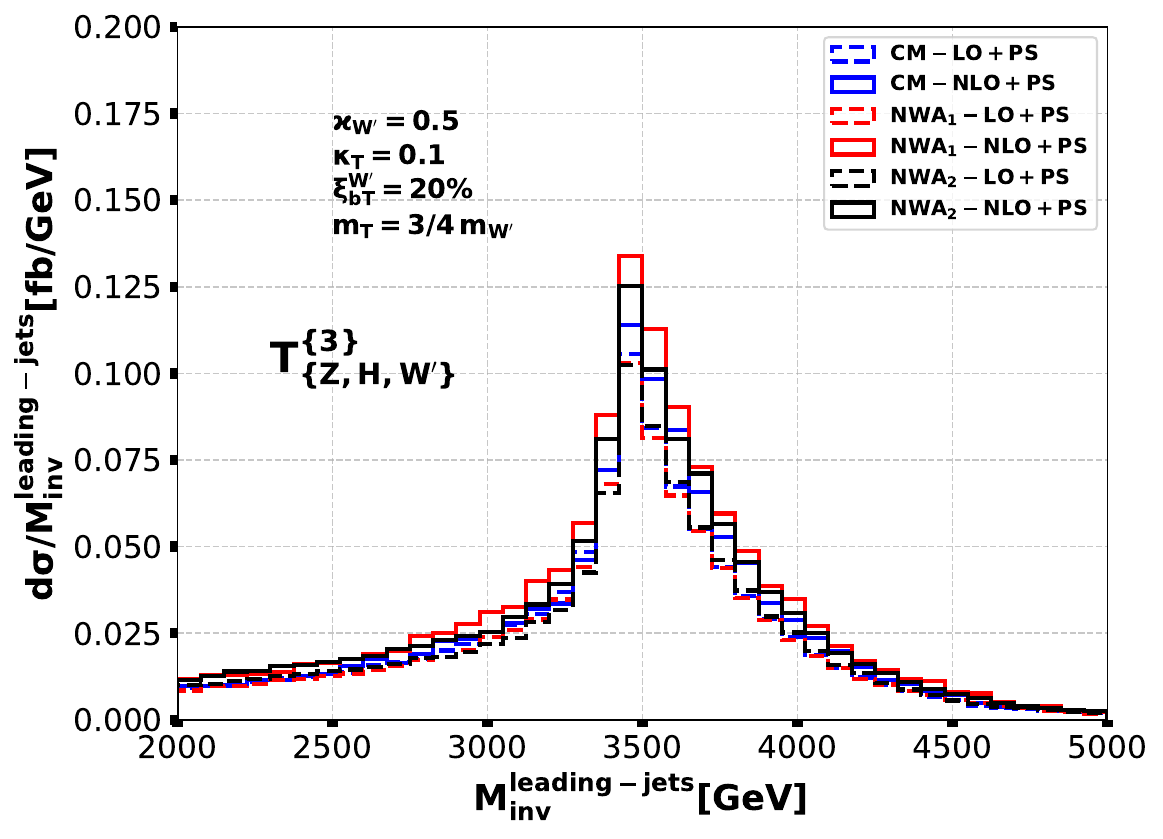}
 \caption{\small LO and NLO leading jets invariant mass differential distribution, for the low-mass case, in DD scenario}
\label{inv-mass-Wp}
\end{figure}

\noindent
Another legitimate question which might be asked: do the 3 approximations: CM, NWA$_1$ and NWA$_2$, give the same differential distribution when parton shower is taken into account? To answer this question, we produced the differential cross section on the invariant mass of the final state \textit{leading jets}, for $m_{\scriptscriptstyle W^{\prime}}=3.5$ TeV, $m_{\scriptscriptstyle T}=1/2\, m_{\scriptscriptstyle W^{\prime}}$ (left panel) and $m_{\scriptscriptstyle T}=3/4\, m_{\scriptscriptstyle W^{\prime}}$ (right panel) at $\sqrt{s}=13\, $ TeV in DD and ${\bf T^{\scriptscriptstyle\{3\}}_{\scriptscriptstyle\{Z,H,W'\}}}$ scenarios, see figure~\ref{inv-mass-Wp}.
\noindent
The invariant mass distribution is produced by using {\tt MadAnalysis}\cite{madanalysis} and {\tt FastJet}\cite{Cacciari:2011ma}
for jet clustering. To reduce the unwanted events, we impose the following cuts on the final-state jets: $p_{\scriptscriptstyle{Tj}}>20$~GeV, $|\eta_{\scriptscriptstyle{j}}|<5$ and $\Delta R_{\scriptscriptstyle{jj}}>0.4$ ($j$ denotes both light-jets and b-jets). We observe that the differential distributions in NWA$_1$, NWA$_2$ and CM are globally similar far from the peaks but differ increasingly as the peaks are approached (particularly at NLO). At LO, the peaks are quite similar, whereas at NLO, the CM exhibits a relatively lower peak compared to those of NLO NWA$_1$ and NWA$_2$. This is because the CM handles the finite width and the off-shell effects. We note that all of these distributions are smeared by the parton-shower (PS), specifically by initial-state radiation (ISR) and final state radiation (FSR). For example, the LO NWA$_2$ distributions would exhibit narrow peaks around $M_{_\text{inv}}^{_\text{leading-jets}}=m_{\scriptscriptstyle W^{\prime}}$ if PS were not applied.

\subsection{Search for $\zp$ boson decaying into vector-like quark}
\label{zp-production}
The first search for heavy vector neutral boson decaying into VLQ at the LHC was performed by CMS collaboration in ref.~\cite{cms7}, where exclusion upper limits ranging between 0.13-10 pb are set on the cross section at 95\% confidence level. In this sub.sec, we investigate the production and decay to VLQ of the $\zp$ boson in the context of our model, considering the benchmark scenarios ${\bf T^{\scriptscriptstyle\{3\}}_{\scriptscriptstyle\{Z,Z'\}}}$ and ${\bf T^{\scriptscriptstyle\{3\}}_{\scriptscriptstyle\{Z,H,Z'\}}}$ at LO and NLO using the 3 approximations employed in the previous subsection (NWA$_1$, NWA$_2$ and CM scheme), and considering the mass ratios: $m_{\scriptscriptstyle{T}}/m_{\scriptscriptstyle{Z^{\prime}}}=1/2$ (low-mass), $m_{\scriptscriptstyle{T}}/m_{\scriptscriptstyle{Z^{\prime}}}=2/3$ (middle-mass) and $m_{\scriptscriptstyle{T}}/m_{\scriptscriptstyle{Z^{\prime}}}=5/6$ (high-mass) as adopted in ref.~\cite{cms7}.\\

\noindent
The technical details are largely similar to those presented in sub.sec. \ref{prod_wp}. Therefore, we will not go into all the details, but we will focus on presenting and analyzing the obtained results in the benchmark scenarios ${\bf T^{\scriptscriptstyle\{3\}}_{\scriptscriptstyle\{Z,Z'\}}}$ and ${\bf T^{\scriptscriptstyle\{3\}}_{\scriptscriptstyle\{Z,H,Z'\}}}$. Here also, the dominant production mechanism of the $\zp$ boson at the LHC (which decays to $T$ and subsequently to $tZ/H$) is the Drell-Yan process: $q\bar{q}\rightarrow\zp\rightarrow \bar{t}\, T + t\, \bar{T} \rightarrow t\, \bar{t}\, Z (H)$. The Born-level, the one-loop and real emission Feynman diagrams associated to this process are similar to those depicted, respectively, in figures~\ref{pp-wp}, \ref{loops} and \ref{reals}, where, without loss of generality, one simply replaces $\wps$ with $\zp$.\\

\noindent 
The ratio of the total width-to-mass of the $\zp$ depends only on the 2 parameters $\vkzp$ and $\xi^{\scriptscriptstyle{Z^{\prime}}}_{\scriptscriptstyle{Tt}}$, it is given by:
\begin{align}
\frac{\Gamma^{\text{ToT}}_{\scriptscriptstyle{Z^{\prime}}}}{m_{\scriptscriptstyle{Z^{\prime}}}}&=\frac{\Gamma[Z^{\prime}\rightarrow T\bar{t}+\bar{T}t]/\xi_{\scriptscriptstyle{Tt}}^{\scriptscriptstyle{Z^{\prime}}}}{m_{\scriptscriptstyle Z^{\prime}}}=\frac{g^2}{4\, \pi}\, \frac{1-2\, s_{\scriptscriptstyle{w}}^2+8/3\, s_{\scriptscriptstyle{w}}^4}{c_{\scriptscriptstyle{w}}^2}\, \frac{\vkzp^2}{1-\xi^{\scriptscriptstyle{Z^{\prime}}}_{\scriptscriptstyle{Tt}}}
\label{GamZpOvM}
\end{align} 
Here also, we note that for $\xi_{{Tt}}^{{Z^{\prime}}} \leq 0.5$ and $\vkzp \leq 1$, $\Gamma_{\scriptscriptstyle{Z^{\prime}}}^{\scriptscriptstyle{\text{ToT}}}/m_{\scriptscriptstyle Z^{\prime}}$ is always less than $5.1\%$ and the values of $\kpzp$ are always smaller than those given in table~\ref{Tab2}, which preserves perturbative unitarity regardless the values of the masses $m_{\scriptscriptstyle T}$ and $m_{\scriptscriptstyle Z^{\prime}}$, see table~\ref{tabkzpPerturba}.
\begin{table*}[h!]
\centering
 \renewcommand{\arraystretch}{1.40}
 \setlength{\tabcolsep}{12.5pt}
 \begin{adjustbox}{width=15cm,height=1.00cm}
 \begin{tabular}{!{\vrule width 1.5pt}l!{\vrule width 1.5pt}l!{\vrule width 1pt}l!{\vrule width 1pt}l!{\vrule width 1.5pt}l!{\vrule width 1pt}l!{\vrule width 1pt}l!{\vrule width 1.5pt}l!{\vrule width 1pt}l!{\vrule width 1pt}l!{\vrule width 1.5pt}l!{\vrule width 1pt}l!{\vrule width 1.5pt}}
 \cline{2-10}
\multicolumn{1}{c!{\vrule width 1.5pt}}{{}}
&\multicolumn{3}{c!{\vrule width 1.5pt}}{{$\bf \vkzp=0.5$}}
&\multicolumn{3}{c!{\vrule width 1.5pt}}{{$\bf \vkzp=1.0$}}
&\multicolumn{3}{c!{\vrule width 1.5pt}}{{$\bf \vkzp=2.0$}}\\
\noalign{\hrule height 1pt}
$\bf \xi^{\scriptscriptstyle{Z^{\prime}}}_{\scriptscriptstyle{Tt}}$&\bf 0.10 &\bf 0.20 &\bf 0.40 &\bf 0.10 &\bf 0.20 &\bf 0.40 &\bf 0.10 &\bf 0.20 &\bf 0.40 \\
\cdashline{1-10}
$\bf \kpzp$&\bf 0.42&\bf 0.30& \bf 0.21& \bf 0.85 &\bf 0.60 &\bf 0.42&\bf 1.69 &\bf 1.20 & \bf 0.85\\
\cdashline{1-10}
$\bf \frac{\Gamma_{\scriptscriptstyle{Z^{\prime}}}^{\scriptscriptstyle{\text{ToT}}}}{m_{\scriptscriptstyle Z^{\prime}}}[\%]$&\bf 0.86 &\bf 0.97 &\bf 1.29 &\bf 3.44 &\bf 3.87 &\bf 5.16 &\bf 13.77 &\bf 15.49 &\bf  20.65 \\
\noalign{\hrule height 1pt}
\end{tabular}
\end{adjustbox}
  \caption{\footnotesize Numerical values of $\kpzp$ for $\vkzp=0.5, 1, 2$ and $\xi^{\scriptscriptstyle{Z^{\prime}}}_{\scriptscriptstyle{Tt}}=0.1, 0.2, 0.4$.}
   \label{tabkzpPerturba}
  \end{table*}

\noindent
We notice that the coupling of $\zp$ to $T$ and $t$ is controlled by the coefficient: $\kzpl=\kappa_{\scriptscriptstyle{Z^{\prime}}}\sqrt{\xi_{\scriptscriptstyle{tT}}^{\scriptscriptstyle{Z^{\prime}}}/\Gamma_{\scriptscriptstyle{tT}}^{\scriptscriptstyle{Z^{\prime}}}}$, cf.~eq.~(\ref{kzpwpcase2}). It should be emphasized that the parameters $\kappa_{\scriptscriptstyle{Z^{\prime}}}$ and $\vkzp$ are not independent in the SSM, as it has been shown in appendix \ref{appA}. So, the free parameters that control the cross sections are: $\vkzp$, the branching ratio $\xi_{\scriptscriptstyle{tT}}^{\scriptscriptstyle{Z^{\prime}}}$, the mass of the $\zp$ boson ($m_{\scriptscriptstyle Z'}$) and the mass of the vector-like quark ($m_{\scriptscriptstyle T}$).
We note also that we assumed, in this section, that $Z'$ decays democratically to up quarks (i.e. $\xi_{\scriptscriptstyle{tT}}^{\scriptscriptstyle{Z^{\prime}}}=\xi_{\scriptscriptstyle{uu}}^{\scriptscriptstyle{Z^{\prime}}}=\xi_{\scriptscriptstyle{cc}}^{\scriptscriptstyle{Z^{\prime}}}=\xi_{\scriptscriptstyle{tt}}^{\scriptscriptstyle{Z^{\prime}}}$). Thus, this implies that in the SSM we must have $\xi_{\scriptscriptstyle{tT}}^{\scriptscriptstyle{Z^{\prime}}}=9.6\%$, see appendix~\ref{appA} for more details.\\

\begin{figure}[h!]
\centering
\includegraphics[width=5.75cm,height=3.75cm]{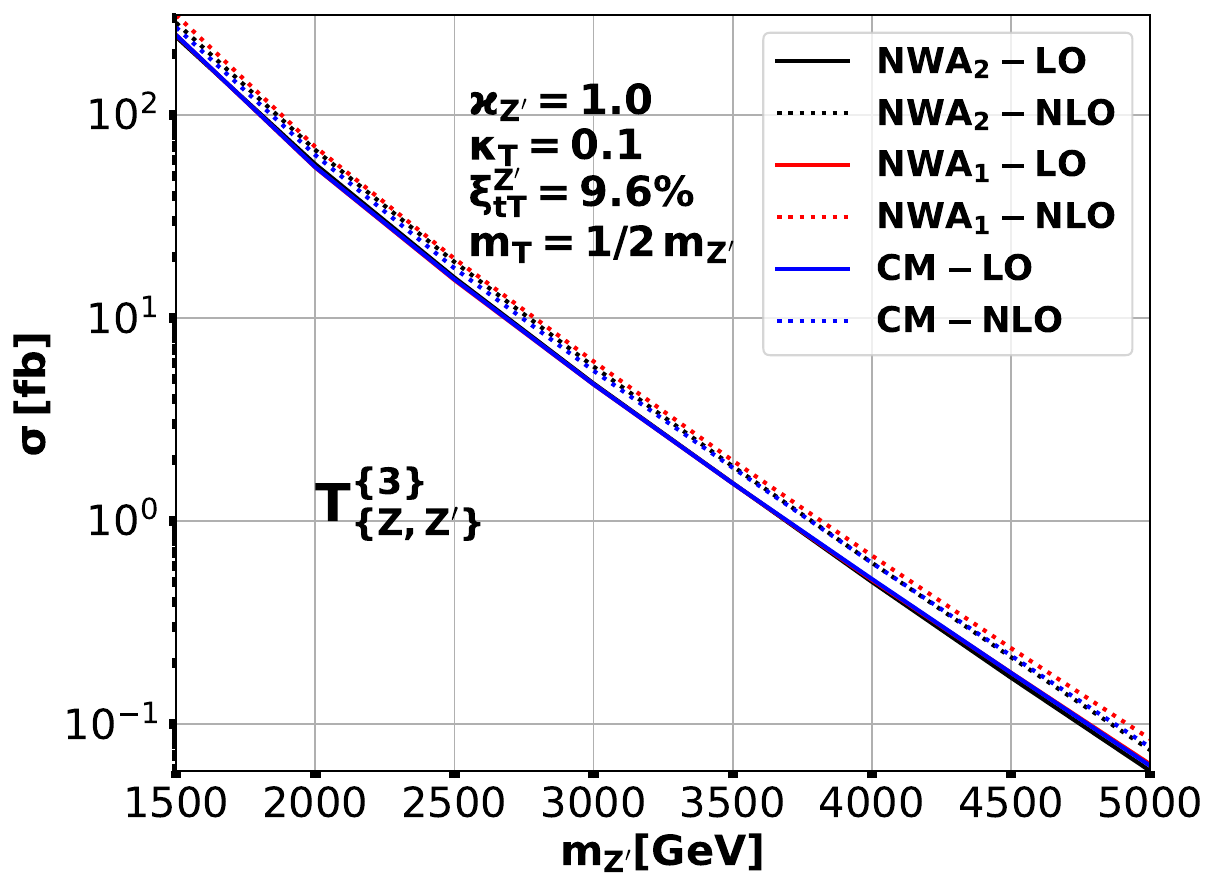}
\includegraphics[width=5.75cm,height=3.75cm]{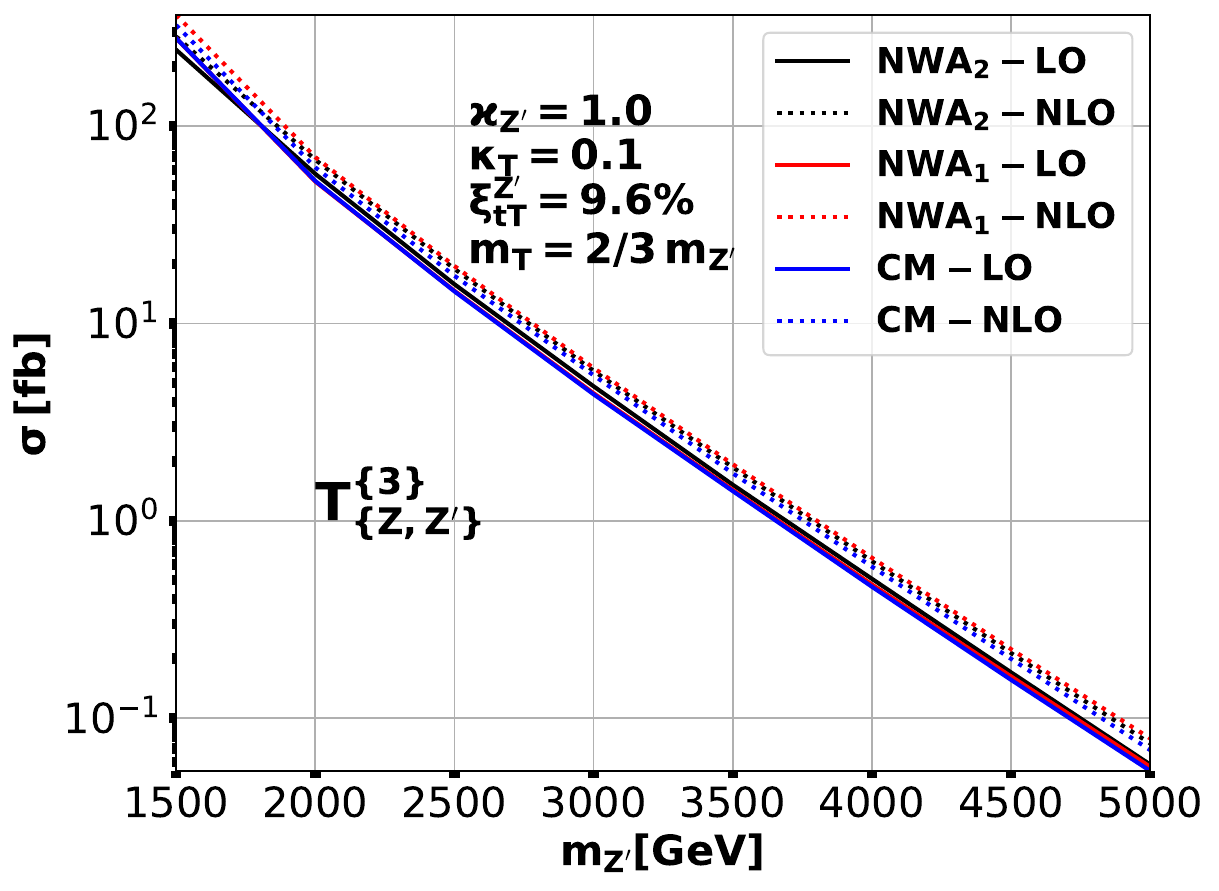}
\includegraphics[width=5.75cm,height=3.75cm]{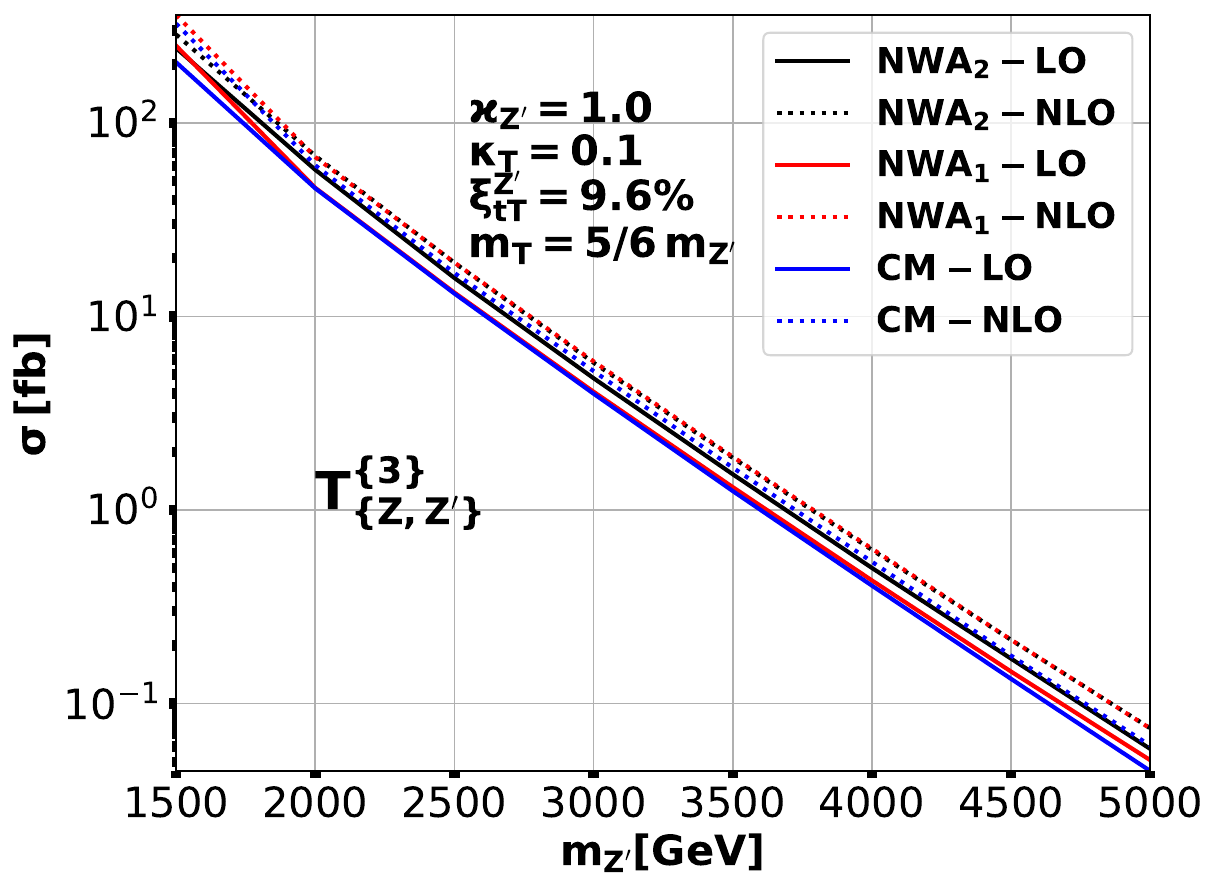}
 \caption{\small Variation of the LO and NLO cross sections as a function of $m_{\scriptscriptstyle Z^{\prime}}$ in the benchmark scenario ${\bf T^{\scriptscriptstyle\{3\}}_{\scriptscriptstyle\{Z,Z^{\prime}\}}}$.}
\label{mZpVarsXsec}
\end{figure}

\noindent 
The mass $m_{\scriptscriptstyle{Z^{\prime}}}$ and the ratio $m_{\scriptscriptstyle T}/m_{\scriptscriptstyle{Z^{\prime}}}$ play a crucial role in controlling the cross section.
In figure~\ref{mZpVarsXsec}, we present the variation of the cross section in term of $m_{\scriptscriptstyle{Z^{\prime}}}$ within the 3 previously discussed approximations,
for: $m_{\scriptscriptstyle{T}}= 1/2\,  m_{\scriptscriptstyle{Z^{\prime}}}, m_{\scriptscriptstyle{T}}= 2/3\, m_{\scriptscriptstyle{Z^{\prime}}}$ and $m_{\scriptscriptstyle{T}}= 5/6\, m_{\scriptscriptstyle{Z^{\prime}}}$. These results are obtained for $\vkzp=1$ (i.e. CSSM) and $\xi^{\scriptscriptstyle{Z^{\prime}}}_{\scriptscriptstyle{Tb}}=9.6\%$ for center of mass energy $\sqrt{s}=13$ TeV, in the benchmark scenario ${\bf T^{\scriptscriptstyle\{3\}}_{\scriptscriptstyle\{Z,Z^{\prime}\}}}$. We observe that all the cross sections (LO and NLO) decrease as the mass of $\zp$ increase and consequently the mass of $T$ increase, as expected. This occurs because the phase space becomes more limited for heavier particles. The same conclusions are obtained for the benchmark scenario ${\bf T^{\scriptscriptstyle\{3\}}_{\scriptscriptstyle\{Z,H,Z^{\prime}\}}}$, the corresponding plots are not included to streamline the presentation. We also note that the scale dependence is extremely reduced at NLO compared to LO as in the case of $\wps$. The plots resemble those in figure~\ref{scalVars} and were excluded to improve the paper's readability.

\noindent
As matter of fact, these plots in figure~\ref{mZpVarsXsec} do not show how significant are the differences between the predictions of NWA$_1$, NWA$_2$ and CM scheme. To illustrate these discrepancies, we give some numerical values of the cross sections for the benchmark scenarios ${\bf T^{\scriptscriptstyle\{3\}}_{\scriptscriptstyle\{Z,Z^{\prime}\}}}$ and ${\bf T^{\scriptscriptstyle\{3\}}_{\scriptscriptstyle\{Z,H,Z^{\prime}\}}}$ in 
tables~\ref{tabVp2Tevzp1} and \ref{tabVp2Tevzp2}, respectively. We observe that the cross sections in NWA$_2$ are insensitive to the ratio $m_{\scriptscriptstyle{T}}/m_{\scriptscriptstyle{Z^{\prime}}}$, which is expected since the VLQ intervene only through the branching ratios $\xi_{\scriptscriptstyle Tt}^{\scriptscriptstyle Z^{\prime}}$ and $\xi_{\scriptscriptstyle Z/H}$. We recall that these quantities are treated as free parameters in our model, i.e. they are fixed for all masses and approximations (LO, NLO , \ldots etc.) as $\xi_{\scriptscriptstyle Tt}^{\scriptscriptstyle Z^{\prime}}=0.096$~(DD to up quarks) and $\xi_{\scriptscriptstyle Z}=1$ (in ${\bf T^{\scriptscriptstyle\{3\}}_{\scriptscriptstyle\{Z,Z^{\prime}\}}}$)  or $\xi_{\scriptscriptstyle Z}=\xi_{\scriptscriptstyle H}=1/2$ (in ${\bf T^{\scriptscriptstyle\{3\}}_{\scriptscriptstyle\{Z,H,Z^{\prime}\}}}$), this justifies why the NWA$_2$ predictions in the benchmark scenarios ${\bf T^{\scriptscriptstyle\{3\}}_{\scriptscriptstyle\{Z,H,Z'\}}}$ and ${\bf T^{\scriptscriptstyle\{3\}}_{\scriptscriptstyle\{Z,Z'\}}}$ are approximately the same. However, in the other approaches (NWA$_1$ and CM), the predictions depend on that ratio due to the phase space restrictions and the kinematics which effect directly the total width of the VLQ in each case.

\begin{table*}[h!]
\boldmath
\centering
 \renewcommand{\arraystretch}{1.40}
 \setlength{\tabcolsep}{10pt}
 \begin{adjustbox}{width=18cm,height=3.0cm}
 \begin{tabular}
 {!{\vrule width 2pt}l!{\vrule width 2pt}l!{\vrule width 2pt}c:c!{\vrule width 2pt}c:c!{\vrule width 2pt}c:c!{\vrule width 2pt}c!{\vrule width 2pt}}
  \cline{1-8}
$\bf m_{\scriptscriptstyle Z^{\prime}}$& $\frac{\Gamma_{\scriptscriptstyle T}^{\scriptscriptstyle\text{ToT}}}{m_{\scriptscriptstyle T}}$ &\multicolumn{2}{c!{\vrule width 2pt}}{{\bf CM}}
&\multicolumn{2}{c!{\vrule width 2pt}}{{\bf NWA$_1$}}
&\multicolumn{2}{c!{\vrule width 2pt}}{{\bf NWA$_2$}}&\multicolumn{1}{c}{{}}\\
\cline{3-9}
 \textbf{[TeV]}& $[\%]$ &$\bf\sigma_{\scriptscriptstyle{\textbf{LO}}}[\textbf{fb}]$& $\bf\sigma_{\scriptscriptstyle{\textbf{NLO}}}[\textbf{fb}]$&$\bf\sigma_{\scriptscriptstyle{\textbf{LO}}}[\textbf{fb}]$& $\bf\sigma_{\scriptscriptstyle{\textbf{NLO}}}[\textbf{fb}]$&$\bf\sigma_{\scriptscriptstyle{\textbf{LO}}}[\textbf{fb}]$& $\bf\sigma_{\scriptscriptstyle{\textbf{NLO}}}[\textbf{fb}]$& $\frac{m_{\scriptscriptstyle T}}{m_{\scriptscriptstyle Z^{\prime}}}$\\
  \noalign{\hrule height 1pt}
  $1.5$ & $0.19$
  &$240.7\, ^{+7.1\%}_{-6.3\%}\, ^{+2.4\%}_{-2.4\%}$
  & $281.2\, ^{+1.6\%}_{-1.9\%}\, ^{+2.4\%}_{-2.4\%}$
  & $240.7\, ^{+7.0\%}_{-6.3\%}\, ^{+2.4\%}_{-2.4\%}$
  & $305.0\, ^{+3.2\%}_{-3.1\%}\, ^{+2.4\%}_{-2.4\%}$
  & $242.0\, ^{+7.0\%}_{-6.3\%}\, ^{+2.4\%}_{-2.4\%}$
  & $284.3\, ^{+2.0\%}_{-2.2\%}\, ^{+2.5\%}_{-2.5\%}$ & \\
\cdashline{3-8}
   $2.5$& $0.53$
   &$15.90\, ^{+10.8\%}_{-9.2\%}\, ^{+3.8\%}_{-3.8\%}$
   & $18.16\, ^{+2.5\%}_{-3.1\%}\, ^{+3.9\%}_{-3.9\%}$
   &$15.78\, ^{+10.7\%}_{-9.1\%}\, ^{+3.8\%}_{-3.8\%}$
   & $20.01\, ^{+3.8\%}_{-4.0\%}\, ^{+3.8\%}_{-3.8\%}$
   & $15.76\, ^{+10.6\%}_{-9.0\%}\, ^{+3.8\%}_{-3.8\%}$
   &$18.85\, ^{+2.6\%}_{-3.2\%}\, ^{+3.9\%}_{-3.9\%}$
   &$\frac{1}{2}$\\
\cdashline{3-8}
  $3.5$& $1.04$
  &$1.567\, ^{+13.4\%}_{-11.1\%}\, ^{+7.7\%}_{-7.7\%}$
  & $1.870\, ^{+3.1\%}_{-4.0\%}\, ^{+7.6\%}_{-7.6\%}$
  & $1.566\, ^{+13.3\%}_{-11.0\%}\, ^{+7.6\%}_{-7.6\%}$
  & $2.014\, ^{+4.4\%}_{-4.9\%}\, ^{+7.4\%}_{-7.4\%}$
  &$1.520\, ^{+13.2\%}_{-11.0\%}\, ^{+7.8\%}_{-7.8\%}$
  & $1.866\, ^{+3.4\%}_{-4.2\%}\, ^{+7.3\%}_{-7.3\%}$
  &\\
 \noalign{\hrule height 1pt}
  $1.5$& $0.34$
  & $233.7\, ^{+7.1\%}_{-6.4\%}\, ^{+2.4\%}_{-2.4\%}$
  &  $283.3\, ^{+3.1\%}_{-3.0\%}\, ^{+2.5\%}_{-2.5\%}$
  & $ 233.3\,^{+7.1\%}_{-6.3\%}\, ^{+2.4\%}_{-2.4\%}$  
  & $ 312.5\,^{+3.6\%}_{-3.4\%}\, ^{+2.4\%}_{-2.4\%}$
  & $243.6\, ^{+7.0\%}_{-6.3\%}\, ^{+2.4\%}_{-2.4\%}$  
  &  $285.1\, ^{+2.2\%}_{-2.4\%}\, ^{+2.5\%}_{-2.5\%}$&\\
\cdashline{3-8}
    $2.5$&$0.94$
    & $15.11\, ^{+10.9\%}_{-9.2\%}\, ^{+3.9\%}_{-3.9\%}$  
    & $18.45\, ^{+3.1\%}_{-3.5\%}\, ^{+3.9\%}_{-3.9\%}$
    & $ 15.07\,^{+10.7\%}_{-9.1\%}\, ^{+3.9\%}_{-3.9\%}$
    &  $ 20.14\,^{+4.5\%}_{-4.5\%}\, ^{+3.8\%}_{-3.8\%}$
    & $15.79\, ^{+10.6\%}_{-9.1\%}\, ^{3.8+\%}_{-3.8\%}$
    & $18.86\, ^{+2.6\%}_{-3.2\%}\, ^{+3.8\%}_{-3.8\%}$ &$\frac{2}{3}$\\
\cdashline{3-8}
  $3.5$& $1.85$
  & $1.442\, ^{+13.6\%}_{-11.2\%}\, ^{+7.5\%}_{-7.5\%}$  
  & $1.780\, ^{+3.8\%}_{-4.5\%}\, ^{+7.1\%}_{-7.1\%}$
  & $ 1.458\,^{+13.3\%}_{-11.0\%}\, ^{+7.6\%}_{-7.6\%}$
  & $ 1.990\,^{+5.1\%}_{-5.3\%}\, ^{+7.7\%}_{-7.7\%}$
  & $1.513\, ^{+13.2\%}_{-11.0\%}\, ^{+8.2\%}_{-8.2\%}$
  & $1.872\, ^{+3.4\%}_{-4.1\%}\, ^{+7.6\%}_{-7.6\%}$ &\\
 \noalign{\hrule height 1pt}
  $1.5$&$0.53$
  &$232.0\, ^{+7.3\%}_{-6.5\%}\, ^{+2.5\%}_{-2.5\%}$  
  & $321.8\, ^{+4.0\%}_{-3.7\%}\, ^{+2.5\%}_{-2.5\%}$
  & $ 231.5\,^{+7.2\%}_{-6.4\%}\, ^{+2.5\%}_{-2.5\%}$
  & $ 350.5\,^{+4.9\%}_{-4.4\%}\, ^{+2.4\%}_{-2.4\%}$
  & $243.1\, ^{+7.0\%}_{-6.3\%}\, ^{+2.4\%}_{-2.4\%}$
  & $283.0\, ^{+2.0\%}_{-2.2\%}\, ^{+2.4\%}_{-2.4\%}$ &\\
\cdashline{3-8}
    $2.5$& $1.47$
    & $14.47\, ^{+11.0\%}_{-9.3\%}\, ^{+4.0\%}_{-4.0\%}$
    &  $18.75\, ^{+4.0\%}_{-4.2\%}\, ^{+4.1\%}_{-4.1\%}$
    & $14.63\,^{+10.7\%}_{-9.2\%}\, ^{+4.0\%}_{-4.0\%}$  
    &  $21.03\,^{+5.5\%}_{-5.3\%}\, ^{+3.9\%}_{-3.9\%}$
    & $15.94\, ^{+10.6\%}_{-9.0\%}\, ^{+3.9\%}_{-3.9\%}$
    & $18.85\, ^{+2.6\%}_{-3.2\%}\, ^{+3.8\%}_{-3.8\%}$ & $\frac{5}{6}$\\
\cdashline{3-8}
  $3.5$& $2.88$
  & $1.331\, ^{+13.7\%}_{-11.3\%}\, ^{+8.4\%}_{-8.4\%}$
  & $1.752\, ^{+5.0\%}_{-5.3\%}\, ^{+7.9\%}_{-7.9\%}$
  & $1.381\,^{+13.3\%}_{-11.1\%}\, ^{+8.2\%}_{-8.2\%}$
  & $1.989\,^{+6.3\%}_{-6.1\%}\, ^{+7.9\%}_{-7.9\%}$
  & $1.532\, ^{+13.2\%}_{-11.0\%}\, ^{+8.1\%}_{-8.1\%}$
  & $1.868\, ^{+3.4\%}_{-4.2\%}\, ^{+7.6\%}_{-7.6\%}$ &\\
 \noalign{\hrule height 1pt}
\end{tabular}
\end{adjustbox}
 \caption{\footnotesize $\sigma_{\scriptscriptstyle{\textbf{LO}}}$ and $\sigma_{\scriptscriptstyle{\textbf{NLO}}}$ for the reaction $pp\rightarrow t\bar{t}Z$ in the benchmark scenario ${\bf T^{\scriptscriptstyle\{3\}}_{\scriptscriptstyle\{Z,Z'\}}}$ (with $\Gamma_{\scriptscriptstyle Z^{\prime}}^{\scriptscriptstyle\text{ToT}}/m_{\scriptscriptstyle Z^{\prime}}=3.42\%$).}
   \label{tabVp2Tevzp1}
  \end{table*} 

\begin{table*}[h!]
\boldmath
\centering
 \renewcommand{\arraystretch}{1.40}
 \setlength{\tabcolsep}{10pt}
 \begin{adjustbox}{width=18cm,height=3.0cm}
 \begin{tabular}
 {!{\vrule width 2pt}l!{\vrule width 2pt}l!{\vrule width 2pt}c:c!{\vrule width 2pt}c:c!{\vrule width 2pt}c:c!{\vrule width 2pt}c!{\vrule width 2pt}}
  \cline{1-8}
$\bf m_{\scriptscriptstyle Z^{\prime}}$& $\frac{\Gamma_{\scriptscriptstyle T}^{\scriptscriptstyle\text{ToT}}}{m_{\scriptscriptstyle T}}$ &\multicolumn{2}{c!{\vrule width 2pt}}{{\bf CM}}
&\multicolumn{2}{c!{\vrule width 2pt}}{{\bf NWA$_1$}}
&\multicolumn{2}{c!{\vrule width 2pt}}{{\bf NWA$_2$}}&\multicolumn{1}{c}{{}}\\
\cline{3-9}
 \textbf{[TeV]}& $[\%]$ &$\bf\sigma_{\scriptscriptstyle{\textbf{LO}}}[\textbf{fb}]$& $\bf\sigma_{\scriptscriptstyle{\textbf{NLO}}}[\textbf{fb}]$&$\bf\sigma_{\scriptscriptstyle{\textbf{LO}}}[\textbf{fb}]$& $\bf\sigma_{\scriptscriptstyle{\textbf{NLO}}}[\textbf{fb}]$&$\bf\sigma_{\scriptscriptstyle{\textbf{LO}}}[\textbf{fb}]$& $\bf\sigma_{\scriptscriptstyle{\textbf{NLO}}}[\textbf{fb}]$& $\frac{m_{\scriptscriptstyle T}}{m_{\scriptscriptstyle Z^{\prime}}}$\\
  \noalign{\hrule height 1pt}
  $1.5$&$0.19$ &$ 241.1\, ^{+7.1\%}_{-6.3\%}\, ^{+2.4\%}_{-2.4\%}$ & $274.0\, ^{+1.6\%}_{-1.9\%}\, ^{+2.4\%}_{-2.4\%}$
  & $239.2\, ^{+7.0\%}_{-6.3\%}\, ^{+2.4\%}_{-2.4\%}$ & $303.3\, ^{+3.2\%}_{-3.1\%}\, ^{+2.4\%}_{-2.4\%}$
  & $244.9\, ^{+7.0\%}_{-6.3\%}\, ^{+2.4\%}_{-2.4\%}$ & $284.8\, ^{+2.0\%}_{-2.2\%}\, ^{+2.5\%}_{-2.5\%}$ &\\
\cdashline{2-8}
   $2.5$&$0.53$ &$15.99\, ^{+10.8\%}_{-9.2\%}\, ^{+3.8\%}_{-3.8\%}$ & $18.58\, ^{+2.5\%}_{-3.1\%}\, ^{+3.9\%}_{-3.9\%}$
   &$15.77\, ^{+10.7\%}_{-9.1\%}\, ^{+3.8\%}_{-3.8\%}$ & $20.09\, ^{+3.8\%}_{-4.0\%}\, ^{+3.8\%}_{-3.8\%}$ 
   & $15.81\, ^{+10.6\%}_{-9.0\%}\, ^{+3.8\%}_{-3.8\%}$ &$18.87\, ^{+2.6\%}_{-3.2\%}\, ^{+3.9\%}_{-3.9\%}$&$\frac{1}{2}$\\
\cdashline{2-8}
  $3.5$&$1.04$ &$1.589\, ^{+13.4\%}_{-11.1\%}\, ^{+7.7\%}_{-7.7\%}$ & $1.912\, ^{+3.1\%}_{-4.0\%}\, ^{+7.6\%}_{-7.6\%}$
  &$1.561\, ^{+13.3\%}_{-11.0\%}\, ^{+7.6\%}_{-7.6\%}$ & $2.022\, ^{+4.4\%}_{-4.9\%}\, ^{+7.4\%}_{-7.4\%}$
  & $1.522\, ^{+13.2\%}_{-11.0\%}\, ^{+7.8\%}_{-7.8\%}$ &$1.863\, ^{+3.4\%}_{-4.2\%}\, ^{+7.3\%}_{-7.3\%}$&\\
 \noalign{\hrule height 1pt}
  $1.5$&$0.34$ &$235.3\, ^{+7.1\%}_{-6.4\%}\, ^{+2.4\%}_{-2.4\%}$ & $283.8\, ^{+3.1\%}_{-3.0\%}\, ^{+2.5\%}_{-2.5\%}$ 
  & $232.3\, ^{+7.1\%}_{-6.3\%}\, ^{+2.4\%}_{-2.4\%}$ & $ 310.3\,^{+3.6\%}_{-3.4\%}\, ^{+2.4\%}_{-2.4\%}$ 
  & $244.9\, ^{+7.0\%}_{-6.3\%}\, ^{+2.4\%}_{-2.4\%}$ & $285.0\, ^{+3.3\%}_{-3.1\%}\, ^{+2.5\%}_{-2.5\%}$ &\\
\cdashline{2-8}
    $2.5$&$0.94$ &$15.41\, ^{+10.9\%}_{-9.2\%}\, ^{+3.9\%}_{-3.9\%}$ & $18.60\, ^{+3.1\%}_{-3.5\%}\, ^{+3.9\%}_{-3.9\%}$ 
    & $15.08\, ^{+10.7\%}_{-9.1\%}\, ^{+3.9\%}_{-3.9\%}$ & $ 20.12\,^{+4.5\%}_{-4.5\%}\, ^{+3.8\%}_{-3.8\%}$ 
    & $15.81\, ^{+10.6\%}_{-9.1\%}\, ^{+3.8\%}_{-3.8\%}$ & $18.88\, ^{+2.6\%}_{-3.2\%}\, ^{3.8+\%}_{-3.8\%}$ &$\frac{2}{3}$\\
\cdashline{2-8}
  $3.5$&$1.85$ &$1.505\, ^{+13.6\%}_{-11.2\%}\, ^{+7.5\%}_{-7.5\%}$ & $1.862\, ^{+3.8\%}_{-4.5\%}\, ^{+7.1\%}_{-7.1\%}$ 
  & $1.464\, ^{+13.3\%}_{-11.0\%}\, ^{+7.6\%}_{-7.6\%}$ & $ 1.983\,^{+5.1\%}_{-5.3\%}\, ^{+7.7\%}_{-7.7\%}$ 
  & $1.521\, ^{+13.2\%}_{-11.0\%}\, ^{+8.2\%}_{-8.2\%}$ & $1.865\, ^{+3.4\%}_{-4.1\%}\, ^{+7.6\%}_{-7.6\%}$ &\\
 \noalign{\hrule height 1pt}
  $1.5$&$0.53$ &$239.5\, ^{+7.3\%}_{-6.5\%}\, ^{+2.5\%}_{-2.5\%}$ & $327.1\, ^{+4.0\%}_{-3.7\%}\, ^{+2.5\%}_{-2.5\%}$ 
  & $ 231.6\,^{+7.2\%}_{-6.4\%}\, ^{+2.5\%}_{-2.5\%}$ & $ 349.7\,^{+4.9\%}_{-4.4\%}\, ^{+2.4\%}_{-2.4\%}$ 
  & $244.2\, ^{+7.0\%}_{-6.3\%}\, ^{+2.4\%}_{-2.4\%}$ & $285.2\, ^{+2.0\%}_{-2.2\%}\, ^{+2.4\%}_{-2.4\%}$ &\\
\cdashline{2-8}
    $2.5$&$1.47$ &$15.59\, ^{+11.0\%}_{-9.3\%}\, ^{+4.0\%}_{-4.0\%}$ & $20.45\, ^{+4.0\%}_{-4.2\%}\, ^{+4.1\%}_{-4.1\%}$ 
    & $14.63\,^{+10.7\%}_{-9.2\%}\, ^{+4.0\%}_{-4.0\%}$ & $20.98\,^{+5.5\%}_{-5.3\%}\, ^{+3.9\%}_{-3.9\%}$ 
    & $15.84\, ^{+10.6\%}_{-9.0\%}\, ^{+3.9\%}_{-3.9\%}$ & $18.86\, ^{+2.6\%}_{-3.2\%}\, ^{+3.8\%}_{-3.8\%}$ & $\frac{5}{6}$\\
\cdashline{2-8}
  $3.5$&$2.88$ &$1.534\, ^{+13.7\%}_{-11.3\%}\, ^{+8.4\%}_{-8.4\%}$ & $1.973\, ^{+5.0\%}_{-5.3\%}\, ^{+7.9\%}_{-7.9\%}$ 
  & $1.384\,^{+13.3\%}_{-11.1\%}\, ^{+8.2\%}_{-8.2\%}$ & $1.990\,^{+6.3\%}_{-6.1\%}\, ^{+7.9\%}_{-7.9\%}$ 
  & $1.521\, ^{+13.2\%}_{-11.0\%}\, ^{+8.1\%}_{-8.1\%}$ & $1.866\, ^{+3.4\%}_{-4.2\%}\, ^{+7.6\%}_{-7.6\%}$ &\\
 \noalign{\hrule height 1pt}
\end{tabular}
\end{adjustbox}
  \caption{\footnotesize $\sigma_{\scriptscriptstyle{\textbf{LO}}}$ and $\sigma_{\scriptscriptstyle{\textbf{NLO}}}$ for the reaction $pp\rightarrow t\bar{t}H+t\bar{t}Z$ in the benchmark scenario ${\bf T^{\scriptscriptstyle\{3\}}_{\scriptscriptstyle\{Z,H,Z'\}}}$ (with $\Gamma_{\scriptscriptstyle Z^{\prime}}^{\scriptscriptstyle\text{ToT}}/m_{\scriptscriptstyle Z^{\prime}}=3.42\%$).}
   \label{tabVp2Tevzp2}
  \end{table*}

\noindent
We note that, in all the results presented in tables~\ref{tabVp2Tevzp1} and \ref{tabVp2Tevzp2}, the widths of the unstable particles $\zp$ and $T$ are narrow ($\Gamma_{_{Z^{\prime}}}^{\text{ToT}}/m_{_{Z^{\prime}}}=3.42\%$ and $\Gamma_{_{T}}^{\text{ToT}}/m_{_{T}}$ varies between $0.19\%-2.88\%$). This ensures the consistency of CM scheme predictions, as explained in section~\ref{sec2}. We observe that the results differ by few percent across the 3 approximations, with NWA$_2$ and CM predictions being much closer, especially in the low and middle-mass cases. Consequently, one can take, with caution, $\sigma_{\scriptscriptstyle \text{NLO}}^{\text{NWA}_2}\approx  \sigma_{\scriptscriptstyle \text{NLO}}^{\text{CM}}$ for these configurations in both benchmark scenarios ${\bf T^{\scriptscriptstyle\{3\}}_{\scriptscriptstyle\{Z,Z'\}}}$ and ${\bf T^{\scriptscriptstyle\{3\}}_{\scriptscriptstyle\{Z,H,Z'\}}}$ (the discrepancy is less than 3\% for most cases).
The situation changes significantly for the high-mass case, $m_{\scriptscriptstyle{T}}/m_{\scriptscriptstyle{\scriptscriptstyle{Z^{\prime}}}}=5/6$. Here, the relative differences are much larger, exceeding the $10\%$ at NLO, which is not negligible. This indicates that off-shell effects cannot be ignored in this regime, rendering the NWA unreliable. Instead, one must employ the CM scheme, cf. appendix~\ref{appC}.\\ 

\noindent
The key conclusion from the previous discussion is that if the mass is close to the decay threshold of the $\zp$ into $Tt$, we recommend using the CM scheme which is, available at LO and NLO for our model. Conversely, if the mass is far from the decay threshold, one may use NWA$_2$ at NLO for quick estimation of the total cross section. However, CM scheme remains the most accurate and reliable for all configurations.\\

\noindent
To conclude this section, we emphasize that this model stands out as a robust candidate for constraining the masses of the $W^{\prime}$ and $\zp$ boson decaying into VLQ. It offers a significant advantage over other models by requiring fewer free parameters (only two in DD scenario) and facilitates the NLO predictions, which provide more precise results. This is achieved through the dedicated {\tt UFO} model {\tt vlQBp}.

%% file: summary.tex
\section{Summary and prospects}
\label{Conc}

\noindent
In the present paper, we have introduced an extension of the SM that includes a singlet VLQ, the heavy gauge bosons $W^{\prime}, Z^{\prime}$ and the scalar $H^{\prime}$, so that their mixing with SM particles is model-independent. VLQs are assumed to mix with SM particles via the Yukawa interaction, whereas we assumed that they mix only with 3rd generation quarks in the phenomenological part of this work. The main idea behind the parametrization of the couplings, in this approach, is to express them in terms of the branching ratios of the new heavy particles and the kinematic functions related to their partial decays. In order to be as general as possible in the derivation of the couplings, we considered two mass hierarchy configurations: {\it (i)} the VLQ is heavier than the supplementary bosons and {\it (ii)} the extra bosons are heavier than the VLQ. To remain as model-independent as possible, we did not include the 3- and 4-vertex of the heavy gauge bosons and  scalars, nor the weak and electromagnetic parts in the covariant derivative associated with the VLQ. The number of the extra free parameters depends on the studied scenario. In benchmark scenarios where the VLQ mixes with 3rd generation quarks via $Z$, $H$ and $W^{\prime}$ or $Z^{\prime}$ (${\bf T^{\scriptscriptstyle\{3\}}_{\scriptscriptstyle\{Z,H,W^{\prime}\}}}$ and ${\bf T^{\scriptscriptstyle\{3\}}_{\scriptscriptstyle\{Z,H,Z^{\prime}\}}}$), this number is five and it can be reduced to three if the NWA is valid, as discussed in section~\ref{sec3}.\\

\noindent
The model couplings are theoretically constrained by the perturbative unitarity procedure, which requires the highest eigenvalues of the partial-wave amplitudes to be less than unity. Restrictive bounds have been set on the parameters normalizing the couplings  for VLQ and heavy bosons masses explorable at the LHC, showing that the perturbative treatment is valid for our phenomenological study. This model is renormalized at one-loop in QCD using the automatic tools {\tt FeynRules/NLCT/FeynArts}. Two versions of the model, in the 4- and 5- flavor-schemes, are generated in {\tt UFO} format and are available for calculation at NLO in the complex mass scheme for narrow widths. This model is validated numerically in several ways: \textit{(i)} comparison with SM counterpart processes, \textit{(ii)} using {\tt MadGraph5} CM scheme automatic check, \textit{(iii)} reproduction of the results with extended version of {\tt loop\_qcd\_qed\_sm\_Gmu} and \textit{(iv)} comparison with existing models, as detailed  in section~\ref{sec2}. We have shown that it provides an excellent approximation to the CM scheme for narrow widths. We plan to extend the model implementation for broad widths in future work. The {\tt FeynRules} inputs and the two versions of the {\tt UFO} models are publicly available at \url{https://github.com/mszidi/VLQ_Wp_Zp_Hp_NLO}.\\

\noindent
In the phenomenological part of this work, we highlighted the case where the masses of the bosons $W^{\prime}$ and $Z^{\prime}$ are larger than the mass of the vector-like quark $T$ ($m_{\scriptscriptstyle{T}}=1/2, 2/3, 3/4\, m_{\scriptscriptstyle{W^{\prime}}}$ and $m_{\scriptscriptstyle{T}}=1/2, 2/3, 5/6\, m_{\scriptscriptstyle{Z^{\prime}}}$). This means that these bosons can decay into a VLQ and a 3rd generation quark, where the VLQ can decay only to SM particles. In the 4FS, we studied the production of the gauge bosons $\wps$ and $\zp$, in proton-proton collision, which then decay to top partner $T$ at center-of-mass energy $\sqrt{s}=13$ TeV, where several benchmark scenarios were explored. We find that for LHC mass range, the perturbative treatment is always available in this model.  We calculated the cross sections at LO and NLO in QCD by adopting 3 approaches: the complex-mass scheme (full calculation), the narrow-width approximation (where $\wps/\zp$ and $T$ are treated as on-shell states), and a mixture of both (where only the $T$ is considered as resonant state).\\

\noindent
We find that the LO predictions from CM scheme, NWA$_1$ and NWA$_2$ are nearly identical, with
acceptable relative differences for both $\wps$ or $\zp$ production, especially if their masses are far from their decay thresholds to VLQ. However, the NLO predictions for NWA$_1$ and NWA$_2$ are
not as close to CM scheme predictions in certain configurations. 
We demonstrated that the relative differences between CM and NWA$_2$ predictions are smaller compared to those between CM and NWA$_1$. For the special case (DD scenario with $\vkwp=0.5$), we showed that the NWA$_2$ and CM scheme predictions are in close agreement at both LO and NLO. However, we found that the discrepancy between CM scheme and NWA$_1$ predictions can become significant (and  large sometimes), when the mass of the gauge boson is close to the decay threshold, particularly in the case of $\zp$ production. Thus, the  NWA$_2$ can be used for a quick estimation of the total cross section, as it is fast and leads to results reasonably comparable to the full calculation (the CM scheme) for the lower- and middle-mass cases. However the CM scheme remains the most accurate method and is therefore recommended for all the mass hierarchies. Fortunately, the full calculation within CM is available in our implementation.\\

\noindent
We have demonstrated that our model is not excluded by {\tt run II} results and that it represents a promising framework for the search of VLQs and heavy gauge bosons. We have shown that for $\kpt=0.1$, the lower $\wps$ mass limit is set between $2.8-3.9$ TeV for $\vkwp$ ranging between 0.3 and 0.8 in DD scenario. We note that this model requires fewer free parameters, only 2 in the DD scenario. Thus, it constitutes a robust candidate for constraining the masses of the $W^{\prime}$ and $Z^{\prime}$ bosons decaying into VLQ. Thanks to its publicly available {\tt UFO} NLO implementation, which directly facilitates accurate NLO predictions. \\

\noindent
This model allows for the existence of a scalar particle $\hp$ capable to interact simultaneously with both SM particles and VLQs. This type of particles play a major role in BSM physics. In this work, we did not include the phenomenological study of such particle at the LHC due to the length of this article. We therefore postpone this part of the work to a separate publication.\\

\noindent
Complete the model by including the missing peaces such as interaction (and self-interaction) between the different gauge bosons and scalars, taking into account the weak and electromagnetic components in the covariant derivative associated with VLQs \ldots, etc. This will restore partial-wave unitarity for processes involving real heavy gauge bosons in the initial or final states, like some processes discussed in appendix~\ref{appB}, and would be a step towards one-loop electroweak renormalization of the model. We aim to achieve this in the near future.

%% file: appendix_A.tex
\section{More detail on the model independent parametrization}
\label{appA}

In this appendix, we derive the model independent parametrization of the mixing of a singlet vector like-quark ($Q$) to ordinary quarks via the SM bosons ($W$, $Z$ and $H$), two new heavy vector-bosons ($\wps$, $\zp$) and a new heavy scalar ($\hp$). 
We recall that only one VLQ of type top ($Q\equiv T$) or bottom ($Q\equiv B$) which is present in this model and not both, and it is assumed to mix only with one quark generation.
 To be more general, we consider all possible mass hierarchies between the VLQs and the extra heavy bosons, on which our parametrization depends~\footnote{Based on several ATLAS and CMS publications, it is more experimentally realistic to assume that $Q$ are lighter than $W', Z'$.
}.
\subsection{Case 1: $m_{\scriptscriptstyle{Q}}>m_{\scriptscriptstyle{B^{\prime}}}+m_{\scriptscriptstyle{q}}$}
\label{appA1}

In this case, the vector-like top quark $Q$ is the heaviest particle in the model, so it can decay to an ordinary quark and one of the bosons $B^{\prime\prime}=W, \wps, Z, \zp, H, \hp$ according to the following decay modes:
\begin{align}
 Q&\rightarrow qH, & Q&\rightarrow qZ, & Q&\rightarrow q^{\prime}W\nonumber\\
 Q&\rightarrow q\hp, & Q&\rightarrow q\zp, & Q&\rightarrow q^{\prime}\wps
\end{align}
where $q$ and $q^{\prime}$ denote, respectively, {\it up} and {\it down}-type quarks of the same generation for $Q\equiv T$, and {\it down} and {\it up}-type for $Q\equiv  B$.
Assuming that only left or right handed coupling chirality (and not both) which contributes, then the partial width for each channel can be written as:
\begin{align}
\Gamma(Q\rightarrow \bpp q)&=\left(\ktbpplr\right)^2 \frac{g^2m_{\scriptscriptstyle{Q}}^3}{64\pi m_{\scriptscriptstyle{W}}^2}\, \, \, \Gamma_{\scriptscriptstyle{B^{\prime\prime}}}(m_{\scriptscriptstyle{Q}}, m_{\scriptscriptstyle{B^{\prime\prime}}}, m_{\scriptscriptstyle{q}})
\label{widthgen}
\end{align}
where $m_{\scriptscriptstyle{B^{\prime\prime}}}$, $m_{\scriptscriptstyle{Q}}$ and $m_{\scriptscriptstyle{q}}$ stand, respectively, for the masses of the boson $\bpp$, the VLQ $Q$ and the SM quark $q$ \footnote{Typically, $m_{\scriptscriptstyle{q}}$ is neglected, except when mixing with the 3\textsuperscript{rd} generation, where it may take the value of $m_{\scriptscriptstyle{t}}$ or $m_{\scriptscriptstyle{b}}$ (in  4FS).}. The parameters $\ktbpplr$ are related to the parameters $\kbpplr$ of the Lagrangian $\call_{\scriptscriptstyle Q}$, cf. eq. (\ref{lQ}), by:
\begin{align}
\ktwlr&=\kwlr &
\ktwplr&=\vkwplr\,\,\,\frac{m_{\scriptscriptstyle{W}}}{m_{\scriptscriptstyle{W^{\prime}}}}& \kthplr&=\khplr\,\,\,\frac{v}{m_{\scriptscriptstyle{Q}}}\nonumber\\
\ktzlr&=\kzlr &\ktzplr&=\vkzplr\,\,\,\frac{m_{\scriptscriptstyle{Z}}}{m_{\scriptscriptstyle{Z^{\prime}}}}&
\kthlr&=\khlr\,\,\,\frac{v}{m_{\scriptscriptstyle{Q}}} & 
\end{align}
and $\Gamma_{\scriptscriptstyle{B^{\prime\prime}}}$ are the kinematic functions, they are given by:
\begin{align}
\Gamma_{\scriptscriptstyle{V}}(m_{\scriptscriptstyle{Q}}, m_{\scriptscriptstyle{V}}, m_{\scriptscriptstyle{q}})&=
\lambda^{\frac{1}{2}}\biggl(1,\frac{m_{\scriptscriptstyle{V}}^2}{m_{\scriptscriptstyle{Q}}^2},\frac{m_{\scriptscriptstyle{q}}^2}{m_{\scriptscriptstyle{Q}}^2}\biggr)\biggl[\biggl(1-\frac{m_{\scriptscriptstyle{q}}^2}{m_{\scriptscriptstyle{Q}}^2}\biggr)^2+\frac{m_{\scriptscriptstyle{V}}^2}{m_{\scriptscriptstyle{Q}}^2}-2\frac{m_{\scriptscriptstyle{V}}^4}{m_{\scriptscriptstyle{Q}}^4}+\frac{m_{\scriptscriptstyle{q}}^2m_{\scriptscriptstyle{V}}^2}{m_{\scriptscriptstyle{Q}}^4}\biggr]/\gamma_{\scriptscriptstyle{V}}\nonumber\\
\Gamma_{\scriptscriptstyle{S}}(m_{\scriptscriptstyle{Q}}, m_{\scriptscriptstyle{S}}, m_{\scriptscriptstyle{q}})&=
\lambda^{\frac{1}{2}}\biggl(1,\frac{m_{\scriptscriptstyle{S}}^2}{m_{\scriptscriptstyle{Q}}^2},\frac{m_{\scriptscriptstyle{q}}^2}{m_{\scriptscriptstyle{Q}}^2}\biggr)\biggl[1+\frac{m_{\scriptscriptstyle{q}}^2}{m_{\scriptscriptstyle{Q}}^2}-\frac{m_{\scriptscriptstyle{S}}^2}{m_{\scriptscriptstyle{Q}}^2}\biggr]/2
\label{kinematic1}
\end{align}
where $V$ and $S$ stand for vector and scalar bosons, $\gamma_{\scriptscriptstyle{W}}=\gamma_{\scriptscriptstyle{W^{\prime}}}=1$, $\gamma_{\scriptscriptstyle{Z}}=\gamma_{\scriptscriptstyle{Z^{\prime}}}=2$ and $\lambda(x,y,z)=x^2+y^2+z^2-2xy-2xz-2yz$.

\noindent
To express the partial decay width of $Q$ into scalars as in the general formula given in eq. (\ref{widthgen}), one has to replace the SM Higgs {\it vev} by $v=2m_{\scriptscriptstyle{W}}/g$.  We note that the {\it vev} of the new scalar ($v^{\prime}$) is hidden in the parameter $\kthplr$, i.e. $\kthplr= (v/v^{\prime})\ktthplr$, where $\ktthplr$ is a free parameter normalizing the coupling $H^{\prime}-Q-q$ as: $\ktthplr\, m_{\scriptscriptstyle Q}/v^{\prime}$. We employ this trick to remain as independent as possible of the BSM model (which might contains other extra neutral or charged scalars).\\

\begin{figure}[h!]
\begin{center}
\subfloat[\label{xiQvars1}]{\includegraphics[width=8.0cm,height=5cm]{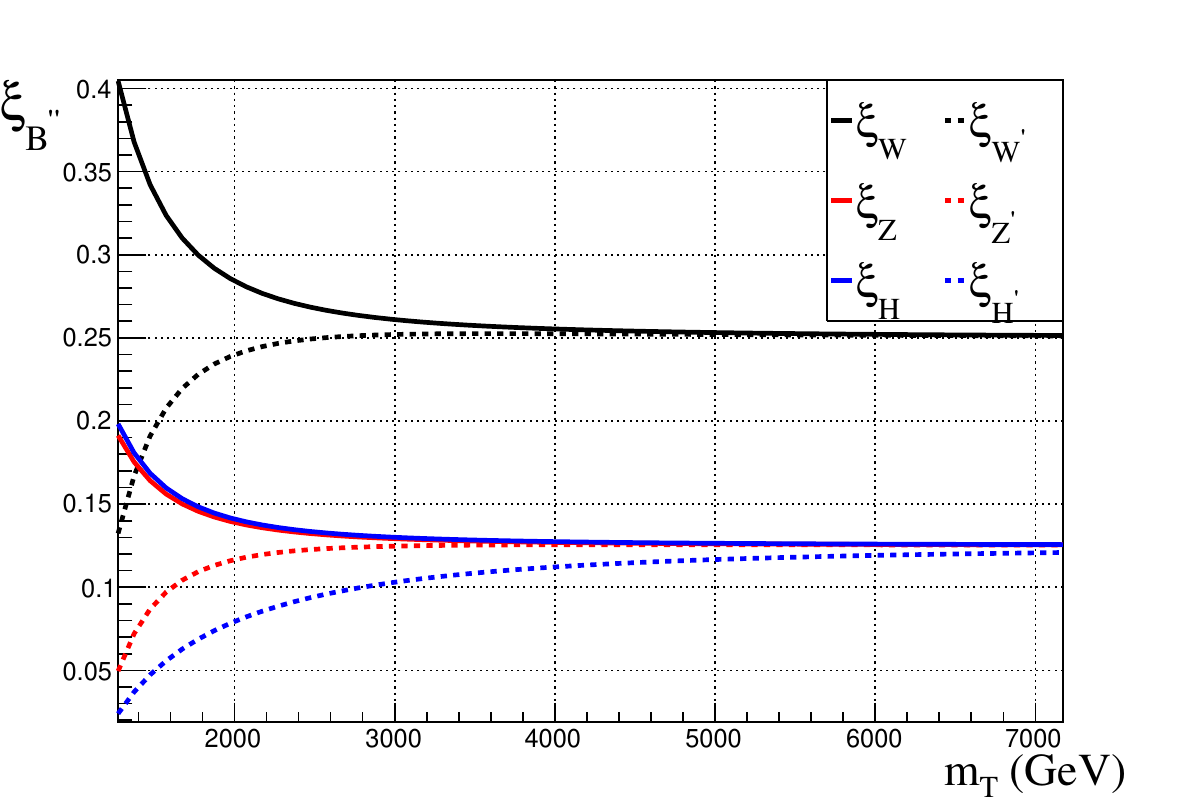}}
\subfloat[\label{xiQvars2}]{\includegraphics[width=8.0cm,height=5cm]{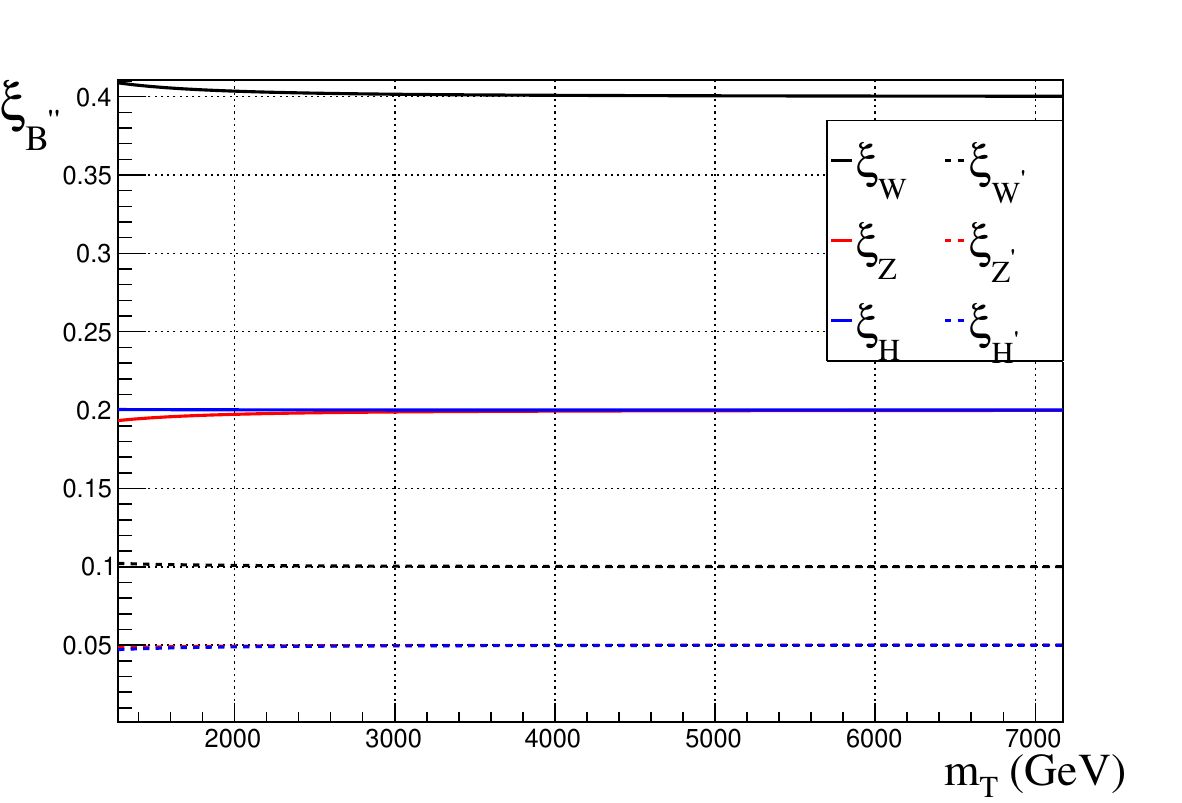}}
\vspace{-0.5cm}
\caption{\small Variation of $\xi_{\scriptscriptstyle B^{\prime\prime}}$ in term of $m_{\scriptscriptstyle{Q}}$ ($Q\equiv T$) for $m_{\scriptscriptstyle{W^{\prime},Z^{\prime},H^{\prime}}}=1000\, \text{GeV}$ and $\ktbpplr=1$ (left) and for $\ktwlr=\ktzlr=\kthlr=1$ and $\ktwplr=m_{\scriptscriptstyle{W}}/m_{\scriptscriptstyle{W^{\prime}}}$, $\ktzplr=m_{\scriptscriptstyle{Z}}/m_{\scriptscriptstyle{Z^{\prime}}}$ and $\kthplr=v/v^{\prime}$ (right).}
\label{brvar}
\end{center}
\end{figure}

\noindent
For mixing with one quark generation, the branching ratio of the decay $Q\rightarrow \bpp q$ is defined as follows:
\begin{align}
\text{Br}(Q\rightarrow B^{\prime\prime}q)&=\frac{(\ktbpplr)^2 \, \, \, \Gamma_{\scriptscriptstyle{B^{\prime\prime}}}(m_{\scriptscriptstyle{Q}}, m_{\scriptscriptstyle{B^{\prime\prime}}}, m_{\scriptscriptstyle{q}})}{\underset{\begin{subarray}{l}\scriptscriptstyle{B^{\prime}=W,W^{\prime},}\\\scriptscriptstyle{Z,Z^{\prime},H,H^{\prime}}\end{subarray}}{\sum}(\ktbplr)^2 \, \, \, \Gamma_{\scriptscriptstyle{B^{\prime}}}(m_{\scriptscriptstyle{Q}}, m_{\scriptscriptstyle{B^{\prime}}}, m_{\scriptscriptstyle{q}})}\equiv\xi_{\scriptscriptstyle{B^{\prime\prime}}}
\label{br}
\end{align}
We observe that in the asymptotic limit ($m_{\scriptscriptstyle{Q}}\rightarrow\infty$) and if all $\ktbpplr$ are equal, the VLQ decays 50\% to charged gauge bosons (25\% to $W$ and 25\% to $\wps$) and 25\% to neutral gauge bosons (12.5\% to $Z$ and 12.5\% to $\zp$) and 25\% to scalars (12.5\% to $H$ and 12.5\% to $\hp$), which is in agreement with Goldstone equivalence theorem\footnote{The Goldstone equivalence theorem states that at high energies, the amplitude of the emission or absorption of longitudinally polarized massive gauge boson is equal to the amplitude of the emission or absorption of the corresponding Goldstone boson. In our case, this translates to: $\xi_{W}=2\,\xi_{Z}=2\,\xi_{H}$ and $\xi_{W^{\prime}}=2\,\xi_{Z^{\prime}}=2\,\xi_{H^{\prime}}$ for $m_{\scriptscriptstyle{Q}}\rightarrow\infty$. }, see sub.fig.~\ref{xiQvars1}. To show this, we set the fraction of the masses on the functions $\Gamma_{\scriptscriptstyle{B^{\prime\prime}}}$ to zero, i.e. $m_{\scriptscriptstyle{q}}^2/m_{\scriptscriptstyle{Q}}^2$,  $m_{\scriptscriptstyle{B^{\prime\prime}}}^2/m_{\scriptscriptstyle{Q}}^2\rightarrow 0$ which is justified for very large $m_{\scriptscriptstyle{Q}}$  ($m_{\scriptscriptstyle{Q}}\gg m_{\scriptscriptstyle{B^{\prime\prime}}}$ and $m_{\scriptscriptstyle{Q}}\gg m_{\scriptscriptstyle{q}}$), we get
\begin{align}
\Gamma_{\scriptscriptstyle{W}}|_{\scriptscriptstyle{m_{\scriptscriptstyle{Q}}\gg}}&\approx\Gamma_{\scriptscriptstyle{W}}^0=1&
\Gamma_{\scriptscriptstyle{W^{\prime}}}|_{\scriptscriptstyle{m_{\scriptscriptstyle{Q}}\gg}}&\approx\Gamma_{\scriptscriptstyle{W^{\prime}}}^0=1&\Gamma_{\scriptscriptstyle{H}}|_{\scriptscriptstyle{m_{\scriptscriptstyle{Q}}\gg}}&\approx\Gamma_{\scriptscriptstyle{H}}^0=1/2\nonumber\\
\Gamma_{\scriptscriptstyle{Z}}|_{\scriptscriptstyle{m_{\scriptscriptstyle{Q}}\gg}}&\approx\Gamma_{\scriptscriptstyle{Z}}^0=1/2&
\Gamma_{\scriptscriptstyle{Z^{\prime}}}|_{\scriptscriptstyle{m_{\scriptscriptstyle{Q}}\gg}}&\approx\Gamma_{\scriptscriptstyle{Z^{\prime}}}^0=1/2&
\Gamma_{\scriptscriptstyle{H^{\prime}}}|_{\scriptscriptstyle{m_{\scriptscriptstyle{Q}}\gg}}&\approx\Gamma_{\scriptscriptstyle{H^{\prime}}}^0=1/2
\label{kinematic2}
\end{align}
It is clear, in this limit, that $\xi_{W,W^{\prime}}=2\,\xi_{Z,Z^{\prime}}=2\,\xi_{H,H^{\prime}}=1/4$. We notice that the latter result is not true if $\ktbpplr$ are not equal even for very large $m_{\scriptscriptstyle{Q}}$. For example, if we take $\ktwlr=\ktzlr=\kthlr=1$ and $\ktwplr=m_{\scriptscriptstyle{W}}/m_{\scriptscriptstyle{W^{\prime}}}$, $\ktzplr=m_{\scriptscriptstyle{Z}}/m_{\scriptscriptstyle{Z^{\prime}}}$ and $\kthplr=v/v^{\prime}$, and if we assume that $m_{\scriptscriptstyle{W}}/m_{\scriptscriptstyle{W^{\prime}}}=m_{\scriptscriptstyle{Z}}/m_{\scriptscriptstyle{Z^{\prime}}}=v/v^{\prime}$, we find that $\xi_{W}=2\,\xi_{Z}=2\,\xi_{H}$ and $\xi_{W^{\prime}}=2\,\xi_{Z^{\prime}}=2\,\xi_{H^{\prime}}$, where 
\begin{align}
\xi_{\scriptscriptstyle{W}}&=\frac{1}{2(1+v^2/v^{\prime\, 2})}&
\xi_{\scriptscriptstyle{W^{\prime}}}&=\frac{v^2/v^{\prime\, 2}}{2(1+v^2/v^{\prime\, 2})}
\end{align}
then for $v^{\prime}=2\, v$, we get $\xi_{W}=2\,\xi_{Z}=2\,\xi_{H}=40\%$ and $\xi_{W^{\prime}}=2\,\xi_{Z^{\prime}}=2\,\xi_{H^{\prime}}=10\%$, see sub.figure~\ref{xiQvars2}.

\noindent
In the left panel of figure~\ref{brvar}, we show the evolution of the branching ratios of $Q$ as a function of $m_{\scriptscriptstyle{Q}}$ for $m_{\scriptscriptstyle{W^{\prime},Z^{\prime},H^{\prime}}}=1$ TeV and equal $\ktbpplr$. We observe that when $m_{\scriptscriptstyle{Q}}$ is of the same order as $m_{\scriptscriptstyle{W^{\prime},Z^{\prime},H^{\prime}}}$, the branching ratios $\xi_{\scriptscriptstyle{W,Z,H}}$ dominate over $\xi_{\scriptscriptstyle{W^{\prime},Z^{\prime},H^{\prime}}}$. However, they become equal for large $m_{\scriptscriptstyle{Q}}$ as expected (i.e. $\xi_{W}=\xi_{W'}$, $\xi_{Z}=\xi_{Z'}$ and $\xi_{H}=\xi_{H'}$). In the right panel, we show the variation of the branching ratio for $\ktwlr=\ktzlr=\kthlr=1$ with $\ktwplr=m_{\scriptscriptstyle{W}}/m_{\scriptscriptstyle{W^{\prime}}}$, $\ktzplr=m_{\scriptscriptstyle{Z}}/m_{\scriptscriptstyle{Z^{\prime}}}$ and $\kthplr=v/v^{\prime}$. Here, we assume $m_{\scriptscriptstyle{W}}/m_{\scriptscriptstyle{W^{\prime}}}=m_{\scriptscriptstyle{Z}}/m_{\scriptscriptstyle{Z^{\prime}}}=v/v^{\prime}$ and $v^{\prime}=2\, v$. For large $m_{\scriptscriptstyle{Q}}$, we find $\xi_{W}=2\,\xi_{Z}=2\,\xi_{H}\approx40\%$ and $\xi_{W^{\prime}}=2\,\xi_{Z^{\prime}}=2\,\xi_{H^{\prime}}\approx10\%$, which is consistent with our expectations.\\

\noindent
The key motivation behind this parametrization is to treat the branching ratios as free parameters rather than as functions of particle masses and couplings, as we done previously, cf. figure~\ref{brvar}. So, the idea is to express the parameters $\ktbpplr$ in terms of $\xi_{\scriptscriptstyle{B^{\prime\prime}}}$ and the kinematic functions $\Gamma_{\scriptscriptstyle{B^{\prime\prime}}}$. To achieve this, we introduce a new free parameter $\kpq$, defined such that $\kpq^2=\sum_{\scriptscriptstyle{B^{\prime}}}(\ktbplr)^2 \, \Gamma_{\scriptscriptstyle{B^{\prime}}}$. This modifies eq.~(\ref{br}) as follows:
\begin{align}
\xi_{\scriptscriptstyle{B^{\prime\prime}}}&=\frac{(\ktbpplr)^2 \, \, \, \Gamma_{\scriptscriptstyle{B^{\prime\prime}}}(m_{\scriptscriptstyle{Q}}, m_{\scriptscriptstyle{B^{\prime\prime}}}, m_{\scriptscriptstyle{q}})}{\kpq^2}
\label{br1}
\end{align}
Then, the parameters $\ktbpplr$ can be easily expressed as:
\begin{align}
 \ktbpplr&=\kpq\sqrt{\frac{\xi_{\scriptscriptstyle{B^{\prime\prime}}}}{\Gamma_{\scriptscriptstyle{B^{\prime\prime}}}}}
 &
 \text{with}
 &&
 \underset{\begin{subarray}{l}\scriptscriptstyle{B^{\prime\prime}=W,W^{\prime},}\\\scriptscriptstyle{Z,Z^{\prime},H,H^{\prime}}\end{subarray}}{\sum}\xi_{\scriptscriptstyle{B^{\prime\prime}}}&=1
 \label{couplingcase1}
\end{align}
The original parameters $\kbpplr$ of the Lagrangian (\ref{lQ}) are then given by:
\begin{align}
\kwlr&=\kpq\sqrt{\frac{\xi_{\scriptscriptstyle{W}}}{\Gamma_{\scriptscriptstyle{W}}}}&
\kwplr&=\kpq\frac{m_{\scriptscriptstyle{W^{\prime}}}}{m_{\scriptscriptstyle{W}}}\sqrt{\frac{\xi_{\scriptscriptstyle{W^{\prime}}}}{\Gamma_{\scriptscriptstyle{W^{\prime}}}}}&\khlr&=\kpq\frac{m_{\scriptscriptstyle{Q}}}{v}\sqrt{\frac{\xi_{\scriptscriptstyle{H}}}{\Gamma_{\scriptscriptstyle{H}}}}\nonumber\\
\kzlr&=\kpq\sqrt{\frac{\xi_{\scriptscriptstyle{Z}}}{\Gamma_{\scriptscriptstyle{Z}}}}&
\kzplr&=\kpq\frac{m_{\scriptscriptstyle{Z^{\prime}}}}{m_{\scriptscriptstyle{Z}}}\sqrt{\frac{\xi_{\scriptscriptstyle{Z^{\prime}}}}{\Gamma_{\scriptscriptstyle{Z^{\prime}}}}}&
\khplr&=\kpq\frac{m_{\scriptscriptstyle{Q}}}{v}\sqrt{\frac{\xi_{\scriptscriptstyle{H^{\prime}}}}{\Gamma_{\scriptscriptstyle{H^{\prime}}}}}
\label{param1}
\end{align}
where $\kpq$ is treated as free parameter which can be determined experimentally.\\

\noindent
As pointed out in refs. \cite{model-indep-vlqs-3,Cacciapaglia:2011fx,model-indep-vlqs-4}, for singlet VLQ only the left-handed mixing angle is the dominant (the right-handed one is suppressed by a factor of $m_q/m_T$). So, we choose to set all the right-handed couplings to zero except those associated to the scalar sector in the case of $Q\equiv T$ which mixes only with 3\textsuperscript{rd} quark generation. In this case, we keep both coupling chiralities since the smaller one is proportional to the top quark mass ($m_t/v$) which is not negligible especially for relatively small $m_{\scriptscriptstyle{T}}$. So, the Lagrangian (\ref{lQ}) describing the interactions of $Q$ with the ordinary quarks via the six bosons is expressed in terms of the 10 independent parameters: $\xi_{\scriptscriptstyle{Z}}, \xi_{\scriptscriptstyle{Z^{\prime}}},\xi_{\scriptscriptstyle{W}}, \xi_{\scriptscriptstyle{W^{\prime}}}, \xi_{\scriptscriptstyle{H}}$, $m_{\scriptscriptstyle{Q}}, m_{\scriptscriptstyle{W^{\prime}}}, m_{\scriptscriptstyle{Z^{\prime}}}, m_{\scriptscriptstyle{H^{\prime}}}$ and $\kpq$ in the general benchmark scenario, denoted by ${\bf Q}^{\scriptscriptstyle\{3\}}_{\scriptscriptstyle\{W, W^{\prime}, Z, Z^{\prime}, H, H^{\prime}\}}$, where all the new hypothetical particles are involved in the model.

\vspace{0.5cm}
\noindent
At first glance, one might think that the couplings of the new heavy bosons with VLQ, cf. eq.~(\ref{param1}), become very large near $Q$ decay thresholds (and diverge at $m_{\scriptscriptstyle{Q}}=m_{\scriptscriptstyle{B^{\prime}}}+m_{\scriptscriptstyle{q}}$, for $B^{\prime}=W^{\prime}, Z^{\prime}, H^{\prime}$) because of the kinematic functions $\Gamma_{\scriptscriptstyle{B^{\prime}}}$, which become very small (and vanishes at $m_{\scriptscriptstyle{Q}}=m_{\scriptscriptstyle{B^{\prime}}}+m_{\scriptscriptstyle{q}}$). Actually, this is not true since in this region also the branching ratios $\xi_{\scriptscriptstyle{B^{\prime}}}$ becomes very small (and vanishes at $m_{\scriptscriptstyle{Q}}=m_{\scriptscriptstyle{B^{\prime}}}+m_{\scriptscriptstyle{q}}$), see the left panel in figure~\ref{brvar}. In practice, the masses of VLQs and the hypothetical heavy bosons are most commonly assumed different, so we rarely approach the dangerous region near the decay threshold. This means that the parametrization is safe for realistic phenomenological studies. However, we can slightly modify it to remain valid even in this problematic region, making it more general.

\noindent
Let's assume that $\Gamma_{\scriptscriptstyle{B^{\prime}}}\propto \xi_{\scriptscriptstyle{B^{\prime}}}$, then we write
\begin{align}
 \xi_{\scriptscriptstyle{B^{\prime}}}&= \alpha_{\scriptscriptstyle Q}^2\, \Gamma_{\scriptscriptstyle{B^{\prime}}}.
 \label{xigam}
\end{align}
where $\alpha_{\scriptscriptstyle Q}$ is a proportionality factor which we assume that it is independent of the ratio $m_{\scriptscriptstyle{B^{\prime}}}/m_{\scriptscriptstyle{Q}}$\footnote{In general, $\alpha_{\scriptscriptstyle Q}$ might be a regular function in $m_{\scriptscriptstyle{Q}}$. However, it can also  be a constant (i.e. independent of $m_{\scriptscriptstyle{Q}}$) if $\kpq$  and $\ktbplr$ are independent of such mass as we will show later on.}. Because eq. (\ref{xigam}) is always satisfied whatever the masses $m_{\scriptscriptstyle Q}$ and $m_{\scriptscriptstyle{B^{\prime}}}$, then it is more practical to write
\begin{align}
 \alpha_{\scriptscriptstyle Q}^2&=\frac{\xi_{\scriptscriptstyle{B^{\prime}}}}{\Gamma_{\scriptscriptstyle{B^{\prime}}}}=\frac{\xi_{\scriptscriptstyle{B^{\prime}}}^{\scriptscriptstyle0}}{\Gamma_{\scriptscriptstyle{B^{\prime}}}^{\scriptscriptstyle0}}\equiv \text{constant}.
 \label{alphaQ}
\end{align}
where $\xi_{\scriptscriptstyle{B^{\prime}}}^{\scriptscriptstyle0}$ and $\Gamma_{\scriptscriptstyle{B^{\prime}}}^{\scriptscriptstyle0}$ are the branching ratio and the kinematic function associated to the heavy boson $B^{\prime}$ taken at the asymptotic limit $m_{\scriptscriptstyle{Q}}\rightarrow\infty$ (i.e. $\alpha_{\scriptscriptstyle Q}^2=1/2$ for ${\bf Q}^{\scriptscriptstyle\{3\}}_{\scriptscriptstyle\{W, W^{\prime}\}}$ and $\alpha_{\scriptscriptstyle Q}^2=1/3$ for ${\bf Q}^{\scriptscriptstyle\{3\}}_{\scriptscriptstyle\{W, W^{\prime}, Z, H\}}$ and so on). We will now express the parameters $\ktbplr$, cf. eq. (\ref{couplingcase1}), as follows: 
\begin{align}
 \ktbplr&=\kpq\sqrt{\frac{\xi_{\scriptscriptstyle{B^{\prime}}}^{\scriptscriptstyle0}}{\Gamma_{\scriptscriptstyle{B^{\prime}}}^{\scriptscriptstyle0}}}.
 \label{couplingcase3}
\end{align}
which clearly show that $\ktbplr$ is finite even for $m_{\scriptscriptstyle{B^{\prime}}}\approx m_{\scriptscriptstyle{Q}}$.\\

 \noindent
 We will now check the validity of equations (\ref{xigam}) and (\ref{couplingcase3}) in the benchmark scenario ${\bf Q}^{\scriptscriptstyle\{3\}}_{\scriptscriptstyle\{W, W^{\prime}, Z, H\}}$ as an example. The branching ratio $\text{Br}(Q\rightarrow W^{\prime}q)$, in this case, can be expressed as follows:
 \begin{align}
\xi_{\scriptscriptstyle{W^{\prime}}}&=\frac{(\ktwplr)^2 \, \, \, \Gamma_{\scriptscriptstyle{W^{\prime}}}(m_{\scriptscriptstyle{Q}}, m_{\scriptscriptstyle{W^{\prime}}}, 0)}{\tilde{\kappa}_{\scriptscriptstyle Q}^2+(\ktwplr)^2 \, \, \, \Gamma_{\scriptscriptstyle{W^{\prime}}}(m_{\scriptscriptstyle{Q}}, m_{\scriptscriptstyle{W^{\prime}}}, 0)}
\label{br2}
\end{align}
where we neglected the masses of the ordinary quarks for simplicity and set
\begin{align}
 \tilde{\kappa}_{\scriptscriptstyle Q}^2&=(\ktwlr)^2 \, \, \, \Gamma_{\scriptscriptstyle{W}}^{\scriptscriptstyle{0}}+(\ktzlr)^2 \, \, \, \Gamma_{\scriptscriptstyle{Z}}^{\scriptscriptstyle{0}}+(\kthlr)^2 \, \, \, \Gamma_{\scriptscriptstyle{H}}^{\scriptscriptstyle{0}}.
\label{kpqt}
\end{align}
In eq. (\ref{kpqt}), we take the kinematic function in the asymptotic limit since the masses of SM particles are very small compared to $m_{\scriptscriptstyle Q}$. So, the parameter $\ktwplr$ is given by:
\begin{align}
 \ktwplr&= \tilde{\kappa}_{\scriptscriptstyle Q}\, \sqrt{\frac{\xi_{\scriptscriptstyle{W^{\prime}}}}{\Gamma_{\scriptscriptstyle{W^{\prime}}}\, (1-\xi_{\scriptscriptstyle{W^{\prime}}})}}
\end{align}
In the following, we will demonstrate that the value of $\ktwplr$ is independent of the ratios between the masses $m_{\scriptscriptstyle Q}$ and $m_{\scriptscriptstyle W^{\prime}}$ by evaluating it in two cases: {\it (i)} when $m_{\scriptscriptstyle Q}$ is much greater than $m_{\scriptscriptstyle W^{\prime}}$ and {\it (ii)} when $m_{\scriptscriptstyle Q}$ and $m_{\scriptscriptstyle W^{\prime}}$ are approximately equal $m_{\scriptscriptstyle Q}\approx m_{\scriptscriptstyle W^{\prime}}$. 
\begin{itemize}
 \item For $m_{\scriptscriptstyle Q}\gg m_{\scriptscriptstyle W^{\prime}}$, the different branching ratios involved in this benchmark scenario are given by:
 \begin{align}
  \xi_{\scriptscriptstyle{W^{\prime}}}&=\xi_{\scriptscriptstyle{W}}=\frac{1}{3}&
  \xi_{\scriptscriptstyle{Z}}&=\xi_{H}=\frac{1}{6}&
  \xi_{\scriptscriptstyle{W^{\prime}}}+\xi_{\scriptscriptstyle{W}}+\xi_{\scriptscriptstyle{Z}}+\xi_{H}&=1.
 \end{align}
then
\begin{align}
 \ktwplr&= \frac{\tilde{\kappa}_{\scriptscriptstyle Q}^{\scriptscriptstyle(1)}}{\sqrt{2}}.
 \label{ktl1}
\end{align}
where we set $\Gamma_{\scriptscriptstyle{W^{\prime}}}\approx\Gamma_{\scriptscriptstyle{W^{\prime}}}^{0}= 1$ and $\tilde{\kappa}_{\scriptscriptstyle Q}^{\scriptscriptstyle(1)}$ is $\tilde{\kappa}_{\scriptscriptstyle Q}$ taken at the asymptotic limit. It can be written in term of $\kpq$ as
\begin{align}
\left(\tilde{\kappa}_{\scriptscriptstyle Q}^{\scriptscriptstyle(1)}\right)^2&=\kpq^2\, \left(\xi_{\scriptscriptstyle{W}}+\xi_{\scriptscriptstyle{Z}}+\xi_{\scriptscriptstyle{H}}\right)\approx \frac{2}{3}\, \kpq^2\equiv \frac{2}{3}\, \left(\kappa_{\scriptscriptstyle Q}^{\scriptscriptstyle (1)}\right)^2.
\end{align}

\item Let's assume that $\xi_{\scriptscriptstyle{W^{\prime}}}\approx 0$, then 
\begin{align}
 \ktwplr&\approx \tilde{\kappa}_{\scriptscriptstyle Q}^{\scriptscriptstyle(2)}\, \sqrt{\frac{\xi_{\scriptscriptstyle{W^{\prime}}}}{\Gamma_{\scriptscriptstyle{W^{\prime}}}}}.
 \label{ktl2}
\end{align}
where $\tilde{\kappa}_{\scriptscriptstyle Q}^{\scriptscriptstyle(2)}\equiv\tilde{\kappa}_{\scriptscriptstyle Q}$ for $\xi_{\scriptscriptstyle{W^{\prime}}}\approx 0$ which can be expressed in term of $\kpq$ as the following: \begin{align}
\left(\tilde{\kappa}_{\scriptscriptstyle Q}^{\scriptscriptstyle(2)}\right)^2&\approx\kpq^2\, \left(\frac{1}{2}+ \frac{1}{4}+\frac{1}{4}\right)= \kpq^2\equiv \left(\kappa_{\scriptscriptstyle Q}^{\scriptscriptstyle (2)}\right)^2.
\end{align}
\end{itemize}

\vspace{0.5cm}
\noindent
Now, we suppose that $\ktwplr$ given in eq.~(\ref{ktl1}) is proportional to the one given in eq. (\ref{ktl2}) but both are finite, so we can write
\begin{align}
 \frac{\tilde{\kappa}_{\scriptscriptstyle Q}^{\scriptscriptstyle(1)}}{\sqrt{2}}=\beta_{\scriptscriptstyle Q}\, \tilde{\kappa}_{\scriptscriptstyle Q}^{\scriptscriptstyle(2)}\, \sqrt{\frac{\xi_{\scriptscriptstyle{W^{\prime}}}}{\Gamma_{\scriptscriptstyle{W^{\prime}}}}}.
 \label{kpq12}
\end{align}
where $\beta_{\scriptscriptstyle Q}$ is a regular function on $m_{\scriptscriptstyle Q}$.\\

\noindent
Solving this equality, expanding the solution around $\xi_{\scriptscriptstyle{W^{\prime}}}\rightarrow 0$ and taking terms up to $\mathcal{O}\left(\xi_{\scriptscriptstyle{W^{\prime}}}\right)$ order, we get
\begin{align}
 m_{\scriptscriptstyle Q}^2&\approx m_{\scriptscriptstyle W^{\prime}}^2+ \sqrt{\frac{2}{3}}\, \frac{\tilde{\kappa}_{\scriptscriptstyle Q}^{\scriptscriptstyle(2)}}{\tilde{\kappa}_{\scriptscriptstyle Q}^{\scriptscriptstyle(1)}}\,  \beta_{\scriptscriptstyle Q}\, m_{\scriptscriptstyle W^{\prime}}^2\, \xi_{\scriptscriptstyle{W^{\prime}}}^{1/2}+\mathcal{O}\left(\xi_{\scriptscriptstyle{W^{\prime}}}\right)
 \approx m_{\scriptscriptstyle W^{\prime}}^2+ \frac{\kappa_{\scriptscriptstyle Q}^{\scriptscriptstyle(2)}}{\kappa_{\scriptscriptstyle Q}^{\scriptscriptstyle(1)}}\, \beta_{\scriptscriptstyle Q}\, m_{\scriptscriptstyle W^{\prime}}^2\,  \xi_{\scriptscriptstyle{W^{\prime}}}^{1/2}+\mathcal{O}\left(\xi_{\scriptscriptstyle{W^{\prime}}}\right)
 \label{mQmWp}
\end{align}
From eq.~(\ref{kpq12}), we find that the kinematic function, in this limit, is given by:
\begin{align}
 \Gamma_{\scriptscriptstyle W^{\prime}}&=2\, \frac{\left(\tilde{\kappa}_{\scriptscriptstyle Q}^{\scriptscriptstyle(2)}\right)^2}{\left(\tilde{\kappa}_{\scriptscriptstyle Q}^{\scriptscriptstyle(1)}\right)^2}\, \beta_{\scriptscriptstyle Q}^2\, \xi_{\scriptscriptstyle{W^{\prime}}}+\mathcal{O}\left(\xi_{\scriptscriptstyle{W^{\prime}}}^{3/2}\right)
 =3\, \frac{\left(\kappa_{\scriptscriptstyle Q}^{\scriptscriptstyle(2)}\right)^2}{\left(\kappa_{\scriptscriptstyle Q}^{\scriptscriptstyle(1)}\right)^2}\, \beta_{\scriptscriptstyle Q}^2\, \xi_{\scriptscriptstyle{W^{\prime}}}+\mathcal{O}\left(\xi_{\scriptscriptstyle{W^{\prime}}}^{3/2}\right).
 \label{gwp0}
\end{align}
which confirm our assumption outlined in eq. (\ref{xigam}). 
Comparing eq. (\ref{xigam}) and eq. (\ref{gwp0}), we find that:
\begin{align}
 \alpha_{\scriptscriptstyle Q}^2&=\frac{1}{3\, \beta_{\scriptscriptstyle Q}^2}\, \frac{\left(\kappa_{\scriptscriptstyle Q}^{\scriptscriptstyle(1)}\right)^2}{\left(\kappa_{\scriptscriptstyle Q}^{\scriptscriptstyle(2)}\right)^2}.
\end{align}
From this equation we see that if $\ktwplr$ and $\kpq$ are the same in the two approximation, i.e. $\beta_{\scriptscriptstyle Q}=1$ and $\kappa_{\scriptscriptstyle Q}^{\scriptscriptstyle(1)}=\kappa_{\scriptscriptstyle Q}^{\scriptscriptstyle(2)}\equiv \kpq$, then $\alpha_{\scriptscriptstyle Q}^2=1/3$ as expected, cf.~eq.~(\ref{alphaQ}) ($ \alpha_{\scriptscriptstyle Q}^2=\xi_{\scriptscriptstyle{W^{\prime}}}^{\scriptscriptstyle0}/\Gamma_{\scriptscriptstyle{W^{\prime}}}^{\scriptscriptstyle0}=1/3$ for  for ${\bf Q}^{\scriptscriptstyle\{3\}}_{\scriptscriptstyle\{W, W^{\prime}, Z, H\}}$).\\

\noindent
Equation (\ref{mQmWp}) states that: for very small $\xi_{\scriptscriptstyle{W^{\prime}}}$, the masses $m_{\scriptscriptstyle Q}$ and $m_{\scriptscriptstyle W^{\prime}}$ are approximately equal provided that $\ktwplr$ is independent of $m_{\scriptscriptstyle Q}$. Conversely, if the the masses $m_{\scriptscriptstyle Q}$ and $m_{\scriptscriptstyle W^{\prime}}$ are approximately the same, the branching ratio will be very small, and the parameter $\ktwplr$ will retain the same value as it is in the asymptotic limit $m_{\scriptscriptstyle Q}\gg m_{\scriptscriptstyle W^{\prime}}$ provided that $\kpq$ is constant, i.e. the parameter $\ktwplr$ is independent of the mass of the vector like quark as expected.\\  

\noindent
While this result can be proven similarly for other benchmark scenarios, we will omit the extensive calculations to avoid an overly technical presentation. Consequently, the following equation summarizes the new form of all the model's couplings (cf. (\ref{param1})) in the case where $m_{\scriptscriptstyle Q}>m_{\scriptscriptstyle W^{\prime}}+m_{\scriptscriptstyle q}$ for the most general benchmark scenario ${\bf Q}^{\scriptscriptstyle\{3\}}_{\scriptscriptstyle\{W, W^{\prime}, Z, Z^{\prime}, H, H^{\prime}\}}$. 
\begin{align}
\kwl&=\kpq\, \sqrt{\frac{\xi_{\scriptscriptstyle{W}}^{\scriptscriptstyle0}}{\Gamma_{\scriptscriptstyle{W}}^{\scriptscriptstyle0}}},&
\kzl&=\kpq\, \sqrt{\frac{\xi_{\scriptscriptstyle{Z}}^{\scriptscriptstyle0}}{\Gamma_{\scriptscriptstyle{Z}}^{\scriptscriptstyle0}}},& 
\ksl&=\kpq\, \frac{m_{\scriptscriptstyle{Q}}}{v}\, \sqrt{\frac{\xi_{\scriptscriptstyle S}^{\scriptscriptstyle0}}{\Gamma_{\scriptscriptstyle S}^{\scriptscriptstyle0}}},\nonumber\\
\kwpl&=\kpq\, \frac{m_{\scriptscriptstyle W^{\prime}}}{m_{\scriptscriptstyle W}}\, \sqrt{\frac{\xi_{\scriptscriptstyle{W}^{\prime}}^{\scriptscriptstyle0}}{\Gamma_{\scriptscriptstyle{W^{\prime}}}^{\scriptscriptstyle0}}},&
\kzpl&=\kpq\, \frac{m_{\scriptscriptstyle Z^{\prime}}}{m_{\scriptscriptstyle Z}}\, \sqrt{\frac{\xi_{\scriptscriptstyle{Z}^{\prime}}^{\scriptscriptstyle0}}{\Gamma_{\scriptscriptstyle{Z^{\prime}}}^{\scriptscriptstyle0}}},&
\ksr&=\kpq\, \frac{m_{\scriptscriptstyle{q}}}{v}\, \sqrt{\frac{\xi_{S}^{\scriptscriptstyle0}}{\Gamma_{\scriptscriptstyle{S}}^{\scriptscriptstyle0}}}. 
\label{kzpwp0}
\end{align}

\begin{figure}[h!]
\includegraphics[width=7cm,height=4cm]{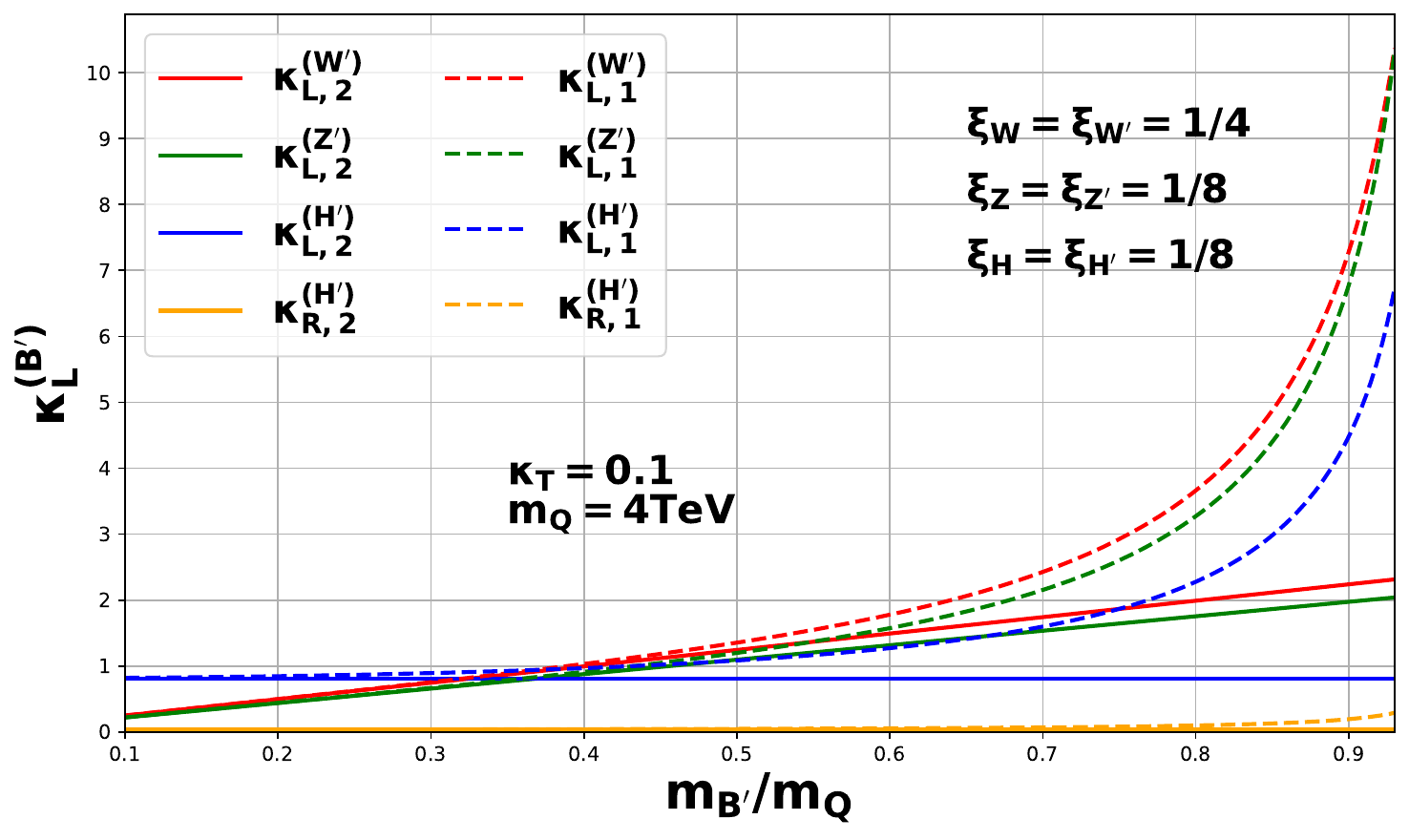}
\caption{\small Variation of the couplings in term of the ratio $m_{\scriptscriptstyle B'}/m_{\scriptscriptstyle Q}$ in the benchmark scenario ${\bf Q}^{\scriptscriptstyle\{3\}}_{\scriptscriptstyle\{W, W^{\prime}, Z, Z^{\prime}, H, H^{\prime}\}}$, where $\kappa_{L/R, 1}^{(B^{\prime})}$ and $\kappa_{L/R, 2}^{(B^{\prime})}$ denote, respectively, the couplings in the original and the modified parametrizations.}
\label{coulingscase1}
\end{figure}

\noindent
In figure~\ref{coulingscase1}, we show the variation of the couplings as a function of the ratio $m_{\scriptscriptstyle B'}/m_{\scriptscriptstyle Q}$, comparing the original parametrization (cf. eq.~\ref{param1}) with the modified parametrization (cf. eq.~(\ref{kzpwp0})) in the  benchmark scenario ${\bf Q}^{\scriptscriptstyle\{3\}}_{\scriptscriptstyle\{W, W^{\prime}, Z, Z^{\prime}, H, H^{\prime}\}}$. This plot clearly demonstrates that both parametrizations yield approximately the same coupling values when $m_{\scriptscriptstyle Q}$ is sufficiently large compared to $m_{\scriptscriptstyle B'}$ (i.e. $m_{\scriptscriptstyle B'}/m_{\scriptscriptstyle Q}<0.5$). However, beyond $m_{\scriptscriptstyle B'}/m_{\scriptscriptstyle Q}=0.5$, the difference starts to become significant until when the couplings become completely different near the decay threshold.\\

\noindent
The two parametrizations options given in eq.~(\ref{param1}) and eq.~(\ref{kzpwp0}) represent two different perspectives. Naturally, when the mass of $Q$ is large compared to the heavy bosons masses, both should yield approximately the same predictions. However, beyond this regime, we cannot definitively say which one is better than the other. Therefore, it is recommended to consider both as 2 versions of the model couplings.  \\

\noindent
We note that the couplings $\kwpl$, $\kzpl$ and $\ksl$ are proportional to the masses $m_{\scriptscriptstyle W^{\prime}}$, $m_{\scriptscriptstyle Z^{\prime} }$ and $m_{\scriptscriptstyle Q}$, respectively (in both parametrization). This proportionality implies that they can become very large, which potentially could invalidate the perturbative treatment at extremely high masses. In section~\ref{pertun} and appendix~\ref{appB}, we establish upper limits on the couplings using perturbative unitarity constraints to ensure that we remain within the perturbative regime.

\subsection{Case 2: $m_{\scriptscriptstyle{Q}}<m_{\scriptscriptstyle{B^{\prime}}}-m_{\scriptscriptstyle{}q}$}
\label{appA2}
\noindent 
In this case, the decay of the VLQ $Q$ to a SM quark and one the heavy bosons ($\bp\equiv \wps$, $\zp$, $\hp$) is kinematically forbidden. So, it can decay only according to the following three modes:
\begin{align}
 Q&\rightarrow qH, & Q&\rightarrow qZ, & Q&\rightarrow q^{\prime}W 
 \end{align}
 {\color{black} where $q$ and $q^{\prime}$ are up or down quarks.}\\

\noindent
Following the same strategy as in the previous section, we show that the parameters normalizing the mixing between $Q$ and SM quark via $W, Z$ and $H$ are given by:
\begin{align}
\kwl&=\kpq\sqrt{\frac{\xi_{\scriptscriptstyle{W}}}{\Gamma_{\scriptscriptstyle{W}}}},& 
\kzl&=\kpq\sqrt{\frac{\xi_{\scriptscriptstyle{Z}}}{\Gamma_{\scriptscriptstyle{Z}}}},&
\khl&=\kpq\frac{m_{\scriptscriptstyle{T}}}{v}\sqrt{\frac{\xi_{\scriptscriptstyle{H}}}{\Gamma_{\scriptscriptstyle{H}}}},&
\khr&=\kpq\frac{m_{\scriptscriptstyle{q}}}{v}\sqrt{\frac{\xi_{\scriptscriptstyle{H}}}{\Gamma_{\scriptscriptstyle{H}}}}. 
\label{klWZH}
\end{align}
with $\underset{\begin{subarray}{l}\scriptscriptstyle{B=}\scriptscriptstyle{Z,W,H}\end{subarray}}{\sum}\xi_{\scriptscriptstyle{B}}=1$.

\vspace{0.25cm}

\noindent
In this case, we suppose that the new heavy bosons can decay into $Q$ then, following our strategy, the parameters describing the mixing between the VLQ and the bosons $\bp$ can be expressed in terms of branching ratio of $\bp$ into $Q$ and the masses of the particles as we will show later on. We remind that the fermions of SM interact with the new bosons according to the SSM (cf. section \ref{sec2}), i.e. they have the same couplings of the SM bosons up to normalization factors.
\subsubsection*{The parameter $\kwpl$}

We suppose that $\wps$ can decay into all the fermions involved in the model, then the partial decay widths of its different decay modes are given by:
\begin{align}
&\Gamma(\wps\rightarrow u_id_j)=N_c(\vkwp)^2\frac{g^2m_{\scriptscriptstyle{W^{\prime}}}}{48\pi},&
&\Gamma(\wps\rightarrow l\nu_l)=(\vtwp)^2\frac{g^2m_{\scriptscriptstyle{W^{\prime}}}}{48\pi},\nonumber\\
&\Gamma(\wps\rightarrow tb)=N_c(\vkwp)^2\frac{g^2m_{\scriptscriptstyle{W^{\prime}}}}{48\pi}\Gamma_{\scriptscriptstyle{tb}}^{\scriptscriptstyle{W^{\prime}}}, &
&\Gamma(\wps\rightarrow Qq^{\prime})=
N_c(\kwpl)^2\frac{g^2m_{\scriptscriptstyle{W^{\prime}}}}{48\pi}\Gamma_{\scriptscriptstyle{Qq^{\prime}}}^{\scriptscriptstyle{W^{\prime}}}.
\label{gamwpffp}
\end{align}
where we neglected the masses of the ordinary fermions (except the top quark), i.e. we set $m_{\scriptscriptstyle{f}}^2/m_{\scriptscriptstyle{W^{\prime}}}^2=0$ ($f\neq t$) since $m_{\scriptscriptstyle{W^{\prime}}}\gg m_{\scriptscriptstyle{f}}$. The kinematic functions involved in eq. (\ref{gamwpffp}) are: 
\begin{align}
\Gamma_{\scriptscriptstyle{tb}}^{\scriptscriptstyle{W^{\prime}}}&=\biggr(1-\frac{m_{\scriptscriptstyle{t}}^2}{m_{\scriptscriptstyle{W^{\prime}}}^2}\biggr)^2\biggl[1+\frac{1}{2}\frac{m_{\scriptscriptstyle{t}}^2}{m_{\scriptscriptstyle{W^{\prime}}}^2}\biggr].\nonumber\\
 \Gamma_{\scriptscriptstyle{Q}{\scriptscriptstyle q^{\prime}}}^{\scriptscriptstyle{W^{\prime}}}&=\lambda^{1/2}\left(1,\frac{m_{\scriptscriptstyle Q}^2}{m_{\scriptscriptstyle W^{\prime}}^2},\frac{m_{\scriptscriptstyle q^{\prime}}^2}{m_{\scriptscriptstyle W^{\prime}}^2}\right)\left[1-\frac{1}{2}\frac{m_{\scriptscriptstyle Q}^2+m_{\scriptscriptstyle q^{\prime}}^2}{m_{\scriptscriptstyle W^{\prime}}^2}-\frac{1}{2}\frac{(m_{\scriptscriptstyle Q}^2-m_{\scriptscriptstyle q^{\prime}}^2)^2}{m_{\scriptscriptstyle W^{\prime}}^4}\right]
 &\approx \frac{1}{2}\, \left(1-\frac{m_{\scriptscriptstyle{Q}}^2}{m_{\scriptscriptstyle{W^{\prime}}}^2}\right)^2\, \left[2+\frac{m_{\scriptscriptstyle{Q}}^2}{m_{\scriptscriptstyle{W^{\prime}}}^2}\right].
 \label{xiwpQq}
\end{align}
In the asymptotic limit (i.e. for $m_{\scriptscriptstyle W^{\prime}}\gg  m_{\scriptscriptstyle Q}, m_{\scriptscriptstyle q}$), we can approximate the kinematic function to:
\begin{align}
 \Gamma_{\scriptscriptstyle{t}{\scriptscriptstyle b}}^{\scriptscriptstyle{W^{\prime}}0}&\approx1,&
 \Gamma_{\scriptscriptstyle{Q}{\scriptscriptstyle q^{\prime}}}^{\scriptscriptstyle{W^{\prime}}0}&\approx 1.
\end{align}
\noindent
The total decay width of the $\wps$ gauge boson is then:
\begin{align}
 \Gamma_{\scriptscriptstyle{W^{\prime}}}^{\text{ToT}}&=3\, \Gamma(\wps\rightarrow l\nu_l)+2\, \Gamma(\wps\rightarrow u_id_j)+\Gamma(\wps\rightarrow tb)+\Gamma(\wps\rightarrow Qq^{\prime})
\end{align}
\noindent

\noindent
The branching fraction of $\wps$ to top VLQ $Q$ and a ordinary quark $q^{\prime}$, in the SSM, is given by:
\begin{align}
\text{Br}(\wps\rightarrow Qq^{\prime})&=\frac{(\kwpl)^2\Gamma_{\scriptscriptstyle{Qq^{\prime}}}^{\scriptscriptstyle{W^{\prime}}}}{(\vtwp)^2+2(\vkwp)^2+(\vkwp)^2\Gamma_{\scriptscriptstyle{tb}}^{\scriptscriptstyle{W^{\prime}}}+(\kwpl)^2\Gamma_{\scriptscriptstyle{Qq^{\prime}}}^{\scriptscriptstyle{W^{\prime}}}}\equiv \xi_{\scriptscriptstyle{Qq^{\prime}}}^{\scriptscriptstyle{W^{\prime}}}
\label{brwp}
\end{align}

\noindent
The main idea of our strategy is to set the denominator of eq.~(\ref{brwp}) equals to $\kpwp^2$ where the later one is treated as a free parameter \footnote{In the SSM, the parameter $\kpwp$ is no longer free; it is constrained by the condition: $\vtwp=\vkwp$, as we will see later.}, i.e. the branching ratio $\xi_{\scriptscriptstyle{Qq^{\prime}}}^{\scriptscriptstyle{W^{\prime}}}$ is expressed as follows:   
\begin{align}
\xi_{\scriptscriptstyle{Qq^{\prime}}}^{\scriptscriptstyle{W^{\prime}}}&=\frac{(\kwpl)^2\Gamma_{\scriptscriptstyle{Qq^{\prime}}}^{\scriptscriptstyle{W^{\prime}}}}{\kpwp^2}
\end{align}
\noindent
Now, we can express the coupling $\kwpl$ in terms of the branching ratio, the kinematic function $\Gamma_{\scriptscriptstyle{Qq^{\prime}}}^{\scriptscriptstyle{W^{\prime}}}$ and the parameter $\kpwp$:
\begin{align}
\kwpl&=\kpwp\sqrt{\frac{\xi_{\scriptscriptstyle{Qq^{\prime}}}^{\scriptscriptstyle{W^{\prime}}}}{\Gamma_{\scriptscriptstyle{Qq^{\prime}}}^{\scriptscriptstyle{W^{\prime}}}}}
\label{nonc-kwpl}
\end{align} 

\noindent
Similarly, the couplings of $\wps$ to SM quarks and lepton are given by:
\begin{align}
\vkwp&=\kpwp\sqrt{\xi_{\scriptscriptstyle{u_id_i}}^{\scriptscriptstyle{W^{\prime}}}}&
\vtwp&=\kpwp\sqrt{3\,\xi_{\scriptscriptstyle{l_il_{\nu_i}}}^{\scriptscriptstyle{W^{\prime}}}}
\label{vkvts}
\end{align}
where we take $\Gamma_{tb}^{W^{'}}\approx 1$ (since $m_{\scriptscriptstyle W^{\prime}}\gg m_{\scriptscriptstyle t}$). We notice that the branching ratios must satisfy:
\begin{align}
\xi_{\scriptscriptstyle{Qq^{\prime}}}^{\scriptscriptstyle{W^{\prime}}}+\sum_{i=1}^{3}\left[\xi_{\scriptscriptstyle{u_id_i}}^{\scriptscriptstyle{W^{\prime}}}+\xi_{\scriptscriptstyle{l_il_{\nu_i}}}^{\scriptscriptstyle{W^{\prime}}}\right]&=1.
\label{brss}
\end{align}

\noindent
As we have already stated in the previous section, eq. (\ref{nonc-kwpl}) is not valid when $m_{\scriptscriptstyle W^{\prime}}\approx m_{\scriptscriptstyle Q}$ because of the threshold singularity caused by the vanishing of $\Gamma_{\scriptscriptstyle{Qq^{\prime}}}^{\scriptscriptstyle{W^{\prime}}}$ in this regions. In our phenomenological study, we never approach this kinematic configuration, so our parametrization remain safe (we consider $m_{\scriptscriptstyle Q}=1/2, \, 2/3,\, 3/4\,  m_{\scriptscriptstyle W^{\prime}}$, cf. section \ref{sec3}). However, for completeness, we can slightly modify this parametrization such that the coupling remains well-defined even when $m_{\scriptscriptstyle W^{\prime}}\approx m_{\scriptscriptstyle Q}$.\\

\noindent
To do so, we replace $\xi_{\scriptscriptstyle{Qq^{\prime}}}^{\scriptscriptstyle{W^{\prime}}}$ and $\Gamma_{\scriptscriptstyle{Qq^{\prime}}}^{\scriptscriptstyle{W^{\prime}}}$ with their asymptotic values (at $m_{\scriptscriptstyle W^{\prime}}\rightarrow\infty$), which is possible under the following requirement:
\begin{align}
\frac{\xi_{\scriptscriptstyle{Qq^{\prime}}}^{\scriptscriptstyle{W^{\prime}}}}{\Gamma_{\scriptscriptstyle{Qq^{\prime}}}^{\scriptscriptstyle{W^{\prime}}}}=\frac{\xi_{\scriptscriptstyle{Qq^{\prime}}}^{\scriptscriptstyle{W^{\prime}0}}}{\Gamma_{\scriptscriptstyle{Qq^{\prime}}}^{\scriptscriptstyle{W^{\prime}0}}}\equiv \text{constant} 
\end{align}

\begin{figure}[h!]
\centering
 \includegraphics[width=8.0cm,height=4cm]{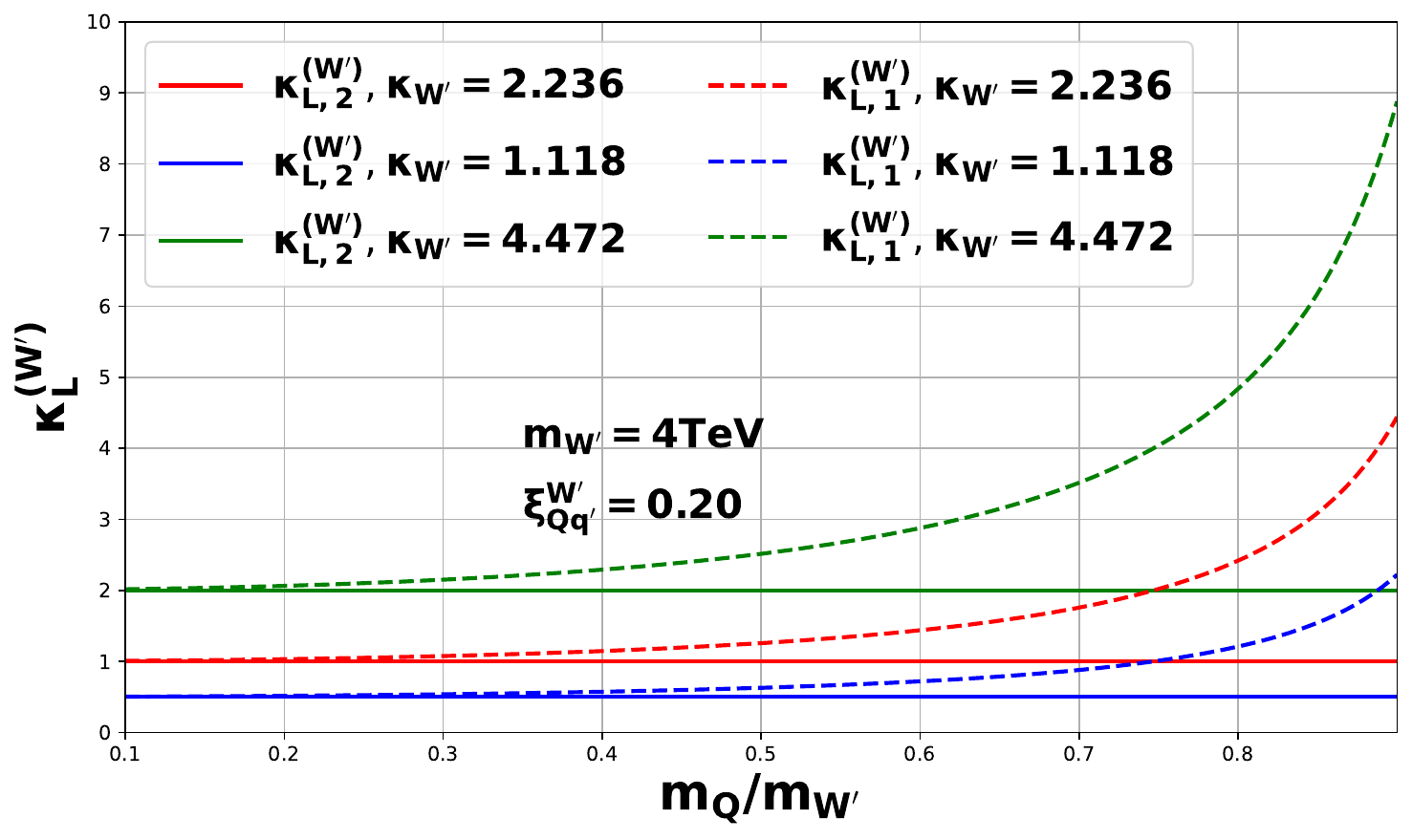}
\includegraphics[width=8.0cm,height=4cm]{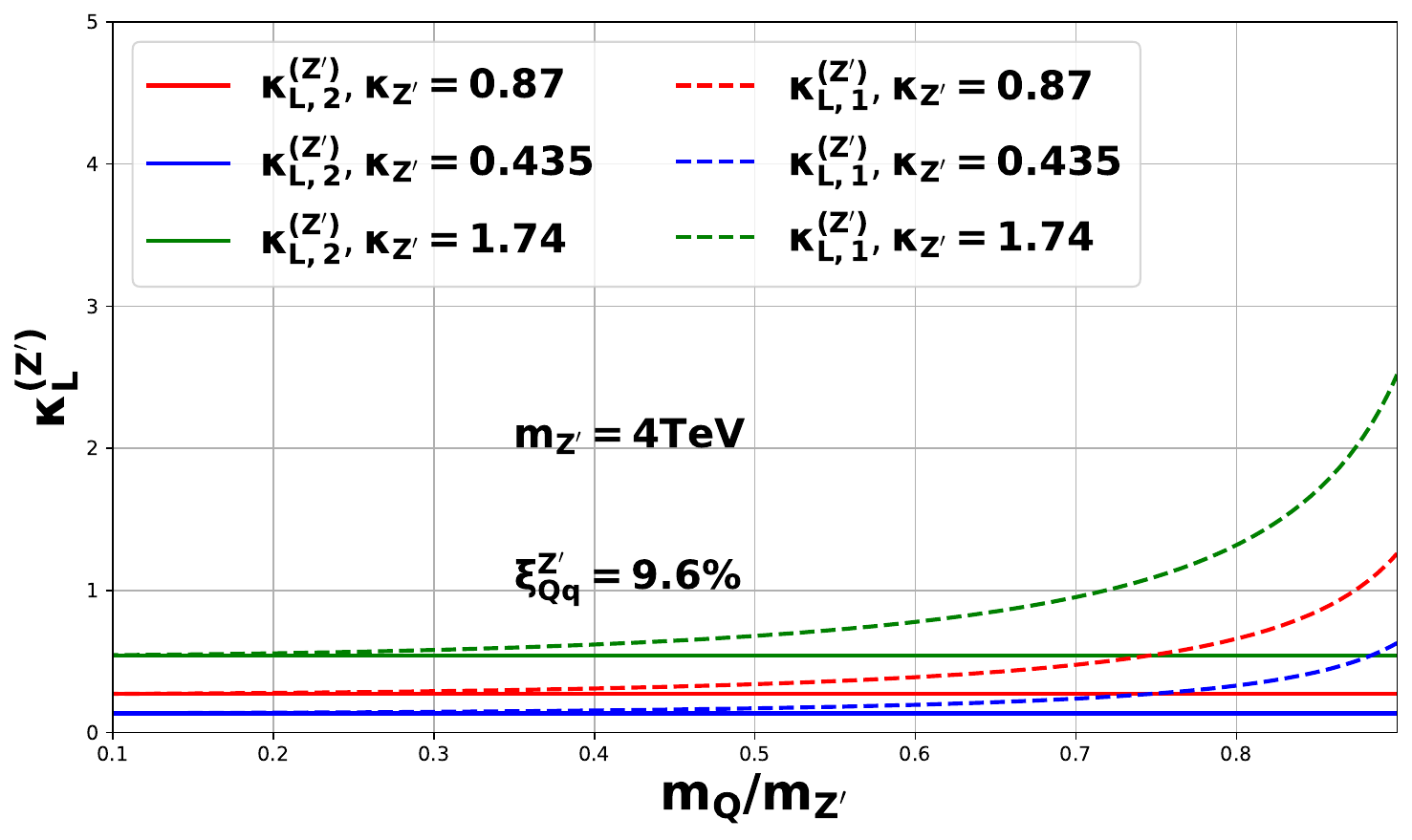}
 \caption{\small Variation of $\kbpl$ as a function of the ratio $m_{\scriptscriptstyle Q}/m_{\scriptscriptstyle B^{\prime}}$ in the original and the modified parametrizations.}
 \label{coulingscase2}
\end{figure}

\noindent
Thus, the coupling of the $\wps$ to $Q$, in the modified parametrization, becomes:
\begin{align}
\kwpl&=\kpwp\sqrt{\frac{\xi_{\scriptscriptstyle{Qq^{\prime}}}^{\scriptscriptstyle{W^{\prime}}0}}{\Gamma_{\scriptscriptstyle{Qq^{\prime}}}^{\scriptscriptstyle{W^{\prime}}0}}}
\label{nonc-kwpl2}
\end{align}
\noindent
In the left panel of figure~\ref{coulingscase2}, we show the variation of the coupling $\kwpl$ in the original parametrization (cf. eq.~\ref{nonc-kwpl}) and the modified parametrization (cf. eq.~\ref{nonc-kwpl2}) in term of the ratio $m_{\scriptscriptstyle Q}/m_{\scriptscriptstyle W^{\prime}}$ for $m_{\scriptscriptstyle W^{\prime}}=4$~TeV. The green, red and blue curves represent, respectively, the coupling for $\vkwp=\vtwp=2, 1, 0.5$. It is clear that the couplings in the two parametrizations are very close when $m_{\scriptscriptstyle W^{\prime}}$ is sufficiently larger than $m_{\scriptscriptstyle Q}$ especially for $\vkwp=\vtwp=0.5$, which implies that the two approaches are approximately equivalent in this configurations.\\

\noindent
According to eq.~(\ref{vkvts}), the parameter $\kpwp$ is not any more free but it is constrained by the branching ratio and the SSM couplings. 
From eq. (\ref{vkvts}), we deduce immediately that in the SSM (i.e. $\vkwp=\vtwp$), the branching ratio into ordinary quarks is 3 times larger the branching ratio into leptons (i.e. $\xi_{\scriptscriptstyle{u_id_i}}^{\scriptscriptstyle{W^{\prime}}}=3\, \xi_{\scriptscriptstyle{l_il_{\nu_i}}}^{\scriptscriptstyle{W^{\prime}}}$). Then eq. (\ref{brss}) becomes:
\begin{align}
 \xi_{\scriptscriptstyle{Qq^{\prime}}}^{\scriptscriptstyle{W^{\prime}}} + 12\, \xi_{\scriptscriptstyle{l_il_{\nu_i}}}^{\scriptscriptstyle{W^{\prime}}}&=1
\end{align}

\noindent
Now, if we assume that $W^{\prime}$ decay democratically to all kind of quarks (i.e. $\xi_{\scriptscriptstyle{Qq^{\prime}}}^{\scriptscriptstyle{W^{\prime}}}=\xi_{\scriptscriptstyle{u_id_i}}^{\scriptscriptstyle{W^{\prime}}}$), we get:
\begin{align}
\xi_{\scriptscriptstyle{Qq^{\prime}}}^{\scriptscriptstyle{W^{\prime}}}=\xi_{\scriptscriptstyle{u_id_i}}^{\scriptscriptstyle{W^{\prime}}}=3/15&\equiv 20\%&
\xi_{\scriptscriptstyle{l_il_{\nu_i}}}^{\scriptscriptstyle{W^{\prime}}}=1/15&\equiv 6.67\%.
\end{align}
\noindent
In the CSSM (i.e. $\vkwp=\vtwp=1$), the parameter $\kpwp$ given by:
\begin{align}
 \kpwp&=\frac{1}{\sqrt{\xi_{\scriptscriptstyle{u_id_i}}^{\scriptscriptstyle{W^{\prime}}}}}\approx 2.236
\end{align}
which fulfill the perturbative unitarity requirement, see subsection \ref{pertun}. \\

\noindent
Let's conclude this subsection by giving the values of the parameters controlling the couplings and the coupling $\kwpl$  for different scenario that we will consider in this work:
\begin{table*}[h!]
\centering
 \renewcommand{\arraystretch}{1.40}
 \setlength{\tabcolsep}{12.5pt}
  \boldmath
 \begin{adjustbox}{width=15cm,height=1.25cm}
 \begin{tabular}{!{\vrule width 1.5pt}l!{\vrule width 1.5pt}l!{\vrule width 1pt}l!{\vrule width 1pt}l!{\vrule width 1.5pt}l!{\vrule width 1pt}l!{\vrule width 1pt}l!{\vrule width 1.5pt}l!{\vrule width 1pt}l!{\vrule width 1pt}l!{\vrule width 1.5pt}l!{\vrule width 1pt}l!{\vrule width 1.5pt}}
 \cline{1-10}
&\multicolumn{3}{c!{\vrule width 1.5pt}}{{$\bf \vkwp=0.5$}}
&\multicolumn{3}{c!{\vrule width 1.5pt}}{{$\bf \vkwp=1.0$}}
&\multicolumn{3}{c!{\vrule width 1.5pt}}{{$\bf \vkwp=2.0$}}
&\multicolumn{2}{c}{{}}\\
\noalign{\hrule height 1pt}
$\bf \xi^{\scriptscriptstyle{W^{\prime}}}_{\scriptscriptstyle{Tb}}$&$0.1$ &$0.2$ &$0.4$ &$0.1$ &$0.2$ &$0.4$ &$0.1$ &$0.2$ &$0.4$& \multicolumn{2}{c!{\vrule width 1.5pt}}{{\bf Kinematic function}} \\
\cdashline{2-10}
\cline{11-12}
$\bf \kpwp$&\bf $1.055$&\bf $1.118$& \bf $1.291$& \bf $2.108$ &\bf $2.236$ &\bf $2.582$&\bf $4.216$ &\bf $4.472$ & \bf $5.16398$& $\bf m_{\scriptscriptstyle W^{\prime}}/m_{\scriptscriptstyle Q}$& $\bf \Gamma_{\scriptscriptstyle Tb}^{\scriptscriptstyle W^{\prime}}$ \\
\noalign{\hrule height 1pt}
$\bf \kwpl$&$0.419$ &$0.629$ &$1.026$ & $0.838$& $1.257$&$2.053$ &$1.676$ & $2.514$& $4.106$& $1/2$&$0.6328$ \\
\cdashline{2-12}
$\bf \kwpl$&$0.543$ & $0.814$&$1.329$ & $1.085$&$1.628$ &$2.659$ &$2.171$ & $3.256$&$5.318$ & $2/3$&$0.3773$\\
\cdashline{2-12}
$\bf \kwpl$&$0.673$ &$1.010$ &$1.649$ & $1.346$&$2.019$ &$3.298$ &$2.692$ &$4.039$ &$6.595$ & $3/4$&$0.2452$ \\
\noalign{\hrule height 1pt}
\end{tabular}
\end{adjustbox}
  \caption{\footnotesize Parameters controlling the couplings $\kwpl$.}
   \label{tabVp1}
  \end{table*}

\subsubsection*{The parameter $\kzpl$}
The partial width of the different modes of the decay of $\zp$ are given by (all the ordinary quarks masses are neglected even the top quark):
\begin{align}
&\Gamma(\zp\rightarrow q\bar q)=N_{\scriptscriptstyle{C}}(\vkzp)^2\frac{g^2m_{\scriptscriptstyle{Z^{\prime}}}}{12\pi c_w^2}\Gamma_{\scriptscriptstyle qq}^{\scriptscriptstyle Z^{\prime}}, &
&\Gamma(\zp\rightarrow l^- l^+)=(\vtzp)^2\frac{g^2m_{\scriptscriptstyle{Z^{\prime}}}}{12\pi c_w^2}\Gamma_{\scriptscriptstyle ll}^{\scriptscriptstyle Z^{\prime}}\nonumber\\
&\Gamma(\zp\rightarrow \nu_l\bar\nu_l )=(\vtzp)^2\frac{g^2m_{\scriptscriptstyle{Z^{\prime}}}}{12\pi c_w^2}\Gamma_{\scriptscriptstyle \nu_l\nu_l}^{\scriptscriptstyle Z^{\prime}}, &
&\Gamma(\zp\rightarrow Q\bar q+\bar Qq)=N_{\scriptscriptstyle{C}}(\kzpl)^2\frac{g^2m_{\scriptscriptstyle{Z^{\prime}}}}{12\pi c_w^2}\Gamma_{\scriptscriptstyle Qq}^{\scriptscriptstyle Z^{\prime}}
\end{align}
with
\begin{align}
&\Gamma_{\scriptscriptstyle ff}^{\scriptscriptstyle Z^{\prime}}=\sqrt{1-4\frac{m_{\scriptscriptstyle{f}}^2}{m_{\scriptscriptstyle{Z^{\prime}}}^2}}\biggl[(g_{\scriptscriptstyle{A}}^{\scriptscriptstyle f})^2\biggl(1-4\frac{m_{\scriptscriptstyle{f}}^2}{m_{\scriptscriptstyle{Z^{\prime}}}^2}\biggr)+(g_{\scriptscriptstyle{V}}^{\scriptscriptstyle f})^2\biggl(1+2\frac{m_{\scriptscriptstyle{f}}^2}{m_{\scriptscriptstyle{Z^{\prime}}}^2}\biggr)\biggr]\approx [(g_{\scriptscriptstyle A}^{\scriptscriptstyle f})^2+(g_{\scriptscriptstyle V}^{\scriptscriptstyle f})^2] \\
 &\Gamma_{\scriptscriptstyle Qq}^{\scriptscriptstyle Z^{\prime}}=\frac{1}{4}\sqrt{\lambda\biggl(1,\frac{m_{\scriptscriptstyle{Q}}^2}{m_{\scriptscriptstyle{Z^{\prime}}}^2},\frac{m_{\scriptscriptstyle{q}}^2}{m_{\scriptscriptstyle{Z^{\prime}}}^2}\biggr)}\biggl[1-\frac{1}{2}\biggl(\frac{m_{\scriptscriptstyle{Q}}^2-m_{\scriptscriptstyle{q}}^2}{m_{\scriptscriptstyle{Z^{\prime}}}^2}\biggr)^2-\frac{1}{2}\frac{m_{\scriptscriptstyle{Q}}^2+m_{\scriptscriptstyle{q}}^2}{m_{\scriptscriptstyle{Z^{\prime}}}^2}\biggr]
 \label{xizpQq}
\end{align}

\noindent
The total decay width of the $\zp$ is then given by:
\begin{align}
 \Gamma_{\scriptscriptstyle{Z^{\prime}}}^{\text{tot}}&=3\, \Gamma(\zp\rightarrow u\bar u)+3\, \Gamma(\zp\rightarrow d\bar d)+3\, \Gamma(\zp\rightarrow l^- l^+)+3\, \Gamma(\zp\rightarrow \nu_l\bar\nu_l )+\Gamma(\zp\rightarrow Q\bar q+\bar Qq)
\end{align}

\noindent
The branching fraction of $\zp$ to $Q\bar{q}$ (or $\bar{Q}q$) in the SSM is expressed as:
\begin{align}
\text{Br}(\zp\rightarrow Q\bar q+\bar Q q)&=\frac{(\kzpl)^2\Gamma_{\scriptscriptstyle{Qq}}^{\scriptscriptstyle Z^{\prime}}}{\kpzp^2}\equiv \xi_{\scriptscriptstyle{Qq}}^{\scriptscriptstyle Z^{\prime}}.
\label{brzp}
\end{align}
with
\begin{align}
\kpzp^2&= \underset{\scriptscriptstyle{f=l,\nu}}{\sum}(\vtzp)^2[(g_{\scriptscriptstyle A}^{\scriptscriptstyle f})^2+(g_{\scriptscriptstyle V}^{\scriptscriptstyle f})^2]+\underset{\scriptscriptstyle{q=u,d}}{\sum}G_{\scriptscriptstyle{q}}(\vkzp)^2[(g_{\scriptscriptstyle A}^{\scriptscriptstyle q})^2+(g_{\scriptscriptstyle V}^{\scriptscriptstyle q})^2]+\Gamma_{\scriptscriptstyle{tt}}^{\scriptscriptstyle Z^{\prime}}+(\kzpl)^2\, \Gamma_{\scriptscriptstyle{Qq}}^{\scriptscriptstyle Z^{\prime}}
\end{align}
with $G_{\scriptscriptstyle{d}}=3$ and $G_{\scriptscriptstyle{u}}=3$.\\

\noindent
We can now write the coupling $\kzpl$ in terms of the branching ratio $\xi_{\scriptscriptstyle{Qq}}^{\scriptscriptstyle Z^{\prime}}$, the kinematic function $\Gamma_{\scriptscriptstyle{Qq}}^{\scriptscriptstyle Z^{\prime}}$, and the new parameter $\kpzp$. Thus, we obtain:
\begin{align}
\kzpl&=\kpzp\sqrt{\frac{\xi_{\scriptscriptstyle{Qq}}^{\scriptscriptstyle{Z^{\prime}}}}{\Gamma_{\scriptscriptstyle{Qq}}^{\scriptscriptstyle{Z^{\prime}}}}}
\label{nonc-kzpl}
\end{align}
Similar to $\kwpl$, to avoid encountering threshold singularity in the expression of $\kzpl$, we can slightly modify this parametrization by replacing the two key elements, $\xi_{\scriptscriptstyle{Qq}}^{\scriptscriptstyle{Z^{\prime}}}$ and $\Gamma_{\scriptscriptstyle{Qq}}^{\scriptscriptstyle{Z^{\prime}}}$, with their values at $m_{\scriptscriptstyle Q}/m_{\scriptscriptstyle W^{\prime}}\rightarrow0$. This is allowed if the following specific condition is satisfied:
\begin{align}
\frac{\xi_{\scriptscriptstyle{Qq}}^{\scriptscriptstyle{Z^{\prime}}}}{\Gamma_{\scriptscriptstyle{Qq}}^{\scriptscriptstyle{Z^{\prime}}}}=\frac{\xi_{\scriptscriptstyle{Qq}}^{\scriptscriptstyle{Z^{\prime}0}}}{\Gamma_{\scriptscriptstyle{Qq}}^{\scriptscriptstyle{Z^{\prime}0}}}\equiv \text{constant} 
\end{align}
So, the coupling in the modified parametrization is given by:
\begin{align}
\kzpl&=\kpzp\sqrt{\frac{\xi_{\scriptscriptstyle{Qq}}^{\scriptscriptstyle{Z^{\prime}}0}}{\Gamma_{\scriptscriptstyle{Qq}}^{\scriptscriptstyle{Z^{\prime}}0}}}&
&\text{with}&
 \Gamma_{\scriptscriptstyle{Qq}}^{\scriptscriptstyle{Z^{\prime}0}}&=\frac{1}{4}.
 \label{nonc-kzplmodif}
\end{align}
In the right panel of figure~\ref{coulingscase2}, we show the variation of the coupling $\kzpl$ in the original parametrization (cf. eq.~(\ref{nonc-kzpl})) and the modified parametrization (cf. eq.~\ref{nonc-kzplmodif}) as a function of the ratio $m_{\scriptscriptstyle Q}/m_{\scriptscriptstyle Z^{\prime}}$ for $m_{\scriptscriptstyle Z^{\prime}}=4$\,~TeV. The green, red and blue curves represent, respectively, the coupling for $\vkzp=\vtzp=2, 1, 0.5$. We observe that the couplings in the two parametrizations are approximately the same for $m_{\scriptscriptstyle Q}/m_{\scriptscriptstyle Z^{\prime}}<0.6$  especially for $\vkzp=\vtzp=0.5$. This means that the two options are approximately equivalent in this regions~\footnote{Anyway, we showed in section 3 that the effects of the ratios $m_{\scriptscriptstyle Q}/m_{\scriptscriptstyle B^{\prime}}$ on the considered cross sections is very limited.}.\\

\noindent
In the same manner, we can express the couplings of the $\zp$ to ordinary quarks and leptons in terms of the associated branchings ratios and kinematic functions, as follows:
\begin{align}
\vkzp&=\kpzp\sqrt{\frac{\xi_{\scriptscriptstyle{u_iu_i}}^{\scriptscriptstyle{Z^{\prime}}}}{\Gamma_{\scriptscriptstyle u_iu_i}^{\scriptscriptstyle Z^{\prime}}}}=\kpzp\sqrt{\frac{\xi_{\scriptscriptstyle{d_id_i}}^{\scriptscriptstyle{Z^{\prime}}}}{\Gamma_{\scriptscriptstyle d_id_i}^{\scriptscriptstyle Z^{\prime}}}}
&
\vtzp&=\kpzp\sqrt{\frac{3\,\xi_{\scriptscriptstyle{l_il_i}}^{\scriptscriptstyle{Z^{\prime}}}}{\Gamma_{\scriptscriptstyle l_il_i}^{\scriptscriptstyle Z^{\prime}}}}=\kpzp\sqrt{\frac{3\,\xi_{\scriptscriptstyle{l_{\nu_i}l_{\nu_i}}}^{\scriptscriptstyle{Z^{\prime}}}}{\Gamma_{\scriptscriptstyle l_{\nu_i}l_{\nu_i}}^{\scriptscriptstyle Z^{\prime}}}}
\label{vkvts2}
\end{align} 
where the branching ratios must satisfy:
\begin{align}
\xi_{\scriptscriptstyle{Qq}}^{\scriptscriptstyle{Z^{\prime}}}+\sum_{i=1}^{3}\left[\xi_{\scriptscriptstyle{u_iu_i}}^{\scriptscriptstyle{Z^{\prime}}}+\xi_{\scriptscriptstyle{d_id_i}}^{\scriptscriptstyle{Z^{\prime}}}+\xi_{\scriptscriptstyle{l_il_{i}}}^{\scriptscriptstyle{Z^{\prime}}}+\xi_{\scriptscriptstyle{l_{\nu_i}l_{\nu_i}}}^{\scriptscriptstyle{Z^{\prime}}}\right]&=1.
\label{brs}
\end{align}
It is evident from eq.~(\ref{vkvts2}), that the parameter $\kpzp$ is not free, but it depends on the SSM parameters ($\vkzp$ or $\vtzp$), the branching ratios and the kinematic functions. Thus, we have:
\begin{align}
\kpzp&=\vkzp\left[\Gamma_{\scriptscriptstyle u_iu_i}^{\scriptscriptstyle Z^{\prime}}/\xi_{\scriptscriptstyle{u_iu_i}}^{\scriptscriptstyle{Z^{\prime}}}\right]^{1/2}=\vtzp\left[\Gamma_{\scriptscriptstyle l_il_i}^{\scriptscriptstyle Z^{\prime}}/\left(3\,\xi_{\scriptscriptstyle{l_{\nu_i}l_{\nu_i}}}\right)\right]^{1/2}
\end{align}
\noindent
If we assume that $\vkzp=\vtzp$ (SSM), we find the following relations between the branching ratios and the kinematic functions of the ordinary fermions:
\begin{align}
 \xi_{\scriptscriptstyle{l_{\nu_i}l_{\nu_i}}}^{\scriptscriptstyle{Z^{\prime}}}&=\xi_{\scriptscriptstyle{l_il_i}}^{\scriptscriptstyle{Z^{\prime}}}\, \frac{\Gamma_{\scriptscriptstyle l_{\nu_i}l_{\nu_i}}^{\scriptscriptstyle Z^{\prime}}}{\Gamma_{\scriptscriptstyle l_il_i}^{\scriptscriptstyle Z^{\prime}}}&
 \xi_{\scriptscriptstyle{u_iu_i}}^{\scriptscriptstyle{Z^{\prime}}}&=3\, \xi_{\scriptscriptstyle{l_il_i}}^{\scriptscriptstyle{Z^{\prime}}}\, \frac{\Gamma_{\scriptscriptstyle u_iu_i}^{\scriptscriptstyle Z^{\prime}}}{\Gamma_{\scriptscriptstyle l_il_i}^{\scriptscriptstyle Z^{\prime}}}&
 \xi_{\scriptscriptstyle{d_id_i}}^{\scriptscriptstyle{Z^{\prime}}}&=3\, \xi_{\scriptscriptstyle{l_il_i}}^{\scriptscriptstyle{Z^{\prime}}}\, \frac{\Gamma_{\scriptscriptstyle d_id_i}^{\scriptscriptstyle Z^{\prime}}}{\Gamma_{\scriptscriptstyle l_il_i}^{\scriptscriptstyle Z^{\prime}}}
\end{align}

\noindent
Now, if we assume that $\xi_{\scriptscriptstyle{Qq}}^{\scriptscriptstyle{Z^{\prime}}}=\xi_{\scriptscriptstyle{u_iu_i}}^{\scriptscriptstyle{Z^{\prime}}}, 2\, \xi_{\scriptscriptstyle{u_iu_i}}^{\scriptscriptstyle{Z^{\prime}}}, 3\, \xi_{\scriptscriptstyle{u_iu_i}}^{\scriptscriptstyle{Z^{\prime}}}$ we get, respectively, $\xi_{\scriptscriptstyle{Qq}}^{\scriptscriptstyle{Z^{\prime}}}=9.60\%, 17.50\%, 24.12\%$.\\

\noindent
In table~\ref{tabVp1zp}, for the sike of comparison, we provide the numerical values of $\kpzp$, $\kzpl$ and $\Gamma_{\scriptscriptstyle Tb}^{\scriptscriptstyle Z^{\prime}}$ for the mass values considered in our phenomenological study ($m_{\scriptscriptstyle T}=\frac{1}{2}, \frac{2}{3}, \frac{5}{6}\, m_{Z^{\prime}}$) in both the SSM and CSSM:

\begin{table*}[h!]
\centering
 \renewcommand{\arraystretch}{1.40}
 \setlength{\tabcolsep}{12.5pt}
  \boldmath
 \begin{adjustbox}{width=13cm,height=1.25cm}
 \begin{tabular}{!{\vrule width 1.5pt}l!{\vrule width 1.5pt}l!{\vrule width 1pt}l!{\vrule width 1pt}l!{\vrule width 1.5pt}l!{\vrule width 1pt}l!{\vrule width 1.5pt}}
 \cline{1-4}
&\multicolumn{1}{c!{\vrule width 1.0pt}}{{$\bf \vkzp=0.5$}}
&\multicolumn{1}{c!{\vrule width 1.0pt}}{{$\bf \vkzp=1.0$}}
&\multicolumn{1}{c!{\vrule width 1.0pt}}{{$\bf \vkzp=2.0$}}
&\multicolumn{2}{c}{{}}\\
\noalign{\hrule height 1pt}
$\bf \xi^{\scriptscriptstyle{Z^{\prime}}}_{\scriptscriptstyle{tT}}$ & $9.6\%$ &$9.6\%$ &$9.6\%$ & \multicolumn{2}{c!{\vrule width 1.5pt}}{{\bf Kinematic function}} \\
\cdashline{2-4}
\cline{5-6}
$\bf \kpzp$&\bf $0.432058$ &\bf $0.864115$ & \bf $1.72823$ &  $\bf m_{\scriptscriptstyle Z^{\prime}}/m_{\scriptscriptstyle Q}$& $\bf \Gamma_{\scriptscriptstyle Tt}^{\scriptscriptstyle Z^{\prime}}$ \\
\noalign{\hrule height 1pt}
$\bf \kzpl$ & $0.337$ & $0.673$ & $1.346$ &  $1/2$ & $0.158$  \\
\cdashline{2-6}
$\bf \kzpl$ & $0.436$ & $0.872$ & $1.744$ &  $2/3$ & $0.094$   \\
\cdashline{2-6}
$\bf \kzpl$ & $0.755$ & $1.509$ & $3.020$ &  $5/6$ & $0.0314$  \\
\noalign{\hrule height 1pt}
\end{tabular}
\end{adjustbox}
  \caption{\footnotesize Some numerical values of the coupling $\kzpl$ for the mass values: $m_{\scriptscriptstyle T}=\frac{1}{2}, \frac{2}{3}, \frac{5}{6}\, m_{Z^{\prime}}$.}
   \label{tabVp1zp}
  \end{table*}

\subsubsection*{The parameter $\khpl$}
Now, let's find the parameter that describes the mixing between $Q$ and $\hp$ in the case where $m_{\scriptscriptstyle H^{\prime}}>m_{\scriptscriptstyle Q}$. The partial decay widths of all possible decay channels of the new heavy scalar are \footnote{We mention that the decays of $\hp$ to gauge bosons or pair of $Q$ are ignored in this approach.}:
\begin{align}
 &\Gamma(\hp\rightarrow t\bar t)=(\vkhp)^2
 \frac{N_{\scriptscriptstyle{C}}}{8\pi}\frac{m_{\scriptscriptstyle{H^{\prime}}}\, m_{\scriptscriptstyle{t}}^2}{v^2}\Gamma_{\scriptscriptstyle{tt}}^{\scriptscriptstyle{H^{\prime}}},&
&\Gamma(\hp\rightarrow Q\bar q+ \bar Qq)=(\khp)^2\frac{N_{\scriptscriptstyle{C}}}{8\pi}\frac{m_{\scriptscriptstyle{H^{\prime}}}\, m_{\scriptscriptstyle{Q}}^2}{v^2}\Gamma_{\scriptscriptstyle{Qq}}^{\scriptscriptstyle{H^{\prime}}}.
\end{align}
with
\begin{align}
 &\Gamma_{\scriptscriptstyle{tt}}^{\scriptscriptstyle{H^{\prime}}}=\biggl(1-4\frac{m_{\scriptscriptstyle{t}}^2}{m_{\scriptscriptstyle{H^{\prime}}}^2}\biggr)^{3/2},&
&\Gamma_{\scriptscriptstyle{Qq}}^{\scriptscriptstyle{H^{\prime}}}=\sqrt{\lambda\biggl(1,\frac{m_{\scriptscriptstyle{Q}}^2}{m_{\scriptscriptstyle{H^{\prime}}}^2},\frac{m_{\scriptscriptstyle{q}}^2}{m_{\scriptscriptstyle{H^{\prime}}}^2}\biggr)}\biggl[\biggl(1+\frac{m_{\scriptscriptstyle{q}}^2}{m_{\scriptscriptstyle{H^{\prime}}}^2}\biggr)\biggl(1-\frac{m_{\scriptscriptstyle{q}}^2+m_{\scriptscriptstyle{Q}}^2}{m_{\scriptscriptstyle{H^{\prime}}}^2}\biggr)-4\frac{m_{\scriptscriptstyle{q}}^2}{m_{\scriptscriptstyle{H^{\prime}}}^2}\biggr]. 
\label{xihpQq}
\end{align}
and
\begin{align}
 \khpl&=\khp \frac{m_{\scriptscriptstyle{Q}}}{v}
\end{align}
where $\khpr$ is neglected.\\

\noindent
The total decay width of $\hp$ is given by:
\begin{align}
\Gamma_{\scriptscriptstyle{H^{\prime}}}^{\text{tot}}&= \Gamma(\hp\rightarrow t\bar t)+\Gamma(\hp\rightarrow Q\bar q+\bar Q q)
\end{align}

\noindent
The branching ratio of $\hp$ into $Q\bar q$ or $\bar Qq$ is defined by:
\begin{align}
 \text{Br}(\hp\rightarrow Q\bar q+\bar Q q)
 &=\frac{(\khp)^2 \Gamma_{\scriptscriptstyle{Qq}}^{\scriptscriptstyle{H^{\prime}}}}{m_{\scriptscriptstyle{t}}^2/m_{\scriptscriptstyle{Q}}^2(\vkhp)^2\Gamma_{\scriptscriptstyle{tt}}^{\scriptscriptstyle{H^{\prime}}}+(\khp)^2\Gamma_{\scriptscriptstyle{Qq}}^{\scriptscriptstyle{H^{\prime}}}}
 \equiv \xi^{\scriptscriptstyle{H^{\prime}}}_{\scriptscriptstyle{Qq}}
 \label{brhp}
\end{align}
We set the denominator of eq. (\ref{brhp}) equal to the square of the new parameter $\kphp$. Then, the parameters $\vkhp$, $\khp$ and $\khpl$ are expressed in terms of the branching ratios, the kinematic functions and the parameter $\kphp$ as follows:
\begin{align}
\color{black} \vkhp&={\color{black}\kphp\, \sqrt{\frac{\xi_{\scriptscriptstyle{tt}}^{\scriptscriptstyle{H^{\prime}}}}{\Gamma_{\scriptscriptstyle{tt}}^{\scriptscriptstyle{H^{\prime}}}}}}&
\khp&=\kphp\, \sqrt{\frac{\xi_{\scriptscriptstyle{Qq}}^{\scriptscriptstyle{H^{\prime}}}}{\Gamma_{\scriptscriptstyle{Qq}}^{\scriptscriptstyle{H^{\prime}}}}}&
\khpl&=\kphp\, \frac{m_{\scriptscriptstyle Q}}{v}\, \sqrt{\frac{\xi_{\scriptscriptstyle{Qq}}^{\scriptscriptstyle{H^{\prime}}}}{\Gamma_{\scriptscriptstyle{Qq}}^{\scriptscriptstyle{H^{\prime}}}}}
\label{nonc-khp}
\end{align} 
Similar to $\kwpl$ and $\kzpl$, to avoid the dangerous region when $m_{\scriptscriptstyle Q}\approx m_{\scriptscriptstyle H^{\prime}}$, we can modify this parametrization by replacing the two key elements $\xi_{\scriptscriptstyle{Qq}}^{\scriptscriptstyle{H^{\prime}}}$ and $\Gamma_{\scriptscriptstyle{Qq^{\prime}}}^{\scriptscriptstyle{H^{\prime}}}$, with their values in the limit $m_{\scriptscriptstyle Q}/m_{\scriptscriptstyle H^{\prime}}\rightarrow0$, while requiring that their ratio remains constant, i.e.
\begin{align}
\frac{\xi_{\scriptscriptstyle{Qq}}^{\scriptscriptstyle{H^{\prime}}}}{\Gamma_{\scriptscriptstyle{Qq}}^{\scriptscriptstyle{H^{\prime}}}}=\frac{\xi_{\scriptscriptstyle{Qq}}^{\scriptscriptstyle{H^{\prime}0}}}{\Gamma_{\scriptscriptstyle{Qq}}^{\scriptscriptstyle{H^{\prime}0}}}\equiv \text{constant} 
\end{align}

\noindent
In the asymptotic limit where $m_{\scriptscriptstyle{H^{\prime}}}\rightarrow\infty$, the kinematic functions $\Gamma_{\scriptscriptstyle{tt}}^{\scriptscriptstyle{H^{\prime}}}$ and $\Gamma_{\scriptscriptstyle{Qq}}^{\scriptscriptstyle{H^{\prime}}}$ tend to unity ($\Gamma_{\scriptscriptstyle{tt}}^{\scriptscriptstyle{H^{\prime}}}=\Gamma_{\scriptscriptstyle{tt}}^{\scriptscriptstyle{H^{\prime}}}\approx 1$). Then, the $H^{\prime}$ decays $m_{\scriptscriptstyle{Q}}^2/(m_{\scriptscriptstyle{t}}^2+m_{\scriptscriptstyle{Q}}^2)\times 100\%$ to VLQ and a 3\textsuperscript{rd} quark and $m_{\scriptscriptstyle{t}}^2/(m_{\scriptscriptstyle{t}}^2+m_{\scriptscriptstyle{Q}}^2)\times 100\%$ to pair of ordinary top quarks, provided that $\khp=\vkhp=1$. So, the two branching ratios are related by $\xi_{\scriptscriptstyle{Qq}}^{\scriptscriptstyle{H^{\prime}}}/\xi_{\scriptscriptstyle{tt}}^{\scriptscriptstyle{H^{\prime}}}=m_{\scriptscriptstyle{Q}}^2/m_{\scriptscriptstyle{t}}^2$, which means that the decay mode $\hp\rightarrow Qq$ is the dominant one since the mass of the VLQ is supposed to be very large compared to the mass of the top quark. For example, for $m_{\scriptscriptstyle{Q}}=10\, m_{\scriptscriptstyle{t}}$, the $\hp$ will decay $99\%$ to $Qq$ and $1\%$ to $t\bar t$.
If we set $m_{\scriptscriptstyle{t}}^2/m_{\scriptscriptstyle{Q}}^2 \approx 0$ in eq. (\ref{brhp}), we see immediately that $\hp$ decay mostly to $Qq$ (i.e. $\xi^{\scriptscriptstyle{H^{\prime}}}_{\scriptscriptstyle{Qq}}\approx 1$ and $\xi^{\scriptscriptstyle{H^{\prime}}}_{\scriptscriptstyle{tt}}\approx 0$), then we get:
\begin{align}
\vkhp& \approx 0&
\khpl& \approx \kphp\, \frac{m_{\scriptscriptstyle Q}}{v}.
\end{align}

\noindent
We conclude this appendix by noting that the couplings of the VLQs to new heavy boson ($\kwpl$, $\kzpl$ and $\khpl$) can be large, which might violate the perturbative unitarity. In the next appendix and section~\ref{sec2}, we theoretically constrain these couplings to ensure the validity of perturbative treatments at the LHC.

%% file: appendix_B.tex
\section{Constraints from perturbative unitarity}
\label{appB}
As we have already noticed in the previous appendix, 
the interaction between the hypothetical particles (new heavy bosons with VLQs) becomes strong in certain configurations (when the ratio $m_{\scriptscriptstyle Q}/m_{\scriptscriptstyle B^{\prime}}$ approaches 1 for example). This can be seen in figure~\ref{coulingscase2}, where such strong interaction may lead to the breakdown of perturbation theory.
In this appendix, we adopt the method developed in ref.~\cite{prtunit3} to set upper bounds on the couplings $\kbpl$ using perturbative unitarity constraints. These bounds translate into relations between the critical masses and the couplings. To do so, we have to calculate the polarized amplitudes for some $2\rightarrow 2$ scattering processes involving SM quarks and VLQ in either initial or finial states, along with extra heavy bosons as real or virtual particles, at very high energy. In the following, we will explore several channels to determine the most restrictive bounds.\\

\noindent
Let us focus on the case where the VLQ is a top partner ($Q\equiv T$). We consider first the processes: $F_{i}\bar{F}_i\longleftrightarrow F_{j}\bar{F}_j$ and $F_{i}\bar{F}_i\longleftrightarrow V_{\alpha}^{\scriptscriptstyle+}V_{\beta}^{\scriptscriptstyle-}, V_{\alpha}^{\scriptscriptstyle0}V_{\beta}^{\scriptscriptstyle0}, S_{\alpha}S_{\beta}, V_{\alpha}^{\scriptscriptstyle 0}S_{\beta}$, with $(F_{\scriptscriptstyle 1,2,3}\equiv q_{_u}, q_{_d}, Q)$, $(V^{\scriptscriptstyle \pm}_{\scriptscriptstyle1,2}\equiv W, W^{\prime})$, $(V^{\scriptscriptstyle0}_{\scriptscriptstyle1,2}\equiv Z, Z^{\prime})$ and $(S_{\scriptscriptstyle1,2}\equiv H, H^{\prime})$. All possible $J=0$ and $J=1$ channels are given by:
\begin{align}
J=0: & \left\{
\begin{aligned}
F_{i+}\bar{F}_{i+},F_{i-}\bar{F}_{i-}.
\end{aligned}
\right.\\
J=1: & \left\{
\begin{aligned}
&F_{i+}\bar{F}_{i-},\\
&F_{i+}\bar{F}_{i+},F_{i-}\bar{F}_{i-}, V^{+}_{\scriptscriptstyle \alpha L}V^{-}_{\scriptscriptstyle \beta L}, {\scriptscriptstyle\frac{1}{\sqrt2}}V^{\scriptscriptstyle 0}_{\scriptscriptstyle \alpha L}V^{\scriptscriptstyle 0}_{\scriptscriptstyle \alpha L}, V^{\scriptscriptstyle 0}_{\scriptscriptstyle \alpha L}V^{\scriptscriptstyle 0}_{\scriptscriptstyle \beta L}, {\scriptscriptstyle\frac{1}{\sqrt2}}S_{\scriptscriptstyle \alpha}S_{\scriptscriptstyle \alpha}, S_{\scriptscriptstyle \alpha}S_{\scriptscriptstyle \beta}, V^{\scriptscriptstyle}_{\scriptscriptstyle \alpha L}S_{\scriptscriptstyle \beta},\\
&F_{i-}\bar{F}_{i+}.
\end{aligned}
\right.
\label{ffbarVV}
\end{align}
\noindent
for $i=1,2,3$ and $\alpha, \beta=1,2$. Here, the subscript $L$ denotes the longitudinal vector bosons, and $q_{_{u/d}}$ refers to the up and down quarks of a given generation, which we neglect their masses in this discussion. \\

\noindent
We have to calculate the polarized matrix elements for 9 scattering processes for $J=0$ (and more than that for $J=1$) in the 6 benchmark scenarios: ${\bf Q}^{\scriptscriptstyle\{3\}}_{\scriptscriptstyle\{W\}}$, ${\bf Q}^{\scriptscriptstyle\{3\}}_{\scriptscriptstyle\{Z\}}$, ${\bf Q}^{\scriptscriptstyle\{3\}}_{\scriptscriptstyle\{H\}}$, ${\bf Q}^{\scriptscriptstyle\{3\}}_{\scriptscriptstyle\{W^{\prime}\}}$, ${\bf Q}^{\scriptscriptstyle\{3\}}_{\scriptscriptstyle\{Z^{\prime}\}}$ and ${\bf Q}^{\scriptscriptstyle\{3\}}_{\scriptscriptstyle\{H^{\prime}\}}$.
To compute the polarized scattering amplitudes associated to these processes, we need the explicit formulae of the helicity spinors involving massive (massless) fermions. For a particle moving in an arbitrary direction, the four possible helicity spinors are given by:
\begin{align}
  u_{\uparrow}&=\sqrt{E+m}
  \left[ 
\begin{array}{c}
\cos(\frac{\theta}{2}) \\
\sin(\frac{\theta}{2}) e^{i \phi} \\
\frac{|\overrightarrow{p}|}{E+m} \cos\left(\frac{\theta}{2}\right) \\
\frac{|\overrightarrow{p}|}{E+m} \sin\left(\frac{\theta}{2}\right) e^{i \phi} 
\end{array} \right] ,&
u_{\downarrow}&=\sqrt{E+m}
\left[
\begin{array}{c}
-\sin(\frac{\theta}{2}) \\
\cos(\frac{\theta}{2}) e^{i \phi} \\
\frac{|\overrightarrow{p}|}{E+m} \sin\left(\frac{\theta}{2}\right) \\
-\frac{|\overrightarrow{p}|}{E+m} \cos\left(\frac{\theta}{2}\right) e^{i \phi} 
\end{array}
\right], \nonumber\\
v_{\uparrow}&=\sqrt{E+m}
  \left[ 
\begin{array}{c}
\frac{|\overrightarrow{p}|}{E+m} \sin\left(\frac{\theta}{2}\right) \\
-\frac{|\overrightarrow{p}|}{E+m} \cos\left(\frac{\theta}{2}\right) e^{i \phi} \\
-\sin(\frac{\theta}{2}) \\
\cos(\frac{\theta}{2}) e^{i \phi} 
\end{array} \right], &
v_{\downarrow}&=\sqrt{E+m}
\left[
\begin{array}{c}
\frac{|\overrightarrow{p}|}{E+m} \cos\left(\frac{\theta}{2}\right) \\
\frac{|\overrightarrow{p}|}{E+m} \sin\left(\frac{\theta}{2}\right) e^{i \phi} \\
\cos(\frac{\theta}{2}) \\
\sin(\frac{\theta}{2}) e^{i \phi} 
\end{array}
\right].
\end{align}

\noindent
Let us denote the 3-momentum of the particle $i$ for $i=1,2,3,4$, in the 2-to-2 process $1+2\rightarrow 3+4$, by $\vec{p}_i$. In the center-of-mass frame (CM), we can fix the polar and the azimuthal angles $\theta$ and $\phi$ associated to the initial and final-state particles as follows:
\begin{align}
 \vec{p}_1:& (\theta=0, \phi=0), & 
 \vec{p}_2:& (\theta=\pi, \phi=\pi), &
 \vec{p}_3:& (\theta=\theta, \phi=0), &  
 \vec{p}_4:& (\theta=\pi-\theta, \phi=\pi). 
\end{align}
\noindent
Since we are only interested on the leading high-energy behaviors ($\sqrt{s}\gg m_{\scriptscriptstyle V}, m_{\scriptscriptstyle S}, m_{\scriptscriptstyle Q}$), we can neglect the transverse components of the massive gauge bosons and retain only the longitudinal ones, which dominants in this regime. Thus, we may disregard the $g_{\mu\nu}$ term in the internal propagators and focus solely on the longitudinal polarization vectors ($\varepsilon^{\mu}_{L}$). This leads to the following approximations:
\begin{align}
 P_{V^{\prime}}^{\mu\nu}(k)&\approx \frac{i}{m_{\scriptscriptstyle V^{\prime}}^2}\, \frac{k^{\mu}\, k^{\nu}}{k^{2}-m^2_{\scriptscriptstyle V^{\prime}}}, & \varepsilon^{\mu}\longrightarrow \varepsilon^{\mu}_{L}(k)&=\frac{k_{\mu}}{m_{\scriptscriptstyle V^{\prime}}}.
\end{align}
\noindent
Here, $P_{V^{\prime}}^{\mu\nu}$ denotes the internal propagator of the vector boson $V^{\prime}$ with 4-momentum $k$ and mass $m_{\scriptscriptstyle V^{\prime}}$.\\

\noindent
In the following, we present the non-vanishing matrix element $\mathcal{T}$ of the polarized scattering amplitudes of the processes $F_{i}\bar{F}_i\rightarrow F_{j}\bar{F}_j$, evaluated in the CM frame in the high-energy limit ($\sqrt{s}\gg m_{\scriptscriptstyle V}, m_{\scriptscriptstyle S}, m_{\scriptscriptstyle Q}$).\\

\noindent
$\bullet$ Processes $q_{_d}\bar{q}_{_d}\longrightarrow q_{_u}\bar{q}_{_u}$, $q_{_u}\bar{q}_{_u}\longrightarrow q_{_d}\bar{q}_{_d}$, $q_{_d}\bar{q}_{_d}\longrightarrow q_{_d}\bar{q}_{_d}$ and $q_{_u}\bar{q}_{_u}\longrightarrow q_{_u}\bar{q}_{_u}$: 
These processes involve $\wps$ in the $t$-channel and $\zp$ and $\hp$ in both the $s$- and $t$-channels. However, the non vanishing polarized amplitudes are proportional to $q_{_u}$ and $q_{_d}$ masses, and can therefore be neglected for very high energies. Thus, their amplitudes are set to zero. \\ 

\noindent
$\bullet$ Processes $Q\bar{Q}\longrightarrow Q\bar{Q}$: In our model, this process is mediated by gluon in the $s$- and $t$-channels. Their contribution is sub-leading at high energies since it is proportional to $\alpha_s$ rather than the VLQ mass, and thus can be neglected \footnote{In this appendix, we ignored contributions from virtual photons and gluons in fermion-fermion scattering \cite{prtunit3}.}. \\

\noindent
$\bullet$ Processes $q\bar{q}\longrightarrow Q\bar{Q}$\, ($Q\bar{Q}\longrightarrow q\bar{q}$): \\

\vspace{-0.25cm}
\begin{table*}[ht!]
\centering
 \renewcommand{\arraystretch}{1.40}
 \setlength{\tabcolsep}{10pt}
  \boldmath
 \begin{adjustbox}{width=16cm,height=2.5cm}
 \begin{tabular}
 {!{\vrule width 1pt}l!{\vrule width 1pt}c!{\vrule width 1pt}c!{\vrule width 1pt}c!{\vrule width 1pt}}
  \noalign{\hrule height 1pt}
 \bf Channels  & \bf Helicity configuration & \bf $\mathcal{T}$ & $\mathcal{T}[m_{q}\rightarrow 0, \ksr\rightarrow0]$  \\
  \noalign{\hrule height 1pt}
& $ + \,\,\ + \longrightarrow - \,\,\ - $ & $ g^2 \kvlsq m_{\scriptscriptstyle q} m_{\scriptscriptstyle Q}/(4\,  c_{\scriptscriptstyle w}^2 m_{\scriptscriptstyle V^{_0}}^2) $& 0 \\
 \bf $t$-channel $V^{0}$
   & $ - \,\,\ - \longrightarrow + \,\,\ + $ & $ g^2 \kvlsq m_{\scriptscriptstyle q} m_{\scriptscriptstyle Q}/(4\,  c_{\scriptscriptstyle w}^2 m_{\scriptscriptstyle V^{_0}}^2) $ & 0 \\
\cdashline{2-4}
   $\mathbf{q_{_u}\bar{q}_{_u}\rightarrow Q\bar{Q}}$& $ + \,\,\ - \longrightarrow - \,\,\ + $ & $ g^2 \kvlsq m_{\scriptscriptstyle q}^2/(4\, c_{\scriptscriptstyle w}^2 m_{\scriptscriptstyle V^{_0}}^2) $ & 0 \\
 & $ - \,\,\ + \longrightarrow + \,\,\ -$ & $ g^2 \kvlsq m_{\scriptscriptstyle Q}^2/(4\, c_{\scriptscriptstyle w}^2 m_{\scriptscriptstyle V^{_0}}^2) $ & $ g^2 \kvlsq m_{\scriptscriptstyle Q}^2/(4\, c_{\scriptscriptstyle w}^2 m_{\scriptscriptstyle V^{_0}}^2) $ \\
\noalign{\hrule height 1pt}
  & $ + \,\,\ + \longrightarrow - \,\,\ - $ & $ -\ksl \ksr\, (-\ksl \ksr)  $& 0\, (0) \\
 \bf $t$-channel $S$
   & $ - \,\,\ - \longrightarrow + \,\,\ + $ & $ -\ksl \ksr\, (-\ksl \ksr)$ & 0\, (0) \\
\cdashline{2-4}
  $\mathbf{q_{_u}\bar{q}_{_u}\rightarrow Q\bar{Q}}$ & $ + \,\,\ - \longrightarrow - \,\,\ + $ & $ \ksrsq\, (\kslsq) $ & 0\, $(\kslsq)$ \\
& $ - \,\,\ + \longrightarrow + \,\,\ -$ & $ \kslsq\, (\ksrsq) $ & $\kslsq$\, (0) \\
\noalign{\hrule height 1pt}
   & $ \mathbf{ + \,\,\ + \longrightarrow - \,\,\ -} $ & $ g^2 \kvpmlsq m_{\scriptscriptstyle q} m_{\scriptscriptstyle Q}/(2\, m_{\scriptscriptstyle V^{_\pm}}^2) $& 0 \\
 \bf $t$-channel $V^{\pm}$
   & $ - \,\,\ - \longrightarrow + \,\,\ + $ & $ g^2 \kvpmlsq m_{\scriptscriptstyle q} m_{\scriptscriptstyle Q}/(2 m_{\scriptscriptstyle V^{_\pm}}^2) $ & 0 \\
\cdashline{2-4}
 $\mathbf{q_{_d}\bar{q}_{_d}\rightarrow Q\bar{Q}}$  & $ + \,\,\ - \longrightarrow - \,\,\ + $ & $ g^2 \kvpmlsq m_{\scriptscriptstyle q}^2/(2\, m_{\scriptscriptstyle V^{_\pm}}^2) $ & 0 \\
  & $ - \,\,\ + \longrightarrow + \,\,\ -$ & $ g^2 \kvpmlsq m_{\scriptscriptstyle Q}^2/(2\, m_{\scriptscriptstyle V^{_\pm}}^2) $ & $ g^2 \kvpmlsq m_{\scriptscriptstyle Q}^2/(2\, m_{\scriptscriptstyle V^{_\pm}}^2) $  \\
 \noalign{\hrule height 1pt}
\end{tabular}
\end{adjustbox}
  \end{table*}
\noindent
The processes $q_{_u}\bar{q}_{_u}\longrightarrow Q\bar{Q}$ and $Q\bar{Q}\longrightarrow q_{_u}\bar{q}_{_u}$ have the same polarized amplitudes, except in the case of $\hp$ and $H$ exchange, where $\mathcal{T}_{+--+}$ and $\mathcal{T}_{-++-}$ are interchanged. \\ 

\noindent
We find that, for the 9 processes considered above, there is no $J=0$ contribution. All the non-vanishing channels have total spin $J=1$ (namely, $\mathcal{T}_{-++-}$ and $\mathcal{T}_{+--+}$), see the table above. Now, let's consider some processes involving gauge bosons and scalars in the final state (or initial state). 

\noindent
$\bullet$ Processes $Q\bar{Q}\longleftrightarrow Z^{\prime}Z^{\prime}, ZZ$:
\begin{table*}[h!]
\centering
 \renewcommand{\arraystretch}{1.40}
 \setlength{\tabcolsep}{10pt}
  \boldmath
 \begin{adjustbox}{width=16cm,height=2.0cm}
 \begin{tabular}
 {!{\vrule width 1pt}l!{\vrule width 1pt}c!{\vrule width 1pt}c!{\vrule width 1pt}c!{\vrule width 1pt}}
  \noalign{\hrule height 1pt}
 \bf Channels  & \bf Helicity configuration & \bf $\mathcal{T}^{Z^{\prime}Z^{\prime}}$ & \bf $\mathcal{T}^{ZZ}$ \\
  \noalign{\hrule height 1pt}
     & $ + \,\,\ + $ & $ - g^{2} \kzplsq m_{\scriptscriptstyle Q} \sqrt{s}\, (1+ \cos\theta) / (8 c_{w}^2 m_{\scriptscriptstyle Z^{\prime}}^2) $&$ - g^{2} \kzlsq m_{\scriptscriptstyle Q} \sqrt{s}\, (1+ \cos\theta) / (8 c_{w}^2 m_{\scriptscriptstyle Z}^2) $  \\
 \bf $t$-channel  & $ - \,\,\ - $ & $  g^{2} \kzplsq m_{\scriptscriptstyle Q} \sqrt{s}\, (1+ \cos\theta) /(8 c_{w}^2 m_{\scriptscriptstyle Z^{\prime}}^2) $& $  g^{2} \kzlsq m_{\scriptscriptstyle Q} \sqrt{s}\, (1+ \cos\theta) /(8 c_{w}^2 m_{\scriptscriptstyle Z}^2) $ \\
\cdashline{2-4}
   & $ + \,\,\ - $ & $ g^{2} \kzplsq m_{\scriptscriptstyle Q}^{2} (1+\cos\theta)/(8c_{w}^2 m_{\scriptscriptstyle Z^{\prime}}^2)\times \sin\theta/(1-\cos\theta)  $&$ g^{2} \kzlsq m_{\scriptscriptstyle Q}^{2} (1+\cos\theta)/(8c_{w}^2 m_{\scriptscriptstyle Z}^2)\times \sin\theta/(1-\cos\theta)  $ \\
   & $ - \,\,\ + $ & $ -g^{2} \kzplsq s\, \sin\theta /(8 c_{w}^2 m_{\scriptscriptstyle Z^{\prime}}^2) $ & $ -g^{2} \kzlsq s\, \sin\theta /(8 c_{w}^2 m_{\scriptscriptstyle Z}^2) $\\
 \noalign{\hrule height 1pt}
 & $ - \,\,\ + $ & $ - g^{2} \kzplsq m_{\scriptscriptstyle Q} \sqrt{s} \, (1-\cos\theta)/(8 c_{w}^2 m_{\scriptscriptstyle Z^{\prime}}^2) $& $ - g^{2} \kzlsq m_{\scriptscriptstyle Q} \sqrt{s} \, (1-\cos\theta)/(8 c_{w}^2 m_{\scriptscriptstyle Z}^2) $ \\
 \bf $u$-channel & $ - \,\,\ - $ & $ g^{2} \kzplsq m_{\scriptscriptstyle Q} \sqrt{s} \, (1-\cos\theta)/(8 c_{w}^2 m_{\scriptscriptstyle Z^{\prime}}^2) $ & $ g^{2} \kzlsq m_{\scriptscriptstyle Q} \sqrt{s} \, (1-\cos\theta)/(8 c_{w}^2 m_{\scriptscriptstyle Z}^2) $\\
\cdashline{2-4}
   & $ + \,\,\ - $ & $ -g^{2} \kzplsq m_{\scriptscriptstyle Q}^{2} \sin\theta\tan(\theta/2)^2 /(8 c_{w}^2 m_{\scriptscriptstyle Z^{\prime}}^2) $& $ -g^{2} \kzlsq m_{\scriptscriptstyle Q}^{2} \sin\theta\tan(\theta/2)^2 /(8 c_{w}^2 m_{\scriptscriptstyle Z}^2) $ \\
   & $ - \,\,\ + $ & $ g^{2} \kzplsq s\, \sin\theta/(8 c_{w}^2 m_{\scriptscriptstyle Z^{\prime}}^2) $ & $ g^{2} \kzlsq s\, \sin\theta/(8 c_{w}^2 m_{\scriptscriptstyle Z}^2) $ \\
 \noalign{\hrule height 1pt}
\end{tabular}
\end{adjustbox}
  \end{table*}
  
\noindent
$\bullet$ Processes $Q\bar{Q}\longleftrightarrow H^{\prime}H^{\prime}, HH$:
\begin{table*}[h!]
\centering
 \renewcommand{\arraystretch}{1.40}
 \setlength{\tabcolsep}{10pt}
  \boldmath
 \begin{adjustbox}{width=16cm,height=2.0cm}
 \begin{tabular}
 {!{\vrule width 1pt}l!{\vrule width 1pt}c!{\vrule width 1pt}c!{\vrule width 1pt}c!{\vrule width 1pt}}
  \noalign{\hrule height 1pt}
 \bf Channels  & \bf Helicity configuration & \bf $\mathcal{T}^{H^{\prime}H^{\prime}}$& \bf $\mathcal{T}^{HH}$  \\
  \noalign{\hrule height 1pt}
     & $ + \,\,\ + $ & $ \khplsq m_{\scriptscriptstyle Q}/\sqrt{s}$ & $ \khlsq m_{\scriptscriptstyle Q}/\sqrt{s}$ \\
 \bf $t$-channel  & $ - \,\,\ - $ & $-\khplsq m_{\scriptscriptstyle Q}/\sqrt{s}$ & $-\khlsq m_{\scriptscriptstyle Q}/\sqrt{s}$ \\
\cdashline{2-4}
   & $ + \,\,\ - $ & $ -\khplsq\cot(\theta/2)$ & $ -\khlsq\cot(\theta/2)$\\
   & $ - \,\,\ + $ & $-\khplsq m_{\scriptscriptstyle Q}^2 \cot(\theta/2)/s $ & $-\khlsq m_{\scriptscriptstyle Q}^2 \cot(\theta/2)/s $ \\
 \noalign{\hrule height 1pt}
     & $ + \,\,\ + $ & $ \khplsq m_{\scriptscriptstyle Q}/\sqrt{s}$ & $ \khlsq m_{\scriptscriptstyle Q}/\sqrt{s}$  \\
 \bf $u$-channel  & $ - \,\,\ - $ & $-\khplsq m_{\scriptscriptstyle Q}/\sqrt{s}$ & $-\khlsq m_{\scriptscriptstyle Q}/\sqrt{s}$ \\
\cdashline{2-4}
   & $ + \,\,\ - $ & $ \khplsq\tan(\theta/2)$ & $ \khlsq\tan(\theta/2)$\\
   & $ - \,\,\ + $ & $\khplsq m_{\scriptscriptstyle Q}^2 \tan(\theta/2)/s $ & $\khlsq m_{\scriptscriptstyle Q}^2 \tan(\theta/2)/s $\\ \noalign{\hrule height 1pt}
\end{tabular}
\end{adjustbox}
  \end{table*}

\noindent
$\bullet$ Processes $Q\bar{Q}\longleftrightarrow W^{+\prime}W^{-\prime},W^{+}W^{-}$:
\begin{table*}[h!]
\centering
 \renewcommand{\arraystretch}{1.40}
 \setlength{\tabcolsep}{10pt}
  \boldmath
 \begin{adjustbox}{width=16cm,height=1.25cm}
 \begin{tabular}
 {!{\vrule width 1pt}l!{\vrule width 1pt}c!{\vrule width 1pt}c!{\vrule width 1pt}c!{\vrule width 1pt}}
  \noalign{\hrule height 1pt}
 \bf Channels  & \bf Helicity configuration & \bf $\mathcal{T}^{ W^{\prime}W^{\prime}}$  &\bf $\mathcal{T}^{ WW}$\\
  \noalign{\hrule height 1pt}
     & $ + \,\,\ + $ & $ -g^{2} \kwplsq m_{\scriptscriptstyle Q} \sqrt{s}\, (1+\cos\theta)/4 m_{\scriptscriptstyle W^{\prime}}^2 $  &$ -g^{2} \kwlsq m_{\scriptscriptstyle Q} \sqrt{s}\, (1+\cos\theta)/4 m_{\scriptscriptstyle W}^2 $\\
 \bf $t$-channel  & $ - \,\,\ - $ & $ g^{2} \kwplsq m_{\scriptscriptstyle Q} \sqrt{s} (1+ \cos\theta) / 4 m_{\scriptscriptstyle W^{\prime}}^2 $ &$ g^{2} \kwlsq m_{\scriptscriptstyle Q} \sqrt{s} (1+ \cos\theta) / 4 m_{\scriptscriptstyle W}^2 $\\
\cdashline{2-4}
   & $ + \,\,\ - $ & $ g^{2} \kwplsq m_{\scriptscriptstyle Q}^{2} (1+\cos\theta)/4 m_{\scriptscriptstyle W^{\prime}}^2\times \sin\theta /(1-\cos\theta) $ &$ g^{2} \kwlsq m_{\scriptscriptstyle Q}^{2} (1+\cos\theta)/4 m_{\scriptscriptstyle W}^2\times \sin\theta /(1-\cos\theta) $\\
   &  $ - \,\,\ + $ & $ -g^{2} \kwplsq s \, \sin\theta / 4 m_{\scriptscriptstyle W^{\prime}}^2  $&$ -g^{2} \kwlsq s \, \sin\theta / 4 m_{\scriptscriptstyle W}^2  $\\
 \noalign{\hrule height 1pt}
\end{tabular}
\end{adjustbox}
  \end{table*}
  
\noindent
We observe that the matrix elements of processes involving gauge bosons grow like $s$ or $\sqrt{s}$ in some channels, thereby violating perturbative unitarity. Such divergences would not occur if the contribution of the interactions (self-interactions) among the various gauge bosons and scalars were included. However, this is beyond the scope of the present paper, as we aim to maintains the maximal independence on the BSM  models. In forthcoming works, we will generalize the present model by including such interactions.
We note that we have not listed all the matrix elements for the processes in eq.~({\ref{ffbarVV}}), as their high-energy behavior is similar to that shown in the 3 preceding tables.\\

\noindent
The previously considered processes did not help to constrains the model for the reasons stated above. Therefore, we will consider the following reactions: $q_{_{u}}\bar{Q}\rightarrow q_{_{u}}\bar{Q}$ and $q_{_{d}}\bar{Q}\rightarrow q_{_{d}}\bar{Q}$~\footnote{We examined numerous processes involving an ordinary quark and a VLQ in the initial or final state (e.g. $\bar{q}Q\rightarrow \bar{q}Q$, $\bar{q}Q\rightarrow q\bar{Q}$  or $qQ\rightarrow qQ$), we found that all lead to similar conclusions.}. Their matrix elements in the high-energy limit ($\sqrt{s}\gg m_{\scriptscriptstyle Q}, m_{\scriptscriptstyle V}, m_{\scriptscriptstyle Q}$) are given in the following tables:\\

\noindent
$\bullet$ Process: $q\bar{Q}\longrightarrow  q\bar{Q}$: the results are shown in the table on the following page.
\vspace{0.5cm}

\begin{table*}[h!]
\centering
 \renewcommand{\arraystretch}{1.40}
 \setlength{\tabcolsep}{10pt}
  \boldmath
 \begin{adjustbox}{width=16cm,height=3.0cm}
 \begin{tabular}
 {!{\vrule width 1pt}l!{\vrule width 1pt}c!{\vrule width 1pt}c!{\vrule width 1pt}c!{\vrule width 1pt}}
  \noalign{\hrule height 1pt}
 \bf Channels  & \bf Helicity configuration & \bf $\mathcal{T}$ & $\mathcal{T}[m_{q_{_u}}\rightarrow 0, \ksr\rightarrow0]$  \\
  \noalign{\hrule height 1pt}
& $ + \,\,\ + \longrightarrow + \,\, + $ & $ -g^2 \kvlsq m_{\scriptscriptstyle q_{_u}}^2/(4\,  c_{\scriptscriptstyle w}^2 m_{\scriptscriptstyle V^{_0}}^2) $& 0 \\
 \bf $s$-channel $V^{0}$
   & $ - \,\,\ - \longrightarrow - \,\,\ - $ & $ -g^2 \kvlsq m_{\scriptscriptstyle Q}^2/(4\,  c_{\scriptscriptstyle w}^2 m_{\scriptscriptstyle V^{_0}}^2) $ & $ -g^2 \kvlsq m_{\scriptscriptstyle Q}^2/(4\,  c_{\scriptscriptstyle w}^2 m_{\scriptscriptstyle V^{_0}}^2) $ \\
\cdashline{2-4}
 $\mathbf{q_{\scriptscriptstyle u}\bar{Q}\longrightarrow  q_{\scriptscriptstyle u}\bar{Q}}$  & $ + \,\, + \longrightarrow - \,\,\ - $ & $ -g^2 \kvlsq m_{\scriptscriptstyle q_{_u}} m_{\scriptscriptstyle Q}/(4\, c_{\scriptscriptstyle w}^2 m_{\scriptscriptstyle V^{_0}}^2) $ & 0 \\
 & $ - \,\,\ - \longrightarrow + \,\,\ +$ & $ -g^2 \kvlsq m_{\scriptscriptstyle q_{_u}} m_{\scriptscriptstyle Q}/(4\, c_{\scriptscriptstyle w}^2 m_{\scriptscriptstyle V^{_0}}^2) $ & 0 \\
\noalign{\hrule height 1pt}
  & $ + \,\,\ + \longrightarrow + \,\,\ + $ & $ -\ksrsq  $& 0 \\
 \bf $s$-channel $S$
   & $ - \,\,\ - \longrightarrow - \,\,\ - $ & $ -\kslsq$ & $-\kslsq$ \\
\cdashline{2-4}
 $\mathbf{q_{\scriptscriptstyle u}\bar{Q}\longrightarrow  q_{\scriptscriptstyle u}\bar{Q}}$  & $ + \,\,\ + \longrightarrow - \,\,\ - $ & $ \ksr \ksl $ & 0 \\
& $ - \,\,\ - \longrightarrow + \,\,\ +$ & $ \ksr \ksl $ & 0 \\
\noalign{\hrule height 1pt}
& $ + \,\,\ + \longrightarrow + \,\, + $ & $ -g^2 \kvpmlsq m_{\scriptscriptstyle q_{_d}}^2/(2\, m_{\scriptscriptstyle V^{_\pm}}^2) $& 0 \\
 \bf $s$-channel $V^{\pm}$
   & $ - \,\,\ - \longrightarrow - \,\,\ - $ & $ -g^2 \kvpmlsq m_{\scriptscriptstyle Q}^2/(2 m_{\scriptscriptstyle V^{_\pm}}^2) $ & $ -g^2 \kvpmlsq m_{\scriptscriptstyle Q}^2/(2 m_{\scriptscriptstyle V^{_\pm}}^2) $ \\
\cdashline{2-4}
 $\mathbf{q_{\scriptscriptstyle d}\bar{Q}\longrightarrow  q_{\scriptscriptstyle d}\bar{Q}}$  & $ + \,\, + \longrightarrow - \,\,\ - $ & $ -g^2 \kvpmlsq m_{\scriptscriptstyle q_{_d}} m_{\scriptscriptstyle Q}/(2 m_{\scriptscriptstyle V^{_\pm}}^2) $ & 0 \\
 & $ - \,\,\ - \longrightarrow + \,\,\ +$ & $ -g^2 \kvpmlsq m_{\scriptscriptstyle q_{_d}} m_{\scriptscriptstyle Q}/(2 m_{\scriptscriptstyle V^{_\pm}}^2) $ & 0 \\
 \noalign{\hrule height 1pt}
\end{tabular}
\end{adjustbox}
  \end{table*} 

\noindent
According to these results, the leading matrix elements at high energy are the $\mathcal{T}_{----}$ helicity amplitudes with total spin $J=0$, corresponding to gauge bosons or scalars in the $s$-channel. The partial-wave amplitude, in the high-energy limit, is defined by:
\begin{align}
a_0&=\frac{1}{32\pi}\int_{-1}^{+1}d\cos(\theta)\, \mathcal{T}(\sqrt{s},\cos(\theta),\{m_i\}).
\label{partAmpl}
\end{align}
where $\{m_i\}$ denotes the set of masses involved in the helicity amplitude $\mathcal{T}$.\\

\noindent
By inserting the different $\mathcal{T}_{----}$ helicity amplitudes into eq.~(\ref{partAmpl}) and integrating over the polar angle, we obtain the partial-wave amplitudes for the 6 benchmark scenarios, as follows:
\begin{align}
&{\bf Q}^{\scriptscriptstyle\{3\}}_{\scriptscriptstyle\{W\}}:\quad a_{0}^{\scriptscriptstyle\{W\}}=-\frac{g^2 \kwlsq}{32\, \pi}\frac{m_{\scriptscriptstyle Q}^2}{m_{\scriptscriptstyle W}^2},&&\qquad\qquad&
&{\bf Q}^{\scriptscriptstyle\{3\}}_{\scriptscriptstyle\{W^{\prime}\}}:\quad a_{0}^{\scriptscriptstyle\{W^{\prime}\}}=-\frac{g^2 \kwplsq}{32\, \pi}\frac{m_{\scriptscriptstyle Q}^2}{m_{\scriptscriptstyle W^{\prime}}^2}, \nonumber\\
&{\bf Q}^{\scriptscriptstyle\{3\}}_{\scriptscriptstyle\{Z\}}:\quad a_{0}^{\scriptscriptstyle\{Z\}}=-\frac{g^2 \kzlsq}{64\, c_{\scriptscriptstyle w}^2\, \pi}\frac{m_{\scriptscriptstyle Q}^2}{m_{\scriptscriptstyle Z}^2},&&&
&{\bf Q}^{\scriptscriptstyle\{3\}}_{\scriptscriptstyle\{Z^{\prime}\}}:\quad a_{0}^{\scriptscriptstyle\{Z^{\prime}\}}=-\frac{g^2 \kzplsq}{64\, c_{\scriptscriptstyle w}^2\, \pi}\frac{m_{\scriptscriptstyle Q}^2}{m_{\scriptscriptstyle Z^{\prime}}^2},\nonumber\\
&{\bf Q}^{\scriptscriptstyle\{3\}}_{\scriptscriptstyle\{H\}}:\quad a_{0}^{\scriptscriptstyle\{H\}}=-\frac{\khlsq}{16\, \pi},&&&
&{\bf Q}^{\scriptscriptstyle\{3\}}_{\scriptscriptstyle\{H^{\prime}\}}:\quad a_{0}^{\scriptscriptstyle\{H^{\prime}\}}=-\frac{\khplsq}{16\, \pi}.
\label{a0}
\end{align}
\noindent
Until now, we have not taken quark colors into account. To include them, we order the non-vanishing channels as: $q_{-}^{R}\bar{Q}_{-}^{\bar{R}}$, $q_{-}^{B}\bar{Q}_{-}^{\bar{B}}$ and $q_{-}^{G}\bar{Q}_{-}^{\bar{G}}$, where the superscripts $R (\bar{R})$, $B(\bar{B})$ and $G(\bar{G})$ denote the quark colors. We first focus on the benchmark scenario ${\bf Q}^{\scriptscriptstyle\{3\}}_{\scriptscriptstyle\{Z^{\prime}\}}$. Since the associated process is of $s$-channel type, all the color-neutral combinations contribute. Thus, the partial wave-amplitude matrix in the basis of color-neutral channels is given by:
\begin{align}
a_{0}^{\scriptscriptstyle\{Z^{\prime}\}}&=-\frac{g^2(\kzpl)^2}{64\, c_{\scriptscriptstyle w}^2\, \pi}\frac{m_{\scriptscriptstyle Q}^2}{m_{\scriptscriptstyle Z^{\prime}}^2}
\begin{pmatrix}
1&1 &1\\
1&1 &1\\
1&1 &1
\end{pmatrix}
\label{partAmplZp}
\end{align}
\noindent
The most restrictive constraints are obtained from the largest eigenvalue of the partial wave-amplitude matrix, cf. eq.~(\ref{partAmplZp}), which is given by:
\begin{align}
\left|\lambda_{0}^{\scriptscriptstyle\{Z^{\prime}\}}\right|&=\frac{g^2(\kzpl)^2}{c_{\scriptscriptstyle w}^2\, \pi}\frac{3}{64}\frac{m_{\scriptscriptstyle Q}^2}{m_{\scriptscriptstyle Z^{\prime}}^2}
\end{align}
Proceeding in the same way, we obtain the absolute values of the largest eigenvalues for the other benchmark scenarios:
\begin{align}
&{\bf Q}^{\scriptscriptstyle\{3\}}_{\scriptscriptstyle\{W\}}:\quad \left|\lambda_{0}^{\scriptscriptstyle\{W\}}\right|=\frac{g^2\kwlsq}{\pi}\frac{3}{32}\frac{m_{\scriptscriptstyle Q}^2}{m_{\scriptscriptstyle W}^2},&&\qquad\qquad&
&{\bf Q}^{\scriptscriptstyle\{3\}}_{\scriptscriptstyle\{W^{\prime}\}}:\quad \left|\lambda_{0}^{\scriptscriptstyle\{W^{\prime}\}}\right|=\frac{g^2 \kwplsq}{\pi}\frac{3}{32}\frac{m_{\scriptscriptstyle Q}^2}{m_{\scriptscriptstyle W^{\prime}}^2}, \nonumber\\
&{\bf Q}^{\scriptscriptstyle\{3\}}_{\scriptscriptstyle\{Z\}}:\quad \left|\lambda_{0}^{\scriptscriptstyle\{Z\}}\right|=\frac{g^2 \kzlsq}{c_{\scriptscriptstyle w}^2\, \pi}\frac{3}{64}\frac{m_{\scriptscriptstyle Q}^2}{m_{\scriptscriptstyle Z}^2},&&&
&{\bf Q}^{\scriptscriptstyle\{3\}}_{\scriptscriptstyle\{Z^{\prime}\}}:\quad \left|\lambda_{0}^{\scriptscriptstyle\{Z^{\prime}\}}\right|=\frac{g^2 \kzplsq}{c_{\scriptscriptstyle w}^2\, \pi}\frac{3}{64}\frac{m_{\scriptscriptstyle Q}^2}{m_{\scriptscriptstyle Z^{\prime}}^2},\nonumber\\
&{\bf Q}^{\scriptscriptstyle\{3\}}_{\scriptscriptstyle\{H\}}:\quad \left|\lambda_{0}^{\scriptscriptstyle\{H\}}\right|=\frac{\khlsq}{\pi}\frac{3}{16},&&&
&{\bf Q}^{\scriptscriptstyle\{3\}}_{\scriptscriptstyle\{H^{\prime}\}}:\quad \left|\lambda_{0}^{\scriptscriptstyle\{H^{\prime}\}}\right|=\frac{\khplsq}{\pi}\frac{3}{16}.
\label{lam0}
\end{align}
\noindent
Thus, instead of deriving constrains from eq.~(\ref{a0}), we use eq.~(\ref{lam0}) to obtain strongest bounds, as done in subsection~\ref{pertun}.\\

\noindent
Finally, we mention that processes involving extra gauge bosons and scalars in the initial and final states (e.g. $\vp_{\!\!1}\vp_{\!\!2}\rightarrow\vp_{\!\!3}\vp_{\!\!4}$, where $\vp_{\!\!i}$ may represent identical or different gauge bosons) are not considered. In reality, such processes require the inclusion of model-dependent interactions, such as self-interactions and interactions between different new heavy gauge bosons and scalars, which lie beyond the scope in this work\footnote{In the Lagrangian of the model (cf. eq. (\ref{lag0})), we do not include pieces describing such interactions.}.

%% file: appendix_C.tex
\section{Narrow width approximation and CM scheme comparison}
\label{appC}
We aim to determine whether the LO and NLO cross sections for the $2\rightarrow3$ processes studied in this work can be approximated either: {\it (i)} to $2\rightarrow2$ production cross sections multiplied the branching ratio of the VLQ (following NWA$_1$), or {\it (ii)} to  $2\rightarrow1$ production cross section times the branching ratios of the heavy gauge boson $\vp=\wps,\zp$ and the VLQ (according to NWA$_2$), under the assumption that the particles $\vp$ and the VLQ  are produced nearly on-shell. Specifically, we want to investigate whether the partonic cross section of the process $q_{_1}\bar{q}_{_2}\rightarrow \bar{q}_{_{_3}} q_{_{_4}}\, \bar{B}$ (for $\bar{B}\equiv Z, H$) can be factorized into {\it production} and {\it decay} components as the following:
\begin{align}
 \hat{\sigma}^{\scriptscriptstyle{q_{_1}\bar{q}_{_2}\rightarrow \bar{q}_{_{_3}}q_{_{_4}}\, \bar{B}}}&\approx \hat{\sigma}^{\scriptscriptstyle{q_{_1}\bar{q}_{_2}\rightarrow \bar{q}_{_{_3}} Q}} \times \text{Br}\left[Q\rightarrow q_{_{_4}} \bar{B}\right]     && \text{NWA$_1$}\nonumber\\
& \approx \hat{\sigma}^{\scriptscriptstyle{q_{_1}\bar{q}_{_2}\rightarrow V^{\prime}}} \times \text{Br}\left[V^{\prime}\rightarrow \bar{q}_{_{_3}} Q\right]\times \text{Br}\left[Q\rightarrow q_{_{_4}} \bar{B}\right]
&& \text{NWA$_2$}
\label{NWA12}
\end{align}

\noindent
We emphasize that for the NWA to be valid, the following conditions must be satisfied \cite{width2,width1,Schwartz2014sze,Fuchs:2014ola}: {(\it i)}~the ratio of the decay width to the mass of the unstable particle must be very small in order to justify the on-shell approximation, {\it (ii)} the internal propagators of the resonant particles must be factorizable from the matrix element, a condition not always satisfied in loop diagrams, ({\it iii}) the interference between resonant and non-resonant contributions should be negligible and ({\it iv}) the production and decay sub-processes must be kinematically accessible and sufficiently far from threshold regions.\\

\noindent
 The key step in proving eq.~(\ref{NWA12}) is to replace the denominators of the squared propagators for the unstable particles ($A\equiv \vp, Q$) in the squared amplitude with their Breit-Wigner forms:
 \begin{align} 
 \frac{1}{(q^2-m_{\scriptscriptstyle A}^2)^2+(\Gamma^{\scriptscriptstyle\text{ToT}}_{\scriptscriptstyle A}\, m_{\scriptscriptstyle A})^2}&\approx \frac{\pi}{\Gamma^{\scriptscriptstyle\text{ToT}}_{\scriptscriptstyle A}\, m_{\scriptscriptstyle A}}\, \delta(q^2-m_{\scriptscriptstyle A}^2)+\mathcal{P}\left[\frac{1}{(q^2-m_{\scriptscriptstyle A}^2)^2}\right]+\mathcal{O}\left(\frac{\Gamma^{\scriptscriptstyle\text{ToT}}_{\scriptscriptstyle A}}{m_{\scriptscriptstyle A}}\right)
\end{align}
\noindent
Here, the delta function encode the on-shell resonance at $q^2=m_{\scriptscriptstyle A}^2$ and the Cauchy principle value $\mathcal{P}$ describes the off-shell effects. The latter one is defined in term of the following distribution:
\begin{align}
 \left<\mathcal{P}\left(\frac{1}{x}\right),\phi(x)\right>&=\lim_{\delta \to\, 0^+} \left(\int_{-\infty}^{-\delta}+\int_{\delta}^{+\infty}\right)\, \frac{\phi(x)}{x}dx
\end{align}
where $\phi(x)$ is an arbitrary differentiable function at $x=0$.\\

\noindent
Since, we are considering configurations where $\Gamma^{\scriptscriptstyle\text{ToT}}_{\scriptscriptstyle A}\ll m_{\scriptscriptstyle A}$\footnote{In section~\ref{sec3}, we showed that the ratios $\Gamma^{\scriptscriptstyle\text{ToT}}_{\scriptscriptstyle V^{\prime}}/m_{\scriptscriptstyle V^{\prime}}$ (for $V^{\prime}\equiv \wps, \zp$) and $\Gamma^{\scriptscriptstyle\text{ToT}}_{\scriptscriptstyle T}/m_{\scriptscriptstyle T}$ are always smaller than $6\%$.}, the principle value can be neglected. Consequently, the denominator of the squared propagator can be replaced by the delta function part, i.e.
\begin{align}
\frac{1}{(q^2-m_{\scriptscriptstyle A}^2)^2+(\Gamma^{\scriptscriptstyle\text{ToT}}_{\scriptscriptstyle A}\, m_{\scriptscriptstyle A})^2}
 &\approx \frac{\pi}{\Gamma^{\scriptscriptstyle\text{ToT}}_{\scriptscriptstyle A}\, m_{\scriptscriptstyle A}}\, \delta(q^2-m_{\scriptscriptstyle A}^2)
\end{align}

\noindent
Let's recall the formula for the partonic cross section with massless initial-state partons:

\begin{align}
 \hat{\sigma}^{2\rightarrow3}&= \frac{1}{2\, \hat{s}}
 \int d\Phi_{3}\, \overline{\sum}|M|^2
 \label{sig2to3}
\end{align}
\noindent
where $\sqrt{\hat{s}}$ is the partonic center-of-mass energy and $d\Phi_{3}$ is the 3-body phase space (PS). The latter one is defined by:
\begin{align}
 d\Phi_{3}&=(2\pi)^{4}\, \delta^{(4)}\left(p_1+p_2-p_3-p_4-p_5\right)
 \frac{d^{3}p_{3}}{(2 \pi)^{3}2 E_{3}}\,
 \frac{d^{3}p_{4}}{(2 \pi)^{3}2 E_{4}}\,
 \frac{d^{3}p_{5}}{(2 \pi)^{3}2 E_{5}}
\end{align}
\noindent
To prove the two approximated formulae in eq.~(\ref{NWA12}) (for NWA$_1$ and NWA$_2$), we need to factorize the PS $d\Phi_{3}$ to production and decay phase spaces. Thus, without any approximation \cite{Fuchs:2014ola}, we can write:
\begin{align}
 d\Phi_{3}&=\left\{
 \begin{array}{l}
 d\Phi_{2}^{(P)} \, \frac{dq_2^2}{2\pi}\, d\Phi_{2}^{(D)}  \equiv d\Phi_{3}^{\text{NWA}_1}\\
\\
 d\Phi_{1}^{(P)} \, \frac{dq_1^2}{2\pi}\, d\Phi_{2}^{(D_1)}\, \frac{dq_2^2}{2\pi}\, d\Phi_{2}^{(D_2)}  \equiv d\Phi_{3}^{\text{NWA}_2}
 \label{PS}
 \end{array}
\right.
\end{align}
where $q_1^2$ and $q_2^2$ are the invariant masses of the unstable particles $\vp$ and $Q$, respectively, and $dq_{1}^2$ and $dq_{2}^2$ are their differentials. The production and decay phase spaces ($d\Phi^{(P)}, d\Phi^{(D)}$) in eq.~(\ref{PS}) are defined by:
\begin{align}
\text{NWA}_1:& \left\{
\begin{aligned}
d\Phi_{2}^{(P)}&=(2\pi)^{4}\, \delta^{(4)}\left(p_1+p_2-p_3-q_2\right)\, \frac{d^{3}p_3}{(2 \pi)^{3}2 E_{3}}\, \frac{d^{3}q_{2}}{(2 \pi)^{3}2 E_{q_2}}\\
d\Phi_{2}^{(D)}&=(2\pi)^{4}\, \delta^{(4)}\left(q_2-p_4-p_5\right)\, \frac{d^{3}p_{4}}{(2 \pi)^{3}2 E_{4}}\, \frac{d^{3}p_{5}}{(2 \pi)^{3}2 E_{5}}
\end{aligned}
\right.
\end{align}
\begin{align}
\text{NWA}_2:& \left\{
\begin{aligned}
d\Phi_{1}^{(P)}&=(2\pi)^{4}\, \delta^{(4)}\left(p_1+p_2-q_1\right)\, \frac{d^{3}q_{1}}{(2 \pi)^{3}2 E_{q_1}}\\
d\Phi_{2}^{(D_1)}&=(2\pi)^{4}\, \delta^{(4)}\left(q_1-p_3-q_2\right)\, \frac{d^{3}p_3}{(2 \pi)^{3}2 E_{3}} \, \frac{d^{3}q_{2}}{(2 \pi)^{3}2 E_{q_2}}\\
d\Phi_{2}^{(D_2)}&=(2\pi)^{4}\, \delta^{(4)}\left(q_2-p_4-p_5\right)\, \frac{d^{3}p_4}{(2 \pi)^{3}2 E_{4}} \, \frac{d^{3}p_5}{(2 \pi)^{3}2 E_{5}}
\end{aligned}
\right.
\end{align}
\subsection{NWA for Leading order}
Let's focus on the process $q(p_1)+\bar{q}^{\prime}(p_2) \rightarrow \wps \rightarrow \bar{b}(p_3)+ t(p_4)+  Z(p_5)$. The demonstration for the other processes can be done in the same way.
The associated Born matrix element (cf.~subsfigure~\ref{qqWp1}), is given by:
\begin{align}
M_{\scriptscriptstyle B}&=i\, \vkwp\, \kwpl\, \kzl\, \frac{g^3}{32\, c_{\scriptscriptstyle{W}}}\,  \bar{v}(p_{\scriptscriptstyle{2}})\gamma_{\mu}(1-\gamma_5)u(p_{\scriptscriptstyle{1}})
\left[\frac{-g^{\mu \nu}+q^{\mu}_{1}q^{\nu}_{1}/m^{\scriptscriptstyle{2}}_{\scriptscriptstyle{W^\prime}}}
{q_{1}^2-m^{\scriptscriptstyle{2}}_{\scriptscriptstyle{W^\prime}}+i\, \lambda}\right]\nonumber\\
&\times\bar{u}(p_{\scriptscriptstyle{4}}) 
\gamma^{\rho}(1-\gamma_5) \left[\frac{\not{\!q}_2+m_{T}}{q_{2}^2-m^{\scriptscriptstyle{2}}_{\scriptscriptstyle{T}}+i\, \lambda}\right] \gamma_{\nu}(1-\gamma_5)
v(p_{\scriptscriptstyle{3}})\, \varepsilon^{*}_{\rho}(p_{\scriptscriptstyle{5}})
\label{mB}
\end{align}
Near resonance, we can consider the virtual particles $\wps$ and $Q\equiv T$ as on-shell particles, see subfigures~\ref{qqWp2},~\ref{qqWp3}. Hence, their associated internal propagators can be approximated by:
\begin{align}
\frac{-g_{\mu \nu}+q_{1\mu}q_{1\nu}/m^{\scriptscriptstyle{2}}_{\scriptscriptstyle{W^\prime}}}
{q_{1}^2-m^{\scriptscriptstyle{2}}_{\scriptscriptstyle{W^\prime}}+i\, \lambda} &\overset{q_1^2\, \approx \, m_{\scriptscriptstyle W^{\prime}}^2}{\xrightarrow{\hspace{2cm}}}\frac{\sum_\text{pol} \varepsilon_{\mu}^{*}(q_1)\varepsilon_{\nu}(q_1)}
{q_{1}^2-m^{\scriptscriptstyle{2}}_{\scriptscriptstyle{W^\prime}}+i\, m_{\scriptscriptstyle{W^\prime}}\Gamma_{\scriptscriptstyle{W^\prime}}}\label{ProPwp}\\
\frac{\not{\!q}_2+m_{T}}{q_{2}^2-m^{\scriptscriptstyle{2}}_{\scriptscriptstyle{T}}+i\, \lambda}&\overset{q_2^2\, \approx\, m_{\scriptscriptstyle T}^2}{\xrightarrow{\hspace{2cm}}}
\frac{\sum_\text{spin} u(q_2)\bar{u}(q_2)}{q_{2}^2-m^{\scriptscriptstyle{2}}_{\scriptscriptstyle{T}}+i\, m_{\scriptscriptstyle{T}}\Gamma_{\scriptscriptstyle{T}}^{\scriptscriptstyle{\text{ToT}}}}\label{ProPQ}
\end{align}
with $q_1=p_1+p_2$, $q_2=p_4+p_5$ and the total widths of $T$ and $\wps$ are denoted by $\Gamma_{\scriptscriptstyle{T}}^{\scriptscriptstyle{\text{ToT}}}$ and $\Gamma_{\scriptscriptstyle{W^{\prime}}}^{\scriptscriptstyle{\text{ToT}}}$, respectively.\\

\noindent
Now, we can factorize $M_{\scriptscriptstyle B}$ into production and decay matrix elements. In the case of NWA$_1$, only the vector-like quark $T$ is taken to be on-shell, so we substitute  the right-hand side of eq.~(\ref{ProPQ}) into eq.~(\ref{mB}). For NWA$_2$, both $\wps$ and $T$ are treated as on-shell particles, requiring the insertion of the right-hand sides of eq.~(\ref{ProPwp}) and eq.~(\ref{ProPQ}) into eq.~(\ref{mB}). This leads to:
\begin{align}
M_{\scriptscriptstyle B}^{\scriptscriptstyle\text{NWA}_1}&=
  M_{\scriptscriptstyle B}^{q\bar{q}^{\prime} \rightarrow \bar{b}T}\frac{1}{q_{2}^2-m^{\scriptscriptstyle{2}}_{\scriptscriptstyle{T}}+i\, m_{\scriptscriptstyle{T}}\Gamma_{\scriptscriptstyle{T}}^{\scriptscriptstyle{\text{ToT}}}}\, M_{\scriptscriptstyle B}^{T\rightarrow tZ}\\
M_{\scriptscriptstyle B}^{\scriptscriptstyle\text{NWA}_2}&=
   M_{\scriptscriptstyle B}^{q\bar{q}^{\prime} \rightarrow W^{\prime}}\, \frac{1}{q_{1}^2-m^{\scriptscriptstyle{2}}_{\scriptscriptstyle{W^{\prime}}}+i\, m_{\scriptscriptstyle{W^{\prime}}}\Gamma_{\scriptscriptstyle{W^{\prime}}}^{\scriptscriptstyle{\text{ToT}}}}\, M_{\scriptscriptstyle B}^{W^{\prime}\rightarrow \bar{b}T}\frac{1}{q_{2}^2-m^{\scriptscriptstyle{2}}_{\scriptscriptstyle{T}}+i\, m_{\scriptscriptstyle{T}}\Gamma_{\scriptscriptstyle{T}}^{\scriptscriptstyle{\text{ToT}}}}\, M_{\scriptscriptstyle B}^{T\rightarrow tZ}
\end{align}
where the Born production and decay matrix elements are defined by:
\begin{align}
\text{Production}&: \left\{
\begin{aligned}
M_{\scriptscriptstyle B}^{q\bar{q}^{\prime} \rightarrow \bar{b}T}&\propto \bar{v}(p_{\scriptscriptstyle{2}})\gamma_{\mu}(1-\gamma_5)u(p_{\scriptscriptstyle{1}})
\left[\frac{-g^{\mu \nu}+q^{\mu}_{1}q^{\nu}_{1}/m^{\scriptscriptstyle{2}}_{\scriptscriptstyle{W^\prime}}}
{q_{1}^2-m^{\scriptscriptstyle{2}}_{\scriptscriptstyle{W^\prime}}+i\, \lambda}\right]\bar{u}(q_2)\gamma_{\nu}(1-\gamma_5)v(p_3)\nonumber\\
M_{\scriptscriptstyle B}^{q\bar{q}^{\prime} \rightarrow W^{\prime}}&\propto \bar{v}(p_2)\gamma_{\mu}(1-\gamma_5)u(p_1) \varepsilon_{\nu}^{*}(q_1)
\end{aligned}
\right.\\
\text{Decay}&: \left\{
\begin{aligned}
M_{\scriptscriptstyle B}^{W^{\prime}\rightarrow \bar{b}T}&\propto \varepsilon_{\nu}(q_1)\, \bar{u}(q_2)\gamma_{\nu}(1-\gamma_5)v(p_{\scriptscriptstyle{3}})\nonumber\\
M_{\scriptscriptstyle B}^{T\rightarrow tZ}&\propto \bar{u}(p_4) \gamma^{\rho}(1-\gamma_5)
u(q_2)\, \varepsilon^{*}_{\rho}(p_{\scriptscriptstyle{5}})
\end{aligned}
\right.
\end{align}
\noindent
In both cases, the squared amplitudes summed over spin and colors and averaged over spin and colors of initial state are given by:
\begin{align}
\overline{|M_{\scriptscriptstyle B}^{\scriptscriptstyle\text{NWA}_1}|}^2&= 
\overline{|M_{\scriptscriptstyle B}^{q\bar{q}^{\prime} \rightarrow \bar{b}T}|}^2\frac{1}{(q_{2}^2-m^{\scriptscriptstyle{2}}_{\scriptscriptstyle{T}})^2 + (m_{\scriptscriptstyle{T}}\Gamma_{\scriptscriptstyle{T}}^{\scriptscriptstyle{\text{ToT}}})^2}\,\, \overline{|M_{\scriptscriptstyle B}^{T\rightarrow tZ}|}^2
\approx 
\overline{|M_{\scriptscriptstyle B}^{q\bar{q}^{\prime} \rightarrow \bar{b}T}|}^2\, \frac{\pi\, \delta\left(q_2^2-m_{\scriptscriptstyle{T}}^2\right)}{\Gamma_{\scriptscriptstyle T}^{\scriptscriptstyle{\text{ToT}}}\, m_{\scriptscriptstyle{T}}} \,\overline{|M_{\scriptscriptstyle B}^{T\rightarrow tZ}|}^2
\end{align}
\begin{align}
\overline{|M_{\scriptscriptstyle B}^{\scriptscriptstyle\text{NWA}_2}|}^2&= 
\overline{|M_{\scriptscriptstyle B}^{q\bar{q}^{\prime} \rightarrow W^{\prime}}|}^2\, \frac{1}{(q_{1}^2-m^{\scriptscriptstyle{2}}_{\scriptscriptstyle{W^{\prime}}})^2+(m_{\scriptscriptstyle{W^{\prime}}}\Gamma_{\scriptscriptstyle{W^{\prime}}}^{\scriptscriptstyle{\text{ToT}}})^2}\,\overline{|M_{\scriptscriptstyle B}^{W^{\prime}\rightarrow \bar{b}T}|}^2\frac{1}{(q_{2}^2-m^{\scriptscriptstyle{2}}_{\scriptscriptstyle{T}})^2+ (m_{\scriptscriptstyle{T}}\Gamma_{\scriptscriptstyle{T}}^{\scriptscriptstyle{\text{ToT}}})^2}\, \overline{|M_{\scriptscriptstyle B}^{T\rightarrow tZ}|}^2\nonumber\\
&\approx 
\overline{|M_{\scriptscriptstyle B}^{q\bar{q}^{\prime} \rightarrow W^{\prime}}|}^2\, \frac{\pi\, \delta\left(q_1^2-m_{\scriptscriptstyle{W^{\prime}}}^2\right)}{\Gamma_{\scriptscriptstyle W^{\prime}}^{\scriptscriptstyle{\text{ToT}}}\, m_{\scriptscriptstyle{W^{\prime}}}} \, \overline{|M_{\scriptscriptstyle B}^{W^{\prime}\rightarrow \bar{b}T}|}^2\,\frac{\pi\, \delta\left(q_2^2-m_{\scriptscriptstyle{T}}^2\right)}{\Gamma_{\scriptscriptstyle T}^{\scriptscriptstyle{\text{ToT}}}\, m_{\scriptscriptstyle{T}}} \, \overline{|M_{\scriptscriptstyle B}^{T\rightarrow tZ}|}^2
\end{align}

\noindent
Let us substitute these squared amplitudes into eq.~(\ref{sig2to3}) and replace the phase space $d\Phi_3$ with the appropriate expression for each case from eq.~(\ref{PS}). This yields:
\begin{align}
\hat{\sigma}^{\scriptscriptstyle\text{NWA}_1}&=\left[\frac{1}{2\hat{s}}\, \int d\Phi_{2}^{(P)} \, \overline{|M_{\scriptscriptstyle B}^{q\bar{q}^{\prime} \rightarrow \bar{b}T}|}^2\right]\, \int \delta\left(q_2^2-m_{\scriptscriptstyle{T}}^2\right)dq_2^2\, \frac{1}{\Gamma_{\scriptscriptstyle T}^{\scriptscriptstyle{\text{ToT}}}}\left[\frac{1}{2m_{\scriptscriptstyle T}}\int  d\Phi_{2}^{(D)}\, \overline{|M_{\scriptscriptstyle B}^{T\rightarrow tZ}|}^2\right]\\
\hat{\sigma}^{\scriptscriptstyle\text{NWA}_2}&=\left[\frac{1}{2\hat{s}}\, \int  d\Phi_{1}^{(P)}\,\overline{|M_{\scriptscriptstyle B}^{q\bar{q}^{\prime} \rightarrow W^{\prime}}|}^2\right] \, \int \delta\left(q_1^2-m_{\scriptscriptstyle{W^{\prime}}}^2\right)dq_1^2\, \frac{1}{\Gamma_{\scriptscriptstyle W^{\prime}}^{\scriptscriptstyle{\text{ToT}}}}\left[\frac{1}{2m_{\scriptscriptstyle W^{\prime}}} \int  d\Phi_{2}^{(D_1)}\, \overline{|M_{\scriptscriptstyle B}^{W^{\prime}\rightarrow \bar{b}T}|}^2\right]\nonumber\\
&\times 
\int \delta\left(q_2^2-m_{\scriptscriptstyle{T}}^2\right)dq_2^2\, \frac{1}{\Gamma_{\scriptscriptstyle T}^{\scriptscriptstyle{\text{ToT}}}}\left[\frac{1}{2m_{\scriptscriptstyle T}}\int  d\Phi_{2}^{(D_2)}\, \overline{|M_{\scriptscriptstyle B}^{T\rightarrow tZ}|}^2\right] 
\end{align}
\noindent
After integration over $q_1^2$ and $q_2^2$ using the delta functions and applying the definition of the partial decay width for $1-$to$-2$ processes, we obtain the famous narrow-width factorization of the cross section:
\begin{align}
 \hat{\sigma}^{\scriptscriptstyle\text{NWA}_1}&=\hat{\sigma}^{q\bar{q}^{\prime} \rightarrow \bar{b}T}\times \frac{\Gamma[T\rightarrow tZ]}{\Gamma_{\scriptscriptstyle T}^{\scriptscriptstyle{\text{ToT}}}}\equiv \hat{\sigma}^{q\bar{q}^{\prime} \rightarrow \bar{b}T}\times\text{Br}[T\rightarrow tZ]\\
\hat{\sigma}^{\scriptscriptstyle\text{NWA}_2}&=\hat{\sigma}^{q\bar{q}^{\prime} \rightarrow W^{\prime}}\times \frac{\Gamma[W^{\prime}\rightarrow \bar{b}T]}{\Gamma_{\scriptscriptstyle W^{\prime}}^{\scriptscriptstyle{\text{ToT}}}}\times \frac{\Gamma[T\rightarrow tZ]}{\Gamma_{\scriptscriptstyle T}^{\scriptscriptstyle{\text{ToT}}}}\equiv \hat{\sigma}^{q\bar{q}^{\prime} \rightarrow W^{\prime}}\times\text{Br}[W^{\prime}\rightarrow \bar{b}T]\times\text{Br}[T\rightarrow tZ]
\label{nwa12LO}
\end{align}

\subsection{NWA for next-to-leading order}
\noindent
At NLO, proving whether or not NWA is valid is not straightforward due the complex topologies involved at the loop level (triangles, boxes, pentagons \ldots etc.). For such diagrams it is not always possible to separate the internal propagators of the unstable particles from the matrix element, especially if these topologies involve particle exchange between the initial and final states.\\

\noindent
In our case, the one-loop amplitude can always be factorized into Born amplitude times higher-order corrections. That is, the virtual matrix element can be expressed as:
\begin{align}
 M_{\scriptscriptstyle V}&= M_{\scriptscriptstyle B}\, \frac{\alpha_s}{2\pi}\, C_{_{\scriptscriptstyle color}}\, A_{\scriptscriptstyle loop}
 \label{factoLoop}
\end{align}
\noindent
where $C_{_{\scriptscriptstyle color}}$ is the color factor and $A_{\scriptscriptstyle loop}$ contains the pure one-loop contributions (i.e. it can be expressed in terms of the four scalar Passarino-Veltman functions).\\

\noindent
The factorization in eq.~(\ref{factoLoop}) holds because all of the following conditions are satisfied: {\it (i)} the tree-level and one-loop Feynman diagrams are all of $s-$channel type only, {\it (ii)} no new Lorentz structure appears at loop level (such as three or four gluons vertices)~\cite{Zidi:2024lid} and {\it (iii)} the color factor factorizes, since every 1-loop graph carries the same factor ($\propto \text{Tr}[T^a\, T^a]$). We have verified eq.~(\ref{factoLoop}) using {\tt FeynArts/FormCalc} tools~\cite{ref5,Hahn:2016ebn}, where we have reduced all one-loop contributions to the basis of Passarino-Veltman scalar functions. It is very important to note that the factorization in eq.~(\ref{factoLoop}) does not guarantee the existence of a generalization for $\hat{\sigma}^{\scriptscriptstyle\text{NWA}_1}$ and $\hat{\sigma}^{\scriptscriptstyle\text{NWA}_2}$ (cf. eq.~(\ref{nwa12LO})) at NLO, due to the quantity $A_{\scriptscriptstyle loop}$, which is not necessary separable. This is particularly true when a virtual boson (gluon for example) connects the initial and final state particles, as occurs in certain boxes and the pentagons diagrams. Fortunately, in our case, these topologies are color-suppressed.
In this paper, we do not provide an analytical proof of the separability nor the loop either the real emission contributions to production and decays factors, as this lies beyond the scope of this paper. However, we will numerically compare the three approximations: NWA$_1$, NWA$_2$ and CM schemes at NLO (and LO), and show when they give approximate predictions and when they differ.

\subsection{NWA and CM scheme comparison}
\subsubsection*{Case 1: Production and decay of $\wps$}
\begin{figure}[h!]
\centering
\includegraphics[width=5.75cm,height=4.25cm]{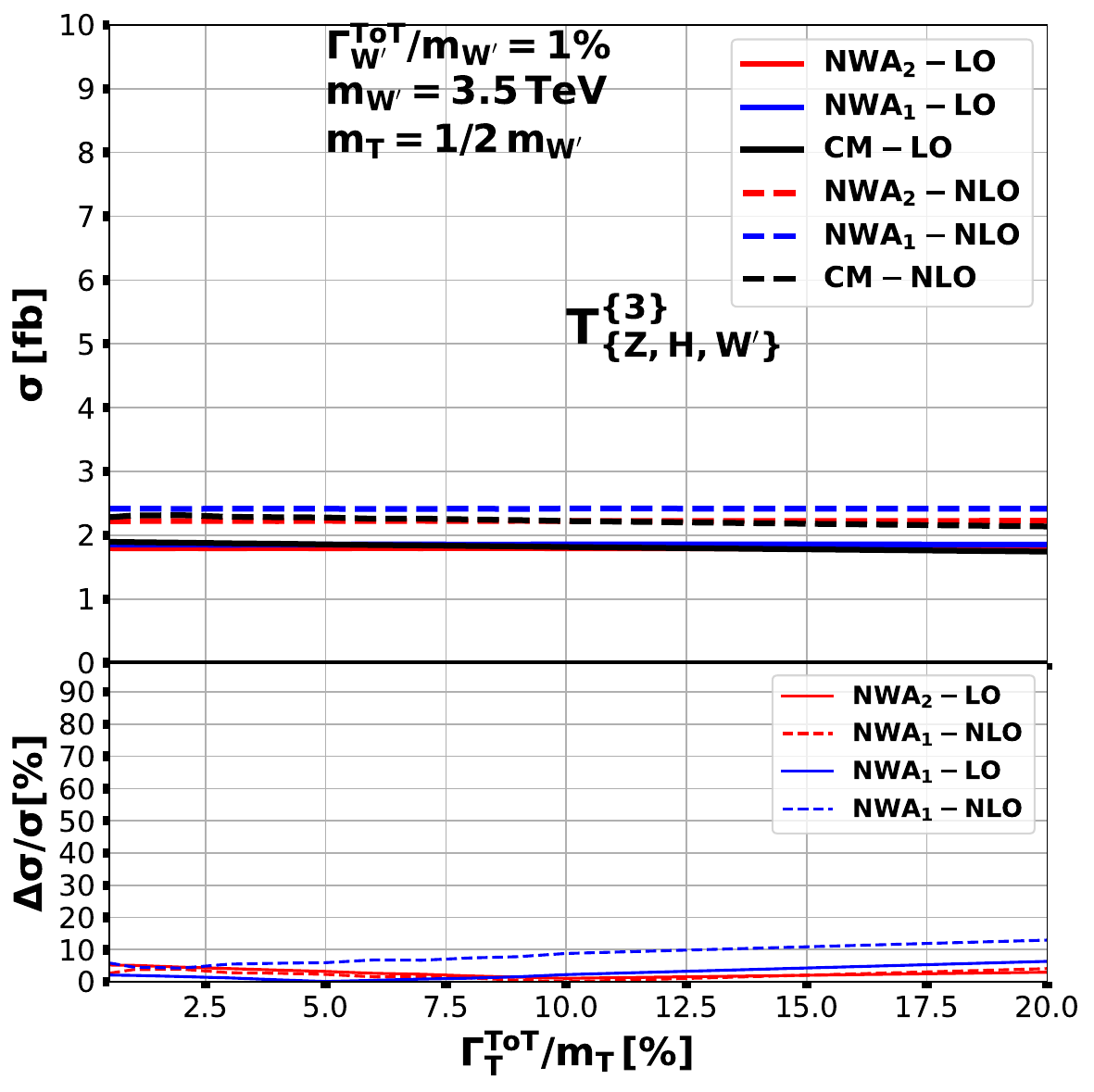}
\includegraphics[width=5.75cm,height=4.25cm]{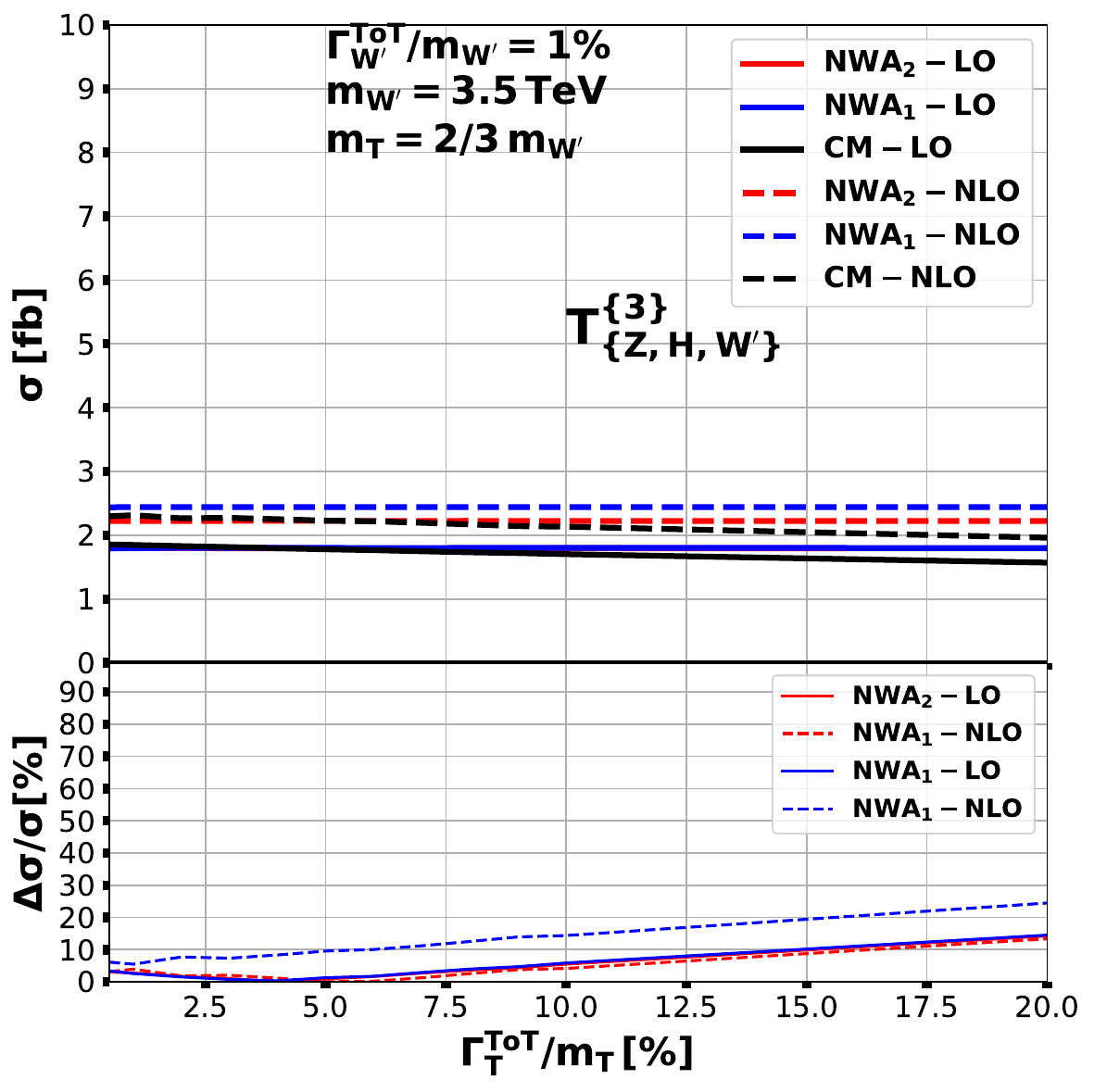}
\includegraphics[width=5.75cm,height=4.25cm]{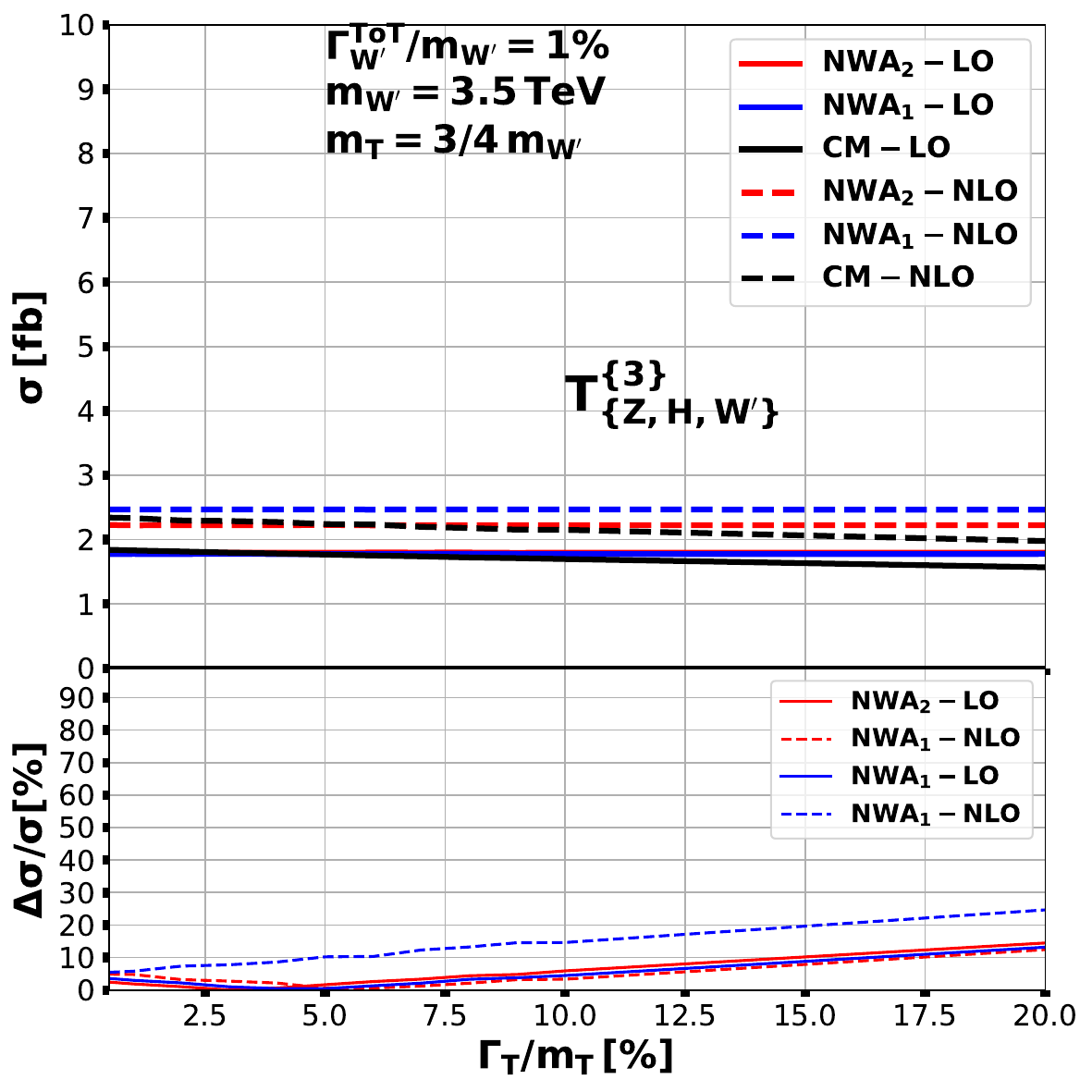}\\
\includegraphics[width=5.75cm,height=4.25cm]{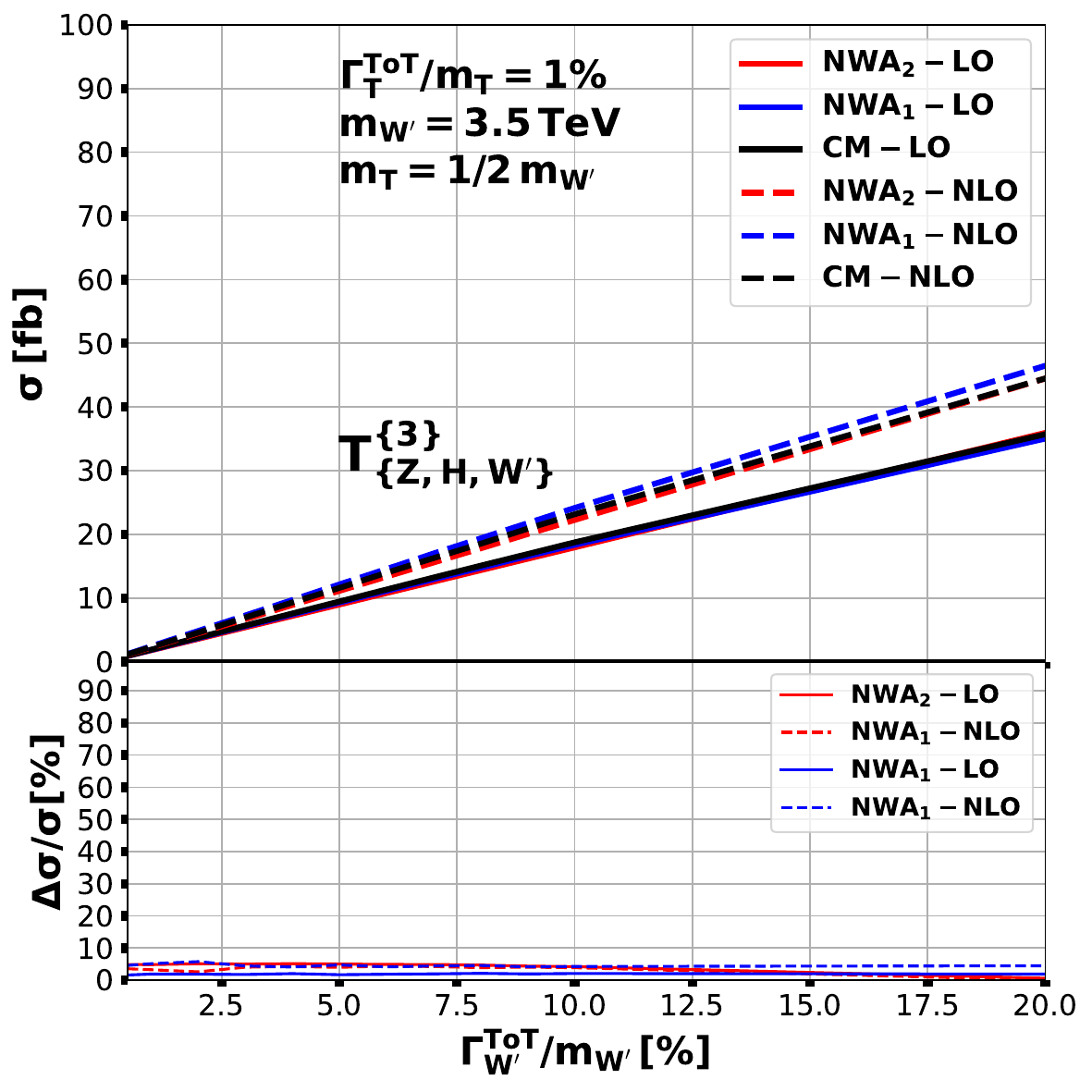}
\includegraphics[width=5.75cm,height=4.25cm]{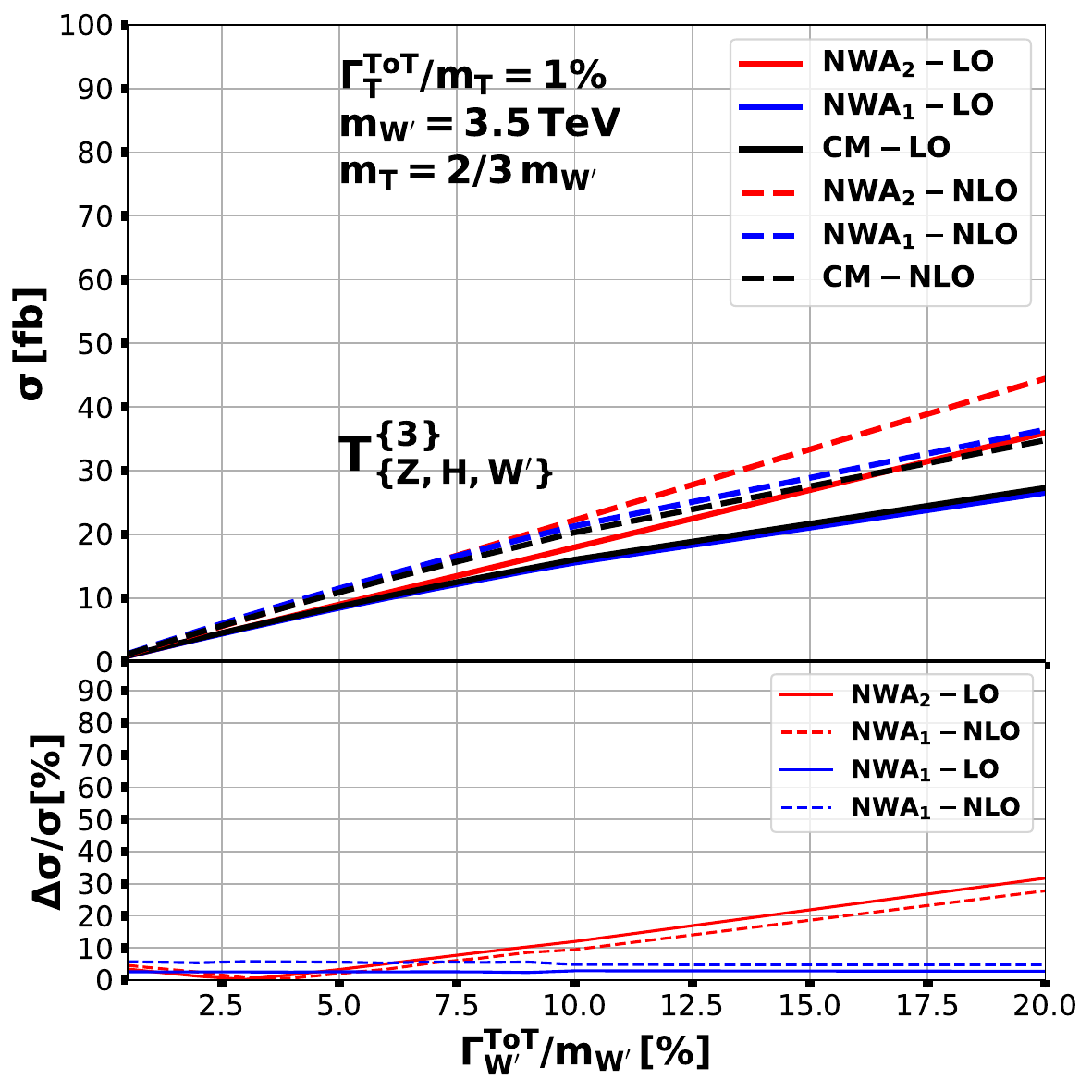}
\includegraphics[width=5.75cm,height=4.25cm]{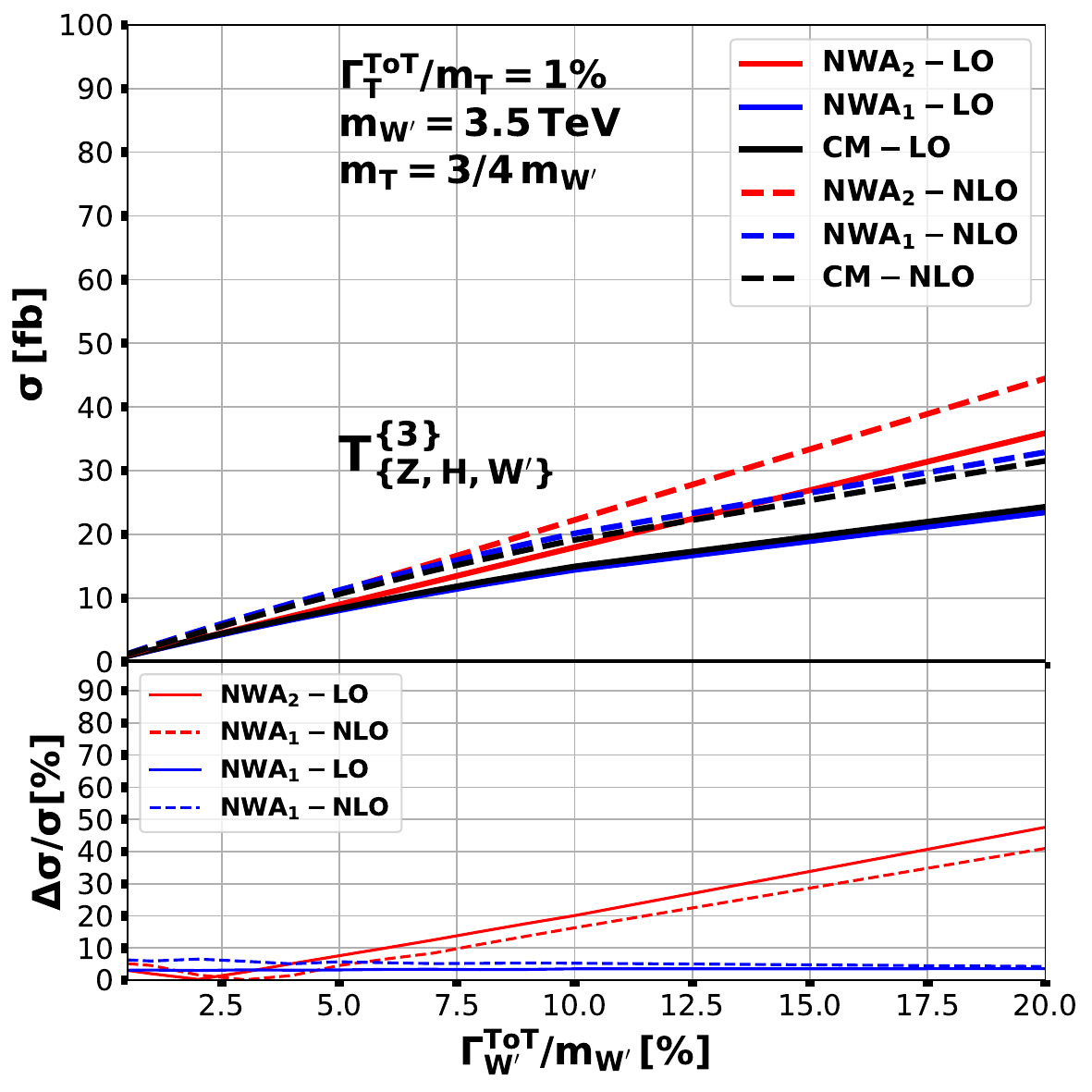}
 \caption{\small Variation of $\sigma$ with $\Gamma_{\scriptscriptstyle T}^{\scriptscriptstyle\text{ToT}}/m_{\scriptscriptstyle T}$ (up) and $\Gamma_{\scriptscriptstyle W^{\prime}}^{\scriptscriptstyle\text{ToT}}/m_{\scriptscriptstyle W^{\prime}}$ (down) at LO and NLO in NWA$_1$, NWA$_2$ and CM.}
\label{XsecVarsGamWpTpV1}
\end{figure}

In figure~\ref{XsecVarsGamWpTpV1}, we show the variation of the hadronic cross sections for the reaction $pp\rightarrow \bar{b}tZ/H+b\bar{t}Z/H$, in the benchmark scenario ${\bf T^{\scriptscriptstyle\{3\}}_{\scriptscriptstyle\{Z,H,W'\}}}$, for $m_{\scriptscriptstyle W^{\prime}} =3.5$~TeV at the center-of-mass energy $\sqrt{s}=13$~TeV, for the low-mass (left panels), middle-mass (central panels) and high-mass (right-panels). The results are shown for the NWA$_1$, NWA$_2$ and CM schemes at both LO and NLO for $\Gamma_{\scriptscriptstyle W^{\prime}}^{\scriptscriptstyle\text{ToT}}/m_{\scriptscriptstyle W^{\prime}}=1\%$ (up panels) and $\Gamma_{\scriptscriptstyle T}^{\scriptscriptstyle\text{ToT}}/m_{\scriptscriptstyle T}=1\%$ (down panels). The sub-panels of each figure represent the relative difference (in percentage) between NWA$_1$ and NWA$_2$ cross sections and CM scheme cross sections, which are defined by: 
\begin{align}
\Delta\tilde{\sigma}_{\text{LO/NLO}}^{\scriptscriptstyle\text{NWA}_1}[\%]&=\frac{\left|\sigma^{\scriptscriptstyle\text{NWA}_1}_{\scriptscriptstyle\text{NLO}}-\sigma^{\scriptscriptstyle\text{CM}}_{\scriptscriptstyle\text{NLO}}\right|}{\sigma^{\scriptscriptstyle\text{CM}}_{\scriptscriptstyle \text{LO/NLO}}}\times 100
&
\Delta\tilde{\sigma}_{\text{LO/NLO}}^{\scriptscriptstyle\text{NWA}_2}[\%]&=\frac{\left|\sigma^{\scriptscriptstyle\text{NWA}_2}_{\scriptscriptstyle\text{NLO}}-\sigma^{\scriptscriptstyle\text{CM}}_{\scriptscriptstyle\text{NLO}}\right|}{\sigma^{\scriptscriptstyle\text{CM}}_{\scriptscriptstyle \text{LO/NLO}}}\times 100
\end{align}
\noindent
We see that for large ratios $\Gamma_{\scriptscriptstyle W^{\prime}}^{\scriptscriptstyle\text{ToT}}/m_{\scriptscriptstyle W^{\prime}}$ and $\Gamma_{\scriptscriptstyle T}^{\scriptscriptstyle\text{ToT}}/m_{\scriptscriptstyle T}$, the NWA predictions deviate from the CM scheme results as expected. However, for smaller ratios, the 3 predictions converge and the discrepancies between them are reduced as illustrated in the sub-panels of figure~\ref{XsecVarsGamWpTpV1}.
To quantify the differences between the predictions of the 3 approaches in the narrow width region, we provide some numerical values of the NLO cross section for $\Gamma_{\scriptscriptstyle W^{\prime}}^{\scriptscriptstyle\text{ToT}}/m_{\scriptscriptstyle W^{\prime}}=1\%,\, 5\%$ and $\Gamma_{_T}^{\scriptscriptstyle\text{ToT}}/m_{_T}=1\%,\, 5\%$, considering both low-mass (i.e. $m_{\scriptscriptstyle T}/m_{\scriptscriptstyle W^{\prime}}=1/2$) and high-mass (i.e. $m_{\scriptscriptstyle T}/m_{\scriptscriptstyle W^{\prime}}=3/4$) cases. These results are summarized in table~\ref{TabGamWpTpNLO}.

\begin{table*}[h!]
\centering
 \renewcommand{\arraystretch}{1.40}
 \setlength{\tabcolsep}{10pt}
 \begin{adjustbox}{width=17.00cm,height=1.5cm}
 \boldmath
 \begin{tabular}
 {!{\vrule width 2pt}l!{\vrule width 2pt}c:c:c:c:c!{\vrule width 2pt}c:c:c:c:c!{\vrule width 2pt}l!{\vrule width 2pt}}
  \cline{1-11}
$\mathbf{\frac{\Gamma_{_{W^{\prime}}}^{\scriptscriptstyle\text{ToT}}}{m_{_{W^{\prime}}}}}$
&\multicolumn{5}{c!{\vrule width 2pt}}{{\bf $\bf 1\%$}}
&\multicolumn{5}{c!{\vrule width 2pt}}{{\bf $\bf 5\%$}}\\
\noalign{\hrule height 1pt}
 $\mathbf{\frac{\Gamma_{_T}^{\scriptscriptstyle\text{ToT}}}{m_{_T}}}$
 & $\bf\sigma_{\scriptscriptstyle{\text{NLO}}}^{\scriptscriptstyle{\text{NWA}_1}}[\text{fb}]$
 & $\bf\Delta\tilde{\sigma}_{\scriptscriptstyle{\text{NLO}}}^{\scriptscriptstyle{\text{NWA}_1}}$
 & $\bf\sigma_{\scriptscriptstyle{\text{NLO}}}^{\scriptscriptstyle{\text{NWA}_2}}[\text{fb}]$
 & $\bf \Delta\tilde{\sigma}_{\scriptscriptstyle{\text{NLO}}}^{\scriptscriptstyle{\text{NWA}_2}}$
 &$\bf\sigma_{\scriptscriptstyle{\text{NLO}}}^{\scriptscriptstyle{\text{CM}}}[\text{fb}]$
 & $\bf\sigma_{\scriptscriptstyle{\text{NLO}}}^{\scriptscriptstyle{\text{NWA}_1}}[\text{fb}]$
& $\bf\Delta\tilde{\sigma}_{\scriptscriptstyle{\text{NLO}}}^{\scriptscriptstyle{\text{NWA}_1}}$
 & $\bf\sigma_{\scriptscriptstyle{\text{NLO}}}^{\scriptscriptstyle{\text{NWA}_2}}[\text{fb}]$
 & $\bf\Delta\tilde{\sigma}_{\scriptscriptstyle{\text{NLO}}}^{\scriptscriptstyle{\text{NWA}_2}}$
 &$\bf\sigma_{\scriptscriptstyle{\text{NLO}}}^{\scriptscriptstyle{\text{CM}}}[\text{fb}]$
 &$\mathbf{\frac{m_{\scriptscriptstyle T}}{m_{\scriptscriptstyle W^{\prime}}}}$\\
\noalign{\hrule height 1pt}
 $\bf 1\%$ & $2.416$ & $\bf 5.31\%$ & $2.226$ & $\bf 2.96\%$ &  $2.294$&  $12.13$ &$\bf 4.38\%$& $11.13$ &$\bf 4.21\%$& $11.62$ &\\
\cdashline{2-11}
   $\bf 5\%$ & $2.416$  & $\bf 7.23\%$ & $2.222$ & $\bf 1.37\%$ & $2.253$ & $12.13$ &$\bf 6.68\%$& $11.12$ &$\bf 2.19\%$& $11.37$&$\mathbf{\frac{1}{2}}$\\
 \noalign{\hrule height 1pt}
  $\bf 1\%$ & $2.468$ & $\bf 7.11\%$ & $2.226$ & $\bf 3.38\%$ & $2.304$ &  $11.26$ &$\bf 6.12\%$&$11.13$  &$\bf 4.90\%$&$10.61$   &\\
 \cdashline{2-11}
 $\bf 5\%$ & $2.467$ & $\bf 10.28\%$ & $2.220$ & $\bf 0.75\%$ &  $2.237$ & $11.26$ &$\bf 9.32\%$& $11.12$ &$\bf 7.96\%$& $10.30$ &$\mathbf{\frac{3}{4}}$\\
 \noalign{\hrule height 1pt}
\end{tabular}
\end{adjustbox}
  \caption{\small NLO cross section for $pp\rightarrow \bar{b}(b)\, t(\bar{t}) Z (H)$  versus $\Gamma_{\scriptscriptstyle W^{\prime}}^{\scriptscriptstyle\text{ToT}}/m_{\scriptscriptstyle W^{\prime}}$ and $\Gamma_{_T}^{\scriptscriptstyle\text{ToT}}/m_{_T}$ for $m_{\scriptscriptstyle W^{\prime}}=3.5$ TeV.}
   \label{TabGamWpTpNLO}
  \end{table*}

\noindent
We observe that the discrepancy between CM and NWA predictions are smaller for the lower-mass case compared to the higher-mass case. The predictions of NWA$_2$ agree more closely with the CM results than those of NWA$_1$, especially for small width-to-mass ratios of $\wps$, as in the case of the DD scenario studied in section~\ref{sec2} ($\Gamma_{\scriptscriptstyle W^{\prime}}^{\scriptscriptstyle\text{ToT}}/m_{\scriptscriptstyle W^{\prime}}=1.09\%$). However, the discrepancy might exceed $7\%$ for NWA$_2$ and $10\%$ for NWA$_1$ in the higher-mass case. This difference arises because in the latter case, we are closer to the decay threshold, while in the lower-mass scenario, we are sufficiently far from  such threshold regions. It is well known that the closer one is to decay threshold, the more significant the off-shell effects become. Therefore, we recommend to either relying on the CM scheme or generalizing the NWA to properly account for off-shell effects, as proposed in ref.~\cite{Fuchs:2014ola}.

\begin{figure}[h!]
\centering
\includegraphics[width=5.75cm,height=4.25cm]{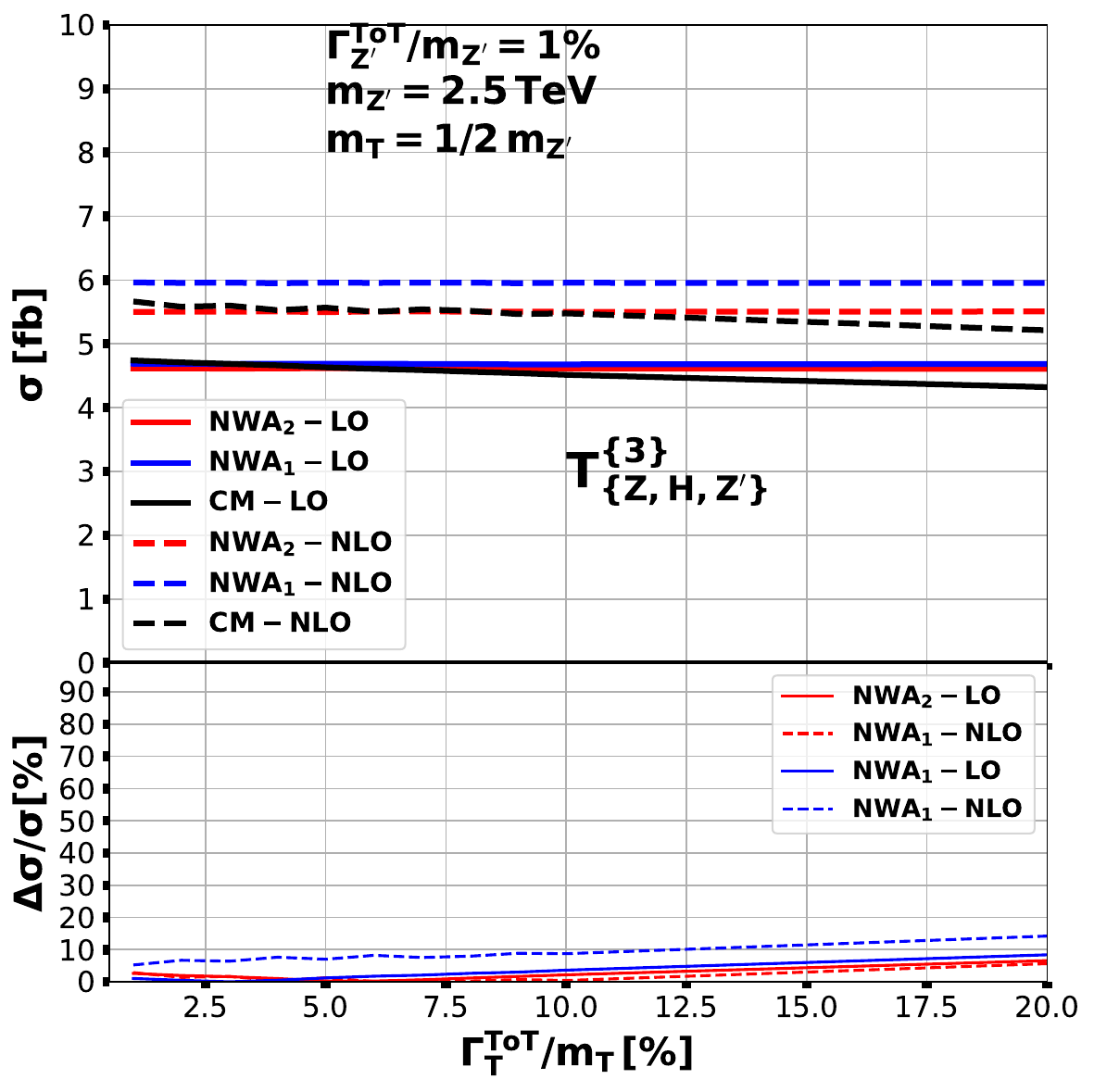}
\includegraphics[width=5.75cm,height=4.25cm]{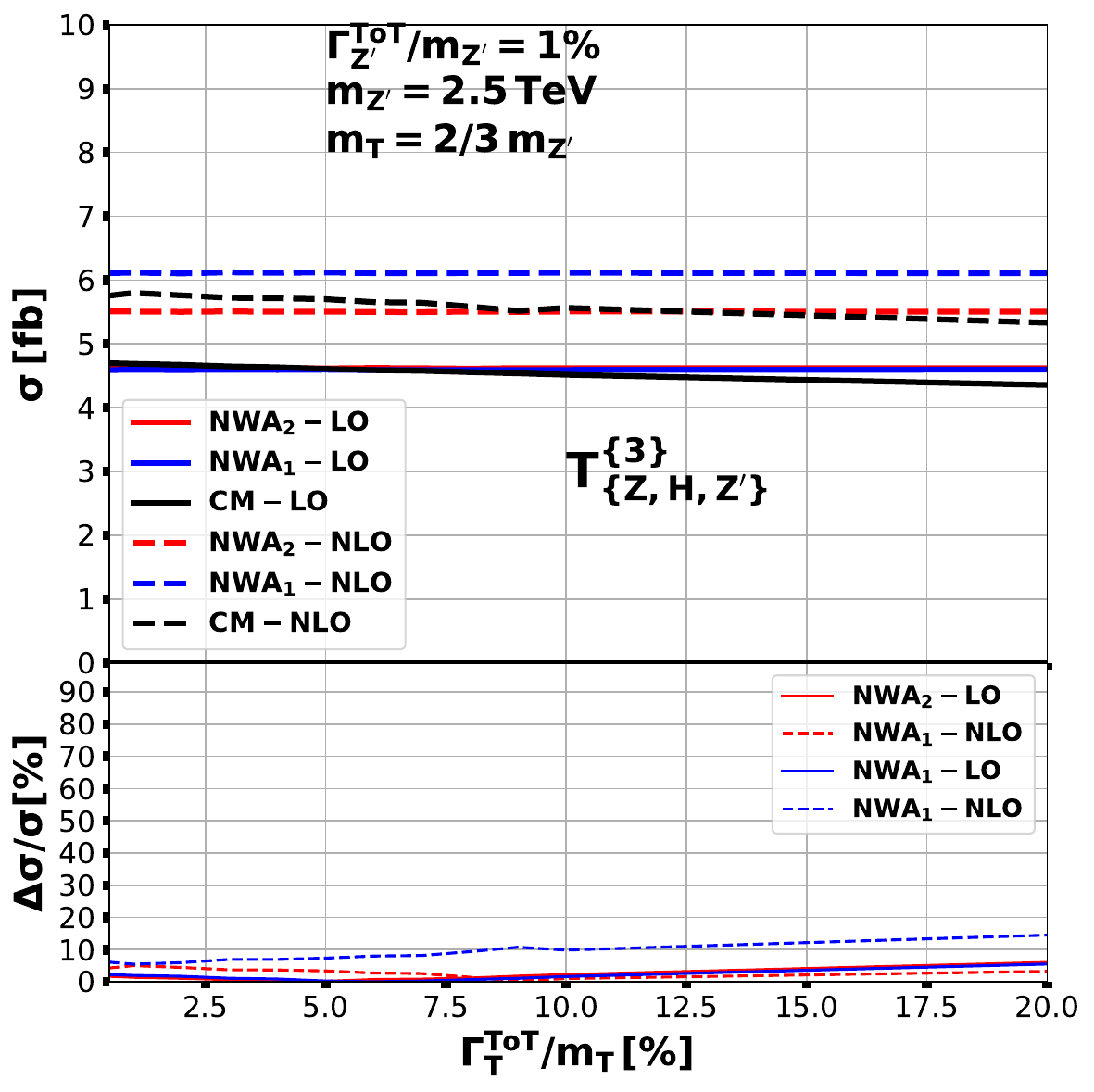}
\includegraphics[width=5.75cm,height=4.25cm]{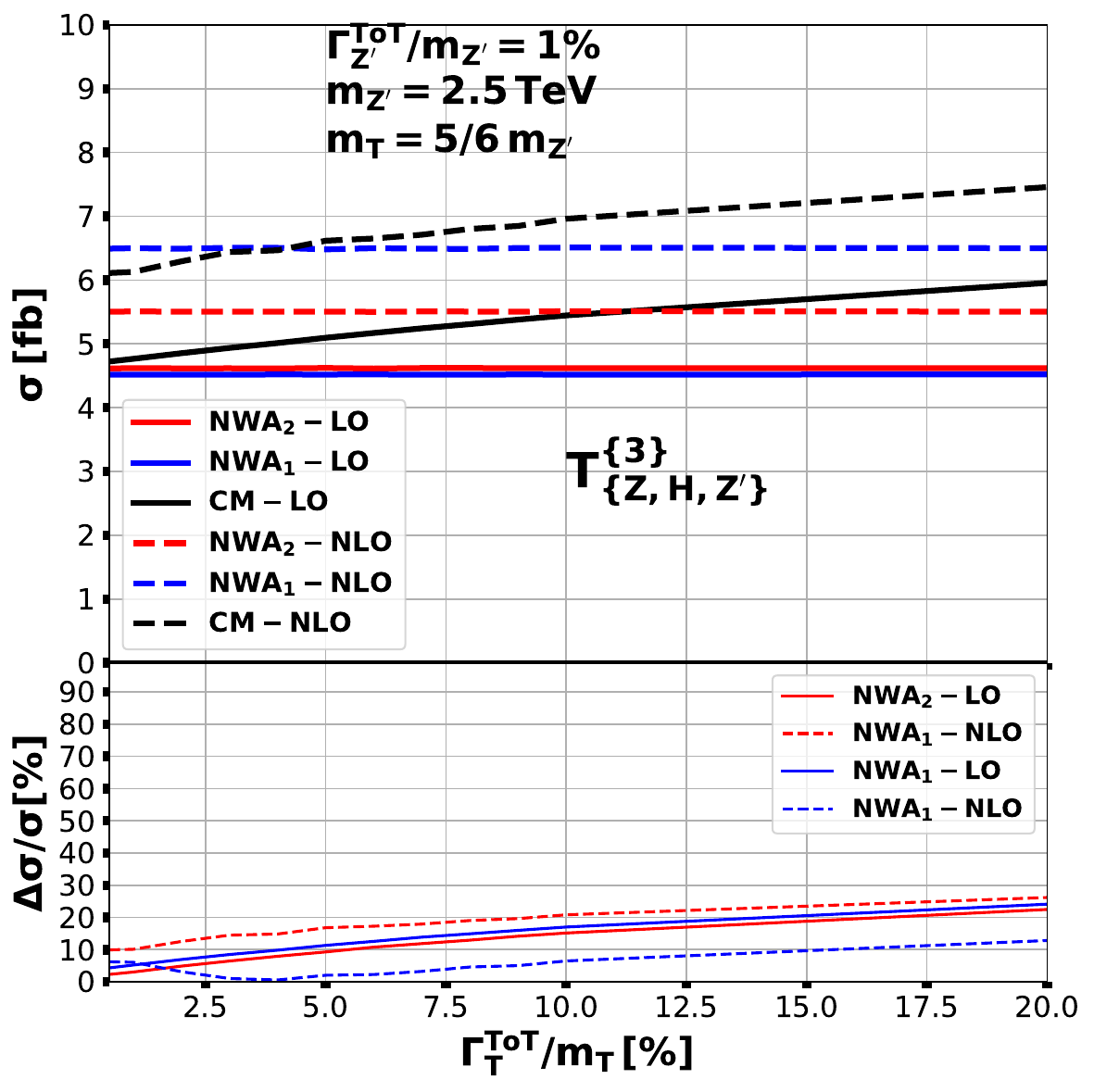}\\
\includegraphics[width=5.75cm,height=4.25cm]{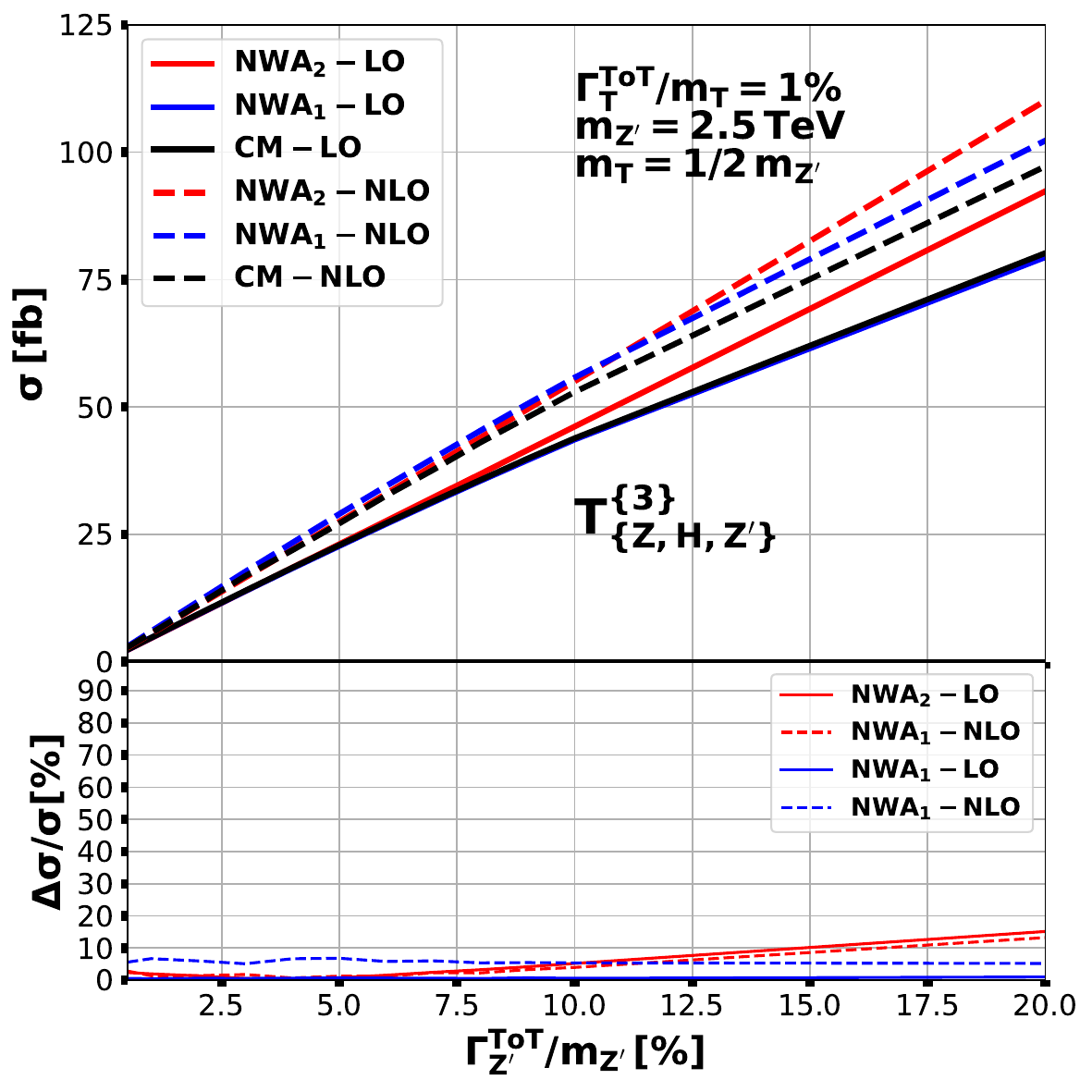}
\includegraphics[width=5.75cm,height=4.25cm]{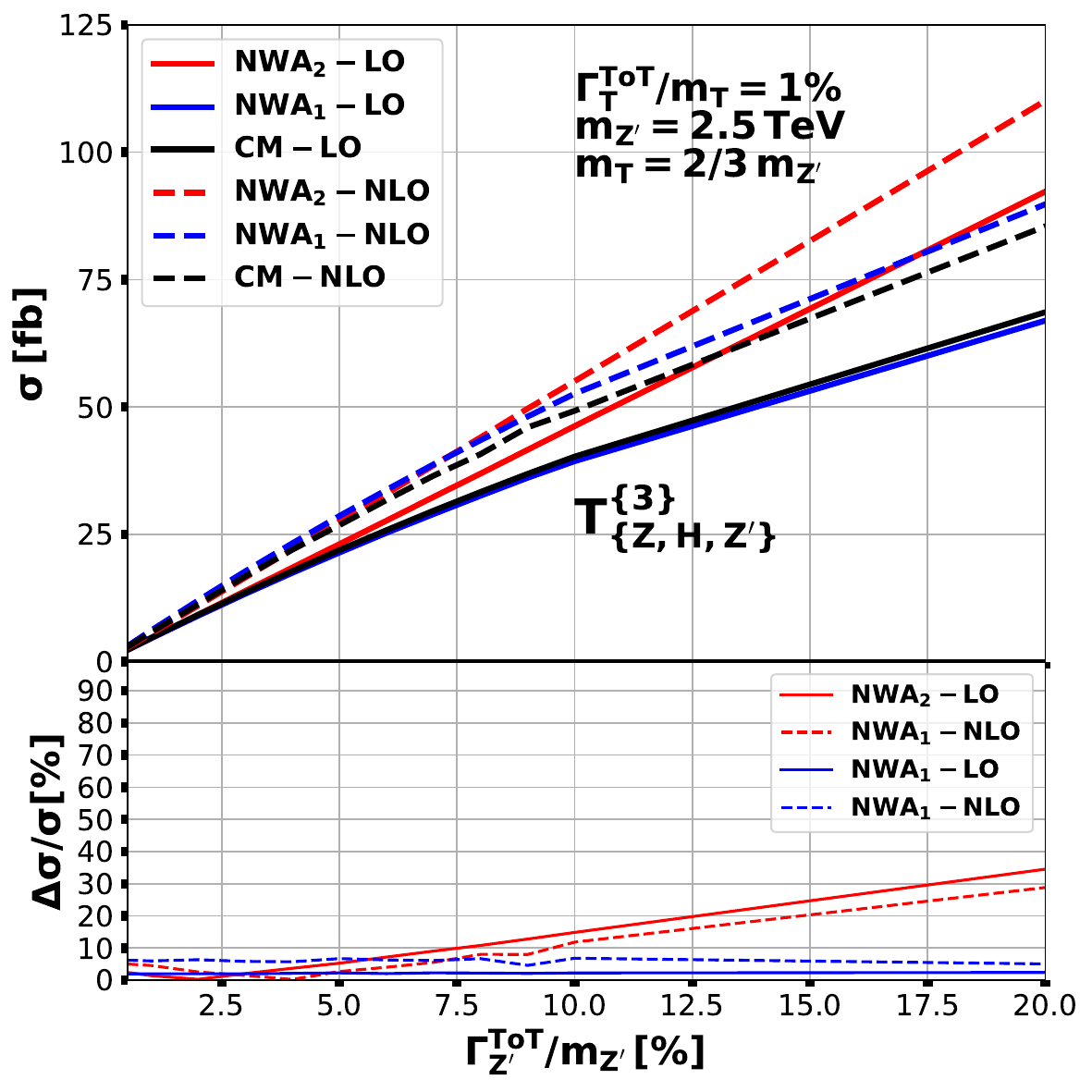}
\includegraphics[width=5.75cm,height=4.25cm]{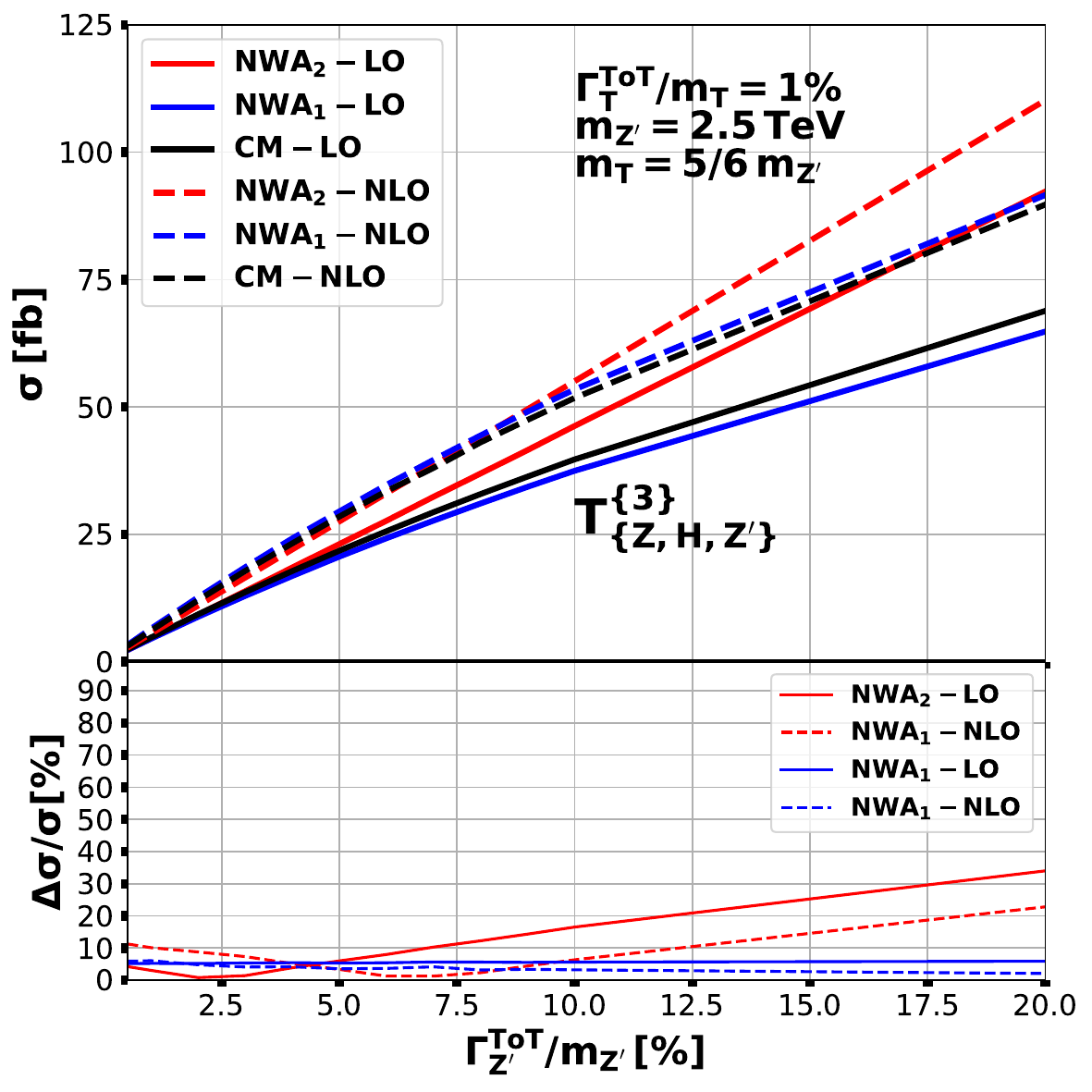}
 \caption{\small Variation of $\sigma_{\scriptscriptstyle\text{LO/NLO}}$ for the reaction $pp\rightarrow \{\zp, T\} \rightarrow t\, \bar{t} Z (H)$ with respect to $\Gamma_{\scriptscriptstyle T}/m_{\scriptscriptstyle T}$ (up) and $\Gamma_{\scriptscriptstyle Z^{\prime}}/m_{\scriptscriptstyle Z^{\prime}}$ (down) at LO and NLO in NWA$_1$, NWA$_2$ and CM (${\bf T^{\scriptscriptstyle\{3\}}_{\scriptscriptstyle\{Z,H,Z'\}}}$ scenario).}
\label{XsecVarsGamZpTp}
\end{figure} 

\subsubsection*{Case 2: Production and decay of $\zp$}
\noindent
In figure~\ref{XsecVarsGamZpTp}, we show the variation of the cross section, for $m_{\scriptscriptstyle{Z^{\prime}}}=2.5\, $ TeV, in terms of $\Gamma^{\text{ToT}}_{\scriptscriptstyle{T}}/m_{\scriptscriptstyle{T}}$ (up panels) and $\Gamma^{\text{ToT}}_{\scriptscriptstyle{Z^{\prime}}}/m_{\scriptscriptstyle{Z^{\prime}}}$ (down panels). In the left, central and right panels respectively, we provide the variation of the cross sections for the low-, middle- and high-mass cases (i.e. $m_{\scriptscriptstyle T}=1/2\, m_{\scriptscriptstyle Z^{\prime}}$, $m_{\scriptscriptstyle T}=2/3\, m_{\scriptscriptstyle Z^{\prime}}$ and $m_{\scriptscriptstyle T}=5/6\, m_{\scriptscriptstyle Z^{\prime}}$). 
 

\noindent
From this figure, we observe that, the relative difference between NWA$_1$ and CM is larger than for NWA$_2$ for the low-mass case, where for the latter the discrepancy is less than $3\%$. Consequently, one can take with caution  $\sigma_{\scriptscriptstyle \text{NLO}}^{\text{NWA}_2}\approx  \sigma_{\scriptscriptstyle \text{NLO}}^{\text{CM}}$.
The situation changes significantly for the high-mass case, $m_{\scriptscriptstyle{T}}/m_{\scriptscriptstyle{\scriptscriptstyle{Z^{\prime}}}}=5/6$ (see the right panels of figure~\ref{XsecVarsGamZpTp}). Here, the relative differences are much larger, where they can exceed the $15\%$ at $\Gamma_{\scriptscriptstyle T}/m_{\scriptscriptstyle T}=10\%$ which is not negligible (red curves in right sub-panels). This indicates that off-shell effects cannot be ignored in this regime, which render the NWA unreliable. Instead, one has to employ the CM scheme. \\

\noindent
We conclude from this appendix that NWA$_2$ can be used for a quick estimation of the cross section, as it is fast and leads to results reasonably comparable to the full calculation (the CM scheme) for the lower-mass case. However the CM scheme remains the most accurate and best method to adopt for all the mass hierarchies.

%% file: appendix_D.tex
\section{More on the validation of the model}
\label{appD}

\noindent
This appendix is a complement of section~\ref{sec2}, where we provide more information on the model validation at NLO for narrow widths within the CM scheme. 
\subsection*{$\bullet$ Validation with the comparison with SM counterpart processes:}
\noindent
Since the LO implementation in the CM is straightforward, we limit ourselves to only the NLO predictions. Note that the SM counterpart processes are generated by making use of the {\tt UFO} model {\tt loop\_qcd\_qed\_sm\_Gmu}, which supports CM scheme calculation. In the following tables~\ref{tabl1AppD} and \ref{tabl2AppD}, eq.~(\ref{KinematicAssumption}) is always satisfied. We remind that the configurations studied here are not physical (for example, they involve giving the $W$ and $Z$ bosons masses different from their measured values). All these exotic replacements are introduced solely for the sake technical comparison.

\begin{table*}[h!]
\centering
 \renewcommand{\arraystretch}{1.40}
 \setlength{\tabcolsep}{10pt}
 \begin{adjustbox}{width=18cm,height=3.5cm}
 \boldmath
 \begin{tabular}{!{\vrule width 1pt}l!{\vrule width 1pt}l!{\vrule width 1pt}l!{\vrule width 1pt}l!{\vrule width 1pt}l!{\vrule width 1pt}l!{\vrule width 1pt}l!{\vrule width 1pt}l!{\vrule width 1pt}}
 \cline{4-8}
 \multicolumn{1}{c}{{}}& \multicolumn{1}{c}{{}}&\multicolumn{1}{c!{\vrule width 1pt}}{{}} 
 &\multicolumn{5}{c!{\vrule width 1pt}}{{{\Large$\mathbf{\sigma_{\scriptstyle NLO}[pb]}$\, for\, $m_{\scriptscriptstyle T}=172.76$ GeV, $m_{W^{\prime}}=80.3790$ GeV and $m_{Z^{\prime}}=91.1876$ GeV}}} 
  \\
 \cline{2-8}
 \multicolumn{1}{c!{\vrule width 1pt}}{{}}&\multicolumn{1}{c!{\vrule width 1pt}}{{\Large\textbf{Model}}} &\multicolumn{1}{c!{\vrule width 1pt}}{{\Large\textbf{Process}}}  & {$\mathbf{\Gamma/m=1\%}$} &{$\mathbf{\Gamma/m=3\%}$} & {$\mathbf{\Gamma/m=6\%}$} &{$\mathbf{\Gamma/m=10\%}$}&{$\mathbf{\Gamma/m=30\%}$}\\
\noalign{\hrule height 1pt}
{\Large$1$} 
&\textbf{vlQBp}&  $p\, p \overset{\textbf{pure QCD}}{\xrightarrow{\hspace{2.0cm}}} T\, \bar{T}$ 
& $6.780\times 10^{2} \pm 5.1\times 10^{-1}$
& $6.784\times 10^{2} \pm 5.0\times 10^{-1}$
& $6.786\times 10^{2} \pm 5.0\times 10^{-1}$
& $6.787\times 10^{2} \pm 5.2\times 10^{-1}$
& $6.793\times 10^{2} \pm 4.9\times 10^{-1}$\\
&\textbf{SM}&  $p\, p \overset{\textbf{pure QCD}}{\xrightarrow{\hspace{2.0cm}}} t\, \bar{t}$ 
& $6.789\times 10^{2} \pm 5.0\times 10^{-1}$
& $6.790\times 10^{2} \pm 5.1\times 10^{-1}$
& $6.789\times 10^{2} \pm 5.1\times 10^{-1}$
& $6.794\times 10^{2} \pm 5.2\times 10^{-1}$
& $6.792\times 10^{2} \pm 5.0\times 10^{-1}$\\
\cline{1-8}
{\Large$2$} 
&\textbf{vlQBp}&  $p\, p \overset{\textbf{virtual}\,\, W^{\prime}}{\xrightarrow{\hspace{2.0cm}}} T\, \bar{b}$ 
&$1.061\times 10^{2} \pm 6.3\times 10^{-1}$
&$1.065\times 10^{2} \pm 3.7\times 10^{-1}$
&$1.062\times 10^{2} \pm 3.6\times 10^{-1}$
&$1.063\times 10^{2} \pm 4.1\times 10^{-1}$
&$1.037\times 10^{2} \pm 3.8\times 10^{-1}$\\
&\textbf{SM}&  $p\, p \overset{\textbf{virtual}\,\, W}{\xrightarrow{\hspace{2.0cm}}} t\, \bar{b}$ 
&$1.071\times 10^{2} \pm 5.0\times 10^{-1}$
&$1.067\times 10^{2} \pm 5.7\times 10^{-1}$
&$1.068\times 10^{2} \pm 1.7\times 10^{-1}$
&$1.073\times 10^{2} \pm 4.0\times 10^{-1}$
&$1.125\times 10^{2} \pm 5.1\times 10^{-1}$\\
\cline{1-8}
{\Large$3$} 
&\textbf{vlQBp}&  $p\, p \overset{\textbf{virtual}\,\, Z^{\prime}}{\xrightarrow{\hspace{2.0cm}}} T\, \bar{t}$ 
&$2.569e+00 \pm 2.3\times 10^{-3}$
&$2.568e+00 \pm 1.3\times 10^{-3}$
&$2.564e+00 \pm 1.6\times 10^{-3}$
&$2.558e+00 \pm 1.8\times 10^{-3}$
&$2.494e+00 \pm 2.2\times 10^{-3}$\\
&\textbf{SM}&  $p\, p \overset{\textbf{virtual}\,\, Z}{\xrightarrow{\hspace{2.0cm}}} t\, \bar{t}$ 
&$2.569e+00 \pm 1.8\times 10^{-3}$
&$2.570e+00 \pm 1.6\times 10^{-3}$
&$2.573e+00 \pm 1.8\times 10^{-3}$
&$2.585e+00 \pm 1.9\times 10^{-3}$
&$2.716e+00 \pm 2.4\times 10^{-3}$\\
\cline{1-8}
{\Large$4$} 
&\textbf{vlQBp}&  $p\, p \overset{\textbf{virtual}\,\, W^{\prime}, T}{\xrightarrow{\hspace{2.0cm}}} \bar{b}\, t\, H$ 
& $1.010\times 10^{-1} \pm 3.0\times 10^{-4}$
& $1.013\times 10^{-1} \pm 4.7\times 10^{-4} $
& $1.008\times 10^{-1} \pm 2.4\times 10^{-4}$
& $1.008\times 10^{-1} \pm 3.3\times 10^{-4}$
& $9.837\times 10^{-2} \pm 2.9\times 10^{-4}$\\
&\textbf{SM}&  $p\, p \overset{\textbf{virtual}\,\, W, t}{\xrightarrow{\hspace{2.0cm}}} \bar{b}\, t\, H$ 
& $1.009\times 10^{-1} \pm 3.6\times 10^{-4}$
& $1.015\times 10^{-1} \pm 4.2\times 10^{-4}$
& $1.010\times 10^{-1} \pm 5.3\times 10^{-4}$
& $1.016\times 10^{-1} \pm 3.0\times 10^{-4}$
& $1.070\times 10^{-1} \pm 4.3\times 10^{-4}$\\
\cline{1-8}
{\Large$5$} 
&\textbf{vlQBp}&  $p\, p \overset{\textbf{virtual}\,\, W^{\prime}, T}{\xrightarrow{\hspace{2.0cm}}} \bar{b}\, t\, Z$ 
& $2.373\times 10^{-1} \pm 1.2\times 10^{-3}$
& $2.369\times 10^{-1} \pm 5.6\times 10^{-4}$
& $2.362\times 10^{-1} \pm 9.7\times 10^{-4}$
& $2.363\times 10^{-1} \pm 5.7\times 10^{-4}$
& $2.285\times 10^{-1} \pm 7.0\times 10^{-4}$\\
&\textbf{SM}&  $p\, p \overset{\textbf{virtual}\,\, W, t}{\xrightarrow{\hspace{2.0cm}}} \bar{b}\, t\, Z$ 
& $2.349\times 10^{-1} \pm 8.8\times 10^{-4}$
& $2.375\times 10^{-1} \pm 1.2\times 10^{-3}$
& $2.380\times 10^{-1} \pm 1.2\times 10^{-3}$
& $2.397\times 10^{-1} \pm 5.6\times 10^{-4}$
& $2.615\times 10^{-1} \pm 1.1\times 10^{-3}$\\
\cline{1-8}
{\Large$6$} 
&\textbf{vlQBp}&  $p\, p \overset{\textbf{virtual}\,\, Z^{\prime}, T}{\xrightarrow{\hspace{2.0cm}}} \bar{t}\, t\, H$ 
& $9.756\times 10^{-3} \pm 1.8\times 10^{-5}$
& $9.766\times 10^{-3} \pm 2.5\times 10^{-5}$
& $9.791\times 10^{-3} \pm 3.5\times 10^{-5}$
& $9.701\times 10^{-3} \pm 2.1\times 10^{-5}$
& $9.420\times 10^{-3} \pm 1.6\times 10^{-5}$\\
&\textbf{SM}&  $p\, p \overset{\textbf{virtual}\,\, Z, t}{\xrightarrow{\hspace{2.0cm}}} \bar{t}\, t\, H$ 
& $9.709\times 10^{-3} \pm 2.2\times 10^{-5}$
& $9.672\times 10^{-3} \pm 2.2\times 10^{-5}$
& $9.716\times 10^{-3} \pm 2.4\times 10^{-5}$
& $9.772\times 10^{-3} \pm 2.8\times 10^{-5}$
& $1.023\times 10^{-2} \pm 2.3\times 10^{-5}$\\
\cline{1-8}
{\Large$7$} 
&\textbf{vlQBp}&  $p\, p \overset{\textbf{virtual}\,\, Z^{\prime}, T}{\xrightarrow{\hspace{2.0cm}}} T\, \bar{t}\, Z$ 
& $7.755\times 10^{-3} \pm 2.2\times 10^{-5}$
& $7.758\times 10^{-3} \pm 1.9\times 10^{-5}$
& $7.774\times 10^{-3} \pm 2.2\times 10^{-5}$
& $7.758\times 10^{-3} \pm 1.8\times 10^{-5}$
& $7.607\times 10^{-3} \pm 2.1\times 10^{-5}$\\
&\textbf{SM}&  $p\, p \overset{\textbf{virtual}\,\, Z, t}{\xrightarrow{\hspace{2.0cm}}} t\, \bar{t}\, Z$ 
& $7.760\times 10^{-3} \pm 1.8\times 10^{-5}$
& $7.824\times 10^{-3} \pm 3.0\times 10^{-5}$
& $7.826\times 10^{-3} \pm 2.1\times 10^{-5}$
& $7.896\times 10^{-3} \pm 3.5\times 10^{-5}$
& $8.654\times 10^{-3} \pm 2.5\times 10^{-5}$\\
\noalign{\hrule height 1pt}
\end{tabular}
\end{adjustbox}
  \caption{\small Comparison of {\tt vlQBp} NLO predictions for some processes with thier SM counterparts in the case of SM-like masses and couplings ($m_{\scriptscriptstyle T}=m_{\scriptscriptstyle t}$, $m_{W^{\prime}}=m_{W}$ and $m_{Z^{\prime}}=m_{Z}$).}
   \label{tabl1AppD}
\end{table*}

\begin{table*}[h!]
\centering
 \renewcommand{\arraystretch}{1.40}
 \setlength{\tabcolsep}{10pt}
 \begin{adjustbox}{width=18cm,height=3.5cm}
 \boldmath
 \begin{tabular}{!{\vrule width 1pt}l!{\vrule width 1pt}l!{\vrule width 1pt}l!{\vrule width 1pt}l!{\vrule width 1pt}l!{\vrule width 1pt}l!{\vrule width 1pt}l!{\vrule width 1pt}l!{\vrule width 1pt}}
 \cline{4-8}
 \multicolumn{1}{c}{{}}& \multicolumn{1}{c}{{}}&\multicolumn{1}{c!{\vrule width 1pt}}{{}} 
 &\multicolumn{5}{c!{\vrule width 1pt}}{{{\Large$\mathbf{\sigma_{\scriptscriptstyle NLO}[pb]}$\, for\, $m_{\scriptscriptstyle T}=1000.0$ GeV, $m_{W^{\prime}}=1205.690$ GeV and $m_{Z^{\prime}}=1367.814$ GeV}}} 
  \\
 \cline{2-8}
 \multicolumn{1}{c!{\vrule width 1pt}}{{}}&\multicolumn{1}{c!{\vrule width 1pt}}{{\Large\textbf{Model}}} &\multicolumn{1}{c!{\vrule width 1pt}}{{\Large\textbf{Process}}}  & {$\mathbf{\Gamma/m=1\%}$} &{$\mathbf{\Gamma/m=3\%}$} & {$\mathbf{\Gamma/m=6\%}$} &{$\mathbf{\Gamma/m=10\%}$}&{$\mathbf{\Gamma/m=30\%}$}\\
\noalign{\hrule height 1pt}
{\Large$1$} 
&\textbf{vlQBp}&  $p\, p \overset{\textbf{pure QCD}}{\xrightarrow{\hspace{2.0cm}}} T\, \bar{T}$ 
& $4.047\times 10^{-2} \pm 3.3\times 10^{-5}$
& $4.048\times 10^{-2} \pm 3.1\times 10^{-5}$
& $4.049\times 10^{-2} \pm 2.9\times 10^{-5}$
& $4.061\times 10^{-2} \pm 3.2\times 10^{-5}$
& $4.062\times 10^{-2} \pm 3.1\times 10^{-5}$\\
&\textbf{SM}&  $p\, p \overset{\textbf{pure QCD}}{\xrightarrow{\hspace{2.0cm}}} t\, \bar{t}$ 
& $4.075\times 10^{-2} \pm 3.1\times 10^{-5}$
& $4.074\times 10^{-2} \pm 2.9\times 10^{-5}$
& $4.072\times 10^{-2} \pm 3.0\times 10^{-5}$
& $4.071\times 10^{-2} \pm 3.1\times 10^{-5}$
& $4.070\times 10^{-2} \pm 2.9\times 10^{-5}$\\
\cline{1-8}
{\Large$2$} 
&\textbf{vlQBp}&  $p\, p \overset{\textbf{virtual}\,\, W^{\prime}}{\xrightarrow{\hspace{2.0cm}}} T\, \bar{b}$ 
&$1.867e+00 \pm 8.6\times 10^{-3}$
&$6.142\times 10^{-1} \pm 2.5\times 10^{-3}$
&$2.968\times 10^{-1} \pm 1.2\times 10^{-3}$
&$1.728\times 10^{-1} \pm 7.8\times 10^{-4}$
&$5.360\times 10^{-2} \pm 2.2\times 10^{-4}$\\
&\textbf{SM}&  $p\, p \overset{\textbf{virtual}\,\, W}{\xrightarrow{\hspace{2.0cm}}} t\, \bar{b}$ 
&$1.861e+00 \pm 8.7\times 10^{-3}$
&$6.100\times 10^{-1} \pm 2.1\times 10^{-3}$
&$2.990\times 10^{-1} \pm 1.2\times 10^{-3}$
&$1.762\times 10^{-1} \pm 8.2\times 10^{-4}$
&$5.787\times 10^{-2} \pm 2.2\times 10^{-4}$\\
\cline{1-8}
{\Large$3$} 
&\textbf{vlQBp}&  $p\, p \overset{\textbf{virtual}\,\, Z^{\prime}}{\xrightarrow{\hspace{2.0cm}}} T\, \bar{t}$ 
& $1.410\times 10^{-4} \pm 9.4\times 10^{-8}$
& $1.409\times 10^{-4} \pm 9.4\times 10^{-8}$
& $1.409\times 10^{-4} \pm 9.5\times 10^{-8}$
& $1.405\times 10^{-4} \pm 8.8\times 10^{-8}$
& $1.373\times 10^{-4} \pm 9.2\times 10^{-8}$ \\
&\textbf{SM}&  $p\, p \overset{\textbf{virtual}\,\, Z}{\xrightarrow{\hspace{2.0cm}}} t\, \bar{t}$ 
& $1.410\times 10^{-4} \pm 9.6\times 10^{-8}$
& $1.411\times 10^{-4} \pm 9.0\times 10^{-8}$
& $1.412\times 10^{-4} \pm 8.8\times 10^{-8}$
& $1.417\times 10^{-4} \pm 9.7\times 10^{-8}$
& $1.497\times 10^{-4} \pm 9.8\times 10^{-8}$\\
\cline{1-8}
{\Large$4$} 
&\textbf{vlQBp}&  $p\, p \overset{\textbf{virtual}\,\, W^{\prime}, T}{\xrightarrow{\hspace{2.0cm}}} \bar{b}\, t\, H$ 
& $7.002\times 10^{-5} \pm 2.7\times 10^{-7}$
& $3.886\times 10^{-5} \pm 2.3\times 10^{-7}$
& $2.940\times 10^{-5} \pm 1.1\times 10^{-7}$
& $2.402\times 10^{-5} \pm 1.3\times 10^{-7}$
& $1.249\times 10^{-5} \pm 1.9\times 10^{-8}$\\
&\textbf{SM}&  $p\, p \overset{\textbf{virtual}\,\, W, t}{\xrightarrow{\hspace{2.0cm}}} \bar{b}\, t\, H$ 
& $6.874\times 10^{-5} \pm 2.8\times 10^{-7}$
& $3.836\times 10^{-5} \pm 1.9\times 10^{-7}$
& $2.976\times 10^{-5} \pm 1.4\times 10^{-7}$
& $2.499\times 10^{-5} \pm 8.8\times 10^{-8}$
& $1.826\times 10^{-5} \pm 1.0\times 10^{-7}$\\
\cline{1-8}
{\Large$5$} 
&\textbf{vlQBp}&  $p\, p \overset{\textbf{virtual}\,\, W^{\prime}, T}{\xrightarrow{\hspace{2.0cm}}} \bar{b}\, t\, Z$ 
&  $2.615\times 10^{-8} \pm 9.6\times 10^{-11}$
&  $2.625\times 10^{-8} \pm 9.6\times 10^{-11}$
&  $2.631\times 10^{-8} \pm 1.0\times 10^{-10}$
&  $2.626\times 10^{-8} \pm 4.3\times 10^{-11}$
&  $2.553\times 10^{-8} \pm 1.4\times 10^{-10}$\\
&\textbf{SM}&  $p\, p \overset{\textbf{virtual}\,\, W, t}{\xrightarrow{\hspace{2.0cm}}} \bar{b}\, t\, Z$ 
& $2.630\times 10^{-8} \pm 7.2\times 10^{-11}$
& $2.615\times 10^{-8} \pm 1.3\times 10^{-10}$
& $2.645\times 10^{-8} \pm 8.3\times 10^{-11} $
& $2.648\times 10^{-8} \pm 9.0\times 10^{-11}$
& $2.893\times 10^{-8} \pm 8.2\times 10^{-11}$\\
\cline{1-8}
{\Large$6$} 
&\textbf{vlQBp}&  $p\, p \overset{\textbf{virtual}\,\, Z^{\prime}, T}{\xrightarrow{\hspace{2.0cm}}} \bar{t}\, t\, H$ 
& $2.878\times 10^{-7} \pm 7.3\times 10^{-10}$
& $2.876\times 10^{-7} \pm 8.6\times 10^{-10}$
& $2.830\times 10^{-7} \pm 8.2\times 10^{-10}$
& $2.770\times 10^{-7} \pm 7.4\times 10^{-10}$
& $2.164\times 10^{-7} \pm 5.4\times 10^{-10}$\\
&\textbf{SM}&  $p\, p \overset{\textbf{virtual}\,\, Z, t}{\xrightarrow{\hspace{2.0cm}}} \bar{t}\, t\, H$ 
& $2.814\times 10^{-7} \pm 8.9\times 10^{-10}$
& $2.801\times 10^{-7} \pm 8.9\times 10^{-10}$
& $2.818\times 10^{-7} \pm 9.1\times 10^{-10}$
& $2.823\times 10^{-7} \pm 8.7\times 10^{-10}$
& $3.011\times 10^{-7} \pm 9.9\times 10^{-10}$\\
\cline{1-8}
{\Large$7$} 
&\textbf{vlQBp}&  $p\, p \overset{\textbf{virtual}\,\, Z^{\prime}, T}{\xrightarrow{\hspace{2.0cm}}} T\, \bar{t}\, Z$ 
& $6.512\times 10^{-9} \pm 2.1\times 10^{-11}$
& $6.504\times 10^{-9} \pm 2.2\times 10^{-11}$
& $6.499\times 10^{-9} \pm 2.4\times 10^{-11}$
& $6.470\times 10^{-9} \pm 2.1\times 10^{-11}$
& $6.407\times 10^{-9} \pm 2.5\times 10^{-11}$\\
&\textbf{SM}&  $p\, p \overset{\textbf{virtual}\,\, Z, t}{\xrightarrow{\hspace{2.0cm}}} t\, \bar{t}\, Z$ 
& $6.547\times 10^{-9} \pm 2.4\times 10^{-11}$
& $6.500\times 10^{-9} \pm 2.2\times 10^{-11}$
& $6.543\times 10^{-9} \pm 2.5\times 10^{-11}$
& $6.599\times 10^{-9} \pm 2.3\times 10^{-11}$
& $7.228\times 10^{-9} \pm 2.2\times 10^{-11}$\\
\noalign{\hrule height 1pt}
\end{tabular}
\end{adjustbox}
  \caption{\small Comparison of {\tt vlQBp} NLO predictions for some processes with their SM counterparts in the case of $m_{\scriptscriptstyle T}=1000.0$ GeV, $m_{W^{\prime}}=1205.690$ GeV and $m_{Z^{\prime}}=1367.814$ GeV (with $m_{\scriptscriptstyle T}=m_{\scriptscriptstyle t}$, $m_{W^{\prime}}=m_{W}$ and $m_{Z^{\prime}}=m_{Z}$).}
   \label{tabl2AppD}
\end{table*}

\noindent
For simplicity, we set the width-to-mass ratios of all involved unstable particles to be equal, denoted $\Gamma/m$ in tables~\ref{tabl1AppD} and \ref{tabl2AppD}.
We recall that the masses of $\wps$ and $\zp$ are chosen such that $s_{\scriptscriptstyle w}$ and $\alpha$ retains their SM values. Thus, the Fermi constant and the Higgs {\it vev}, in tables~\ref{tabl2AppD}, are changed, while always satisfying:
\begin{equation}
G_{\scriptscriptstyle F}=\frac{\sqrt{2}}{8}\frac{4\pi\alpha}{s_{\scriptscriptstyle w}^2\, m_{\scriptscriptstyle W}^2},\qquad
v=2\, m_{\scriptscriptstyle W}\, \sqrt{\frac{s_{\scriptscriptstyle w}^2}{4\pi\alpha}}, \qquad
s_{\scriptscriptstyle w}^2=1-\frac{m_{\scriptscriptstyle W}^2}{m_{\scriptscriptstyle Z}^2}, \qquad
\text{with}\,\,\,  m_{\scriptscriptstyle W}=m_{\scriptscriptstyle W^{\prime}}\, \, \text{and}\, \, m_{\scriptscriptstyle Z}=m_{\scriptscriptstyle Z^{\prime}}.
\label{fermiConst}
\end{equation}

\noindent
From tables~\ref{tabl1AppD} and \ref{tabl2AppD}, we observe a good agreement between the predictions of our model and the SM counterpart cross sections at NLO, up to width-to-mass ratios of $10\%$ for the unstable particles. Various other cases have been studied, including ones with heavier masses (e.g.~$m_{\scriptscriptstyle T}=3000$ GeV, $m_{\scriptscriptstyle Z^{\prime}}=4000$ GeV and $m_{\scriptscriptstyle W^{\prime}}=3525.87$ GeV), they are omitted for brevity. All of them led to the same conclusions. Thus, our model provides an excellent approximation of the CM scheme at NLO for narrow widths. For broad widths, the discrepancy becomes more significant, see the last column of each table, meaning that this model cannot be used for such configurations. We aim to extend the dedicated model {\tt vlQBp} to handle broad widths consistently at NLO in future work.

\subsection*{$\bullet$ Validation by reproducing the results with and extended version of {\tt loop\_qcd\_qed\_sm\_Gmu}:}

\noindent
The UV counterterms and the $R_2$ rational terms associated with the VLQ $T$, which are required for one-loop QCD calculation, are similar to those associated to the standard top quark. Therefore, we adapted the UV and $R_2$ counterterms from the top quark for the VLQ $T$ and added them to the SM {\tt UFO} model {\tt loop\_qcd\_qed\_sm\_Gmu}, which supports the CM scheme at NLO. We note that the ingredients associated with the $T-W-b$ and $T-H-t$ vertices were added manually in the modified UFO model. The extra heavy gauge bosons ($W^{\prime}$ and $Z^{\prime}$) are deliberately not included in this modified version to minimize changes to the original {\tt loop\_qcd\_qed\_sm\_Gmu} model. Instead, the standard gauge  bosons ($W$ and $Z$) can serve as proxies for $W^{\prime}$ and $Z^{\prime}$ in processes involving only a single gauge boson, such as $pp\rightarrow\{W^{\prime}, T\}\rightarrow t\bar{b}H+\bar{t}bH$. \\

\noindent
We have calculated the LO and NLO cross sections of the processes $pp\rightarrow\{W^{\prime}\}\rightarrow T\bar{b}+\bar{T}b$ and $pp\rightarrow\{W^{\prime}, T\}\rightarrow t\bar{b}H+\bar{t}bH$ (involving virtual $W^{\prime}$ and/or $T$ propagators) using both our dedicated model {\tt vlQBp} and the modified {\tt loop\_qcd\_qed\_sm\_Gmu} model. As noted, the SM $W$ serves as $W^{\prime}$ in the extended {\tt loop\_qcd\_qed\_sm\_Gmu} for these two processes. It is important to note that the Fermi constant and the mass and width of the ordinary $Z$ boson had to be adjusted to  ensure that the values of $\alpha$ and $s_{_{w}}$ remained unchanged, cf. eq.~(\ref{fermiConst}). The LO and NLO results for  $pp\rightarrow\{W^{\prime}, T\}\rightarrow t\bar{b}H+\bar{t}bH$, in the CM scheme, are presented in table~\ref{tabloopqcd} (3rd and 4th columns) and show excellent agreement between the predictions of the extended {\tt loop\_qcd\_qed\_sm\_Gmu} and {\tt vlQBp} at both LO and NLO.\\

\noindent
We also found excellent agreement between the models for NWA$_1$ LO and NLO cross sections (i.e. the process $pp\rightarrow\{W^{\prime}\}\rightarrow T\bar{b}+\bar{T}b$). The LO and NLO results are presented in the 5th column of table~\ref{tabloopqcd} (for the extended {\tt loop\_qcd\_qed\_sm\_Gmu} model) and the 4 column of table~\ref{tabVp2Tev} (for {\tt vlQBp} model). This validation method is not universally applicable, particularly for processes including two gauge bosons (such as $pp\rightarrow\{W^{\prime}, T\}\rightarrow t\bar{b}Z+\bar{t}bZ$). Currently, it is only used for the two processes mentioned above. An extension that includes both extra gauge bosons is required to validate other reactions.

\begin{table*}[h!]
\boldmath
\centering
 \renewcommand{\arraystretch}{1.40}
 \setlength{\tabcolsep}{10pt}
 \begin{adjustbox}{width=18cm,height=2.0cm}
 \begin{tabular}
 {!{\vrule width 2pt}l!{\vrule width 2pt}l!{\vrule width 2pt}c:c!{\vrule width 1pt}c:c!{\vrule width 2pt}c:c!{\vrule width 2pt}c!{\vrule width 2pt}}
  \cline{1-8}
$\bf m_{\scriptscriptstyle W^{\prime}}$& $\frac{\Gamma_{\scriptscriptstyle T}^{\scriptscriptstyle\text{ToT}}}{m_{\scriptscriptstyle T}}$ &\multicolumn{2}{c!{\vrule width 1pt}}{{{\tt \bf CM: vlQBp}}}
&\multicolumn{2}{c!{\vrule width 2pt}}{\bf CM: extended {\tt\bf loop\_qcd\_qed\_sm\_Gmu}}
&\multicolumn{2}{c!{\vrule width 2pt}}{\bf NWA$_1$: extended {\tt\bf loop\_qcd\_qed\_sm\_Gmu}}
\\
\cline{3-9}
 \textbf{[TeV]}& $[\%]$ &
 $\bf\sigma_{\scriptscriptstyle{\textbf{LO}}}[\textbf{fb}]$& 
 $\bf\sigma_{\scriptscriptstyle{\textbf{NLO}}}[\textbf{fb}]$&
 $\bf\sigma_{\scriptscriptstyle{\textbf{LO}}}[\textbf{fb}]$&
 $\bf\sigma_{\scriptscriptstyle{\textbf{NLO}}}[\textbf{fb}]$&
 $\bf\sigma_{\scriptscriptstyle{\textbf{LO}}}[\textbf{fb}]$&
 $\bf\sigma_{\scriptscriptstyle{\textbf{NLO}}}[\textbf{fb}]$&
 $\frac{m_{\scriptscriptstyle T}}{m_{\scriptscriptstyle W^{\prime}}}$\\
  \noalign{\hrule height 1pt}
  $2.5$ & $0.53$& $ 22.58 \pm 6.2\times 10^{-2}$ & $26.31 \pm 2.4\times 10^{-1}$
  & $22.59 \pm 6.2\times 10^{-2}$ & $25.78 \pm 1.3\times 10^{-1}$ 
  & $22.27 \pm 9.1\times 10^{-2}$ & $28.32 \pm 2.4\times 10^{-1}$
  &\\
\cdashline{2-8}
   $3.5$& $1.04$ & $2.080 \pm 6.1\times 10^{-3}$ & $2.485 \pm 1.8\times 10^{-2}$
   & $2.070 \pm 6.1\times 10^{-3}$ & $2.484 \pm 7.2\times 10^{-3}$
   & $2.035 \pm 9.0\times 10^{-3}$ & $2.599 \pm 1.9\times 10^{-2}$
  &$\frac{1}{2}$\\
\cdashline{2-8}
  $4.0$& $1.36$ & $0.626 \pm 1.9\times 10^{-3}$ & $0.766 \pm 5.8\times 10^{-3}$
  & $0.624 \pm 2.0\times 10^{-3}$ & $0.754 \pm 3.8\times 10^{-3}$
  & $0.615 \pm 2.9\times 10^{-3}$ & $0.799 \pm 6.5\times 10^{-3}$
 &\\
 \noalign{\hrule height 1pt}
  $2.5$&$0.94$ & $22.19 \pm 6.3\times 10^{-2}$ & $25.75 \pm 3.0\times 10^{-1}$
  & $22.26 \pm 6.3\times 10^{-2}$ & $26.75 \pm 1.2\times 10^{-1}$
  & $21.63 \pm 8.9\times 10^{-2}$ & $28.98 \pm 2.5\times 10^{-1}$
  &\\
\cdashline{2-8}
    $3.5$& $1.85$ & $2.021 \pm 5.8\times 10^{-3}$ & $2.505 \pm 1.9\times 10^{-2}$
    & $2.025 \pm 5.9\times 10^{-3}$ & $2.553 \pm 1.3\times 10^{-2}$
    & $1.963 \pm 8.7\times 10^{-3}$ & $2.635 \pm 2.1\times 10^{-2}$
    &$\frac{2}{3}$\\
\cdashline{2-8}
  $4.0$& $2.41$ & $0.607 \pm 1.9\times 10^{-3}$ & $0.754 \pm 4.7\times 10^{-3}$
  & $0.608 \pm 1.9\times 10^{-3}$ & $0.762 \pm 3.8\times 10^{-3}$
  & $0.586 \pm 2.8\times 10^{-3}$ & $0.807 \pm 6.0\times 10^{-3}$
  &\\
 \noalign{\hrule height 1pt}
  $2.5$& $1.19$ & $22.18 \pm 6.3\times 10^{-3}$ & $26.66 \pm 3.2\times 10^{-1}$
  & $22.31 \pm 6.3\times 10^{-2}$ & $27.81 \pm 7.7\times 10^{-2}$
  & $21.61 \pm 8.8\times 10^{-2}$ & $29.13 \pm 2.3\times 10^{-1}$
  &\\
\cdashline{2-8}
    $3.5$& $2.34$ & $2.027 \pm 5.8\times 10^{-3}$ & $2.510 \pm 1.7\times 10^{-2}$
    & $2.026 \pm 5.8\times 10^{-3}$ & $2.567 \pm 9.8\times 10^{-3} $
    & $1.941 \pm 8.6\times 10^{-3}$ & $2.677 \pm 2.0\times 10^{-2}$
 & $\frac{3}{4}$\\
\cdashline{2-8}
  $4.0$& $3.05$ & $0.608 \pm 1.9\times 10^{-3}$ & $0.775 \pm 7.2\times 10^{-3}$
  & $0.606 \pm 1.9\times 10^{-3}$ & $ 0.784 \pm 4.4\times 10^{-3}$
  & $0.578 \pm 2.8\times 10^{-3}$ & $0.811 \pm 6.1\times 10^{-3}$
 &\\
 \noalign{\hrule height 1pt}
\end{tabular}
\end{adjustbox}
  \caption{\footnotesize Hadronic cross section for the reaction $pp\rightarrow\{W^{\prime}, T\}\rightarrow t\bar{b}H+\bar{t}bH$ in the benchmark scenario ${\bf T^{\scriptscriptstyle\{3\}}_{\scriptscriptstyle\{H,W'\}}}$ and $pp\rightarrow\{W^{\prime}, T\}\rightarrow T\bar{b}+\bar{T}b$, with $\Gamma_{\scriptscriptstyle W^{\prime}}^{\scriptscriptstyle\text{ToT}}/m_{\scriptscriptstyle W^{\prime}}=1.09\%$, $\xi_{{Tb}}^{{W^{\prime}}}=20\%$ and $\vkwp = 0.5$ (DD scenario).}
   \label{tabloopqcd}
  \end{table*}

\subsection*{$\bullet$ Validation by {\tt MadGraph5} automatic CM scheme check:}
\noindent
A useful check of the consistency of the implementation of the CM scheme is provided within {\tt MG5\_aMC@NLO}. By running the command {\tt check cms}, it compares the squared amplitude calculated in the CM scheme against the NWA. This is done by scaling all relevant couplings and widths by the parameter $0\leq\lambda\leq 1$ for kinematic configuration where all the resonances are off-shell. If the CM implementation is consistent, the difference between CM scheme and NWA results divided by $\lambda$ (denoted $\Delta$) must tend to constant when $\lambda$ approaches zero. For more details, see \url{https://cp3.irmp.ucl.ac.be/projects/madgraph/wiki/ComplexMassScheme}.\\

\begin{figure}[h!]
\includegraphics[width=8cm,height=6cm]{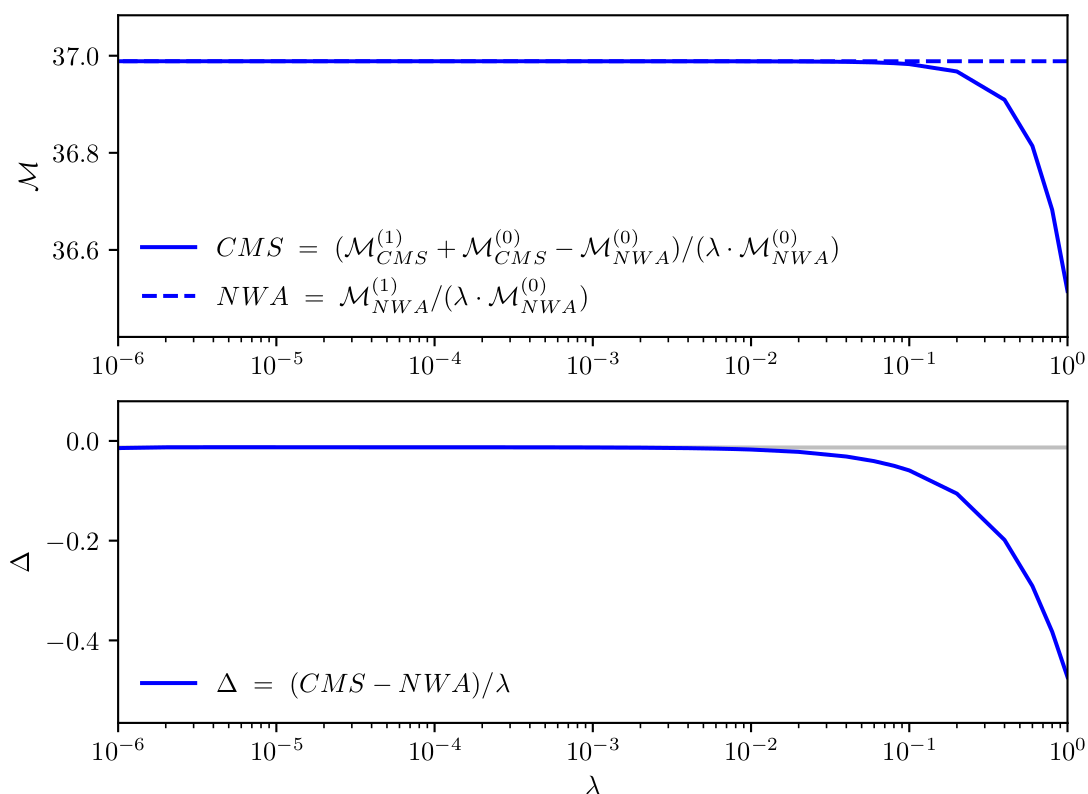}\qquad
\includegraphics[width=8cm,height=6cm]{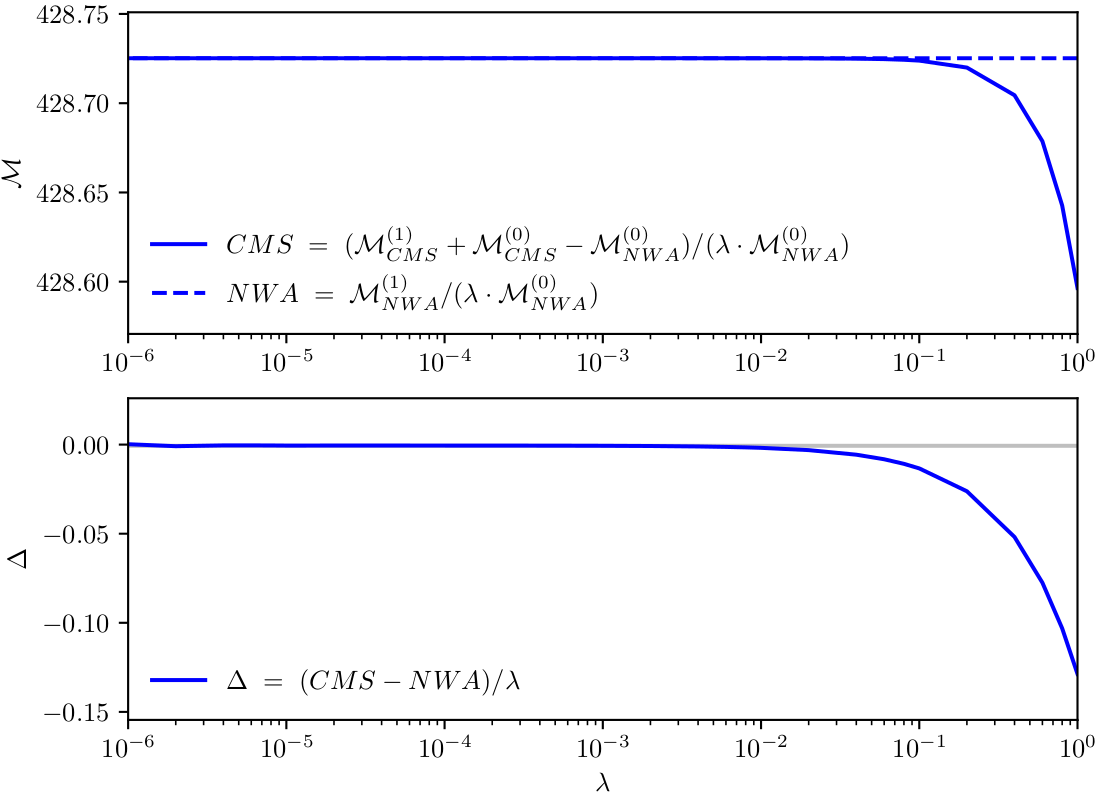}
\caption{\small {\tt MadGraph5} automatic CM scheme check at one-loop in QCD. The left panels are associated to the process $q\bar{q}^{\prime}\rightarrow \{W^{\prime}, T\}\rightarrow t\bar{b} Z(H)$ for the $\wps$-resonance, and the right panels are associated to the process $q\bar{q}^{\prime}\rightarrow \{Z^{\prime}, T\}\rightarrow t\bar{t} Z (H)$ for $T$-resonance.}
\label{cmsTab58}
\end{figure}

\noindent
We performed this test for all processes studied in this paper, using multiple kinematic configurations with narrow widths. They all successfully passed the test. Two representative examples are shown in figure~\ref{cmsTab58}. The upper panels display the variation of the virtual squared amplitudes in QCD, while the lower panels show the corresponding $\Delta$ for the sub-processes: $q\bar{q}^{\prime}\rightarrow \{W^{\prime}, T\}\rightarrow t\bar{b} Z/H$ with $2m_{\scriptscriptstyle T}=m_{\scriptscriptstyle W^{\prime}}=2.5$ TeV (left panels) and $q\bar{q}\rightarrow \{Z^{\prime}, T\}\rightarrow t\bar{t} Z/H$ with $2m_{\scriptscriptstyle T}=m_{\scriptscriptstyle Z^{\prime}}=1.5$ TeV (right panels). These plots clearly show that for large $\lambda$, the effect of the finite width is significant, where the CM scheme and NWA predictions deviate. The deviation $\Delta$ becomes progressively smaller as $\lambda$ decreases, until it tends to a constant (close to zero), confirming the consistency of the model implementation in the CM scheme. We note that this test was also performed at LO, where it led to the same conclusions.    